\documentclass[12pt]{article}
\usepackage{amssymb}
\usepackage{amsmath}
\usepackage{amscd}
\usepackage{mathrsfs}
\usepackage{amssymb}
\usepackage{dsfont}
\usepackage{slashed,bbold}
\usepackage{feynmf}
\usepackage{graphicx}
\usepackage[pdftex,bookmarks=true,pdfauthor={},pdftitle={},pagebackref=false]{hyperref}
\usepackage{youngtab}
\usepackage{color}
\usepackage{mdframed}
\usepackage{breakurl}
\usepackage{cite}
\hypersetup{colorlinks,urlcolor=blue,citecolor=blue,linkcolor=blue}

\numberwithin{equation}{section}

\newcommand{\xdownarrow}[1]{%
  {\left\downarrow\vbox to #1{}\right.\kern-\nulldelimiterspace}
}

\def\Dsl{\,\raise.15ex\hbox{/}\mkern-13.5mu D}
\def\dsl{\raise.15ex\hbox{/}\kern-.57em\partial}

\begin{document}

\thispagestyle{empty}

\begin{flushright}
\begin{tabular}{r}
CERN-TH.6736/92 \\
FTUAM-38/92
\end{tabular}
\end{flushright}

\vspace{4cm}

\renewcommand*{\thefootnote}{\fnsymbol{footnote}}

\begin{center}

{\Large \bf Topics in String Theory and Quantum Gravity}\footnote{Based  
on the lectures given by L. \'Alvarez-Gaum\'e at Les
Houches
Summer School on Gravitation and Quantizations, July 6-31, 1992. Published in
{\em Gravitation and Quantizations}, Proceedings of the 1992 Les Houches Summer
School, eds. B.~Julia and J.~Zinn-Justin, North-Holland 1995.}

\renewcommand*{\thefootnote}{\arabic{footnote}}
\setcounter{footnote}{0}

\vspace{1.cm}

L. \'Alvarez-Gaum\'e\footnote{Currently also at 
Simons Center for Geometry and Physics, State University of New York, Stony Brook, NY, USA.
Email: {\tt Luis.Alvarez-Gaume@cern.ch}.} \\
{\it Theory Division CERN \\ CH-1211 Geneva 23 \\ Switzerland}
\vspace{0.4cm}

and

\vspace{0.4cm}

M.~\'A. V\'azquez-Mozo\footnote{Present address: Departamento de F\'{\i}sica Fundamental,
Universidad de Salamanca, E-37008 Salamanca, Spain. Email:~{\tt vazquez@usal.es}.} \\
{\it Departamento de F\'{\i}sica Te\'orica C-XI \\
Universidad Aut\'onoma de Madrid \\
E-28049 Madrid, Spain}

\end{center}

\vspace{5cm}

\begin{tabular}{l}

CERN-TH.6736/92 \\
November 1992 (v1)/March 2026 (v2)

\end{tabular}

\newpage

\setcounter{page}{1}

\tableofcontents

\newpage

\renewcommand*{\thefootnote}{\arabic{footnote}}
\setcounter{footnote}{0}

\section{General Introduction}

The goal of the 1992 Les Houches Summer School was to bring together
practitioners of different research areas concerned with the
quantization
of gravity (see~\cite{Oxford-symp} and references therein). In this
context, the present lectures partly represent the viewpoint of
string theorists (for details and references see, e.g., 
refs.~\cite{Green-Schwarz-Witten,Kaku,Kaku2,Lust-Theisen}). The most ambitious
approach to the problem of quantizing gravity is certainly embodied by superstring theory
which addresses, together with the problem of quantizing the gravitational field, its
unification with the other three fundamental
forces. In fact, superstring theory is rich enough to account for many of the features of the
standard model of particle physics. This is not accidental. String
theory can be seen as the culmination of several decades of effort dedicated to
incorporate gravity and the other
known interactions within the same framework. 

After nearly a decade of renewed interest in string
theory, it seems natural to recap how it has come to
be one of the leading candidates for a unified theory of all four fundamental interactions. 
We still do not know the basic physical or geometrical
principles underlying string theory. There has been some progress in the
formulation of a string field theory (see the lecture by B. Zwiebach
in this school~\cite{Zwiebach_lectures},
as well as
ref.~\cite{SFT} and references therein) but we are still far from having a
complete theory of string fields. Important progress in string theory
has been made by imposing consistency requirements which worked quite well in ordinary
field theories, and particularly in the standard model. These conditions, when
applied to the unification of gravity with particle physics, lead to very strong
constraints, which, for the time being, seem to be fulfilled only by string
theory.

One of the basic features of the standard model, responsible for its
successes, is its renormalizability. This means that after
a finite number of parameters are specified, we have a rather successful
machinery to calculate and explain a large number of high- and low-energy
phenomena. This predictability that we have in renormalizable quantum field theory is one of
the desired features we would like to find also in a quantum theory of
gravity.

The standard model also has several properties that 
any theory trying to go beyond it should explain: the chiral nature of the
families of quarks and leptons with respect to the standard model gauge group
$\mbox{SU}(3)\times \mbox{SU}(2)\times \mbox{U}(1)$, the origin of the gauge interactions, the
source of symmetry breaking, the vanishing of the cosmological
constant, etc. One of the more popular ``beyond the standard model'' avenues is to
study theories incorporating supersymmetry~\cite{SUSY}, 
a central property of superstring theory. Unfortunately, there is no
evidence at present that supersymmetry is realized in
nature. Nevertheless, if current or future accelerators would find evidence
of any supersymmetric partners to the known elementary particles, the
temptation to extrapolate and believe that superstring theory
plays a central role in the force unification
would be nearly overwhelming.

When looking for a predictive theory including both quantum
gravity and the chiral structure of the low energy degrees of
freedom, one seems to be left just with string theory. These two
requirements alone pose very stringent constraints on candidate
theories. As we will see in the next section, various proposals
including
combinations of supergravity \cite{SUGRA} and Kaluza-Klein theories
\cite{KK},
do not satisfy any of these coditions. If, on the other hand, we focus on the
quantization of the gravitational field alone without including other
interactions, there has been some substantial progress in recent
years. This is reviewed in the lectures by A.~Ashtekar at this School~\cite{Ashtekar_lectures}.

The outline of the present lectures is as follows. In section~\ref{sec:QFTapproach} we summarize
various approaches to quantize gravity within the framework of quantum field theory and unify it with
the remaining interactions. 
The difficulties encountered in applying standard
field theory techniques to perturbative computations including graviton
loops are briefly analyzed. Kaluza-Klein theories are also presented and, as we will see,
not only are afflicted with
uncontrollable infinities, but present the additional problem that chiral
fermions are hard to obtain at low energies. We also study briefly the
conceptual problems encountered in the quantization of fields in the
presence of external gravitational backgrounds. 

This section concludes
with a collection of general remarks concerning the Euclidean and
Hamiltonian approach to the quantization of gravity. We will argue that, if taken
seriously, the naive approach of summing over all possible
topologies
and geometries meets with formidable mathematical difficulties.
The use of semiclassical methods based on instantons, which
could in principle provide some insights into quantum-gravitationally
induced processes, is also presented (and criticized). It should be said that this first section paints
a rather negative portrait of the field-theoretical attempts to understanding
quantum gravity, and needless to say, this view was not shared by many
of the others lecturers at the School.

In section~\ref{sec:anomalies} we come to the study of the constraints imposed by
having consistent gauge and gravitational interactions
among chiral fermions. This brings us to the analysis of anomalies,
both local and global. The general conceptual problem of
computating local anomalies in various dimensions is presented, including some
powerful mathematical techniques that we apply, among other
examples, to the Green-Schwarz anomaly cancelation mechanism
\cite{Green-Schwarz} that opened the way to the formulation of the
heterotic string \cite{Pricenton-quartet}. We also briefly discuss the
issue of global anomalies (anomalies with respect to
transformations not continuously connected to the identity), 
and in particular Witten's $\mbox{SU}(2)$ anomaly
\cite{WittenSU(2),Witten-anom}. The aim of this section is to exhibit the
difficulties in having anomaly-free chiral theories containing gravitational
and gauge interactions. In the case of the Green-Schwarz mechanism, for example, we
are left with just~$\mbox{SO}(32)$,~$\mbox{E}_{8}\times
\mbox{E}_{8}$, and~$\mbox{E}_{8}\times \mbox{U}(1)^{248}$ as possible gauge groups. 
This severely restricts the spectrum of low-energy
excitations (i.e., massless states). Global anomalies, on the other hand, also
play a central role in determining the spectrum of possible
superstring
theories. In section~\ref{sec:fermionic_strings} we analyze in detail how the absence of
global world-sheet diffeomorphism anomalies (modular invariance), a
purely two-dimensional condition, restricts the space-time spectrum in
string theory.

In section~\ref{sec:bosonic_string} we begin our study of string theory,
restricting our attention first to the bosonic string. Its quantization 
is presented discussing the difference between critical and
non-critical strings and how the spectrum contains a graviton and, 
unfortunately, a tachyon as well. 
We then present a
general analysis of string perturbation theory and discuss the operator
formalism proposed in ref.~\cite{Alvarez-Gaume-Gomez-Moore-Vafa}, (for other
approaches, see refs.~\cite{Op-form-2-1,Op-form-2-2}). This approach emphasizes
the geometrical
nature of first quantized string theory, where string scattering amplitudes
and physical-state conditions are beautifully reflected in the
properties of the moduli space of Riemann surfaces. This geometric
formulation of string scattering amplitudes provides a neat understanding
of the origin of infinities. For the bosonic string in particular it will become clear how the source of
infinities is the tachyon appearing in the spectrum. We have also
included a brief section on target-space duality, to exhibit a property of
strings not shared by field theories. This symmetry renders further
evidence that string theory has a fundamental length despite string interactions 
being purely local, something that one
cannot have in field theory. This fundamental length may
be the ultimate reason for the finiteness of superstrings amplitudes. We
also analyze how the Einstein equations follow from the two-dimensional
consistency requirement of conformal invariance, and fully determines the classical
equations of motion.

Section~\ref{sec:fermionic_strings} is devoted to the study of 
supersymmetric strings. We see
how the tachyon is projected out from the spectrum using the GSO projection \cite{GSO},
and present its geometric interpretation
displaying the interplay between modular invariance and the
spectrum of the theory. The most remarkable example of supersymmetric strings is the heterotic
string, that we summarize paying special attention to the important
role of modular invariance. Two more topics are discussed before closing the section.
The first is a brief summary of the finite temperature behavior of fermionic strings where,
again, the modular group plays an interesting role. We present the
property
of temperature duality, consisting in the canonical partition function having 
the same value at temperatures~$T$ and~${\rm const.}/T$, and some of its riddles.
Secondly, we provide a brief status report on the finiteness of superstring
perturbation theory and what is known about its behavior at high orders.

Finally, in section~\ref{sec:developments} we summarize some active areas of current
research in string theory, as well as in string or string-inspired
black-hole physics, and present our conclusions. 

There are obviously many aspect of string theory left
out of these lectures. Most notably, the subject of classical solutions
to string theory, which started with the Calabi-Yau \cite{Calabi-Yau}
and orbifold solutions \cite{orbifolds}, and string phenomenology
\cite{phenomenology}. Both topics bloomed into rather large subjects but
have not been included. The focus of the School was centered on the
issue of quantum gravity, so it seemed more appropriate to leave these
subjects to a different occasion and to more competent lecturers. Our aim here
is to show how far we have gone with a rather simple set of
requirements. A lot of work is still necessary to clarify and explore
many aspects of string theory. Perhaps its most outstanding open
problem (as of any other theory of gravity) is the
vanishing of the cosmological constant, an issue where string theory provides
no clue.

Needless to say, many attendants to the school did not share the
moderately optimistic opinion conveyed in these lectures. This was a constant source 
of useful (and often entertaining) discussions. We still believe
that many of the ingredients and properties of string theory will be
contained in the correct theory of quantum gravity. It is however too early to
tell whether this expectation will be fulfilled.

\paragraph{Acknowledgements.} 
It is a pleasure to thank the organizers
of the 1992 Les Houches Summer School B. Julia and J. Zinn-Justin for the opportunity to present
these lectures and for creating a very productive and inspiring
enviroment. In preparing these lectures we have benefitted from
discussions with many colleagues. We would like to thank E.~\'Alvarez,
J.~L.~F.~Barb\'on,  E.~Br\'ezin, A.~Connes, C.~G\'omez, L.~Ib\'a\~nez, 
S.~Jain, W.~Lerche, J.~Louis, and M.~A.~R.~Osorio for many conversations
about string theory. The reviews by E.~\'Alvarez
\cite{EAlvarez1,EAlvarez2} were
particularly useful in the preparation of section~\ref{sec:QFTapproach}. Finally, we also
thank B.~Julia for the careful reading of a preliminary version of these lectures
and his very useful comments.

\subsection{A note on this revised version} 

In the over three decades passed since the first version of this review 
was posted on the arXiv,
string theory has undergone some momentous developments transoforming it
almost beyond recognition: among others, the second
superstring revolution, bringing about D-branes and M-theory, 
and the formulation of the AdS/CFT correspondence with its multiple spinoffs. 
Despite this fact, we have refrained from updating the content of the lectures. 
It would have meant either writing a completely different review or
expanding the original version to a medium-size book.
But, together with these practical reasons, our decision is also based on what we should call archeological 
grounds. As they stand, the lectures somehow reflect 
the zeitgeist of the years between the first and the second superstring revolutions.
This is particularly glaring in the big emphasis given to closed over open strings, 
an attitude that would radically change with the advent of D-branes. But also in the tacit optimism 
about supersymmetry and the heterotic string as the right path towards a reliable theory of everything. 
The contemporary 
reader might find interesting and amusing to judge to what extend these expectations 
have been confirmed or dashed.

All these considerations notwithstanding, the present version incorporates significant changes and improvements.  
Besides carrying out a substantial editing, fixing a large number of typos, and improving the layout, 
we have implemented the many corrections prompted by Bernard Julia's 
critical reading of the 
original manuscript, included in the 
published version that still remains unavailable online. 
In addition, all figures have also been included, something that back in 1992 were well beyond the
technical skills of the authors. For historical reasons, some outdated terminology has been 
preserved, most notably $R$-duality or space-time duality
for what shortly afterwards became universally known as $T$-duality. Finally, a few references 
that should have been originally quoted but were overlooked have now been included. 

We hope this new version of an ancient review might still
be of some use to early-stage students looking for a compact introduction to the very basics 
of string theory, as well as to the other topics presented in the following pages. 

\newpage

\section{Field-theoretical approach to quantum gravity}
\label{sec:QFTapproach}

General relativity (GR) and quantum field theory (QFT)
are among the most impressive achievements in 20th century science.
One of the first predictions of GR was the centennial precession of the
orbit of Mercury. The relativistic contribution is (see, e.g.,
ref.~\cite{Weinberg})
\begin{align}
\Delta\phi^{\rm th}=43.03 \,\, \mbox{s/century},
\end{align}
while the observed value is
\begin{align}
\Delta\phi^{\rm obs}=43.11\pm 0.45 \,\, \mbox{s/century}.
\end{align}
Even gravitational radiation, also a prediction of GR, has been
indirectly detected by the observation of the decay of the orbit
of the binary pulsar PSR $1913+16$ \cite{Taylor-pulsar}.

On the other hand, QFT allows to calculate observables at microscopic
scales, often with a very high precision. To give an example, the
predicted
value in quantum electrodynamics~(QED) for the anomalous magnetic moment of
the electron is~\cite{Itzykson-Zuber}
\begin{align}
a_{e}^{\rm th}=1.159652359(282) \times 10^{-12},
\end{align}
where uncertainties are given by those in the value of the fine-structure 
constant. Experimentally it is found that
\begin{align}
a_{e}^{\rm exp}=1.159652410(200) \times 10^{-12}.
\end{align}

To obtain a more complete fundamental picture of the physical world we
would like to put together QFT and GR. There are still many obstacles
to achieve this goal. It is also sufficiently clear that such a
synthesis should provide profound insights into many of the riddles of
the standard model (SM) of strong, weak, and electromagnetic interactions.
Most notably, the origin of the chiral nature of the quarks
and leptons families and possibly also the origin of mass. Studying the running
of the coupling constants for the three factors 
$\mbox{SU}(3)\times \mbox{SU}(2)\times \mbox{U}(1)$ of the SM gauge
group (denoted by~$\alpha_{3},\alpha_{2},\alpha_{1}$
respectively), it is found with some assumption about the degrees of
freedom between present experimental energies and the Planck scale (the
desert hypothesis) and extrapolating from their present
experimental values that the three of them converge to a common value at a scale 
between~$10^{16}$-$10^{17}$~GeV, where quantum gravitational effects can certainly
not be neglected. Unification of GR and QFT is in fact the most coveted Holy
Grail in theoretical physics in the last quarter of the 20th century.

\subsection{Linearized gravity}

QFT has been successfully applied to both strong and electroweak
interaction, successfully described in terms of local gauge field
theories. It is precisely this
local gauge symmetry that ensures that the SM is
perturbatively
renormalizable~\cite{'tHooft1}, i.e. there is an effective
decoupling of the high-energy degrees of freedom from the low energy
predictions. Thus, infinities appearing in the perturbative expansion
can be absorbed in redefinitions of parameters of the theory that
are experimentally measurable.

Gauge invariance and renormalizability seem to be two of the cornerstones of
our understanding of present high energy physics. One then may wonder why
not try the same program with GR. We begin with the
Einstein-Hilbert action
\begin{align}
S_{\rm EH}=-\frac{1}{2\kappa^{2}}\int d^{4}x\sqrt{-g} R,
\end{align}
and make a decomposition of the metric into a background
$\widehat{g}_{\mu \nu}$ and a dynamical part $h_{\mu \nu}$:
\begin{align}
g_{\mu \nu}=\widehat{g}_{\mu \nu}+ \kappa h_{\mu \nu}.
\end{align}
By doing so, 't Hooft and Veltman \cite{'tHooft-Veltman} showed that
four-dimensional pure gravity has a finite one-loop $S$-matrix,
although it is not one-loop renormalizable off-shell.
Considering gravity coupled to a scalar field, the one-loop counterterm
in the Lagrangian is the sum of two terms, the first one
proportional to the equation of motion and the second one given by
\begin{align}
{\cal L}_{c}^{1}=\frac{1}{\epsilon}\sqrt{-\widehat{g}}\frac{203}{80}
\widehat{R}^{2},
\end{align}
where $\epsilon=4-d$ is the usual parameter in dimensional
regularization. In the case of pure gravity we should have $\widehat{R}=0$,
so this counterterm vanishes and the theory is finite at one-loop.
When considering however the coupling to matter, the counterterm is
divergent (since now we have~$\widehat{R}\neq 0$).

The finiteness of the one-loop $S$-matrix for pure gravity was a rather
surprising result, since Einstein theory is non-renormalizable by
naive power counting. At two loops the situation changes radically.
Goroff and Sagnotti~\cite{Goroff-Sagnotti} were
able to prove that Einstein pure gravity diverges at the two-loop level, and
that the divergent part of the on-shell effective action is
\begin{align}
\Gamma^{(2)}=\frac{209}{2880(4\pi)^{4}}\frac{1}{\epsilon}
\int d^{4}x\sqrt{-g} R^{\alpha \beta}_{\;\;\;\;\gamma \delta} R^{\gamma
\delta}_{\;\;\;\;\rho \sigma} R^{\rho \sigma}_{\;\;\;\;\alpha \beta}.
\end{align}
Thus, the program that allowed the construction of a reliable quantum field theory for
both the strong and the electroweak interactions fails when applied to~GR.

Insisting in that renormalizability (or
predictability) should hold in the presence of gravitational
interactions results in strong constraints on
the possible quantum theories describing gravity.
If the theory includes other fields, then
subtle cancellations must exist among different contributions.
Following this line of thought, the first proposed candidate for an operative theory of quantum 
gravity was
supergravity~(SUGRA) \cite{SUGRA}.

\subsection{Supergravity}

In SUGRA theories supersymmetry is local. Since the
commutator of two supercurrents generates the energy-momentum tensor,
gauging supersymmetry inevitably leads to a theory
containing gravity. As in ordinary supersymmetry, in~$N=1$ supergravity
\cite{N=1}
each particle of mass $m$ and spin $s$ has associated with it another
one with the same mass and spin $s\pm\frac{1}{2}$. Thus, the
graviton, which has spin $2$, comes accompanied by the gravitino field
with $s=\frac{3}{2}$. In principle one can also construct theories with
extended supergravity, $N>1$. Since it is not known how to consistently
quantize an interacting theory containing a massless spin-$\frac{5}{2}$
field, the requirement that the highest helicity state be the graviton
imposes the constraint~$N\leq 8$ \cite{N<9}.

The ultraviolet properties of SUGRA perturbation theory improve with respect
to those of pure gravity due to the cancelation of divergences between
bosons and fermions in the same supermultiplet. Although
non-renormalizable by power counting, it is legitimate to ask whether
the cancelation of infinities will render the
SUGRA $S$-matrix finite for some $1\leq N\leq 8$, without the need to renormalize. Even if the
off-shell Green functions may be plaged with infinities, a presumably
weaker requirement than renormalizability is to impose finiteness of the
scattering amplitudes.

As an example, let us consider $N=1$, $d=4$ SUGRA described by the
Lagrangian \cite{N=1}
\begin{align}
{\cal L}=-\frac{1}{2\kappa^{2}}eR-
\frac{1}{2}\epsilon^{\mu\nu\sigma\lambda}
\overline{\psi}_{\mu}\gamma_{5}\gamma_{\nu}D_{\sigma}\psi_{\lambda},
\end{align}
where $\psi_{\mu}$ is the gravitino field, $e=\det(e_{\mu}^{\;\;a})$ the
determinant of the vierbein, and $D_{\mu}$ the covariant derivative
\begin{align}
D_{\mu}=\partial_{\mu}+\frac{1}{2}\omega_{\mu}^{\;\;mn}\sigma_{mn},
\end{align}
with $\omega_{\mu}^{\;\;mn}$ the spin connection and
$\sigma_{mn}=\frac{1}{4}[\gamma_{m},\gamma_{n}]$. The
Lagrangian ${\cal L}$ is invariant under the local supersymmetry
transformations
\begin{align}
\delta
e_{\mu}^{\;\;m}
&=\frac{1}{2}\kappa\overline{\epsilon}(x)\gamma^{m}\psi_{\mu}(x),
\nonumber \\[0.2cm]
\delta\omega_{\mu}^{\;\;mn}&=0,
\\[0.2cm]
\delta \psi_{\mu}&=D_{\mu}\epsilon(x).
\nonumber 
\end{align}
In the quantum theory, besides the graviton and the gravitino,
we have to include a coordinate spin-$1$ ghost and its
corresponding partner, a spin-$\frac{1}{2}$ ghost field
\cite{Freedman-Das}.

It can be shown that on-shell amplitudes in $N=1$ SUGRA are finite up to
two-loops~\cite{N=1-finite}, since the possible counterterms in
the action are zero on-shell. At three-loops, however, there is a counterterm which
does not vanish even on-shell \cite{N=1-3-l} (see also the second
reference in \cite{N=1-finite}). Unless the coefficient
multiplying it is zero, $N=1$ SUGRA would be three-loop
divergent. Whether this happens or not, is still an open question. The computation to settle 
the question
is a formidable task that nobody has undertaken yet. This notwithstanding, since there
exists a possible counterterm, one would tend to be rather pessimistic
about its outcome.

Given that $N=1$ SUGRA is expected to diverge at three-loops, one has to
examine the case for
a finite theory in $N$-extended supergravity theories. For
some time $N=8$ SUGRA in four dimensions was a candidate for a theory of
``everything''. This theory can be formulated as $N=1$ SUGRA in
eleven dimensions \cite{N=1-d=11} which, after dimensional reduction,
becomes
$N=8$ SUGRA in four dimensions \cite{Cremmer-Julia}. In the
eleven-dimensional theory we have an elfbein, $e_{\mu}^{\;\;m}$,
together with a gravitino, $\psi_{\mu}$, and an antisymmetric tensor of
rank three, $A_{\mu\nu\sigma}$ (it is easy to check that the number of
propagating bosonic and fermionic degrees of freedom is the same). The
mechanism of dimensional reduction consists in making the fields
depend only on the four-dimensional coordinates
$x^{0},\ldots,x^{3}$ and not on the remaining ones
ones $x^{4},\ldots,x^{10}$. At the end of the procedure, we find the spectrum
in~$d=4$ shown in table~\ref{SUGRA_spec}.
\begin{table}[t]
\begin{center}
\begin{tabular}{ccc}
\hline
\# of fields    & spin  &	 \\
\hline
\\[-0.3cm]
1  & 2	& graviton \\[0.1cm]
8  & $\frac{3}{2}$  & gravitinos  \\[0.1cm]
28 & 1	  &  \\[0.1cm]
56 & $\frac{1}{2}$  &  \\[0.1cm]
70 & 0	 &  \\
\hline
\end{tabular}
\end{center}
\caption{$N=8$~SUGRA spectrum in~$d=4$.}
\label{SUGRA_spec}
\end{table}
Incidentally, it is not possible to construct SUGRA theories in dimensions $d>11$
\cite{Nahm} since there is no way of matching bosonic and
fermionic degrees of freedom.
Would it be possible to build simple SUGRA in $d>11$,
implementing dimensional reduction we would have $N>8$ SUGRA in $d=4$,
thus violating so the upper bound for the number of supersymmetric partners
(see above).

Although by symmetry arguments $N=8$ SUGRA is expected to be finite up
to seven loops, there are possible
on-shell countertems at higher loops \cite{Kallosh}. This means that the theory could be
divergent at large
orders if there is no miraculous cancellation of the coefficients in
front of the counterterms. Naively, this seems to hint that supergravity is not the
correct place to look for a finite theory of quantum gravity. In fact,
for $N$-extended SUGRA there are possible non-zero on-shell counterterms
beyond the
$(N-1)$-loop when $N\geq 3$, and beyond two-loops for~$N=1,2$.

\subsection{Kaluza-Klein theories}

During the 1970s and early 1980s there was a good deal of
activity in
Kaluza-Klein theories. The main idea goes
back to the 1920s when Kaluza and Klein \cite{Kaluza-Klein} tried to
unify Einstein's general relativity and Maxwell's theory by formulating
GR in $M_{4}\times S^{1}$, with $M_{4}$ the ordinary
four-dimensional space-time and $S^{1}$ a circle of very small radius
(typically of the order of the Planck lenght, $\ell_{P}\sim 1.6\times
10^{-33}$
cm). Introducing coordinates $(x^{\mu},\phi)$, we take the
Kaluza-Klein ansatz for
the metric tensor in five dimensions
\begin{align}
g_{AB}(x,\phi)=\left(
\begin{array}{cc}
g_{\mu \nu}(x)+e^{2}\kappa^{2}\sigma(x)A_{\mu}(x)A_{\nu}(x) &  e\kappa\sigma(x) A_{\mu}(x)  \\[0.2cm]
e\kappa \sigma(x)A_{\mu}(x)  & \sigma(x)
\end{array}
\right),
\label{U(1)KK}
\end{align}
where we have retained only the zero-modes in the fifth coordinate
$\phi$ (this is equivalent to restricting ourselves to massless modes
in $M_{4}$). Substituting eq.~\eqref{U(1)KK} in the five-dimensional
Einstein-Hilbert action, we
obtain the gravitational action in four dimensions plus the Maxwell
kinetic term $(F_{\mu \nu})^{2}$ for the four-dimensional vector field
$A_{\mu}$, as well as the kinetic term for the scalar~$\sigma$.

This setup can be generalized to include non-Abelian gauge theories
\cite{Cho-Freund}. To do so, we consider the Einstein-Hilbert action in a
$(4+n)$-dimensional space-time  $M_{4}\times B$, where $B$ is a compact
internal manifold having $G$ as its group of isometries. Considering
coordinates
$(x^{\mu},\phi_{i})$ (with $\mu=0,1,2,3$ and $i=1,\ldots,n$), the Kaluza-Klein
ansatz for the metric tensor in~$M_{4}\times B$ is
\begin{align}
g_{AB}(x,\phi)=\left(
\begin{array}{cc}
\displaystyle g_{\mu \nu}(x)+\sum_{a,b=1}^{{\rm dim\,}G}\gamma_{ij}(\phi)\xi^{ia}(\phi)\xi^{jb}(\phi)
A_{\mu}^{a}(x)A_{\nu}^{b}(x)	  &  \displaystyle\sum_{a=1}^{{\rm dim\,} G} A_{\mu}^{a}(x)\xi_{i}^{a}(\phi)
\\
\displaystyle\sum_{a=1}^{{\rm dim\,} G} A_{\mu}^{a}(x)\xi_{i}^{a}(\phi) &  \displaystyle \gamma_{ij}(\phi)
\end{array}
\right),
\label{KK-n}
\end{align}
with $\xi_{i}^{a}$ ($a=1,\ldots,{\rm dim\,} G$) the Killing vectors
associated with the symmetry group $G$
\begin{align}
T^{a}:\phi_{i} \longrightarrow \phi_{i}+\xi_{i}^{a}(\phi)\epsilon^{a}.
\end{align}
Substituting \eqref{KK-n} into
\begin{align}
S_{\rm EH}=-\frac{1}{2\kappa^{2}}\int d^{4+n}x \sqrt{-g} R^{(4+n)},
\end{align}
we obtain in particular the Einstein-Hilbert action in four dimensions
as well as the
Yang-Mills action $\sum_{a}(F_{\mu \nu}^{a})^{2}$, invariant under the
gauge transformation
\begin{align}
(x^{\mu},\phi_{i}) &  \longrightarrow  
\left( x^{\mu},\phi_{i}+\sum_{a}\epsilon^{a}(x)\xi_{i}^{a}(\phi) \right),
\\[0.2cm]
A_{\mu}^{a}(x) &  \longrightarrow  A_{\mu}^{a}(x)+D_{\mu}\epsilon^{a}(x),
\end{align}
where $\epsilon^{a}(x)$ are the gauge parameters and $D_{\mu}$ is the
covariant derivative with respect to the gauge field $A_{\mu}^{a}$.

This illustrate the main tenet of Kaluza-Klein theories: to represent gauge
invariance in four-dimensional space-time as resulting from the
group of isometries of an internal $n$-dimensional manifold, a
beautiful geometrical idea indeed. Of course,
the problem is to construct realistic Kaluza-Klein theories, i.e.
theories with in which the isometry group~$G$ of the internal manifold contains the SM gauge group
$\mbox{SU}(3)\times \mbox{SU}(2) \times \mbox{U}(1)$. This condition imposes a lower bound
on the dimension of the internal space \cite{K-K-Witten1}. The space with symmetry group~$G$ of
lowest dimension is always a homogeneous space $G/H$, 
with $H\subset G$ a maximal subgroup of $G$. In our 
case~$G=\mbox{SU}(3)\times \mbox{SU}(2) \times \mbox{U}(1)$, while the largest 
suitable subgroup
is $H=\mbox{SU}(2)\times \mbox{U}(1)\times \mbox{U}(1)$. This means that the minimal
dimension of a manifold with symmetry group  $\mbox{SU}(3)\times \mbox{SU}(2) \times
\mbox{U}(1)$ is $12-5=7$. This means that the minimal dimension of the Kaluza-Klein space-time 
is~$11$,
precisely the maximal dimension for SUGRA (see above).

Unfortunately, Kaluza-Klein theories also have serious problems. Being essentially Einstein gravity 
in dimensions higher than four, one of them is non-renormalizability. But perhaps their
main difficulty is the impossibility of reproducing the chiral nature
of our four-dimensional world. To see this, consider a massless Dirac
fermion in $4+n$ dimensions. It obeys the Dirac equation
\begin{align}
\Dsl \psi\equiv\sum_{i=1}^{4+n}\Gamma^{i}D_{i}\psi=0,
\end{align}
which we rewrite as
\begin{align}
\Dsl ^{(4)}\psi+\Dsl ^{({\rm int})}\psi=0,
\label{4+int}
\end{align}
with $\Dsl ^{(4)}=\sum_{i=1}^{4} \Gamma^{i}D_{i}$ and
$\Dsl ^{({\rm int})}=\sum_{i=5}^{4+n} \Gamma^{i}D_{i}$. It is easy to see
from \eqref{4+int} that the eigenvalues of $\Dsl ^{({\rm int})}$ give the
masses of the four-dimensional fermions. Since the non-zero eigenvalues
will be of the order of $1/R$, with $R$ the typical length of the
internal manifold (i.e. of the order of the Planck length), we conclude that the
phenomenologically relevant four-dimensional fermions correspond to the
zero-modes of~$\Dsl ^{({\rm int})}$. However, in order to reproduce the
SM, these fermions should be chiral in four dimensions,
which means that the representations of the gauge group 
coupling to their left- and right-handed components
should be different. This
cannot be achieved using Dirac fermions in $4+n$ dimensions, as can
be seen by the following argument due to Lichnerowicz
\cite{Lichnerowicz,K-K-Witten2}: squaring
the Dirac operator~$i\Dsl ^{({\rm int})}$ we have
(dropping the label ``int'' for simplicity)
\begin{align}
(i\Dsl )^{2}=-\sum_{i=5}^{4+n} D_{i}D^{i} + \frac{1}{4}R.
\end{align}
where~$R$ is the scalar curvature of the internal manifold. Thus, if $R>0$ everywhere on~$B$, 
and since $-\sum_{i} D_{i}D^{i}$ is positive definite,
the Dirac operator cannot have zero-modes. Furthermore,
a theorem by Lawson and Yau \cite{Lawson-Yau} states that on any compact
space $B$ with
non-Abelian symmetry group~$G$, there is always a $G$-invariant metric
with positive curvature. The upshot is that we cannot obtain massless fermions in four
dimensions from Dirac
fermions in $4+n$ dimensions. 

What can we say about Rarita-Schwinger
fields?
After all, in $N=8$ SUGRA we do not start with Dirac spinors
but rather with
Rarita-Schwinger fields in eleven dimensions, giving rise after dimensional
reduction to four-dimensional Dirac fermions. The problem of
obtaining four-dimensional chiral spin-$\frac{1}{2}$ fermions from
Rarita-Schwinger
fields in higher dimension has been studied by Witten, who also carried
out the general analysis of whether chirality could be generated from
Kaluza-Klein theories. Here we are essentially summarizing his arguments
\cite{K-K-Witten2}.
By using the Atiyah-Hirzebruch theorem
\cite{Atiyah-Hirzebruch}, he showed that the zero-modes of
the Rarita-Schwinger operator on any homogeneous space~$G/H$ are in a real
representation leading to vector-like quantum numbers in four
dimensions. It is therefore imposible to reproduce a chiral low-energy
field theory in four dimensions implementing the Kaluza-Klein program on
eleven-dimensional SUGRA.

The only way to avoid Witten's result without renouncing to the
Kaluza-Klein ideas is to consider internal
manifolds that are neither coset spaces nor compact
\cite{Wetterich}. Another way to bypass the problem of generating
chiral fermions is including gauge fields in the higher-dimensional
theory \cite{gauge-d>4}. This last proposal, however, spoils
the most relevant
aim of Kaluza-Klein theories, namely, the unification of gravity with
Yang-Mills theories in a geometrical framework. One needs other principles
to justify the presence of gauge symmetries in higher dimensions. String
theory is the only candidate generating at the same time extra
gauge symmetries, chirality, and a likely consistent quantum theory of
the gravitational field.

\subsection{Quantum field theory and classical gravity}
\label{sec:QFT_curved_spaces}

Instead of being too ambitious and starting directly with quantum gravity, 
we could learn some useful lessons studying
QFT in the presence of a classical gravitational
background. This subject is reviewed in the
lectures by R. Wald~\cite{Wald_lectures}, and
we will limit our considerations here to some general remarks. QFT on
curved space-times exhibits many interesting features on its own right
\cite{Birrel-Davies}. In particular, many of the
properties holding in Minkowski space-time either do not apply or
change radically on a curved background. One important example concerns the
vacuum state, which in Minkowskian QFT has a number of well-known properties:
\begin{itemize}
\item[-]
It is unique.

\item[-]
There is a well-defined concept of localized excitations
(particles).

\item[-]
It determines the symmetries of the world and their realizations.
\end{itemize}
On the other hand, in the presence of a gravitational field the vacuum state loses the
absolute meaning it has in flat space. The reason is that strong
or rapidly varying gravitational fields can produce
particles. It is easy to see that this makes the concept of
particle ambiguous (i.e. observer-dependent). Let us consider a region of the
space-time with a very strong gravitational field iwhere
a static observer will see that particles are created in pairs. 
However, a free-falling observer due to the equivalence principle
will not observe (locally) any gravitational field, so for him/her there is
no creation of particles at all.
This means that what the free-falling observer identifies as the
vacuum state of her/his QFT is not a vacuum at all for the static observer.

There are a number of elements that we require in order to build a QFT
in a fixed gravitational background~\cite{Gibbons}:
\begin{itemize}
\item[-]
A Hilbert space of states ${\cal H}$.

\item[-]
A classical space-time $({\cal M},g_{\mu\nu})$.

\item[-]
Fields operators acting on ${\cal H}$.

\item[-]
A set of canonical commutation relations obeyed by these fields.

\item[-]Wave equations.

\item[-]
Rules for constructing ``particles'' observables and Fock basis of
${\cal H}$.

\item[-]
Regularization and renormalization schemes to make expectation
values like $\langle T_{\mu\nu} \rangle$ finite.

\end{itemize}
In constructing a QFT in curved background, on the other hand, we face some important
issues. First of all, since we are performing the quantization of the theory
in a given background metric~$g_{\mu\nu}$, it is natural
to ask about the back-reaction of the QFT on the space-time geometry.
To be more precise, when solving the quantum theory we usually impose
the consistency condition (in~$G=1$ units)
\begin{align}
R_{\mu\nu}-\frac{1}{2}g_{\mu\nu}\big(R-2\Lambda\big)=8\pi\langle
T_{\mu\nu}\rangle.
\label{Einstein}
\end{align}
In general our energy-momentum tensor will not verify \eqref{Einstein},
so it is necessary to proceed \`a la Hartree-Fock: we quantize the
theory in the presence of a ``test'' background metric
$g_{\mu\nu}^{(0)}$, and evaluate the vacuum expectation value of
the energy-momentum tensor $\langle T_{\mu\nu} \rangle$. Then we
solve \eqref{Einstein} with $\langle T_{\mu\nu} \rangle$ and
obtain a first-corrected metric $g_{\mu\nu}^{(1)}$. Repeating the
process we get a series of metrics $\{g_{\mu\nu}^{(0)},g_{\mu\nu}^{(1)},
\ldots,g_{\mu\nu}^{(N)},\ldots\}$ which, hopefully, should eventually
converge to a self-consistent solution $g_{\mu\nu}$.

A second problem is that, in order to solve wave equations, we need our
background space-time to be globally hyperbolic \cite{Hawking-Ellis}
(meaning that there exists a global Cauchy hypersurface~$\Sigma$). 
When there are singularities or Cauchy horizons we would also need
to specify certain (unclear) boundary conditions.

Moreover, in ordinary QFT the field's divergent vacuum energies can be
subtracted without any further ado since it amounts
to a redefinition of the origin of energies.
In the presence of gravity,
vacuum energies contribute to the value of the
cosmological constant. Once symmetry breaking occurs, the
quantum contributions are
unacceptably large. In view of this, the smallness of the cosmological
constant is probably the most mysterious fact with no
explanation from any theory of quantum gravity, including string theory.

Let us now consider fields on a curved background. Since we have no
invariant notion of time, it seems natural to work in the Heisenberg
picture where the states are time-independent. The existence of a
Cauchy hypersurface, $\Sigma$, allows the introduction of an inner product in 
field space: given two classical fields $\Phi_{1}$, $\Phi_{2}$ 
their inner product $(\Phi_{1},\Phi_{2})$ is given by
\begin{align}
(\Phi_{1},\Phi_{2})=-\int_{\Sigma}
d\Sigma^{\mu}J_{\mu}(\Phi_{1},\Phi_{2}),
\label{product}
\end{align}
where $J_{\mu}(\Phi_{1},\Phi_{2})$ is defined by
\begin{align}
J_{\mu}(\phi_{1},\phi_{2})=i(\overline{\phi}_{1}\nabla_{\mu}\phi_{2}-
\overline{\phi}_{2}\nabla_{\mu}\phi_{1}),
\end{align}
for two scalar fields and
\begin{align}
J_{\mu}(\psi_{1},\psi_{2})=\overline{\psi}_{1}\gamma_{\mu}\psi_{2},
\end{align}
when $\psi_{1}$, $\psi_{2}$ are spinors. It is easy to check 
using the equations of motion that both currents are conserved
($\nabla_{\mu}J^{\mu}=0$), so the product \eqref{product} does not depend
on the particular choice of the Cauchy hypersurface, as long as
the fields vanish at infinity.

For simplicity, we consider the case of a real massless
scalar
field $\phi(x)$ \cite{DeWitt}. Let~$\{u_{i}(x)\}$ be a complete set of
solution of its wave
equation $g^{\mu\nu}\nabla_{\mu}\nabla_{\nu}\phi(x)=0$. Its 
general solution can then be written as
\begin{align}
\phi(x)=\sum_{i}u_{i}(x)a_{i}+\overline{u}_{i}(x)\overline{a}_{i}.
\end{align}
We assume moreover that $\{u_{i}(x)\}$ is an orthonormal set with
respect to the inner product~\eqref{product}. If $\xi^{\mu}(x)$ is a timelike
Killing field, we can define
\begin{align}
H(\xi)=\int_{\Sigma} d\Sigma^{\nu}T_{\mu\nu}\xi^{\mu},
\label{Hamiltonian}
\end{align}
where $\Sigma$ is again a Cauchy hypersurface and $T_{\mu\nu}$ is the
energy-momentum tensor of the scalar field. This quantity determines the field evolution
with respect to the time coordinate defined by~$\xi^{\mu}$. 
Indeed, the
Poisson bracket of (\ref{Hamiltonian}) with the field $\phi(x)$ is easy
to evaluate, with the result
\begin{align}
\{H(\xi),\phi(x)\}={\cal L}_{\xi}\phi(x),
\end{align}
where ${\cal L}_{\xi}$ is the Lie derivative with respect to the Killing
vector $\xi$. Since ${\cal L}_{\xi}$ is an anti-Hermitian operator with
respect
to the inner product, we will choose the $u_{i}(x)$'s to be
their eigenstates
\begin{align}
{\cal L}_{\xi}u_{i}(x)=-i\kappa u_{i}(x),
\label{eigenstates}
\end{align}
with $\kappa$ real. Next, we group the modes $u_{i}(x)$ into two
classes: when $\kappa>0$ we say that the associated mode is a
positive-frequency solution, while for $\kappa<0$ we have a negative-frequency solution. From
(\ref{eigenstates}) it is easy to see that $u_{i}(x)$ and
$\overline{u}_{i}(x)$
both have the same frequencies with opposite sign. Upon
quantization,
we associate creation/annihilation operators with positive/negative frequency
solutions, thus
\begin{align}
\widehat{\phi}(x)=\sum_{i}\Big[ u_{i}(x)a_{i}+\overline{u}_{i}(x)a_{i}^{\dagger}\Big],
\end{align}
with the vacuum state defined by $a_{i}|0\rangle=0$, for every $a_{i}$.
It is important to stress that the classification of the modes into
positive and negative frequency, and the definition of
the vacuum state, has been done with respect to a given Killing field
$\xi^{\mu}(x)$. Once the choice of the Killing field is made one has
introduced the notion of particle. A specially interesting case are
the so-called ``sandwich'' space-times composed by
three regions: two of them,~$M^{-}$ and~$M^{+}$, are static and correspond,
respectively, to the past and future. Between them we have the
third region, $M^{0}$, in which there is a time varying gravitational
field. In this situation we have different notions of particle
appropriate to particular measurement processes at different times. 
The concept of particle is therefore observer dependent.

The ambiguity in the concept of particle stems from its global nature.
We may speculate that, since we always have (by the equivalence
principle) free falling observers, it should be possible to define
particles only by their localized effects and
the dispersion relation $p^{2}=m^{2}$. This would lead to a ``bundle''
of ground states, which may not have a global section in the presence of
horizons.

We illustrate the ambiguity of the notion of particle with
a well-known example due to Unruh~\cite{Unruh}. Let us
consider a particle detector consisting of a quantum system
in its ground state in
the absence of interaction with an external field.
After it interacts, the system gets excited and
is not in its fundamental state any longer. The
presence of particles could then be detected by checking whether our detector is 
in an excited state.

Imagine now the detector moving along a world line
$x^{\mu}(\tau)$, where $\tau$ is the proper time and suppose
that it interacts with the scalar field
$\phi(x)$ through the Lagrangian
\begin{align}
{\cal L}_{\rm int}=c m(\tau)\phi[x^{\mu}(\tau)],
\end{align}
where $m(\tau)$ is the monopole moment of the detector. Furthermore, let us
assume
that we are in Minkowski space and that the field $\phi(x)$ is in its
ground state $|0_{M}\rangle$. As we will see, for a general trajectory the detector will
not remain in its ground state with energy $E_{0}$, but will jump to
an excited state with $E>E_{0}$, while the field will also be in some excited 
state $|\psi\rangle$. For $|c|\ll 1$ the amplitude for this
process to occur is
\begin{align}
{\cal A}=ic\langle
E,\psi|\int_{-\infty}^{+\infty}d\tau\,m(\tau)\phi[x(\tau)]
|0_{M},E_{0}\rangle.
\end{align}
Here $m(\tau)$ can be rewritten as
\begin{align}
m(\tau)=e^{iH_{0}\tau}m(0)e^{-iH_{0}\tau},
\end{align}
with $H_{0}$ the detector's Hamiltonian. Since $H_{0}|E\rangle=E|E\rangle$,
the amplitude ${\cal A}$ becomes
\begin{align}
{\cal A}=
ic\langle E|m(0)|E_{0}\rangle\int_{-\infty}^{+\infty}d\tau\,
e^{i(E-E_{0})\tau} \langle\psi|\phi[x(\tau)]|0_{M}\rangle.
\end{align}
When $|\psi\rangle=|k\rangle$ is a state with a single particle with
momentum~$k$, we can use the expression of
$\phi(x)$ in terms of Minkowski modes
\begin{align}
\phi(x)=\int\frac{d^{3}k'}{\sqrt{16\pi^{3}\omega'}}\left(
a_{k}^{\dagger}e^{-i\vec{k}'\cdot\vec{x}+i\omega't}+ \mbox{h.c.}\right),
\end{align}
with $\omega=\sqrt{k^{2}+m^{2}}$, to write
\begin{align}
\langle k|\phi(x)|0\rangle=\frac{1}{\sqrt{2(2\pi)^{3}\omega}}
e^{-i\vec{k}\cdot\vec{x}+i\omega t}.
\end{align}
For an inertial observer ${\bf x}={\bf x}_{0}+
{\bf v}\tau(1-v^{2})^{-1/2}$, and performing the integration we obtain
\begin{align}
{\cal A}=\frac{1}{\sqrt{4\pi\omega}}e^{-i\vec{k}\cdot\vec{x}_{0}}
\delta\big(E-E_{0}+(\omega-{\bf k}\cdot{\bf v})(1-v^{2})^{-1/2}\big).
\end{align}
Since $E>E_{0}$ and ${\bf k}\cdot{\bf v}\leq kv<\omega$, the
argument of the $\delta$-function is positive and the
amplitude vanishes. For a more complicated trayectory $x^{\mu}(\tau)$, however,
the probability that after interaction the system is not in its ground state is given by
\begin{align}
{\cal P}=c^{2}\sum_{E}\left|\langle E|m(0)|E_{0}\rangle\right|^{2}
{\cal F}(E-E_{0}),
\end{align}
where
\begin{align}
{\cal F}(E-E_{0})=\int_{-\infty}^{+\infty} d\tau
\int_{-\infty}^{+\infty} d\tau'\,e^{-iE(\tau-\tau')}
G^{+}\big(x(\tau),x(\tau')\big).
\end{align}
Hence, the particle absorption per unit time is
\begin{align}
{\cal W}=c^{2}\sum_{E}\left|\langle E|m(0)|E_{0}\rangle
\right|^{2}\int_{-\infty}^{+\infty} d(\Delta\tau)
e^{-i(E-E_{0})\Delta\tau}G^{+}(\Delta\tau).
\end{align}
In the particular case of a 
uniformly accelerated detector, the world line is given by
\begin{align}
x&=y=0 \nonumber, \nonumber \\[0.2cm]
z&=\alpha\cosh\left({\tau\over \alpha}\right), \\[0.2cm]
t&=\alpha \sinh\left({\tau\over \alpha}\right),
\nonumber 
\end{align}
and the probability per unit time takes the form
\begin{align}
{\cal W}=\frac{c^{2}}{2\pi}\sum_{E}(E-E_{0})\frac{
\left|\langle E|m(0)|E_{0}\rangle \right|^{2}}{
e^{2\pi\alpha (E-E_{0})}-1}.
\end{align}
Thus, an observer moving with constant acceleration detects a flux of
scalar massless particles whose spectrum corresponds to a thermal bath 
with temperature
\begin{align}
T=\frac{1}{2\pi\alpha}.
\end{align}
For the inertial observer, on the other hand, the field $\phi(x)$ is in its
vacuum state $|0_{M}\rangle$ and no particles are detected. 
This clearly shows how, when we depart from Minkowski space, the
concepts of vacuum and particle states become observer dependent.

An alternative way to understand how the notion of no-particle state changes
in curved backgrounds is to use Bogoliubov transformations (for more
details and references see~\cite{Gibbons}). We have seen above
that upon choosing a time-like Killing field $\xi^{\mu}(x)$ we
can write any solution of the wave equation in terms
of a set of positive frequency solutions~$u_{i}(x)$ with respect to $\xi^{\mu}(x)$,
and their complex conjugates $\overline{u}_{i}(x)$. However,
many curved backgrounds do not
have time-like Killing fields, or the Killing fields are not time-like
everywhere.
In the first case we usually deal with vector fields satisfying the
Killing condition only asymptotically, which demands a $S$-matrix
formulation.

Let us consider two complete sets of solutions of the
the massless scalar field wave equation,~$\{u_{i}(x)\}$ and $\{w_{i}(x)\}$.
Associated
with them we have the corresponding creation and annihilation
operators $\{a_{i},a_{i}^{\dagger}\}$ and
$\{b_{i},b_{j}^{\dagger}\}$. Since both sets of modes define bases for
the space of solutions to the wave equation, we can expand one set in terms of
the other
\begin{align}
w_{i}(x)&=\sum_{j} \alpha_{ij}u_{j}(x)+\beta_{ij}\overline{u}_{j}(x),
\nonumber \\[0.2cm]
\overline{w}_{i}(x)&=\sum_{j}
\overline{\beta}_{ij}u_{j}(x)+\overline{\alpha}_{ij}\overline{u}_{j}(x).
\end{align}
Taking into account that
\begin{align}
\phi(x)=\sum_{i}\Big[a_{i}u_{i}(x)+a_{i}^{\dagger}\overline{u}_{i}(x)\Big]=
\sum_{i}\Big[b_{i}w_{i}(x)+b_{i}^{\dagger}\overline{w}_{i}(x)\Big],
\end{align}
we find that both sets of creation-annihilation operators are
related by the Bogoliubov transformations
\begin{align}
b_{i}&=\sum_{j}\overline{\alpha}_{ij}a_{j}-\overline{\beta}_{ij}a_{j}^{\dagger},
\nonumber \\[0.2cm]
b_{i}^{\dagger}&=\sum_{j}-\beta_{ij}a_{j}+\alpha_{ij}a_{j}^{\dagger}.
\end{align}

We apply next these relations to a space-time with two
asymptotically flat regions, that we label as {\it in} and {\it out}. The
corresponding
set of modes $u_{i}^{\rm in}(x)$ and $u_{i}^{\rm out}(x)$ are related to one another by
\begin{align}
u_{i}^{\rm out}(x)=\sum_{j}\alpha_{ij}u_{j}^{\rm in}(x)+
\beta_{ij}\overline{u}_{j}^{\rm in},
\end{align}
and by applying the previous result we obtain:
\begin{align}
a^{\rm out}_{i}&=\sum_{j}\overline{\alpha}_{ij}a^{\rm in}_{j}-
\overline{\beta}_{ij}a_{j}^{\rm in\dagger}, \nonumber \\[0.2cm]
b_{i}^{\rm out \dagger}&=
\sum_{j}-\beta_{ij}a_{j}^{\rm in}+\alpha_{ij}a_{j}^{\rm in \dagger},
\label{in-out}
\end{align}
while the corresponding {\it in} and {\it out} vacuum states 
are respectively defined by
\begin{align}
a_{i}^{\rm in}|0_{\rm in}\rangle &= 0 \hspace{1cm} \forall i, \nonumber \\[0.2cm]
a_{i}^{\rm out}|0_{\rm out}\rangle &= 0 \hspace{1cm} \forall i,
\end{align}
Thus, using eq.~(\ref{in-out}) we find that if there are some $\beta_{ij}\neq 0$, the {\it in} vacuum
is not a vacuum state for {\it out} particles, rather
\begin{align}
n_{i}^{\rm out}=\langle 0_{\rm in}|a_{i}^{\rm out \dagger}a_{i}^{\rm out}| 0_{in}
\rangle= \sum_{i}|\beta_{ij}|^{2}.
\end{align}
The total number of {\it out} particles in the {\it in} vacuum is then
given by
\begin{align}
N^{+}=\sum_{j} n_{j}^{\rm out}={\rm tr\,}\beta\beta^{\dagger}.
\end{align}
With this we show that if our system is in the {\it in} vacuum state, an 
{\it out} observer would find the quantum state to contain a total number of particles
given by ${\rm tr\,}\beta\beta^{\dagger}$. In practical terms, the problem
is to find the Bogoliubov transformation between the two asymptotic
regions. Although Bogoliubov
transformations are also applied in ordinary QFT, they are ubiquitous in the presence of 
physically interesting
gravitational backgrounds.

One of the most spectacular results in the
study of quantum fields in curved space-times
was found by Hawking in $1975$~\cite{Hawking1}.
He found that a Schwarzschild black hole emits
particles as if it were a black body with temperature\footnote{For details
and references on Hawking radiation, see the lectures by R. Wald at this school~\cite{Wald_lectures}.}
\begin{align}
T_{\rm Hawking}=\frac{1}{8\pi M},
\label{eq:hawking_temp}
\end{align}
where~$M$ is the black hole mass.
Since the temperature is inversely proportional to the mass, and the
total amount of energy radiated per unit time and per unit area is
proportional to the fourth power of
the temperature, we see that, assuming the
semiclassical approach all the way, the black hole evaporates
completely in a time proportional to $M_{0}^{3}$, with $M_{0}$ the
initial mass. 
From a thermodynamical viewpoint it is quite counterintuitive
that the black hole heats up as it radiates.

Black hole evaporation rises important conceptual issues (for a
recent review see ref.~\cite{Preskill}). As explained, Hawking found that the radiation
emited by a black hole is thermal. If this is the case, the radiated particles are
in a mixed quantum state.
On the other hand, in classical GR no information from inside the horizon can escape to the
asymptotic region. It would seem therefore natural that when quantizing a field theory 
in a black hole background one should trace over the states inside the
horizon. This produces a mixed state. Thus, if the
radiation is exactly
thermal during the whole evaporation, we are forced to conclude that a matter system in a
pure state collapsing to form a black hole would end up, after evaporation, 
in a mixed quantum state.
In other words, although we know
the initial quantum state of the system we cannot predict what the final state
would be. This clearly violates the laws of quantum mechanics.

Of course, in the previous discussion it was assumed that the
semiclassical derivation of the particle emission is valid during
the whole process of black hole evaporation.
Nevertheless, when the mass of the black hole becomes close to the Planck mass~$M_{\rm Pl}\sim
10^{19}\,\mbox{GeV}$
the semiclassical description breaks down. The prediction of what
happens after that point is impossible without a reliable theory of
quantum gravity. One can imagine at least four possibilities
\cite{Hawking2}:
\begin{itemize}
\item[-]
The evaporations produces a naked singularity of negative mass,
which persists.

\item[-]
The evaporation slows down and stops, leaving a remnant black hole
with a mass of the order of the Planck mass.

\item[-]The black hole evaporates completely, but all the information
about the initial state is encoded in the radiation escaping to
infinity.

\item[-]The black hole evaporates completely, taking with it all the
information about its initial state and conserved quantities, except
those coupled to long range fields (mass, charge, and angular momentum).
\end{itemize}

There are some arguments \cite{Hawking2,Preskill} in favour of
the fourth posibility. If this were the case, we would be left with the puzzles pointed
out above. Consider a past asymptotic region where
our system is described by some density operator, $\rho_{-}$, defined in the
Fock space of a free field, and a future asymptotic region in
which the system's density
operator is $\rho_{+}$. Hawking \cite{Hawking2} proposed to use a
superscattering operator $\$$ relating the density operators in the
two asymptotic regions
\begin{align}
\rho_{+\;\;B}^{\;\;A}=\$^{A\;\;\;\; D}_{\;\;BC}\rho_{-\;\;D}^{\;\;C}.
\end{align}
The $\$$-operator should satisfy several conditions:
first, it has to map initial density operators of unit trace into final
density operators which are positive semi-definite and of unit trace.
This implies that
\begin{align}
\$^{C\;\;\;\;B}_{\;\;CA}=\delta_{A}^{\;\;B}.
\end{align}
Moreover, in order $\$$ to map Hermitian operators into Hermitian operators, we require
\begin{align}
\$^{A\;\;\;\;D}_{\;\;BC}=\$^{B\;\;\;\;D}_{\;\;AC}.
\end{align}

In ordinary (Minkowski) QFT we have the $S$-operator mapping states
in the asymptotic past into states in the asymptotic future. This would be 
a particular case where the $\$$-operator admits a factorization as the tensor 
product $S\otimes S^{\dagger}$
\begin{align}
\$^{A\;\;\;\;D}_{\;\;BC}=S^{A}_{\;\;C}\overline{S}_{B}^{\;\;D}.
\label{factorization}
\end{align}
With this form for the superscattering operator, a system in a
pure state remains in a pure state, as required by the
laws
of quantum mechanics. This is however not the case for a general
$\$$. In fact, it is possible to show that for the
$\$$-operator to have the form (\ref{factorization})
the Green functions of the theory must satisfy
asymptotic completeness,
i.e. the asymptotic {\it in} and {\it out} states span the
Hilbert space of the theory. Hawking has argued that the axiom of
asymptotic completeness is not verified in quantum gravity. This
has the consequence that quantities not coupled to long range
fields are not conserved (e.g., global $\mbox{U}(1)$ charges
\cite{Hawking2}). A further analysis and criticism of these arguments
can be found in ref.~\cite{Banks-Peskin-Susskind}.

There has been some recent developments in the study of black-hole
evaporation. On the one hand, the classical no-hair theorems have been
reanalyzed \cite{Kraus-Wilczek} and found to
be substantially modified by quantum fields. Furthermore, some simple
renormalizable theories of two-dimensional gravity coupled to dilatons,
and containing many of the interesting features of four-dimensional
black holes, have recently been proposed \cite{Callan-Giddings}. In these
models one can study the collapse of matter to form a black hole and its
subsequent evaporation. Unfortunately, the final stages of this process
are not yet accesible to the approximations used in
ref.~\cite{Callan-Giddings}.

\subsection{Euclidean approach to quantum gravity}

In ordinary QFT we are faced with the computation of path integrals of
the form \begin{align}
Z=\int {\cal D}\phi \,e^{iS[\phi]},
\label{field-fpi}
\end{align}
with~$S[\phi]$ the classical action. However, the oscillatory nature of
the exponential inside the integral makes
the expression ill-defined. The problem can be
ameliorated by performing a Wick rotation to Euclidean
space: an analytic continuation of real
Minkowski time to pure imaginary values,~$t\rightarrow -it$, such that
eq.~(\ref{field-fpi}) is rewritten as
\begin{align}
Z_{E}=\int {\cal D}\phi \,e^{-S_{E}[\phi]},
\label{eucl-field}
\end{align}
where $S_{E}[\phi]$ is the Euclidean action. Since in most instances
this Euclidean action is positive definite,
eq.~(\ref{eucl-field}) is better defined since the contribution of field
configurations with large actions are damped by the
exponential factor. Once the path integral is evaluated in Euclidean
space we return to Minkowski signature by undoing the analytical continuation.

In Minkowski space QFT this procedure is justified because the 
Wightman axioms~\cite{Glimm-Jaffe}
guarantee that the analytic continuation can be performed.
In fact, switching to Euclidean space has also many other applications in QFT.
For example, when studying quantum systems
at finite temperature we need to calculate the canonical partition
function
\begin{align}
Z(\beta)={\rm tr\,}e^{-\beta H},
\end{align}
with $\beta=1/T$ and $H$ the Hamiltonian of the system. This trace can
be represented as a path integral~\cite{Ramond} in Euclidean space
where time is compactified to a circle whose length equals the inverse
temperature. The thermal partition function is then given by
\begin{align}
Z(\beta)=\int_{\mathbb{R}^{3}\times S^{1}_{\beta}} {\cal D}\phi
\,e^{-S_{E}[\phi]}.
\end{align}
In addition, for bosonic fields we have to asume periodic boundary conditions
in the compactified dimension, while fermions
are antiperiodic.
Euclidean field theory is also the arena where
Yang-Mills instantons are formulated \cite{Eguchi-Gilkey-Hanson}.

Let us come next to gravity. There we may expect to do something similar with the
general theory of relativity, namely, to perform an analytical continuation
from Minkowski to Euclidean signature, so that a path
integral formulation of quantum gravity would render better results than
the methods described in previous sections. Once
physical observables are computed in Euclidean space, we would return to the physical signature.

This program is however not so easy carry out because of the difficulties emerging 
from the very beginning. For starters, when working with curved spaces, even
globally hyperbolic ones, there are no axioms ensuring that the analytic
continuation in the time coordinate can be done. Moreover, when performing the analytic
continuation we might find that
singularities disappear. This is the case, e.g., of the
Schwarzschild metric \begin{align}
ds^{2}=\left(1-\frac{2m}{r}\right)dt^{2}-
\left[\frac{dr^{2}}{1-\frac{2m}{r}}+
r^{2}\big(d\theta^{2}+\sin{\theta}d\phi^{2}\big)\right].
\end{align}
When analytically continuing time into Euclidean space, we miss
the region inside the horizon~$r<2m$, and are left only with the
exterior part~$r>2m$ and the horizon in which the Killing vector
$\partial_{t}$ vanishes.

Even if we would decide not to worry about this problem and continue ahead, we 
will soon find more trouble. Unlike Yang-Mills theories,
the gravitational action is unbounded from below. This is because the
integrand is linear in the scalar curvature and the action can be made
arbitrarily negative. Let us consider the action for the
Euclidean gravitational
field coupled to matter in a four-dimensional Riemannian manifold $M$ with metric~$g_{ab}$
and boundary $\partial M$ (see \cite{Hawking-cent} and references
therein)\footnote{The Euclidean metric has signature~$(+,+,+,+)$
and~$\kappa=8\pi G$.}
\begin{align}
S[g]&=-\frac{1}{16\pi G}\int_{M}d^{4}x\sqrt{g}\big(R-2\Lambda\big)-
\frac{1}{8\pi G}\int_{\partial M}d^{3}x\sqrt{h}\big(K-K^{0}\big) 
-\int_{M}d^{4}x\sqrt{g}\,{\cal L}_{\rm matt},
\label{gravity-action}
\end{align}
where $K$ is the trace of the second fundamental form on $\partial
M$ (i.e., the extrinsic curvature) and~$h_{ab}$ is the metric on the
three-dimensional boundary. The third term represents the
action of the matter fields. Let us consider now the metric
\begin{align}
\widetilde{g}_{ab}=\Omega^{2}g_{ab},
\end{align}
with $\Omega(x)$ a real function. The scalar curvature for this new metric is written in 
terms of the one for~$g_{ab}$ as
\begin{align}
\widetilde{R}= \Omega^{-2}R-6\Omega^{-3}\Box \Omega,
\end{align}
while the extrinsic curvature is
\begin{align}
\widetilde{K}=\Omega^{-1}K+3\Omega^{-2}n^{a}\nabla_{a}\Omega,
\end{align}
with $n^{a}$ the unitary normal vector on the boundary $\partial M$. Using these
expressions we rewrite the gravitational part of the action~(\ref{gravity-action}) 
evaluated on the rescaled metric~$\widetilde{g}_{ab}$ as
\begin{align}
S[\widetilde{g}]&=-\frac{1}{16\pi
G}\int_{M}d^{4}x\sqrt{g}\Big(\Omega^{2}R+6g^{ab}\nabla_{a}\Omega
\nabla_{b}\Omega-2\Lambda\Omega^{4}\Big) \nonumber\\[0.2cm]
&-\frac{1}{8\pi G}\int_{\partial M} d^{3}x\sqrt{h}\Omega^{2}\big(K-K^{0}\big).
\end{align}
Now, by taking a rapidly varying~$\Omega(x)$ we can make the action arbitrarily negative. Since in
principle we should integrate over all possible conformal factors,
this implies that the path integral over 
$\Omega(x)$ diverges.
Gibbons, Hawking, and Perry \cite{Gibbons-Hawking-Perry} gave a
prescription for the evaluation of the functional integral over the
conformal factor: separate all possible four-dimensional metrics into
conformal
classes and pick within each class a metric with vanishing
scalar curature. For a given~$g_{ab}$ this could be accomplishe
by solving the equation
\begin{align}
\left(\Box-\frac{1}{6}R\right)\Omega(x)=0,
\end{align}
so $\widetilde{g}_{ab}=\Omega^{2}g_{ab}$ has zero scalar curvature.
Once this is done, we integrate over the conformal factor and over
all conformal classes, choosing a particular analytic continuation of the
integral over~$\Omega$ to render it well-defined. Since in
this prescription we consider
metrics with zero scalar curvature, the gravitational action is enterely
governed by the surface term, and therefore by the boundary conditions on
$\partial M$. Furthermore, 
physical gravitational fields vanish asymptotically at
infinity, so in realistic cases we deal with metrics that are
asymptotically flat. This fact led Gibbons, Hawking, and Perry to
propose the positive action conjecture \cite{Gibbons-Hawking-Perry},
according to which the gravitational action is non-negative for any
asymptotically Euclidean and positive-definite metric with~$R=0$. This
conjecture was finally proved in $1979$ by Schoen and Yau
\cite{Schoen-Yau}.

Let us assume that we have successfully dealt with
the unboundness of the gravitational action. We can then
think about applying Polyakov's prescription, which provided such good
results in two-dimensional gravity \cite{Polyakov}. It states that in
calculating the partition function (or any correlation function) we should sum over
all topologies, which in the case of two-dimensional closed orientable surfaces
are classified symply by their genus
\begin{align}
Z_{d=2}=\sum_{g=0}^{\infty} \int_{{\cal I}_{g}} {\cal D}g
\,e^{-S_{\rm E}[g]},
\end{align}
where the path integral is performed at fixed topology. 
Applying a similar expansion to four-dimensional gravity, we would write
\begin{align}
Z_{d=4}=\sum_{\rm topologies} \int_{\rm fix.\, top.} {\cal D}g \,e^{-S_{\rm E}[g]}.
\end{align}
However, carrying out this program in four dimensions meets with a basic
difficulty: there exists no algorithm to decide whether two
four-dimensional manifolds are homeomorphic. We will briefly review
the set of theorems leading to this result (see ref.~\cite{EAlvarez1}
and references therein for full details).

The starting point is the fact (known as Markov's theorem) that any
finitely generated group $G$ can be the fundamental group of a
four-dimensional, smooth, compact, and connected manifold. Saying that the
group
\begin{align}
G=\langle a_{1},\ldots,a_{n} |R_{1},\ldots,R_{k}\rangle,
\end{align}
is finitely
generated simply means that it is generated
by a finite number of elements $\{a_{i}\}$ subject to a finite
number of relations $\{R_{j}\}$. The second important ingredient we need
is a result first posed by Dehn and
proved by Novikov, known in group theory as the word problem:
given a family of finitely generated groups $\{G_{k}\}$, with
$k=1,2,\ldots$, there is no algorithmic way of distinguishing
any trivial group in this family.

It is possible to show now the impossibility of classifying topologies in
four dimensions. Let us construct for the family $G_{k}$ a set of
manifolds $M_{k}$ such that $\pi_{1}(M_{k})=G_{k}$ 
(possible by Markov's theorem).
Were we able
to say whether a given manifold is homeomophic to one of the manifolds in
the family, we could also decide that
their fundamental groups are equal. Thus, by checking which elements
of $\{M_{k}\}$ are homeomorphic to a given manifold $M_{0}$
with trivial fundamental group, we would be able to
say which elements of $\{G_{k}\}$ are trivial. But we saw that this cannot be done.

Although the previous result implies that it is impossible to classify
four-dimensional topologies in general, we can still try to classify
topologies within some subclass of four-dimensional manifolds. For example,
we can restricts ourselves to the sector with trivial fundamental
group,~$\pi_{1}(M)=0$. Given a manifold~$M$, let us begin by defining the Kirby-Siebenmann 
invariant~$\alpha(M)\in \mathbb{Z}_{2}$ such that $\alpha(M)=0$ if the manifold
$M\times S^{1}$ admits a smooth structure and $\alpha(M)=1$ if it does
not. A theorem by M. Freedman states that if $M$ is a compact, connected
manifold  with trivial fundamental group, then it is classified by it 
second homology group,~$H^{2}(M,\mathbb{Z})$, and the Kirby-Siebenmann
invariant, $\alpha(M)$.

Given a four-dimensional manifold, we can define the intersection form
$\Omega$ as the symmetric bilinear form $\Omega:H^{2}(M,\mathbb{Z})\times H^{2}(M,\mathbb{Z})
\rightarrow \mathbb{Z}$
\begin{align}
\Omega(\beta_{1},\beta_{2})=\int_{M}\beta_{1}\wedge \beta_{2},
\end{align}
for any~$\beta_{1},\beta_{2}\in H^{2}(M,\mathbb{Z})$.
This intersection form can be used to classify simply connected, i.e.~$\pi_{1}(M)=0$, compact
four-dimensional manifolds into homeomorphism classes. Two of such
manifolds are homotopy equivalent if and only if their intersection forms are
isomorphic (Whitehead's theorem). Introducing the notion of signature
$\sigma(\Omega)$ for the intersection form $\Omega$ as the number of
its positive eigenvalues
minus the negative ones, it can be proved that, whenever~$\Omega$ 
is even, its signature is automatically divisible by eight. The concept of
signature is important since there is a one-to-one correspondence
between the set of simply-connected compact, four-dimensional manifolds
and the pairs $\langle \Omega,\alpha \rangle$, where $\alpha$ is the
Kirby-Siebenmann invariant defined earlier 
and~$\sigma(\Omega)/8\equiv\alpha\,\,(\mbox{mod}\,2)$  if $\Omega$ is even.
Thus, the pairs $\langle \Omega,\alpha
\rangle$ constitute the space of parameters for four-dimensional
compact topological manifolds in the sector with~$\pi_{1}(M)=0$.

GR is invariant under diffeomorphisms. Thus,
when performing the path integral at fixed
topology, we should integrate over all diffeomorphism classes without
counting each class more than once. This means that, 
in order to carry out the path integration, we have to classify
four-dimensional manifolds into diffeomorphism classes. This can be done by invoking a theorem by
Donaldson: given a closed four-dimensional, smooth, and
simply connected
manifold with an intersection form $\Omega$ which is positive definite, then
$\Omega$ is diagonalizable over the integers,~$\Omega=\mbox{diag\,}(1,\ldots,1)$. 
In other words,
a manifold with intersection form
$\Omega$ does not exist smoothly for any non-zero, positive definite
$\Omega$. This implies the existence of different classes of
differentiable structures classified in part by Donaldson's
invariants. There are, e.g., exotic
differentiable
structures in $\mathbb{R}^{4}$, the so called $\mathbb{R}_{\rm fake}^{4}$.

In view of the problems encountered, we should
conclude that carrying out the Polyakov approach in 
four-dimensional gravity to the end seems to be impossible.
Nevertheless, one could
impose some kind of  restriction in the set of manifolds we integrate
over based in some yet unknown physical principle. For example,
to integrate over simply connected 
manifolds~[$\pi_{1}(M)=0$]. At any rate, it should be said that the arguments 
presented above are maybe too naive. They are
based
on a semiclassical analysis of the problem of quantizing gravity. In the
functional integral we will certainly have contributions coming from
singular metrics. Once the conditions that the manifolds and
geometries should be continuous and smooth are relaxed, the objections presented in this section
lose much of their strength. The physical conditions implied by considering more general 
manifolds are unknown.

\subsection{Canonical quantization of gravity}

Instead on considering general manifolds, in this approach
we restrict our attention to space-times with
topology $\Sigma_{3}\times\mathbb{R}$, where $\Sigma_{3}$ is a compact
three-manifold interpreted as a
constant-time hypersurface ``folliating'' the whole space-time.
Assuming the existence of a global time
coordinate allows the definition of a Hamiltonian and a canonical
quantization of the space-time\footnote{A thorough presentation of
canonical gravity is found in the lectures by A. Ashtekar at this school~\cite{Ashtekar_lectures}.}. 
The corresponding Fock space would not
be
equivalent to that obtained from path-integral quantization, because in
the latter case we integrate over all four-dimensional manifolds, including those
whose topology is not $\Sigma_{3}\times\mathbb{R}$. 

It is clear that 
in carrying out a canonical quantization of gravity, we are interested in the classification of
three-dimensional manifolds. The determination of the classical
phase space is ``simpler'' now, as long as one believes Thurston's
geometrization program \cite{Thurston}. In fact, every
three-dimensional manifold can be included in an exhaustive list with a
countable infinite number of parameters. We cannot be completely happy, however, 
because of a potential problem with infinite overcounting.

There are several steps in the classification of three-dimensional
manifolds. First we can consider genus $g$ Heegard splitting
(for an application of this technique, see ref.~\cite{Dijkgraaf-Witten})
which consist in decomposing a three-manifold into two ``handle
bodies'' of genus $g$ by cutting along a Riemman surface $\Sigma_{g}$.
After that, the boundaries of the two pieces are identified by a
diffeomorphism not connected with the identity.
A problem,
nevertheless, arises in considering the fundamental group
$\pi_{1}(M)$
of a three-dimensional manifold, because although finitely
generated there is no algorithmic way of knowing whether a given group
is the fundamental group of a three-manifold.

We would like to find a set of topological invariants
that could be used to classify three-manifolds along the same lines as
two-dimensional
compact orientable manifolds are classified by the Euler invariant (i.e. the genus).
The search of these invariants for the
three-dimensional manifolds is a matter of current investigation
(Jones-Witten
invariants, Casson invariants,...). A particularly interesting attempt of classification is the 
one carried out by Thurston~\cite{Thurston} based on the
geometrization conjecture: every compact, orientable three-manifold can
be cut by disjoint embedded two-spheres and tori into pieces which,
after gluing three-balls to all boundary spheres, admit geometric
structures. That the manifold admits a geometric structure simply means
that for every pair of points $x,y \in M$ there are two isometric
neighborhoods~$U_{x},U_{y} \subset M$.

One possible way to dispose of these problems is by considering that
at short distances whaat we have is a topological field theory~\cite{Witten-Topological}. 
In this case, below some critical length, the
theory is in a symmetrical phase depending only on some well
defined topological invariants, so the phase space is perfectly well-characterized. 
In the-low energy phase the invariance under
diffeomorphisms is spontaneously broken by the vacuum, which is only
Lorentz invariant, and the
graviton is the corresponding Goldstone boson. This idea is an old one
and can be traced back to the
early seventies \cite{Isham-Salam-Strathdee}.

The group of $C^{\infty}$-diffeomorphisms of a smooth compact manifold
$M$, $\mbox{Diff}(M)$, has some peculiar properties \cite{Milnor}. It is
a Lie
group whose Lie algebra is given by the set $\mbox{Vect}(M)$ of smooth
vector fields on $M$ closed under Lie brackets.
This group is very different from ordinary Lie groups, since in general it is not
possible to get any element in
the connected component of $\mbox{Diff}(M)$ as the result of the
exponential
map acting on an element of $\mbox{Vect}(M)$. However, there always exists a set
$v_{1},\ldots ,v_{k} \in \mbox{Vect}(M)$ such that any element $f\in
\mbox{Diff}(M)$ can be written as
\begin{align}
f=\exp(v_{1})\circ \exp(v_{2}) \circ \ldots \circ \exp(v_{k}).
\end{align}

For our purposes, we are interested in studying the topological structure of the set of
all metrics modulo the group of diffeomorphism. The topology of this
space is intimately connected to the topology of $\mbox{Diff}(M)$. For
instance
\begin{align}
\pi_{1}[\mbox{Metrics}(M)/\mbox{Diff}(M)] \cong \pi_{0}[\mbox{Diff}(M)],
\end{align}
that is to say, the simply connectness of
$\mbox{Metrics}(M)/\mbox{Diff}(M)$ is determined
by the connectness of $\mbox{Diff}(M)$, and similarly for higher
homotopy groups. Considering the group $\mbox{Diff}(S^{n})$, Smale proved
that \cite{Smale}
\begin{align}
\pi_{k}[\mbox{Diff}(S^{n})] \cong \pi_{k}[\mbox{O}(n+1)],
\end{align}
for $n=1,2$ and all $k$. Smale himself conjectured that this
relation also holds for
$n=3$, namely
\begin{align}
\pi_{k}[\mbox{Diff}(S^{3})] \cong \pi_{k}[\mbox{O}(4)].
\end{align}
The conjecture was finally proved by Hatcher \cite{Hatcher}.

As a final speculation, one can entertain the idea that
in the right theory maybe one is allowed to make deformations wilder
than homeomorphisms, so that there is some kind of quantum equivalence
of otherwise distinct classical geometrical structures. This happens, for
example, in string theory in the context of $R$-duality~\cite{duality} 
and mirror symmetry~\cite{mirror}. For example a string
moving on a circle of radius $R$ is equivalent to a string on a circle
of radius~$1/R$ (see sec.~\ref{sec:R-duality}). One also finds that
topologically very different Calabi-Yau manifolds (Ricci-flat, compact
K\"{a}hler manifolds) lead to the same physical theory (mirror symmetry).
This is properly known to hold only for string theory. This short
section has hopefully convinced the reader that canonical quantum gravity 
meets with
rather severe mathematical difficulties if we insist on having smooth
manifolds and geometries. Very important progress towards a proper
understanding of canonical quantum gravity has been achieved however with the
work initiated by Ashtekar and coworkers \cite{Ashtekar_lectures,Ashtekar}.

\subsection{Gravitational instantons}

In Yang-Mills theory, instantons appear as classical solutions to the
Euclidean field equations. From the
point of view of Minkowski space these solutions are interpreted as
tunneling between vacuum states that are not topologically equivalent.
If Euclidean gravity makes any sense we could expect to find
gravitational instantons representing tunneling between
inequivalent vacua. Some properties of these gravitational instantons
are \cite{Eguchi-Hanson}:
\begin{itemize}
\item[-]
They describe gravitational fields which are localized in
Euclidean space-time.
\item[-]
These solutions approach an asymptotically locally Euclidean
vacuum
metric at infinity. This means that although the
metric at infinity is locally flat, the space globally is not
topologically equivalent
to flat space-time. This is equivalent to self-dual Yang-Mills
self-dual solutions being pure-gauge at infinity.
\item[-]
They have non-trivial topological quantum numbers.
\end{itemize}
All these properties are very similar to those of 
Yang-Mills instantons~\cite{Eguchi-Gilkey-Hanson}.
The basic reason to expect self-dual solutions in
four-dimensional Euclidean space is that the holonomy group
is~$\mbox{SU}(2)\times \mbox{SU}(2)$, the self-duality condition being 
that the curvature takes values only on
one~$\mbox{SU}(2)$ factor. Searching for
regular manifolds
with instanton-like solutions, we find that the surface $K3$ is the only
compact, regular, and simply connected manifold without boundary 
admitting a non-trivial self-dual curvature. Allowing for solutions with
singular points increases the number of candidates.

Let us illustrate our discussion of gravitational instantons by considering
some particular examples~\cite{Eguchi-Hanson}. We begin with the
following metric ansatz
\begin{align}
ds^{2}=U({\bf x})^{-1}(d\tau^{2}+ \boldsymbol{\omega}\cdot d{\bf x})
+U({\bf x})d{\bf x}\cdot d{\bf x},
\label{instanton}
\end{align}
which is self-dual as long as
\begin{align}
\nabla \times \boldsymbol{\omega}=\nabla U({\bf x}).
\end{align}
The general solution to the vacuum~$R_{\mu\nu}=0$ can be written as
\begin{align}
U({\bf x})=\epsilon+\sum_{i=1}^{k}\frac{2M_{i}}{|{\bf x}-{\bf x}_{i}|},
\label{eq:U_grav_inst}
\end{align}
with $\epsilon$ an integration constant. In order to make
(\ref{instanton}) well behaved and free of singularities, we have to
choose $M_{i}=M$ and to make $\tau$ periodic with period $8\pi M/k$.

We take $\epsilon=1$, so
the action is $S=4\pi kM^{2}$. If~$k=1$ we have the
Euclidean Schwarzschild solution, the only 
gravitational instanton which has found applications so far. It appears in the
nucleation rate of black holes on a thermal gas of gravitons at a given
temperature~\cite{Gross-Perry-Yaffe}.
For~$k>1$, on the other hand, we have 
multi-Taub-NUT metrics, which are
asymptotically locally flat at spatial infinity~$|{\bf x}|\rightarrow
\infty$. In this region these instantons look like
$S^{2}\times(S^{1}/\mathbb{Z}_{k})$, where~$S^{2}$ has a radius growing 
as $|x|\rightarrow\infty$ like $|x|$ and~$S^{1}/\mathbb{Z}_{k}$ approaches a
fixed radius. The fiber identification implied by the
$\mathbb{Z}_{k}$-modding makes the physical interpretation of these
solutions very difficult.

Taking~$\epsilon=0$ in eq.~\eqref{eq:U_grav_inst} we have that the action is zero
for all values of $k$. For $k=1$ we recover flat space-time
while for $k=2$ we have the Eguchi-Hanson metric, in both cases
modulo coordinate transformations. Asymptotically, these instantons look
like $S^{3}/\mathbb{Z}_{k}$. Again their physical interpretation is
unclear.

There is a very important difference between Yang-Mills and
gravitational instantons. In the Yang-Mills case, instanton-mediated
amplitudes are suppressed by factors of the
order~$\exp{(-4\pi^{2}/g^{2})}$, the exponential of minus the instanton action. For
asymptotically locally Euclidean instantons (like
multi-Eguchi-Hanson instantons), however, the action vanishes so
the amplitudes are not suppressed exponentially, but only 
by power-like terms arising from the quantization of zero modes and moduli.
Unfortunately, no physical process has been described so far requiring
the use of these instantons.

With this we conclude our quick overview of some of the strategies to
quantize gravity within the framework of QFT.

\section{Consistency conditions: anomalies}
\label{sec:anomalies}

\subsection{Generalities about anomalies}
\label{eq:anomalies_generalities}

It is well known result in classical field theory that the
presence of a continuous symmetry in the action leads to the existence
of
a conserved current associated with it (Noether's theorem). However, under certain conditions
classical symmetries may not be conserved after quantization.
Conservation laws may get modified by terms proportional to the Planck constant
\begin{align}
\nabla_{\mu}\langle j^{\mu}(x) \rangle = O(\hbar).
\end{align}
There is simple rationale for this. Currents $j^{\mu}(x)$ are composite
operators and therefore naively ill-defined. They need to be regularized
and trouble might arise when
regularization becomes incompatible with some classical symmetry. 
If this is the case, the quantum theory does not
necessarily preserve the classical symmetry after renormalization.

By definition an anomaly is the breakdown of a classical symmetry by
quantum corrections. In QFT there are well-known examples of anomalies. For
example, massless $\phi^{4}$ scalar field theory is invariant under
scale transformations at the classical level. The process of
regularization and renormalization requires however
the introduction an energy scale $\mu$. As a result, the
renormalized coupling constant $\lambda_{R}$ becomes a function of the
energy at which experiments are carried out and the field itself acquires an anomalous scaling dimension.
In short, classical scale invariance has been broken by quantum corrections.

Perhaps the most celebrated example of a QFT anomaly is the
Adler-Bell-Jackiw (ABJ) anomaly of QED, which is behind 
the decay of the neutral pion~$\pi^{0}\rightarrow 2\gamma$~\cite{ABJ}. It originates 
in a triangle diagram with
two vector currents and one axial vector current
\begin{align*}
\centerline{\includegraphics[scale=0.30]{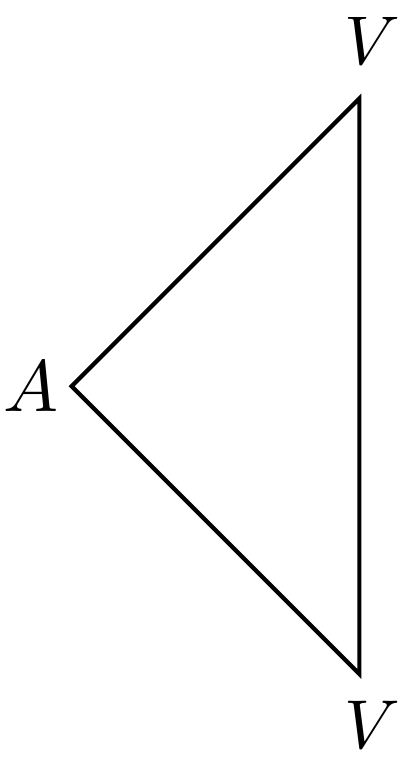}}
\end{align*}
Imposing
Bose symmetry and conservation of the vector currents results in 
the axial vector current $A_{\mu}(x)=\overline{\psi}(x)\gamma_{5}
\gamma_{\mu} \psi(x)$ being not conserved in massless QED.
Instead, we find the anomalous Ward identity
\begin{align}
\partial_{\mu} A^{\mu}(x)=-\frac{\alpha\hbar}{16\pi^{2}} \epsilon_{\mu
\nu \sigma \lambda} F^{\mu \nu}(x)F^{\sigma \lambda}(x),
\label{anomaly1}
\end{align}
where~$\alpha$ is the fine-structure constant,~$F^{\mu \nu}(x)$ 
the electromagnetic field strength, and~$\epsilon_{\mu \nu \sigma \lambda}$ the completely antisymmetric
Levi-Civita tensor. This anomaly is not dangerous at all since it affects
a global current. As mentioned above, it is in fact the reason why the 
neutral pion decays into two photons~\cite{Pi-0}.

Much more worrying is the existence of anomalies in local gauge
symmetries. The reason is that very important issues such as
renormalizability or unitarity are based upon the preservation of the
gauge symmetry at the quantum level. Thus, gauge
anomalies
jeopardize the consistency of the theory. The best example of the type of danger
posed by gauge anomalies is found 
in the SM based in the gauge group~$\mbox{SU}(3)\times \mbox{SU}(2)\times
\mbox{U}(1)$. As we will see below, anomalous
contribution can arise from a triangle diagram with three $V-A$
currents in the vertices coupled to gauge fields \cite{V-A}:
\begin{align*}
\centerline{\includegraphics[scale=0.30]{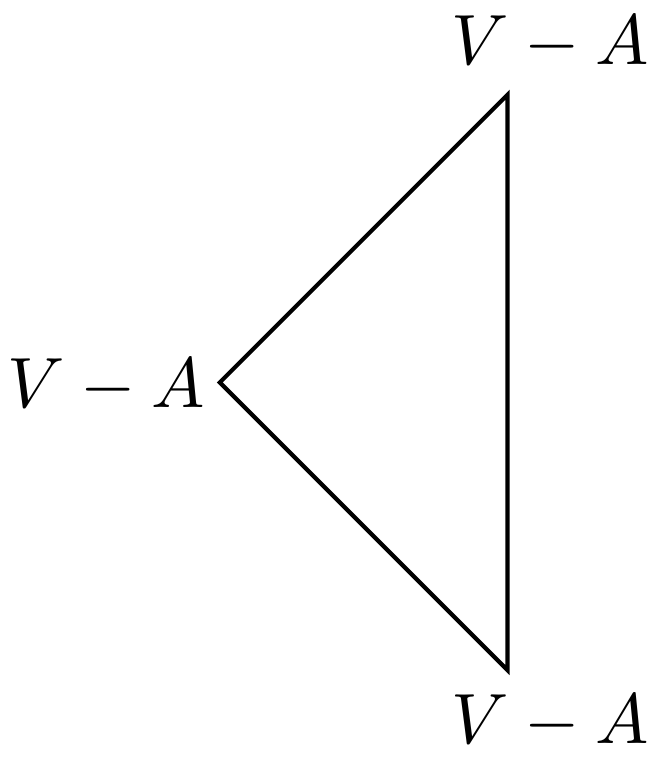}}
\end{align*}
This contributes to the Ward identity for the~$V-A$ current, so if
after summing over all fermion species running in the fermionic loop we get a
non-zero contribution gauge invariance will be anomalous. Applying
Feynman rules to calculate a diagram with one fermion loop 
and~$N$ vertices coupled to external gauge fields, we find that the result
is proportional to an overall group
theoretical factor of the form
\begin{align}
\sum_{T_{L}}{\rm str}(T_{L}^{a_{1}}\ldots T_{L}^{a_{N}})-
\sum_{T_{R}}{\rm str}(T_{R}^{a_{1}}\ldots T_{R}^{a_{N}}).
\label{eq:group_theory_factorStr}
\end{align}
Here~$T_{L}^{a}$ and $T_{R}^{a}$ are respectively the generators of the
representation carried by the left- and right-handed fermions,
and ${\rm str}$ is the symmetrized trace (symmetrization is
needed because of Bose symmetry). For non-chiral theories in which left
and right-handed fermions transform under the same representation of the
gauge group this factor is always zero, and consequently there are no
anomalies. The problem arises when the theory is chiral (as the real world
is) because in this case the contribution of the diagram in non-zero, 
unless we have a cancellation when summing over all fermion species in the loop.

Let us consider the SM. The quantum numbers of a single
family with respect to the~$\mbox{SU}(3)\times \mbox{SU}(2)\times \mbox{U}(1)$ gauge group are
\begin{align}
(3,2)_{1/6} \oplus (\overline{3},1)_{-2/3} \oplus (\overline{3},1)_{1/3} \oplus
(1,2)_{-1/2} \oplus (1,1)_{1},
\label{SM}
\end{align}
where the first entry in the parenthesis indicates the representation 
of~$\mbox{SU}(3)$, the second one that of~$\mbox{SU}(2)$ and the subindex the weak
hypercharge $Y$. It can be seen that all gauge anomalies cancel for these representations. 
In the case of three hypercharge generators, we see
that the contribution
of the leptons to the group theory factor~\eqref{eq:group_theory_factorStr} 
is exactly canceled by that of the quarks within a single family
(notice that only left-handed fermions carry non-zero hypercharge)
\begin{align}
\sum_{\rm leptons}Y^{3}_{L}&=2\times
\left(-\frac{1}{2}\right)^{3}+1=\frac{3}{4}, \nonumber \\[0.2cm]
\sum_{\rm quarks}Y^{3}_{L}&=3\times\left[2\times\left(\frac{1}{6}\right)^{3}
+\left(-\frac{2}{3}\right)^{3}+\frac{1}{3}\right]=-\frac{3}{4}.
\end{align}
It can be easily checked that the group theory factors for
all other combinations of the generators of $\mbox{SU}(3)\times \mbox{SU}(2)\times
\mbox{U}(1)$ also vanish, so the~SM is
anomaly free.

On general grounds, we only expect gauge anomalies in chiral theories or,
equivalently, in parity-violating theories. This is because in a non-chiral 
theory it is always possible to construct a fermionic mass term in the
Lagrangian that is gauge invariant. This
allows the use of Pauli-Villars regularization, preserving
gauge invariance. The resulting quantum theory is therefore free from gauge anomalies.

\subsection{Spinors in $2n$ dimensions}
\label{sec:spinors}

Since the existence of anomalies is intimately related to the presence of
chiral fermions it is worthwhile to review briefly the properties of spinors in 
$2n$~dimensions before plunging into the study of
anomalies\footnote{Chiral theories exist only in even-dimensional
space-times.}\cite{LAG1}.
The starting point is the $2n$-dimensional Clifford algebra
\begin{align}
\{\Gamma^{a},\Gamma^{b}\}=2\eta^{ab},
\label{Clifford}
\end{align}
with $a,b=0,1,\ldots,2n-1$ and $\eta_{ab}$ the flat metric with
signature $(s,t)$
\begin{align}
\eta_{ab}={\rm diag}(\,\overbrace{1,\ldots,1}^{t},\overbrace{-1,\ldots,-1}^{s}\,).
\end{align}
A Dirac spinor field is an object transforming under an infinitesimal~$\mbox{SO}(t,s)$ 
transformation according to
\begin{align}
\delta \psi=-\frac{1}{2}\epsilon_{mn}\Sigma^{mn} \psi,
\label{eq:fermion_trans_general}
\end{align}
with $\epsilon_{mn}=-\epsilon_{nm}$ the parameters of the transformation and
\begin{align}
\Sigma^{mn}=-\frac{1}{4} [\Gamma^{m},\Gamma^{n}].
\label{spinorial}
\end{align}
The Dirac matrices~$\Gamma^{m}$ are chosen to satisfy the Hermiticity properties
\begin{align}
(\Gamma^{1})^{\dagger}&=\Gamma^{1},\ldots,
(\Gamma^{t})^{\dagger}=\Gamma^{t}, \nonumber \\[0.2cm]
(\Gamma^{t+1})^{\dagger}&=-\Gamma^{t+1},\ldots,(\Gamma^{t+s})^{\dagger}=
-\Gamma^{t+s}.
\end{align}
Using eq.~(\ref{Clifford}) it can be checked that $\Sigma^{mn}$ indeed
satisfies the commutation relations of the Lie algebra of~$\mbox{SO}(t,s)$.

Since we are working in even dimension, we can define an element
$\overline{\Gamma}$ of the Clifford algebra analog to the
$4$-dimensional $\gamma_{5}$ chirality matrix
\begin{align}
\overline{\Gamma}=\alpha\Gamma^{0}\Gamma^{1}\ldots\Gamma^{2n-1}.
\label{eq:Gammm5gen}
\end{align}
It anticommutes with all Dirac matrices
\begin{align}
\{\overline{\Gamma},\Gamma^{m}\}=0, \hspace{2cm} m=0,\ldots,2n-1,
\label{gamma51}
\end{align}
and satisfies
\begin{align}
\overline{\Gamma}^{2}=1, \hspace{1cm} \overline{\Gamma}^{+}=\overline{\Gamma}.
\label{gamma52}
\end{align}
These last identities can be translated into conditions over the 
parameter~$\alpha$ introduced in eq.~\eqref{eq:Gammm5gen}
\begin{align}
\alpha^{2}=(-1)^{\frac{s-t}{2}}, \hspace{1cm}
\alpha^{*}=(-1)^{\frac{s-t}{2}}\alpha.
\end{align}
{}From (\ref{gamma51}) we also deduce that $[\overline{\Gamma},\Sigma^{ab}]=0$.
This, together with eq.~(\ref{gamma52}), suffices to construct the chiral
projectors
\begin{align}
P_{\pm}=\frac{1}{2}(1\pm \overline{\Gamma}).
\end{align}
They split the spinor~$\psi$, whose transformation is given in eq.~\eqref{eq:fermion_trans_general},
into two irreducible representations of $\mbox{SO}(t,s)$
\begin{align}
\psi_{\pm} = P_{\pm}\psi,
\end{align}
where $\psi_{+}$ (resp.~$\psi_{-}$) is a positive (resp. negative) chirality
Weyl spinor.

It is a known fact that the Clifford algebra (\ref{Clifford}) has a
unique faithful representation of dimension $2^{n}$. Thus, the
representations supplied by
\begin{align}
\Gamma^{m},\hspace*{0.5cm} (\Gamma^{m})^{*},\hspace*{0.5cm}-(\Gamma^{m})^{*},
\hspace*{0.5cm}(\Gamma^{m})^{T},\hspace*{0.5cm}
-(\Gamma^{m})^{T},
\end{align}
are all equivalent. In particular, there exists a matrix~$B$ such that
\begin{align}
(\Sigma^{mn})^{*}=B\Sigma^{mn}B^{-1}.
\label{eq:B_oper_cliff_def}
\end{align}
The charge conjugate spinor~$\psi^{c}$ is defined by means of the antilinear
operator $C$
\begin{align}
\psi^{c}=C\psi\equiv B^{-1}\psi^{*}.
\end{align}
It can be easily seen that $\psi$ and $\psi^{c}$ transform in the same way
under $\mbox{SO}(s,t)$ transformations
\begin{align}
\delta \psi^{c}=B^{-1} \delta \psi^{*}=
-\frac{1}{2}\epsilon_{mn}[B^{-1}(\Sigma^{mn})^{*}B]B^{-1}\psi^{*}=
-\frac{1}{2}\epsilon_{mn}\Sigma^{mn}\psi^{c}.
\end{align}
On the other hand, if $\psi$ carries
a gauge index and transforms in the~$R(G)$ representations, its charge conjugate
spinor transforms in the complex conjugate representation
$\overline{R}(G)$.

Notice that whenever the matrix~$B$ defined in eq.~\eqref{eq:B_oper_cliff_def} can be found such that
$C^{2}=1$, it is possible to construct two Majorana projectors
\begin{align}
\frac{1}{2}(1\pm C),
\end{align}
such that
\begin{align}
\psi_{M}=\frac{1}{2}(1+C)\psi,
\end{align}
and
\begin{align}
\psi_{\overline{M}}=\frac{1}{2}(1-C)\psi,
\end{align}
are Majorana and anti-Majorana fields, respectively satisfying~$\psi_{M}^{c}=\psi_{M}$ 
and~$\psi_{\overline{M}}^{c}=-\psi_{\overline{M}}$.

Let us now see under what circunstances the Weyl and Majorana
conditions can be simultaneously imposed. We calculate the
commutator of $\overline{\Gamma}$ and $C$, with the result
\begin{align}
C\overline{\Gamma}=(-1)^{\frac{s-t}{2}}\overline{\Gamma}C.
\label{eq:[C,Gamma]}
\end{align}
This means that if $(s-t)/2$ is odd a Weyl spinor and its charge conjugate
have oposite chiralities, while for $(s-t)/2$ even their chiralities are the same. If we
consider now the case of Minkowski space, $s=2n-1$, $t=1$, it can be 
seen that chirality and helicity are equivalent notions. Indeed, the
massless Dirac equation in momentum space can be written as
\begin{align}
\Gamma^{0}\psi(p)=\Gamma^{2n-1}\psi(p),
\label{Dirac}
\end{align}
where we took coordinate so the particle
moves along the $x^{2n-1}$~direction and used the massless dispersion relation~$p^{0}=p^{2n-1}$. 
By definition, the helicity operator in
$2n$ dimensions is
\begin{align}
h=\Sigma^{12}\Sigma^{34}\ldots\Sigma^{2n-3,2n-2}=
\left(\frac{i}{2}\right)^{n-1}\Gamma^{1}\Gamma^{2}\ldots\Gamma^{2n-2}.
\end{align}
Applying $h$ to $\psi(p)$, we arrive at
\begin{align}
h\psi(p)=
\left(\frac{i}{2}\right)^{n-1}\Gamma^{1}\Gamma^{2}\ldots\Gamma^{2n-2}
\psi(p)=\left(\frac{i}{2}\right)^{n-1}\overline{\Gamma}\psi(p),
\end{align}
after using the Dirac equation (\ref{Dirac}). From eq.~\eqref{eq:[C,Gamma]} we also learn
that in Minkowski space charge conjugation flips helicity if~$d=4k$,
while for $d=4k+2$ it does not.

In order to further investigate the possibility of finding $B$ such that
$C^{2}=1$, let us consider~$B$ as the matrix intertwining~$\Gamma^{m}$ with $-(\Gamma^{m})^{*}$
\begin{align}
-(\Gamma^{m})^{*}=B\Gamma^{m} B^{-1}.
\end{align}
Thus, we write
\begin{align}
\Gamma^{m}&=-B^{*}(\Gamma^{m})^{*}B^{-1} \nonumber \\[0.2cm]
&=(B^{*}B)\Gamma^{m}(B^{*}B)^{-1}.
\end{align}
This implies that $B^{*}B$ commutes with all the  $\Gamma$'s. Since the Dirac matrices
build an irreducible representation of the Clifford algebra
(\ref{Clifford}), Schur's lemma implies
\begin{align}
B^{*}B=\epsilon \mathbb{1},
\label{B-condition}
\end{align}
with $|\epsilon|=1$. In fact, taking the complex conjugate of the last
expression one can see, after computing the trace, that
$\epsilon$ must be real, so $\epsilon =\pm 1$.

On the other hand, the standard charge conjugation matrix
relates~$\Gamma^{m}$ and $-(\Gamma^{m})^{T}$
\begin{align}
(\Gamma^{m})^{T}=-C \Gamma^{m} C^{-1}.
\label{eq:GammaT=-CGammaC-1}
\end{align}
Since in Minkowski space
$(\Gamma^{m})^{\dagger}=\Gamma^{0}\Gamma^{m}\Gamma^{0}$ we can write, after
normalizing properly
\begin{align}
B^{\dagger}B=\mathbb{1}.
\end{align}
Combining this result with eq.~(\ref{B-condition}), we arrive at
\begin{align}
B^{T}=\epsilon B,
\end{align}
and
\begin{align}
C=B\Gamma^{0}.
\end{align}

To find how the value of~$\epsilon $ depends on the dimension, we
work in an explicit basis for~$d=2n$
\begin{align}
\Gamma^{0}&=\sigma_{x}\otimes \mathbb{1}\otimes\ldots \otimes \mathbb{1},
\nonumber \\[0.2cm]
\Gamma^{1}&=i\sigma_{y}\otimes \mathbb{1}\otimes\ldots\otimes \mathbb{1},
\nonumber \\[0.2cm]
\Gamma^{2}&=i\sigma_{3}\otimes \sigma_{x}\otimes\ldots\otimes \mathbb{1},
\nonumber \\[0.2cm]
\Gamma^{3}&=i\sigma_{3}\otimes \sigma_{y}\otimes\ldots\otimes \mathbb{1},
 \\[0.2cm]
&\vdots 
\nonumber \\[0.2cm]
\Gamma^{2n-2}&=
i\overbrace{\sigma_{3}\otimes\ldots\otimes\sigma_{3}}^{n-1}
\otimes\sigma_{x}\otimes \mathbb{1}\otimes\ldots \otimes \mathbb{1},
\nonumber\\[0.2cm]
\Gamma^{2n-1}&=i\sigma_{3}\otimes\ldots\otimes\sigma_{3}\otimes
\sigma_{y}\otimes \mathbb{1}\otimes\ldots\otimes \mathbb{1},
\nonumber
\end{align}
and
\begin{align}
\overline{\Gamma}=\sigma_{3}\otimes\sigma_{3}\otimes\ldots\otimes\sigma_{3}
\end{align}
where $\sigma_{x}$, $\sigma_{y}$, and $\sigma_{3}$ are the
two-dimensional Pauli matrices. The charge
conjugation operator satisfying~\eqref{eq:GammaT=-CGammaC-1} is given by:
\begin{align}
C=\sigma_{y}\otimes\sigma_{x}\otimes\sigma_{y}\otimes\ldots,
\end{align}
where the number of $\sigma_{y}$'s in the tensor product is $k$ for $d=4k$ and $k+1$ for $d=4k+2$.
We find then the following values for $\epsilon$
\begin{align}
\epsilon =+1 \hspace{2cm} d=2,4 \hspace{5mm} \mbox{mod} \hspace{2mm} 8,
\nonumber \\[0.2cm]
\epsilon =-1 \hspace{2cm} d=0,6 \hspace{5mm} \mbox{mod} \hspace{2mm} 8.
\end{align}
The result of our calculation is that the Majorana condition can only be imposed in
$d=2,4 \hspace{2mm} \mbox{mod} \hspace{2mm} 8$, since only in this case
$C^{2}=(BB^{*})^{-1}=1$. Moreover, since we found above 
that~$[\overline{\Gamma},C]=0$ only for $d=4k+2$, Weyl and Majorana conditions
can be simultaneously imposed only when $d=2\,\,\,\mbox{mod}\,\,\,8$. This includes 
in particular~$d=2$ and~$d=10$, two cases that as we will see below are of special interest in 
superstring theory.

\subsection{When can we expect to find anomalies?}
\label{eq:expect}

Up to here we have encountered two kinds of anomalies. The first is called the singlet anomaly
and affects global currents
(this is exemplified by the ABJ anomaly). As we already said, anomalies of
this type not only do not jeopardize the consistency of the theory, but
they might be crucial to explain some physical
processes (for example, the $\pi^{0} \rightarrow 2\gamma$ decay). The second
class
are gauge anomalies, leading to a breakdown of gauge
invariance at the quantum level. Contrary to the first case, these
anomalies are very dangerous as its presence renders the theory
inconsistent.

However, besides these we can also find a third type:
gravitational anomalies~\cite{G-A,LAG-Witten}. They appear in theories
coupled to gravity and lead to a breakdown of diffeomorphism invariance in the quantum theory.
Unless the gravitational anomaly cancels after summing over all chiral species in the
spectrum, the theory cannot be consistently coupled to
gravity.

Gauge and gravitational anomalies can be further classified into
two groups. Local gauge (resp. gravitational) anomalies are those
in which the breakdown of gauge (resp. diffeomorphism) invariance 
is restricted to transformations in the connected component
of the identity. They can therefore be spotted in perturbation theory. This is the
case, for example, of the potential gauge anomaly in the standard
model that we discussed at the beginning of this section.
But we can also have global gauge (resp. gravitational) anomalies. In this case,
although the theory is invariant under infinitesimal gauge
transformations (resp. diffeomorphisms), it may not be so under transformations
that cannot be continuously deformed into the identity. The most celebrated
example of this type is Witten's~$\mbox{SU}(2)$ anomaly (see sec.~\ref{sec:GSandWSU2} below).

We look now at the quantum conservation of the
gauge current from the viewpoint of the quantum effective action
of fermions coupled to a classical external gauge field~\cite{LAG-Witten}. It is defined by
\begin{align}
e^{-\Gamma[A_{\mu})]_{\rm eff}}=\int {\cal D}\psi {\cal D}\overline{\psi}
\exp{\left[-\int d^{d}x \,\overline{\psi}\,i \Dsl
\left(\frac{1-\overline{\Gamma}}{2}\right)\psi\right]},
\label{effective}
\end{align}
where $\Dsl=\Gamma^{\mu}(\partial_{\mu}+iA_{\mu}^{a}\lambda^{a})$ is
the Dirac operator and~$A_{\mu}^{a}$ the non-Abelian external gauge field. 
It is easy to show that under an infinitesimal gauge transformation
\begin{align}
\delta A_{\mu}^{a}=D_{\mu}\epsilon ^{a},
\end{align}
the effective action changes by
\begin{align}
\delta \Gamma_{\rm eff}=tr\int d^{d}x\, (D_{\mu}\epsilon^{a}) \frac{\delta
\Gamma_{\rm eff}}{\delta A_{\mu}^{a}}=-{\rm tr\,}\int d^{d}x\, \epsilon^{a} D_{\mu}
\left(\frac{\delta \Gamma_{\rm eff}}{\delta A_{\mu}^{a}} \right).
\label{eq:var_effective_action1}
\end{align}
On the other hand, taking a functional derivative of eq.~(\ref{effective}) with respect
to the gauge field, we find
\begin{align}
\frac{\delta \Gamma_{\rm eff}}{\delta A_{\mu}^{a}}&=
-e^{\Gamma[A_{\mu}^{a}]_{\rm eff}}\int {\cal D}\psi {\cal D}\overline{\psi}
\left[\overline{\psi}\Gamma^{\mu}\lambda^{a}\left(\frac{1-\overline{\Gamma}}{2}
\right) \psi\right] e^{-S(\psi,\overline{\psi},A)}
\nonumber
\\[0.2cm]
&=-\left \langle \overline{\psi}
\Gamma^{\mu}\lambda^{a}\left(\frac{1-\overline{\Gamma}}{2}\right) \psi
\right \rangle.
\end{align}
Plugging this result into eq.~\eqref{eq:var_effective_action1} we express the infinitesimal
variation of the fermion effective action in terms of the quantum conservation of the gauge current
\begin{align}
\delta \Gamma_{\rm eff}&={\rm tr\,}\int d^{d}x\, \epsilon^{a} D_{\mu}
\left \langle \overline{\psi}
\Gamma^{\mu}\lambda^{a}\left(\frac{1-\overline{\Gamma}}{2}\right) \psi
\right \rangle \nonumber \\[0.2cm]
&={\rm tr\,}\int d^{d}x\, \epsilon^{a} D_{\mu}\langle j_{a}^{\mu} \rangle .
\end{align}
Thus, the non-invariance of the fermion effective action under infinitesimal gauge
tranformations implies a breakdown in the conservation of the expectation value of the
current~$\langle j_{a}^{\mu} \rangle$.

A similar calculation can be done for gravitational anomalies. Now, we
perform an infinitesimal coordinate tranformation $x^{\mu}\rightarrow
x^{\mu}+\xi^{\mu}(x)$ inducing a change in the metric tensor
\begin{align}
\delta g_{\mu \nu} =D_{\mu}\xi_{\nu}+D_{\nu}\xi_{\mu},
\end{align}
where~$D_{\mu}$ the curved space covariant derivative.
A calculation similar to the one for the gauge case presented above gives the 
variation of the fermion effective action\footnote{Due to the coupling of the fermions to gravity,
the effective action now depends also on the metric tensor.}
\begin{align}
\delta \Gamma_{\rm eff}=\int d^{d}x \sqrt{g}\, \xi_{\mu}D_{\nu} \langle
T^{\mu \nu} \rangle,
\end{align}
where $\langle T^{\mu \nu}\rangle$ is the expectation value of the
energy-momentum tensor of the matter field and~$D_{\mu}$ is the
covariant derivative. Again, the failure in the
conservation of the energy-momentum tensor is signalled by a non-invariance of the
fermion effective action under diffeomorphisms. 

Anomalies are expected only in parity-violating theories. 
Otherwise we can always construct gauge invariant
fermion mass terms and regularize the theory using the Pauli-Villars method.
Besides, anomalies are  associated
only with the imaginary part of the fermion effective action. To see this, suppose the
fermions transform in a complex representation~$R(G)$ of the gauge group. If the theory is chiral,
there is no evident
way to regularize the theory while preserving gauge invariance. Thus, the
fermion effective action $\Gamma[A]_{R}$ is complex and not necessarily
gauge invariant. Now, let us consider that fermions transform in the complex conjugate
representation~${\overline R}(G)$. In this case, the same discussion presented above 
leads to the conclusion that the effective action~$\Gamma[A]_{\overline R}$ [which is
the complex conjugate of $\Gamma[A]_{R}$] is also potentially not gauge invariant.

Considering now fermions transforming in the real representation
$R(G)\oplus{\overline R}(G)$, we find that it is possible to construct a gauge-invariant 
mass term. We can then use Pauli-Villars regularization and construct a quantum theory 
that is anomaly-free. Since the
path integral is Gaussian in the fermion fields, the
effective action for this theory is simply  $\Gamma[A]_{R}
+\Gamma[A]_{\overline R}=2\,{\rm Re\,}\Gamma[A]_{R}$. We conclude therefore that the real part of
$\Gamma[A]_{R}$ is always gauge invariant, or can be made gauge invariant by
adding appropriate local counterterms. As a consequence, anomalies
only appear in the imaginary part of the fermion effective action.

The perturbative analysis of anomalies proceeds by computing
one-loop diagrams where fermions in the loop are
coupled to external gauge fields,~$A_{\mu}^{a}$, or gravitons,~$h_{\mu \nu}$, through the chiral
current or the energy-momentum tensor
\begin{align}
\overline{\psi}T^{a}\Gamma_{\mu}P_{+}\psi &\,\,\,\longleftrightarrow \,\,\, A_{\mu}^{a},
\nonumber \\[0.2cm]
\overline{\psi}\big(D_{\mu}\Gamma_{\nu}+D_{\nu}\Gamma_{\mu}\big)\psi &\,\,\,\longleftrightarrow\,\,\,
h_{\mu \nu}.
\end{align}
These diagrams, appearing in the perturbative analysis of the Ward
identities, have their parity-preserving and parity-violating parts. The
anomaly comes from the latter, since, just like in the case of the fermion effective 
action, the former can be regularized 
without breaking gauge invariance.

The impossibility of defining a parity-violating amplitude satisfying
all physical principles implies that there two possible definitions of the
amplitude \cite{Bardeen-Zumino}. The first, which is the hardest one to obtain
from a computational
point of view, is to define the parity-violating part of the amplitude
preserving Bose symmetry in the external lines. The anomaly so obtained
is called the consistent anomaly because it satisfy the Wess-Zumino
consistency condition \cite{Wess-Zumino}.

A second alternative (called the Adler-Rosenberg method~\cite{ABJ,Rosenberg}) 
is to consider the same diagram but with a single axial-vector current in one of the
vertex and vector currents in the remaining ones. The anomaly is then calculated
by imposing vector current conservation and Bose symmetry on the
external vector lines. This form of the anomaly, called the covariant anomaly,
does not satisfy the Wess-Zumino consistency conditions, but it is much easier to
compute. There exists a standard formalism interpolating
between consistent and covariant form of the anomaly~\cite{LAG-Ginsparg1,LAG1}. Thus, we
can calculate the covariant anomaly (which is easier) and obtain from it
the consistent one.

We now determine what diagrams are potentially anomalous. Let us
consider a $(k+1)$-poligon with $k$ external vector currents and one 
axial-vector current. Since the amplitude is parity violating, it should contain a
Levi-Civita tensor $\epsilon_{\mu_{1}\ldots \mu_{2n}}$ (remember that we work in 
$2n$~dimensions). Since we want to check the conservation of the axial-vector current,
the polarization vector in the axial-vector channel has to be taken proportional to the
incoming momentum $P_{\mu}$ of this vertex. We also have the momenta entering the other
vertices~$p_{1},\ldots,p_{k}$ and their corresponding polarization 
vectors~$\epsilon_{1}^{\mu},\ldots,\epsilon_{k}^{\mu}$. We have to saturate the~$2n$
indices of the Levi-Civita tensor and at the same time preserve momentum 
conservation,~$P^{\mu}+\sum p_{i}^{\mu}=0$. The only way to accomplish this 
is setting~$k=n$, so the simplest potentially anomaoloous diagram
is a $(n+1)$-polygon. The amplitude associated to this diagram then has the structure
\begin{align}
\left[\sum_{T_{L}}\,{\rm str\,}(T_{L}^{a_{1}}\ldots
T_{L}^{a_{n+1}})-\sum_{T_{R}}\,{\rm str\,}(T_{R}^{a_{1}}\ldots
T_{R}^{a_{n+1}})\right]A(p,\epsilon),
\end{align}
where~$A(p,\epsilon)$ contains all dependence on the external momenta and polarization vectors.

As for the group theory prefactor,
instead of~${\rm str\,}(T^{a_{1}}\ldots T^{a_{n+1}})$ it is easier,
and completely equivalent, to evaluate ${\rm tr\,}H^{n+1}$, where~$H$ is any
element of the gauge group Lie algebra. For $d=4k$ we would have
to compute~${\rm tr\,}H^{2k+1}$ and ${\rm tr\,}H^{2k+2}$ for $d=4k+2$. Let us consider first
that fermions are in a real or pseudoreal representation of the
gauge group. If this is the case,~$H$ satisfies
\begin{align}
H^{T}=-S^{-1}HS,
\end{align}
for some $S=\pm S^{T}$ ($+1$ for real representations and $-1$ for
pseudoreal). Then, for $d=4k$ we find
\begin{align}
{\rm tr\,}H^{2k+1}={\rm tr\,}(H^{2k+1})^{T}
={\rm tr\,}(-S^{-1}HS)^{2k+1}=-{\rm tr\,}H^{2k+1}=0,
\end{align}
meaning that in $d=4k$ there is no gauge anomaly whenever the gauge group only has real
or pseudoreal representations. Thus, chiral theories
with gauge groups~$\mbox{SO}(2n+1)$, $\mbox{G}_{2}$, $\mbox{F}_{4}$, 
$\mbox{E}_{7}$, and $\mbox{E}_{8}$ are
always anomaly free. The only potentially dangerous groups are~$\mbox{SO}(2n)$, $\mbox{SU}(N)$, 
and~$\mbox{E}_{6}$.

So much for gauge anomalies. In the case of
gravitational anomalies we find again that the first potentially anomalous diagram
in~$d=2n$ dimensions is a polygon with $n+1$~vertices~\cite{LAG-Witten}. The same
reasoning as in the gauge case applies here, keeping in mind that 
the polarization vector of the graviton is a second rank symmetric tensor. 
An interesting question to answer is when
pure gravitational anomalies appear. We know that in four-dimensions 
particles and antiparticles running in the loop have 
opposite helicity, since $\{\overline{\Gamma},C\}=0$ in~$d=4$ [see eq.~\eqref{eq:[C,Gamma]}].
Gravitons, however, do not distinguish between particles and antiparticles and its coupling
to fermions ``looks'' vector-like. The same happens when $d=4k$, for 
which~$\overline{\Gamma}$ and $C$ also anticommute. The
conclusion is that there are no pure gravitational anomalies when the space-time dimension
is a multiple of four.

This is not the case in $d=4k+2$. Now $[\overline{\Gamma},C]=0$ so particles
and antiparticles have the same helicity. The gravitational interaction is 
genuinely chiral and we can expect anomalies in the conservation of the energy-momentum 
tensor. Pure gravitational anomalies can only
appear in $4k+2$ dimensions. 

\begin{figure}[t]
\centerline{\includegraphics[scale=0.40]{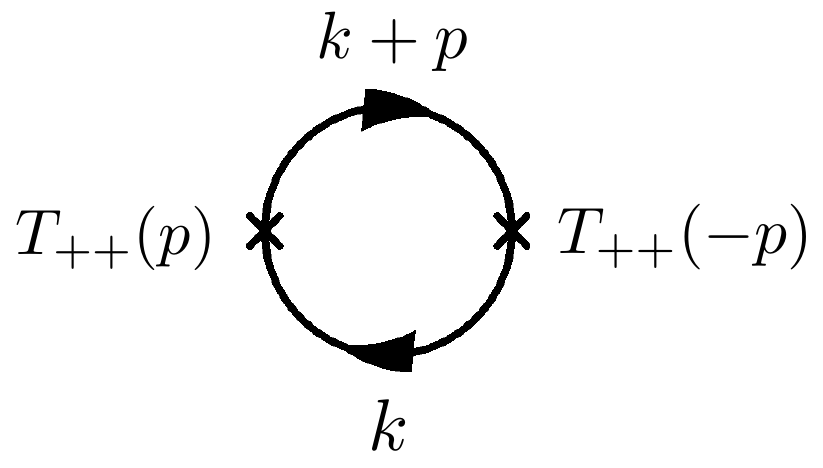}}
\caption{The gravitational anomaly in two dimensions}
\label{fig1}
\end{figure}

As an example, let us compute the
gravitational anomaly for a spin-{1/2} fermion in two dimensions
\cite{LAG-Witten}. We consider a weak gravitational fields so
the metric tensor can be written as a small perturbation of Minkowski 
space-time,~$g_{\mu\nu}=\eta_{\mu\nu}+h_{\mu\nu}$. Working at linear order, the 
graviton field~$h_{\mu\nu}$ couples to the fermion field through its
energy-momentum tensor,~$\Delta {\cal L}=
-\frac{1}{2}h^{\mu\nu}T_{\mu\nu}$. Considering a chiral fermion
satisfying~$\overline{\Gamma} \psi=-\psi$, it can be seen that the fermionic energy-momentum 
tensor has a single non-vanishing component~$T_{++}$, where we are using two-dimensional
light-cone coordinates. Moreover, we are interested in the two-point
function
\begin{align}
U(p)=\int d^{2}x \langle \Omega |T\big[T_{++}(x)T_{++}(0)\big]
|\Omega\rangle e^{ipx},
\end{align}
which, assuming the naive conservation of the energy-momentum tensor,~$\partial_{-}T_{++}(x)=0$, 
satisfies the Ward identity
\begin{align}
p_{-}U(p)=0,
\end{align}
so $U(p)=0$ for all $p_{-}\neq 0$. This implies by analyticity that $U(p)=0$ for all $p_{-}$, 
and this is impossible for the two-point
function of a Hermitian operator. This means that the energy-momentum tensor Ward identity
must be anomalous, which can be explicitly calculated
from fig.~\ref{fig1} with the result~\cite{LAG-Witten}
\begin{align}
p_{-}U(p)=\frac{i}{24\pi}p_{+}^{3}.
\label{eq:2D_grav_anomaly}
\end{align}
Given this value of~$U(p)$ we can write the effective action to second
order in the metric perturbation $h_{\mu\nu}$ by simply coupling each vertex in
fig.~\ref{fig1} to $-\frac{1}{2}h_{--}$, namely
\begin{align}
\Gamma_{\rm eff}=-\frac{1}{192\pi}\int d^{2}p \frac{p_{+}^{3}}{p_{-}}
h_{--}(p)h_{--}(-p),
\label{effective-grav}
\end{align}
where $h_{--}(p)$ is the Fourier transform of $h_{--}(x)$. Implementing a 
general coordinate transformation, we see that eq.~\eqref{effective-grav} does
not remain invariant. Still, we have the freedom of 
adding local counterterms to the fermion effective and see whether the result
is diffeomorphism invariant. In fact, writing the most
general counterterm of the appropriate dimension
\begin{align}
\Delta \Gamma_{\rm local}&= \int
d^{2}p\Big[Ap_{+}^{2}h_{--}(p)h_{+-}(-p)+Bp_{+}p_{-}h_{+-}(p)h_{+-}(-p)
\nonumber
\\[0.2cm]
&+Cp_{+}p_{-}h_{++}(p)h_{--}(-p)+Dp_{-}^{2}h_{++}(p)
h_{+-}(-p)\Big],
\end{align}
it can be seen that there is no choice of the constants~$A$, $B$, $C$, and~$D$
making~$\Gamma_{\rm eff}+\Delta \Gamma_{\rm local}$ invariant under coordinate
transformations.

The situation is quite different if when considerin a Dirac fermions in $1+1$
dimensions. The fermion effective action in this case is obtained by 
adding to~(\ref{effective-grav}) its parity transformed, to write
\begin{align}
\Gamma_{\rm eff}=-\frac{1}{192\pi}\int
d^{2}p\left[\frac{p_{+}^{3}}{p_{-}}h_{--}(p)h_{--}(-p)+
\frac{p_{-}^{3}}{p_{+}}h_{++}(p)h_{++}(-p)\right].
\label{effective2}
\end{align}
This action is not invariant under general coordinate
transformations. However, there exists a local couterterm that we can add to 
get the diffeomorphism-invariant action
\begin{align}
\overline{\Gamma}=-\frac{1}{192\pi}\int d^{2}p \frac{R(p)R(-p)}{p_{+}p_{-}},
\label{eq:inv2Doneloopgravaction}
\end{align}
where $R(p)=p_{+}^{2}h_{--}+p_{-}^{2}h_{++}-2p_{+}p_{-}h_{+-}$ is the
linearized scalar curvature.

Using the result~\eqref{eq:inv2Doneloopgravaction} we can compute the trace anomaly of the energy
momentum tensor of a Dirac fermion in two dimensions. Since the theory does not have 
any mass scale, the trace of the energy-momentum tensor
is zero classically,~$T^{\mu}_{\mu}=2T_{+-}=0$. Although~$h_{+-}$ does not appear in the effective 
action~(\ref{effective2}), once we add the local counterterm to restore
diffeomorphism invariance we get a dependence on this graviton field component
and a non-zero result for the expectation
value
\begin{align}
\langle 2T_{+-}(p) \rangle=-2\frac{\delta \overline{\Gamma}}{\delta h_{+-}(-p)}=
-\frac{1}{24\pi}R(p),
\end{align}
implying that classical conformal invariance is broken by quantum
corrections.

\subsection{The Atiyah-Singer index theorem and the computation of
anomalies}

In the last two decades sophisticated tools borrowed from topology
have been applied to the computation of anomalies. 
One that has been specially useful is the Atiyah-Singer index theorem
\cite{Atiyah-Singer,Atiyah-Pattodi-Singer} and its generalization
for families of
elliptic operators~\cite{Atiyah-Singer-2}.
The first one allows a neat computation of singlet anomalies,
while the second one has been applied to the analysis of both local and
global gauge and gravitational anomalies.

Here we are not going to give the full details on how to compute anomalies using index theorems,
which the reader can find in a number of reviews (see, e.g, refs.~\cite{LAG-Ginsparg1,LAG1}). Instead,
we will present the basic results and how to apply them. Since
the problem of computing anomalies is related to the
index theorem for the Dirac operator (as well as that for
Rarita-Schwinger operator), we start by reviewing its main features.

The index of an operator ${\cal O}$ is defined as the
difference between the dimensions of its kernel and that of
its adjoint
\begin{align}
{\rm ind\,}{\cal O}= {\rm dim\,}{\rm ker\,}{\cal O}-{\rm dim\,}{\rm ker\,}{\cal O}^{\dagger}.
\end{align}
The Atiyah-Singer index theorem relates the index of an elliptic
operator with the topological properties of the manifold on which it is
defined. For obvious reasons, we are going to focus on the particular case in which 
${\cal O}$ is the Weyl operator, $D_{+}=i\Dsl P_{+}$.

Let~$S_{+}\otimes E$ and
$S_{-}\otimes
E$ be two vector bundles over a manifold~$M$, where~$S_{\pm}$ is the space of spinors of
positive and negative chirality and~$E$ is the space
carrying the representation of the gauge group. We introduce
the gauge connection one-form $A=A_{\mu}dx^{\mu}$ over $E$, transforming as
\begin{align}
A \rightarrow g^{-1}(A+d)g,
\end{align}
and its curvature $F$
\begin{align}
F=dA+A^{2}\;,
\end{align}
where, for simplicity, wedge product between forms are omitted (i.e. $A\wedge A=A^{2}$). In
the presence of gravity, besides the
gauge bundle, we also have the vector bundles of orthonormal or coordinates frames. In
the first case, covariant derivatives are written in terms of the spin 
connection~$\omega^{m}_{\,\,\,\,n}=\omega_{\mu \,\,\,\,n}^{\,\,\,m}dx^{\mu}$, while the curvature two-form has an 
expression similar to the gauge field strength,~$R^{m}_{\,\,\,n}=d\omega^{m}_{\,\,\,n}+
\omega^{m}_{\,\,\,\,p}\omega^{p}_{\,\,\,\,n}$. 
The Dirac operator $\Dsl$ is given explicitly by
\begin{align}
\Dsl =
\Gamma^{\mu}\left(\partial_{\mu}+\frac{1}{2}\omega_{\mu\,mn}\Sigma^{mn}
+A_{\mu}^{a}T^{a}\right).
\label{dir-op}
\end{align}
The corresponding Weyl operators are defined by~$D_{\pm}=i \Dsl
P_{\pm}$ mapping
\begin{align}
D_{\pm}:S_{\pm}\otimes E \longrightarrow S_{\mp}\otimes E.
\label{eq:D_pm->S_pmxE}
\end{align}
and verify $D_{+}^{\dagger}=D_{-}$. Using a basis where~$\overline{\Gamma}$ is
diagonal, the Dirac operator~\eqref{dir-op} can be written as
\begin{align}
i\Dsl = \left(
\begin{array}{cc}
0     &     D_{-}  \\
D_{+} &     0
\end{array}
\right).
\label{eq:iD=(0D-D+0)}
\end{align}

Let us focus on the index of the operator Weyl operator $D_{+}$
\begin{align}
{\rm ind\,}D_{+}={\rm dim\,}{\rm ker\,}D_{+}-{\rm dim}\,{\rm ker\,}D_{-}.
\end{align}
Using the Atiyah-Singer index
theorem, it can be expressed in terms of the topological invariants of the
bundle. Before doing so, however, we need to introduce some elementary notions
from the theory of characteristic classes~\cite{LAG-Ginsparg1,LAG1,Nash}. Let $\Omega$ be some
matrix-valued
two-form taking values in the Lie algebra of some group $G$ and let
$P(\Omega)$ be an invariant polynomial, i.e. a polynomial in $\Omega$
satisfying
\begin{align}
P(g^{-1}\Omega g)=P(g),
\end{align}
for any element $g$ of $G$. We are going to consider the particular case where~$\Omega$ 
is either the gauge field strength or the curvature two-form. Thus, 
the invariant polynomial~$P(\Omega)$ satisfies two important properties:
\begin{itemize}
\item[-]
$P(\Omega)$ is a closed form, i.e.~$dP(\Omega)=0$.

\item[-]
The integrals of $P(\Omega)$ over the manifold define topological invariants.

\end{itemize}
To prove these two properties it is enough to look at monomials of the
form~$P_{m}(\Omega)\equiv {\rm tr\,}(\Omega^{m})$, since these are the building blocks
of any other invariant polynomial. It is straightforward to show
that~$P_{m}(\Omega)$ is closed.
\begin{align}
dP_{m}=m\,{\rm tr\,}(d\Omega\,\Omega^{m-1})=m\,{\rm tr\,}(D\Omega \, \Omega^{m-1})=0,
\end{align}
where~$D$ is the covariant derivative and we used the
Bianchi identity
\begin{align}
D\Omega=d\Omega+\Omega \omega-\omega \Omega=0,
\end{align}
where $\omega$ is the corresponding connection one-form.

The second property is not as evident as the first one. We need to
prove that
the integral of the polynomial~$P_{m}$ over~$M$ is independent of the
connection~$\omega$. Consider a one-parameter family
interpolating between two given connections~$\omega_{0}$ and
$\omega_{1}$
\begin{align}
\omega_{t}=\omega_{0}+t(\omega_{1}-\omega_{0}),
\end{align}
with $0\leq t \leq 1$, and the associated family of curvatures
\begin{align}
\Omega_{t}=d\omega_{t}+\omega^{2}_{t}.
\end{align}
We evaluate the derivative of $\Omega_{t}$ with respect to
the parameter $t$ that, for simplicity, we denote by a dot
\begin{align}
\frac{\partial P_{m}(\Omega_{t})}{\partial
t}&=m\,{\rm tr\,}\left(\dot{\Omega}_{t}
\,\Omega^{m-1}_{t}\right) \nonumber \\[0.2cm]
&=-m\,{\rm tr\,}\left(D_{t}
\dot{\omega}_{t}
\Omega^{m-1}_{t}\right)  \label{top-invariant}\\[0.2cm]
&= m\,d\,{\rm tr\,}\left(
\dot{\omega}_{t}\Omega^{m-1}_{t}\right),
\nonumber
\end{align}
after applying
\begin{align}
\dot{\Omega}_{t}=d\dot{\omega}_{t}+\dot{\omega}_{t}\omega_{t}+
\omega_{t}\dot{\omega}_{t}.
\end{align}
Now, integrating over~$t$ from~$0$
to~$1$ in eq.~(\ref{top-invariant}), we find
\begin{align}
P_{m}(\Omega_{1})-P_{m}(\Omega_{0})&=m\,d\int
dt\,{\rm tr}(\dot{\omega}_{t}\Omega_{t}^{m-1}) \nonumber \\[0.2cm]
&=dQ_{2n-1}^{0}(\omega_{t},\Omega_{t}),
\label{Chern-Simons}
\end{align}
where $Q_{2n-1}^{0}$ is called the Chern-Simons form.
If we now integrate the previous expression over any closed $2m$-dimensional submanifold
$M_{2m} \subset M$, we have that the right-hand side vanishes a arrive at the result
\begin{align}
\int_{M_{2m}} P_{m}(\Omega_{1}) = \int_{M_{2m}} P_{m}(\Omega_{0}).
\end{align}
Thus, the integral of $P_{m}(\Omega)$ over $M_{2m}$ is independent of the
connection and therefore is a topological invariant as we wanted to prove. 
Furthermore, it is easy to show that this integral is also invariant under
deformations of $M_{2m}$. Let us consider a deformation~$M_{2m}'$ of the submanifold~$M_{2m}$
and let $B_{2m+1}$ be a $(2m+1)$-dimensional manifold  such that~$\partial B_{2m+1}=
M_{2m}-M_{2m}'$. Since $dP_{m}(\Omega)=0$, the
Stokes theorem implies that
\begin{align}
0=\int_{B_{2m+1}} dP_{m}({\Omega}) = \int_{M_{2m}} P_{m}(\Omega)-
\int_{M_{2m}'} P_{m}(\Omega).
\end{align}

We have all the basic ingredients now to introduce some well-known characteristic classes. 
Let us first consider a bundle with group~$\mbox{U}(n)$, so the
curvature $\Omega$ is a Hermitian matrix of two-forms. The
total Chern class is defined as
\begin{align}
c(\Omega)=\det\,\left(1+\frac{i}{2\pi}\Omega\right).
\label{eq:total_chern_class}
\end{align}
This polynomial can be expanded as a sum
\begin{align}
c(\Omega)=1+c_{1}(\Omega)+c_{2}(\Omega)+\ldots,
\end{align}
where~$c_{i}(\Omega)$ is a $2i$-form called the $i$th Chern class. To
write closed expressions for the Chern classes, we use the
following
trick: since the matrix $\Omega$ is Hermitian, it can be formally
diagonalize to write
\begin{align}
\frac{i}{2\pi}\Omega=\left(
\begin{array}{ccc}
x_{1}	&	 &	\\
	& \ddots &	\\
	&	 &  x_{n}
\end{array}
\right)\; ,
\end{align}
where the eigenvalues $x_{i}$ are two-forms. Then, the
total Chern class~\eqref{eq:total_chern_class} is given by
\begin{align}
c(\Omega)=\prod_{i=1}^{n} (1+x_{i})=1+\sum_{i=1}^{n} x_{i} +
\sum_{i<j} x_{i}x_{j} +\ldots +\prod_{i=1}^{n} x_{i}.
\end{align}
Each of the terms in this expansion can be easily identified as
\begin{align}
c_{1}(\Omega)&=\sum_{i=1}^{n} x_{i}=\frac{i}{2\pi}\, {\rm tr\,}\Omega,
\nonumber
\\[0.2cm]
c_{2}(\Omega)&=\sum_{i<j}
x_{i}x_{j} = \left(\frac{i}{2\pi}\right)^{2}\left[({\rm tr\,}\Omega)^{2}-
{\rm tr\,}(\Omega^{2})\right],
\nonumber
\\[0.2cm]
&\vdots 
\nonumber
\\[0.2cm]
c_{n}(\Omega)&=\prod_{i=1}^{n} x_{i} =
\det\,\left(\frac{i}{2\pi}\Omega\right),
\end{align}
so the Chern classes have simple expressions in terms of~$\Omega$. 

The total
Chern class behaves nicely with respect to the direct sum (Whitney
sum) of bundles: if $E$ and $F$ denote two $\mbox{U}(n)$ bundles, the total Chern
class of their direct sum~$E\oplus F$ is
\begin{align}
c(E\oplus F)=c(E)c(F).
\end{align}
Unfortunately, this behavior of the total Chern class with respect to
direct sums is not maintained for tensor
products. Associated with a $\mbox{U}(n)$ bundle we can define another useful polynomial, the Chern
character
\begin{align}
{\rm ch}(\Omega)={\rm tr\,}e^{\frac{i}{2\pi}\Omega}=
{\rm ch}_{0}(\Omega)+{\rm ch}_{1}(\Omega)+\ldots,
\label{Ch.1}
\end{align}
where ${\rm ch}_{i}(\Omega)$ is $i$th Chern character. They are identified by expanding the exponential
\begin{align}
{\rm tr\,}e^{\frac{i}{2\pi}\Omega}=\sum_{k=0}^{\infty}
\frac{1}{n!}{\rm tr\,}\left(\frac{i}{2\pi}\Omega\right)^{k}.
\end{align}
Notice that this sum always contains a finite number of terms. Since~$\Omega^{k}$ is
a $2k$-form, it vanishes whenever $2k>{\rm dim\,}M$. We then find the expression of the $j$th Chern character
\begin{align}
{\rm ch}_{0}(\Omega)&=2n, \nonumber \\[0.2cm]
{\rm ch}_{j}(\Omega)&=\frac{1}{j!}\left(\frac{i}{2\pi}\right)^{j}\,{\rm tr\,}
\Omega^{j},
\hspace{1cm} 2\leq 2j \leq \,{\rm dim\,} M.
\label{Ch.2}
\end{align}
A convenient property of the total Chern character is that it is well-behaved with respect to both
direct sums and tensor products of bundles
\begin{align}
{\rm ch}(E\oplus F)&={\rm ch}(E)+{\rm ch}(F),
\nonumber
\\[0.2cm]
{\rm ch}(E\otimes F)&={\rm ch}(E){\rm ch}(F).
\end{align}

Up to now we have defined invariant polynomials for~$\mbox{U}(n)$ bundles in
which the curvature is a Hermitian matrix of two-forms.
We also want to consider $\mbox{SO}(n)$ bundles,  as
in Riemannian geometry. We then define the total Pontrjagin class
\begin{align}
p(\Omega)=\det\left(1+\frac{1}{2\pi}\Omega\right)=
1+p_{1}(\Omega)+p_{2}(\Omega)+\ldots,
\end{align}
where $p_{j}(\Omega)$ is the $j$th Pontrjagin class.
The curvature $\Omega$ is an antisymmetric matrix of two-forms that
cannot be diagonalized by a similarity transformation. Nevertheless, it can
be brought to a skew-diagonal form
\begin{align}
\frac{1}{2\pi}\Omega=\left(
\begin{array}{ccccc}
0      &   x_{1} &  &  &   \\
-x_{1} &    0 &  &  &	\\
       & &    0   & x_{2} &   \\
       & & -x_{2} &   0   &   \\
       & &  &  &  \ddots
\end{array}
\right).
\label{eq:skew_eigenvalues_Omega}
\end{align}
The antisymmetry of $\Omega$ implies that only even-degree polynomials
in $\Omega$ are different from zero, so~$p_{j}(\Omega)$ is a~$4j$-form.
The total Pontrjagin class can be written in terms of the formal
eigenvalues as
\begin{align}
p(\Omega)=\prod_{i}(1+x_{i}^{2})=1+\sum_{i} x_{i}^{2}+\sum_{i<j}
x_{i}^{2}x_{j}^{2}+\ldots
\end{align}

We are ready now to formulate the Atiyah-Singer index theorem for the
Weyl operator
\begin{align}
D_{+}=i\Gamma^{\mu}\left(\partial_{\mu}+
\frac{1}{2}\omega_{\mu\,mn}\Sigma^{mn}+A_{\mu}\right)P_{+}.
\end{align}
The theorem states that
\begin{align}
{\rm ind\,}D_{+}=\int_{M}\left[\widehat{A}(M){\rm ch}(F)\right]_{\rm vol},
\end{align}
where the subindex {\rm vol} indicates that we only retain
from the product the volume form and~$\widehat{A}(M)$, called the Dirac 
genus of~$M$ (or the $A$-roof genus),
is defined in terms of
the two-form skew-eigenvalues $x_{i}$ by
\begin{align}
\widehat{A}(M)=\prod_{a} \frac{x_{a}/2}{\sinh{(x_{a}/2)}}\;.
\end{align}
It can be rewritten in terms of the invariant polynomials
${\rm tr}(R^{m})$, with $R$ the curvature two-form of the manifold~$M$
\begin{align}
\widehat{A}(M)=1+\frac{1}{(4\pi)^{2}}\frac{1}{12}\,{\rm tr\,}R^{2}+
\frac{1}{(4\pi)^{4}}\left[\frac{1}{288}({\rm tr\,}R^{2})^{2}+
\frac{1}{360}\,{\rm tr\,}R^{4}\right]+\ldots
\end{align}
This equation, together with the expansion of the total Chern character
(\ref{Ch.1}) and (\ref{Ch.2}), leads to an explicit expression for the
index of $D_{+}$. For example, in $d=4$ dimensions we have
\begin{align}
{\rm ind\,}D_{+}=-\frac{1}{2(2\pi)^{2}}\int_{M}\left({\rm tr\,}F^{2}
-\frac{r}{24}\,{\rm tr\,}R^{2}\right), \hspace{1cm} \mbox{for~$d=4$},
\label{eq:indD+d=4}
\end{align}
with $r$ the dimension of the vector bundle.

The analysis of anomalies also involves the calculation of
the index of the Rarita-Schwinger operator. Here we have to be careful, however, because ghost fields
are needed in order to remove unphysical degrees of freedom. The
constraint
\begin{align}
k_{\mu}\psi^{\mu}=0,
\end{align}
together with the invariance
\begin{align}
\psi^{\mu}\rightarrow \psi^{\mu}+k^{\mu}\chi,
\end{align}
remove two spin-$\frac{1}{2}$ degrees of freedom of the same chirality,
while using the constraint
\begin{align}
\Gamma^{\mu}\psi_{\mu}=0,
\end{align}
gets rid of an additional spin-$\frac{1}{2}$ degree of freedom, this time with the opposite
chirality. To find the index of the Rarita-Schwinger operator we
subtract the contribution of a~spin-$\frac{1}{2}$ field from that of
spin-$\frac{3}{2}$ with an additional vector index, namely
\begin{align}
{\rm ind\,}\Dsl_{3/2}=
\int_{M}\left[\widehat{A}(M)\left({\rm tr\,}e^{iR/2\pi}-1\right){\rm ch}(F)\right]_{\rm vol}.
\end{align}
Finally, when studying anomalies in ten dimensions (needed
when analyzing the low-energy field theory of superstrings) we also may
expect anomalies from antisymmetric tensor fields whose field
strength is self-dual or anti-self-dual. The index theorem for these fields reads
\begin{align}
{\rm ind\,}\left(iD_{A}\right)=\frac{1}{4}\int_{M} [L(M)]_{\rm vol},
\end{align}
where $L(M)$ is the Hirzebruch polynomial defined by
\begin{align}
L(M)\equiv 2^{n}\prod_{a} \frac{x_{a}/2}{\tanh{(x_{a}/2)}} .
\end{align}

We apply now all the technology introduced to the problem of computing anomalies,
beginning with the use of the Atiyah-Singer index theorem to recover the
single chiral~$\mbox{U}(1)$ anomaly studied in sec.~\ref{eq:anomalies_generalities}. 
To do this we use Fujikawa's approach~\cite{Fujikawa} which begins with the Euclidean fermion
one-loop effective action
\begin{align}
e^{-\Gamma[A]_{\rm eff}}=\int {\cal D}\psi {\cal D}\overline{\psi}
\exp{\left(-\int d^{2n}x\sqrt{g}\,\overline{\psi}\,i\Dsl \psi\right)},
\label{5-action}
\end{align}
with $\Dsl=\Gamma^{\mu}(\partial_{\mu}+A_{\mu})$
and we have to keep in mind that in Euclidean space~$\psi$ and~$\overline{\psi}$ are
independent variables. The classical action is
invariant under global chiral transformation
\begin{align}
\psi' &= e^{i\alpha\overline{\Gamma}}\psi, \nonumber\\[0.2cm]
\label{chiral-trans}
\overline{\psi}' &= \overline{\psi}e^{i\alpha \overline{\Gamma}},
\end{align}
and Noether's theorem leads to the classically conserved current
\begin{align}
j_{5}^{\mu}=\overline{\psi}\,\overline{\Gamma}\Gamma^{\mu} \psi.
\end{align}

We can implement the change of variables defined in eq.~\eqref{chiral-trans} on the 
quantum effective action~\eqref{5-action}, with an $x$-dependent $\alpha$. 
For the classical action, we find
\begin{align}
\int (dx)\,\overline{\psi}'\,i\Dsl\,\psi' =
\int (dx)\,\overline{\psi}\,i\Dsl\,\psi+ \int (dx)\,\alpha(x)
\nabla_{\mu}j_{5}^{\mu},
\end{align}
where~$(dx)\equiv d^{2n}x\sqrt{g}$ denotes the volume element and
$\nabla_{\mu}=\partial_{\mu}+\omega_{\mu}$ is the covariant derivative.
Here we have to be careful, since the change~\eqref{chiral-trans} may
induce a nontrivial Jacobian~${\cal J}$ in the functional integration measure.
Taking this into account, we find
\begin{align}
\int {\cal D}\psi' {\cal D}\overline{\psi}'& \exp{\left(\int
(dx)\,\overline{\psi}'\,i\Dsl\,\psi'\right)} \nonumber \\
&=\int {\cal D}\psi {\cal D}\overline{\psi}\,{\cal J}\,\exp{\left(\int
(dx)\,\overline{\psi}\,i\Dsl\,\psi+
\int (dx)\,\alpha(x)\nabla_{\mu}j_{5}^{\mu}\right)}.
\label{eff.act.j5}
\end{align}
Were the Jacobian a constant, we could simply expand the exponential on the right-hand side
to first order in~$\alpha(x)$ to obtain the conservation
of the axial current, $\langle \nabla_{\mu}j_{5}^{\mu}\rangle=0$.
This is, howver, not necessarily the case. To compute ${\cal J}$,
we begin by expanding~$\psi$ and~$\overline{\psi}$ in a basis of eigenfunctions of
the Dirac operator
\begin{align}
i\Dsl \psi_{n} = \lambda_{n} \psi_{n},
\end{align}
to write
\begin{align}
\psi&=\sum_{n}a_{n}\psi_{n},
\nonumber \\[0.2cm]
\overline{\psi}&=\sum_{n}\overline{b}_{n}\psi^{\dagger}_{n},
\end{align}
where $a_{n}$ and $b_{n}$ are Grassmann parameters.
The integration measure is simply $\prod_{n}d\overline{b}_{n}da_{n}$, while
the action can be recast as
\begin{align}
\int(dx)\,\overline{\psi}\,i\Dsl\,\psi=\sum_{n} \lambda_{n}
\overline{b}_{n}a_{n}.
\end{align}
In this basis, the Jacobian is computed to be
\begin{align}
{\cal J}=\exp{\left[-2i\sum_{n}\langle
\psi_{n}|\alpha(x)\overline{\Gamma}|\psi_{n}\rangle\right]}\simeq
1-2i\sum_{n}\langle \psi_{n}|\alpha(x)\overline{\Gamma}|\psi_{n}\rangle,
\label{Jacobian}
\end{align}
where the inner product is defined as
\begin{align}
\langle \psi|\alpha(x)\overline{\Gamma}|\psi \rangle=
\int (dx) \alpha(x)\psi^{\dagger}(x)\overline{\Gamma}\psi(x).
\label{inner}
\end{align}
A basic problem of the result~\eqref{Jacobian} is that it diverges, so 
we have to regularize somehow. This can be done by inserting a Gaussian
cut-off
\begin{align}
-2i\sum_{n}\int (dx)
\alpha(x)\psi^{\dagger}_{n}(x)&\overline{\Gamma}\psi_{n}(x)
e^{-\frac{\lambda^{2}_{n}}{M^{2}}}=-2i\sum_{n}\langle
\psi_{n}|\alpha(x)\overline{\Gamma}
e^{-\frac{{\not D}^{2}}{M^{2}}}|\psi_{n}\rangle.
\label{preanomaly}
\end{align}

An important point here is that the positive and negative eigenvalues of $i\Dsl$
are paired. Since~$\{ \Dsl ,\overline{\Gamma}\}=0$,
for any eigenfunction $\psi_{n}$ with $\lambda_{n}>0$ we have 
that~$\overline{\Gamma}\psi_{n}$ is also an eigenfunctions with opposite 
eigenvalue,~$\Dsl \overline{\Gamma}\psi_{n}=-\lambda_{n}\overline{\Gamma}\psi_{n}$. 
Since eigenvectors with different eigenvalues are orthogonal, this implies
\begin{align}
\langle \psi_{n}|\overline{\Gamma}|\psi_{n}\rangle=0,
\end{align}
provided $i\Dsl\psi_{n}\neq 0$. Thus, 
the asymmetry in the spectrum of the Dirac operator is restricted
to its zero modes. In the limit
$\partial_{\mu}\alpha(x)\rightarrow 0$, eq.~(\ref{preanomaly}) only receives
contributions from the zero modes
\begin{align}
\alpha\lim_{M\rightarrow \infty}\langle\psi_{n}|\overline{\Gamma}
e^{-\frac{{\not D}^{2}}{M^{2}}}|\psi_{n}\rangle=\alpha
\sum_{\rm zero\,\,modes} \langle\psi_{n}|\overline{\Gamma}|\psi_{n}\rangle.
\end{align}
The sum on the right-hand side gives the number of zero modes of $D_{+}$ minus
the number of zero modes of $D_{-}=D_{+}^{\dagger}$ and is therefore equal to the index of the Weyl operator.
Applying the Atiyah-Singer index theorem, we obtain the final result for the integrated anomaly
\begin{align}
\int (dx) \langle \nabla_{\mu}j_{5}^{\mu}\rangle = 2\int_{M}
[\widehat{A}(M){\rm ch}(F)]_{\rm vol},
\end{align}
This expression of the axial anomaly in valid 
in $2n$~dimensions in the presence of a gravitational background. 
In the four-dimensional case, we can use the explicit expressions  
given in eq.~\eqref{eq:indD+d=4} to write
\begin{align}
\int_{M}(dx) \langle \nabla_{\mu} j_{5}^{\mu} \rangle=
-\frac{1}{(2\pi)^{2}}
\int_{M}\left({\rm tr\,}F^{2}-\frac{r}{24}\,{\rm tr\,}R^{2}\right).
\end{align}
This results agrees with eq.~(\ref{anomaly1}) in the flat space-time limit,~$R=0$.

We are now ready to face the computation of local gauge and
gravitational anomalies using the Atiyah-Singer index theorem for
families of operators. Once again we start with the Euclidean fermion effective action
\begin{align}
e^{-\Gamma[A]_{\rm eff}}=\int {\cal D}\psi {\cal D}\overline{\psi}
\exp{\left(-\int d^{2n}x\,\overline{\psi}\,iD_{+}\,\psi\right)},
\label{eq:effaction_general_2n+2}
\end{align}
where~$D_{+}=\Dsl P_{+}$. There is however a problem in making sense this
functional integral since, as we saw in eq.~\eqref{eq:D_pm->S_pmxE}, the operator~$D_{+}$ does not map the space
of positive chirality spinors into itself.
This means that there is no well-posed eigenvalue problem,
$D_{+}\psi=\lambda \psi$, and no natural definition of the
functional integral in terms of the determinant of $D_{+}$. To solve this problem,
we are going to introduce a new operator~$i\widehat{D}$ with a well defined eigenvalue problem and whose
(properly regularized) determinant can be identified with the functional integral of interest
\begin{align}
\int {\cal D}\psi {\cal D}\overline{\psi}
\exp{\left(-\int d^{2n}x\,\overline{\psi}\,iD_{+}\,\psi\right)}=\det\,\widehat{D}.
\label{eq:detD=effaction_general_2n+2}
\end{align}
Moreover, since we know that the anomaly
comes from the imaginary part of the effective action (see
above), $|\det\,\widehat{D}|$ has to be gauge invariant. We also have to make
sure that the perturbative expansion for $\widehat{D}$ is the same as that
for~$D_{+}$.

We are going to work with the operator~\cite{LAG-Ginsparg2}
\begin{align}
\widehat{D}=\left(
\begin{array}{cc}
  0	    &	D_{+} \\
 \dsl_{-} &   0
\end{array}
\right),
\end{align}
which satisfies all the properties listed above. For example, 
\begin{align}
|\det\,\widehat{D}|^{2}=\det\,(\widehat{D}^{\dagger}D)=
\det(\dsl_{+}\dsl_{-})\,\det(D_{+}D_{-})
\end{align}
where $\det(\dsl_{+}\dsl_{-})$ does not depend of the gauge fields
and therefore is gauge invariant. Since $\det(D_{+}D_{-})=\det(i\Dsl)$ 
[see eq.~\eqref{eq:iD=(0D-D+0)}], we conclude that~$|\det\,\widehat{D}|$ is gauge
invariant.

The computation of gauge and gravitational anomalies proceed by using the
Atiyah-Singer index theorem for a Dirac operator in $2n+2$ dimensions~\cite{LAG-Ginsparg2}. 
Let us choose appropriate boundary conditions to compactify effectively the $2n$-dimensional 
Euclidean space-time to~$S^{2n}$, 
and consider a one-parameter family of gauge
transformations
$g(\theta,x):S^{1}\times S^{2n} \rightarrow G$ with boundary
conditions $g(0,x)=g(2\pi,x)=1$. We then define the
family of connections
\begin{align}
A^{\theta}\equiv g^{-1}(\theta,x)(A+d)g(\theta,x),
\end{align}
where $A$ is a reference connection chosen such that
the associated Dirac operator has no zero modes,~$\det i\Dsl(A)\neq 0$.
Defining the corresponding family of operators $\widehat{D}(A^{\theta})$ and
using that~$|\det\widehat{D}(A^{\theta})|$ is invariant, we can
write
\begin{align}
\det\widehat{D}(A^{\theta})=\left[\det i\Dsl(A)\right]^{\frac{1}{2}}
e^{iw(A,\theta)},
\end{align}
showing that, as expected, the anomaly is associated with the phase of the determinant.
Combining eqs.~\eqref{eq:effaction_general_2n+2} and~\eqref{eq:detD=effaction_general_2n+2}
we relate~$\det\widehat{D}$ to fermion effective action, whose variation under an infinitesimal
change in the $\theta$~parameter is given by
\begin{align}
\delta\Gamma[A^{\theta}]_{\rm eff}=-i\frac{\partial w(A,\theta)}{\partial \theta}
\delta\theta,
\end{align}
Since~$A$ and~$A^{\theta}$ are gauge equivalent, this relation provides connection of the gauge anomaly 
with the derivative of $w(\theta,A)$. The function~$\exp{[iw(\theta,A)]}$ defines a map~$S^{1}\rightarrow S^{1}$
classified by its winding
number
\begin{align}
\frac{1}{2\pi}\int_{0}^{2\pi}d\theta\,\frac{\partial
w(\theta,A)}{\partial \theta}=m.
\label{eq:winding_number_general}
\end{align}

\begin{figure}[t]
\centerline{\includegraphics[scale=0.40]{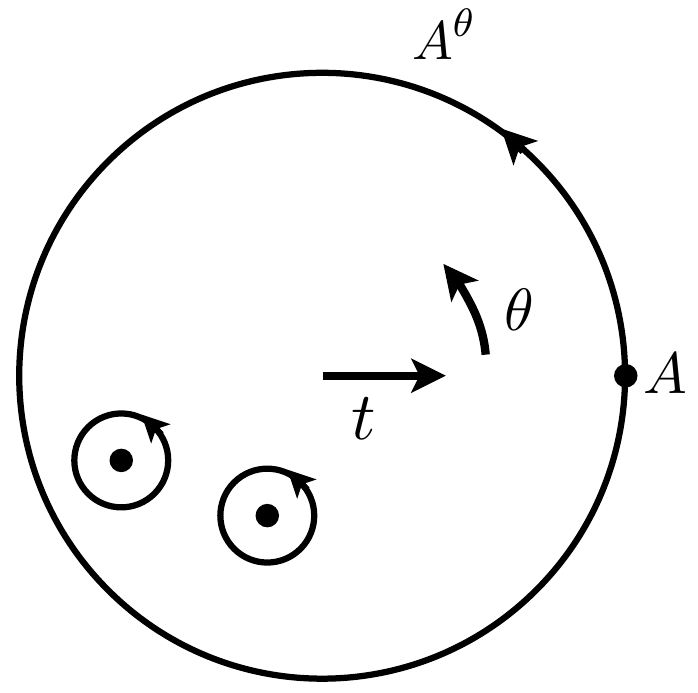}}
\caption{The disc ${\cal D}$ which parametrizes the two-parameter family
of gauge fields $A^{t,\theta}$ with polar coordinates $t,\theta$}
\label{fig-4}
\end{figure}

The anomaly is thus identified with the winding number density,
and it is obtained from the index theorem.
Let us define the two-parameter family of connections
\begin{align}
A^{t,\theta}=tA^{\theta},
\end{align}
where $t\in [0,1]$. This defines a disc ${\cal D}$ in the space
of connections whose boundary is~$A^{\theta}$ (see fig.~\ref{fig-4}).
Unlike~$\det D(A^{\theta})$, the determinant of $D(A^{t,\theta})$ may
vanish in a number of points inside the disk, 
since $A^{t,\theta}$ and $A$ are not related by a gauge
transformation for $0\leq t <1$. Thus, deforming the boundary~$S^{1}=\partial\mathcal{D}$ 
towards the interior, the total winding number equals the signed sum of 
all winding numbers associated with each of
the internal zeroes of the determinant. Moreover, it is possible to
show that these winding numbers are equal to $\pm 1$, coinciding with
the chirality of the Dirac operator eigenfunction whose eigenvalue vanishes at
that point~\cite{LAG-Ginsparg2}. These arguments
lead to identifying the winding
number~\eqref{eq:winding_number_general} 
with the index of the Dirac operator~$\Dsl_{2n+2}\equiv D(A^{t,\theta})$ 
defined on~$S^{2n}\times {\cal D}$
\begin{align}
{\rm ind\,}\Dsl_{2n+2}=\frac{1}{2\pi}\int_{0}^{2\pi}d\theta\,\frac{\partial
w(\theta,A)}{\partial \theta}\;.
\label{localdensity}
\end{align}

To write an explicit expression of the index appearing in the left hand side of this equation, we could
to use the Atiyah-Pattodi-Singer index theorem \cite{Atiyah-Pattodi-Singer}, a generalization
of the Atiyah-Singer theorem to manifolds with a boundary. Here, however,
we proceed in a different way. Let us consider the closed manifold~$S^{2}\times S^{2n}$
and slice~$S^{2}$ into its two hemispheres $S^{2}_{\pm}$. We set coordinates $(t,\theta)$ in the 
upper hemisphere and consider the gauge field
\begin{align}
{\cal A}(x,t,\theta)=A^{t,\theta}+tg^{-1}d_{\theta}g,
\end{align}
where~$d_{\theta}$ denotes the exterior differentiation with respect to $\theta$, i.e.
$d_{\theta}=d\theta \partial/\partial \theta$.
In the lower hemisphere, on the other hand, we take coordinates $(s,\theta)$ while the
gauge field is simply given by ${\cal A}(x,s,\theta)=A(x)$. We see that at the equator~$s=t=1$ 
both connections are related by a gauge transformation and we have a well-defined bundle
over~$S^{2n}\times S^{2}$. Applying now the Atiyah-Singer index theorem to this manifold, we have
\begin{align}
{\rm ind\,}[i\Dsl_{2n+1}({\cal A})]&=
\int_{S^{2}\times S^{2n}} [{\rm ch\,}({\cal
F})]_{\rm vol}  \nonumber \\[0.2cm]
&=\frac{1}{(n+1)!}\left(\frac{i}{2\pi}\right)^{n+1}
\int_{S^{2}\times S^{2n}}{\rm tr\,}({\cal F}^{n+1}),
\end{align}
where ${\cal F}$
\begin{align}
{\cal F}=(d+d_{\theta}+d_{t}){\cal A}+{\cal A}^{2},
\label{eq:AStoAPS1}
\end{align}
is the curvature associated with the connection~${\cal A}$
and~$d_{t}$ is defined analogously to~$d_{\theta}$.

Next, we split the integral in~\eqref{eq:AStoAPS1} into two integrals over~$S^{2n}\times S_{\pm}^{2}$.
Using then eq.~\eqref{Chern-Simons} with~$\omega_{0}=0$, we find that ${\rm tr\,}{\cal
F}^{n+1}=dQ_{2n+1}({\cal A})$ where $Q_{2n+1}({\cal A})$ is the
$2n+1$-Chern-Simons form. Thus, using the Stokes theorem, we write
the index of $\Dsl_{2n+2}$ as the difference
\begin{align}
{\rm ind\,}[i\Dsl_{2n+1}({\cal A})]&=
\frac{1}{(n+1)!}\left(\frac{i}{2\pi}\right)^{n+1}
\left[\int_{S^{1}\times S^{2n}} Q_{2n+1}({\cal A})\Big|_{t=1}
\right. \nonumber \\[0.2cm]
&-\left. \int_{S^{1}\times S^{2n}} Q_{2n+1}({\cal
A})\Big|_{s=1}\right].
\end{align}
To connect with the anomaly we need the local
winding density $d_{\theta}w(\theta,A)$. Let us notice that
the second integral vanishes, because there~$\mathcal{A}=A$ and we would need 
a~$d\theta$ to soak up the integral over~$S^{1}$. Then we
write
\begin{align}
{\rm ind\,}[i\Dsl_{2n+1}({\cal A})]=
\frac{1}{(n+1)!}\left(\frac{i}{2\pi}\right)^{n+1}
\int_{S^{1}\times S^{2n}} Q_{2n+1}(A^{\theta}+\widehat{v},F^{\theta}),
\end{align}
where $\widehat{v}=g^{-1}d_{\theta}g$ and $F^{\theta}$ is the gauge
curvature associated with $A^{\theta}$. It is now
straightforward to get the local winding density
\begin{align}
id_{\theta}w(A,\theta)=\frac{i^{n+2}}{(2\pi)^{n}(n+1)!}\int_{S^{2n}}
Q^{1}_{2n}(\widehat{v},A^{\theta},F^{\theta}),
\label{eq:gauge_anomaly_winding}
\end{align}
where $Q^{1}_{2n}(\widehat{v},A^{\theta},F^{\theta})$ is the term 
in~$Q_{2n+1}(\widehat{v},A^{\theta},F^{\theta})$ linear in $\widehat{v}$.
This form of the anomaly verifies the Wess-Zumino consistency
conditions. 

The $2n$-form on the right-hand side of eq.~\eqref{eq:gauge_anomaly_winding}
appears in the so-called Stora-Zumino descent equations~\cite{Zumino83}, that we derive next. 
Let us consider a
family of gauge transformations~$g(\theta^{\alpha},x)$ parametrized by
a collection of angles $\theta^{\alpha}$ and define
\begin{align}
\overline{A}&=g^{-1}(A+d)g, \nonumber \\[0.2cm]
\widehat{v}&=g^{-1}\delta g\;,
\end{align}
where~$\delta$ is the differential with respect to the parameters
$\theta^{\alpha}$. The following identity, known in the literature as the Russian formula, is easy to prove
(see, e.g.,~\cite{LAG-Ginsparg1})
\begin{align}
(d+\delta)Q_{2n+1}(\overline{A}+\widehat{v},\overline{F})=dQ_{2n+1}(\overline{A},\overline{F}),
\label{N-parameter}
\end{align}
with~$\overline{F}$ the curvature associated with $\overline{A}$. On the other hand,
the Chern-Simons form~$Q_{2n+1}(\overline{A}+\widehat{v},\overline{F})$ admits the following expansion in 
powers of $\widehat{v}$
\begin{align}
Q_{2n+1}(\overline{A}+\widehat{v},\overline{F})=
Q_{2n+1}^{0}(\overline{A},\overline{F})+Q_{2n}^{1}(\widehat{v},\overline{A},\overline{F})+\ldots
+Q_{0}^{2n+1},
\end{align}
where the superscript of the $Q$'s indicates its order in~$\widehat{v}$. Substituting it into
eq.~(\ref{N-parameter}), we arrive at the Stora-Zumino descent equations
\begin{align}
\delta Q_{2n+1}^{0}+dQ_{2n}^{1}&=0, \nonumber \\[0.2cm]
\delta Q_{2n}^{1} + dQ_{2n-1}^{2} &=0, \nonumber \\[0.2cm]
&\vdots \nonumber \\
\delta Q_{1}^{2n}+dQ_{0}^{2n+1} &= 0, \\[0.2cm]
\delta Q_{0}^{2n+1}&=0.
\nonumber 
\end{align}
In fact, the second identity is equivalent to the Wess-Zumino consistency
condition.

In the previous analysis, we have confined our attention to gauge anomalies. 
We would like to find a similar procedure to compute gravitational and mixed anomalies.
To do this, we need the Atiyah-Singer index theorem for
families of elliptic operators \cite{Atiyah-Singer-2}.

Let us consider a family of elliptic operators $D_{p}$ with $p$ lying in
some manifold $P$. As long as the
operator varies smoothly over $P$, the index of $D_{p}$ is constant over the manifold $P$, 
since it is invariant under continuous deformations. Nonetheless, this does not have to be true for
${\rm dim\,}{\rm ker\,}D_{p}$ and ${\rm dim\,}{\rm ker\,}D_{p}^{\dagger}$ separately, since 
${\rm ker\,}D_{p}$ and ${\rm ker\,}D_{p}^{\dagger}$ are not necessarily 
well-defined vector bundles over the parameter space $P$. However, in the context
of $K$-theory for $P$ \cite{Atiyah}, we can have a well-defined
vector bundle over $P$ defined by
\begin{align}
{\rm Ind\,}D_{p}={\rm ker\,}D_{p} \ominus {\rm ker\,}D_{p}^{\dagger},
\end{align}
where~${\rm ind\,}D_{p}={\rm ch}_{0}({\rm Ind\,}D_{p})$.
Using the Atiyah-Singer index theorem for families of operators, we
write~\cite{Alvarez-Singer-Zumino}
\begin{align}
{\rm ch}({\rm Ind\,}D)=\int_{M_{2n}}\widehat{A}(Z){\rm ch}(V),
\label{A-Sfamily}
\end{align}
where $Z$ is a fiber space with base $P$ and fiber $M_{2n}$, and $V$ is
the vector bundle. In the case of pure gravity, $V$ is of 
geometric origin: e.g.,~$V=0$ for a spin-$\frac{1}{2}$ fermion $V=0$ and
$V=TM_{2n}$, the tangent bundle of $M_{2n}$ for a Rarita-Schwinger field. 
In the case of mixed anomalies $V$ also contains the gauge bundle. Actually,
our calculation of the gauge anomalies presented above used a version of
the Atiyah-Singer index theorem for a family of operators, 
where~$M_{2n}=S^{2n}$,~$P=S^{2}$, and the vector bundle was constructed using
the gauge transformation~$g(\theta,x)$.

We now turn to gravitational anomalies. In presence of gravity,
we have to make sure that the quantum theory is invariant under
such classical symmetries as coordinate diffeomorphisms
and local Lorentz transformations. In principle, we then may expect  
Einstein and Lorentz anomalies. In the
first case a breakdown of diffeomorphism invariance leads to
a non-conservation of the energy-momentum tensor
\begin{align}
\langle\nabla_{\mu\nu}T^{\mu}\rangle\neq 0.
\end{align}
Lorenz anomalies, on the other hand, are signaled by a non-zero
variation of the effective action under 
local Lorentz transformation, $\delta
e^{m}_{\,\,\,\,\mu}=\alpha^{m}_{\,\,\,\,n}e^{n}_{\,\,\, \mu}$,
\begin{align}
\delta_{\alpha}\Gamma_{\rm eff}&=-\int (dx)\,\alpha^{m}_{\,\,\,\,n}e^{n}_{\,\,\,\mu}
\frac{\delta \Gamma_{\rm eff}}{\delta e^{m}_{\,\,\,\,\mu}} \nonumber \\[0.2cm]
&=-\int
(dx)\,e\,\alpha^{mn}\langle T_{mn} \rangle.
\end{align}
Since for a Lorentz transformation $\alpha_{(ab)}=0$, we have that $\delta_{\alpha}\Gamma_{\rm eff}\neq
0$ implies $\langle T_{[ab]} \rangle \neq 0$. It can be shown however
that Einstein and Lorentz anomalies~\cite{Bardeen-Zumino,LAG-Ginsparg1} are related. 
It is always possible to add a local term to the action to switch from one type of anomaly to the
other. Here, we prefer to think of gravitational anomalies in terms of Einstein anomalies.

After these general remarks we proceed to give the prescription for the computation of
gravitational anomalies. Following the same argument leading
to the gauge anomaly, the gravitational anomaly is expressed in terms of 
a local winding number density obtained
from the characteristic polynomial (\ref{A-Sfamily}) of some
$(2n+2)$-dimensional Dirac operator. For
spin-$\frac{1}{2}$ fermions, the characteristic polynomial is given by
\begin{align}
I_{1/2}=\prod_{i} \frac{x_{i}/2}{\sinh{(x_{i}/2)}},
\end{align}
while for a Rarita-Schwinger field, we have
\begin{align}
I_{3/2}=\prod_{i}\frac{x_{i}/2}{\sinh{(x_{i}/2)}}
\Big[{\rm tr\,}\big(e^{R/2\pi}-1\big)+4k-3\Big].
\end{align}
Finally, the result for the self-dual antisymmetric tensor is
\begin{align}
I_{\rm s.d.}=-\frac{1}{8}\prod_{i}\frac{x_{i}}{\tanh{x_{i}}},
\end{align}
where the minus sign indicates the bosonic nature of the
antisymmetric field. In all previous expressions $x_{i}$ are the skew-eigenvalues of $R/2\pi$
defined [see eq.~\eqref{eq:skew_eigenvalues_Omega}]. To compute mixed anomalies,
we have to multiply these polynomials by ${\rm ch}({\cal F})$, where ${\cal F}$
is the field strength of the $(2n+2)$-dimensional gauge field.

\subsection{Examples: the Green-Schwarz cancellation mechanism and Witten's
$\mbox{SU(2)}$ global anomaly}
\label{sec:GSandWSU2}

As a first application of the technology we have presented, we discuss the
non-trivial cancellation of gravitational anomalies for some
supergravity theories~\cite{LAG-Witten}. In two dimensions the
relevant characteristic four-form polynomials are
\begin{align}
I_{1/2}&=-\frac{1}{24}p_{1}, \nonumber \\[0.2cm]
I_{3/2}&=\,\,\frac{23}{24}p_{1},  \\[0.2cm] 
I_{\rm s.d.}&=-\frac{1}{24}p_{1}.
\nonumber
\end{align}
There are many ways to cancel anomalies. Notice in particular that 
in~$d=2$ a ``self-dual antisymmetric tensor field'' is just a right-moving scalar.
The identity $I_{\rm s.d.}=I_{1/2}$ then reflects its equivalence to a positive
chirality fermion because of
two-dimensional bosonization~\cite{Coleman}. 

On the other hand, in six dimensions we find
\begin{align}
I_{1/2}&=\frac{1}{5760}(7p_{1}^{2}-4p_{2}),\nonumber \\[0.2cm] 
I_{3/2}&=\frac{1}{5760}(275p_{1}^{2}-980p_{2}), \\[0.2cm] 
I_{\rm s.d.}&=\frac{1}{5760}(16p_{1}^{2}-112p_{2}), 
\nonumber
\end{align}
and the minimal cancellation occurs for
\begin{align}
21I_{1/2}-I_{3/2}+8I_{\rm s.d.}=0.
\end{align}
Thus, we must have $21$~Weyl spin-$\frac{1}{2}$ fermions, $1$~gravitino
with the opposite helicity, and $8$~self-dual antisymmetric tensors. 

Finally, in $d=10$ we have
\begin{align}
I_{1/2}&=\frac{1}{967680}(-31p_{1}^{3}+44p_{1}p_{2}-16p_{3}), \nonumber \\[0.2cm]
I_{3/2}&=\frac{1}{967680}(225p_{1}^{3}-1620p_{1}p_{2}+7920p_{3}), \\[0.2cm]
I_{\rm s.d.}&=\frac{1}{967680}(-256p_{1}^{3}+1664p_{1}p_{2}-7936p_{3}),
\nonumber
\end{align}
and the minimal solution is found to be \cite{LAG-Witten}
\begin{align}
-I_{1/2}+I_{3/2}+I_{\rm s.d.}=0.
\end{align}
The gravitational anomaly then cancels when we have a spin-$\frac{1}{2}$ Weyl
fermion, a gravitino field of opposite chirality, and a self-dual tensor
field. These are exactly the fields appearing in the~$N=2$ chiral supergravity multiplet in
$d=10$~\cite{LAG-Witten}.

As a further example, we discuss the mechanism of
anomaly cancelation in type-$I$ superstring theory found by Green and
Schwarz~\cite{Green-Schwarz} (here we follow the presentation
of ref.~\cite{Green-Schwarz-West}). The low-energy field theory of type-$I$
superstrings is $N=1$ Yang-Mills coupled to $N=1$ SUGRA in ten
dimensions. The $N=1$ SUGRA multiplet contains a graviton, 
a left-handed Weyl-Majorana gravitino,
a right-handed Weyl-Majorana dilatino, an antisymmetric
tensor, and a dilaton. In addition, the super Yang-Mills
multiplet includes a non-Abelian gauge field and its
Weyl-Majorana gaugino, both in the adjoint representation of the
gauge group~$G$. Shifting all gravitational anomalies to Lorentz anomalies, 
we obtain the twelve-form characteristic polynomial:
\begin{align}
I_{12}&=-\frac{1}{15}{\rm tr\,}F^{6}+\frac{1}{24}{\rm tr\,}F^{4}\,{\rm tr\,}R^{2}
\nonumber \\[0.2cm]
&-\frac{1}{960}{\rm tr\,}F^{2}\left[5({\rm tr\,}R^{2})^{2}+4{\rm tr\,}R^{4}\right]+
\frac{N-496}{7560}{\rm tr\,}R^{6} \\[0.2cm]
&+\left(\frac{N-496}{5760}+\frac{1}{8}\right)
{\rm tr\,}R^{4}\,{\rm tr\,}R^{2}+\left(\frac{N-496}{13824}+\frac{1}{32}\right)
({\rm tr\,}R^{2})^{3}, \nonumber
\end{align}
where traces over gauge indices are in the adjoint representation
and $N$ is the dimension of the gauge group. This polynomial determines
the anomalies of the low-energy field theory of
type-$I$ superstrings.

The key observation behind the cancellation of anomalies discovered by Green and
Schwarz is that all anomalies in the type-I superstring can be cancelled by adding 
an appropriate term to the action
provided $I_{12}$ admits the factorization
\begin{align}
I_{12}=({\rm tr\,}R^{2}+k\,{\rm tr\,}F^{2})I_{8},
\label{eq:GS_factorization}
\end{align}
where $I_{8}$ is an
invariant eight-form constructed in terms of $F$ and $R$. 
An obstruction to this is
the presence of terms proportional to~${\rm tr\,}R^{6}$
and~${\rm tr\,}F^{6}$. An obvious way to get rid of the first trace is to restrict to 
gauge groups with dimension~$N=496$. As to~${\rm tr\,}F^{6}$, we search for groups in which
this trace is not an independent Casimir
\begin{align}
{\rm tr\,}F^{6}=\alpha\,({\rm tr\,}F^{2})\,({\rm tr\,}F^{4})+\beta({\rm tr\,}F^{2})^{3}.
\end{align}
In fact, the factorization~\eqref{eq:GS_factorization} exists for~$k=-1/30$ and
\begin{align}
{\rm tr\,}F^{6}=\frac{1}{48}({\rm tr\,}F^{2})\,({\rm tr\,}F^{4})-
\frac{1}{14400}({\rm tr\,}F^{2})^{3},
\label{eq:factorizationtrF6}
\end{align} 
so $I_{8}$ is given by
\begin{align}
I_{8}&=\frac{1}{24}{\rm tr\,}F^{4}-\frac{1}{7200}({\rm tr\,}F^{2})^{2}-\frac{1}{240}
({\rm tr\,}F^{2})\,({\rm tr\,}R^{2})
+\frac{1}{8}{\rm tr\,}R^{4}+\frac{1}{32}({\rm tr\,}R^{2})^{2}.
\end{align}
There are only three possible groups with dimension~$N=496$ and 
satisfying eq.~\eqref{eq:factorizationtrF6}: $\mbox{SO}(32)$, $\mbox{E}_{8}\times \mbox{E}_{8}$ or 
$\mbox{E}_{8}\times \mbox{U}(1)^{248}$. It is a remarkable feature of the Green-Schwarz mechanism
that by simply imposing the cancellation of gauge and gravitational
anomalies the possible gauge groups are narrowed to just three possible choices, of which
type-I superstring theory can only accommodate~$\mbox{SO}(32)$. 
Soon after the Green-Schwarz anomaly cancellation was discovered, 
the heterotic string~\cite{Pricenton-quartet} was formulated with both~$\mbox{SO}(32)$ 
and $\mbox{E}_{8}\times
E_{8}$ gauge groups. It is however the latter theory that seems to have better prospects to make contact
with the SM at low energies.

After all local anomalies have been properly cancelled, we still have to make
sure that the theory is not afflicted by global gauge or gravitational
anomalies. As it was explained above, global anomalies lead to a breakdown of
gauge or diffeomorphism invariance under transformations that do not lie
in the connected component of the identity.
The best known example of this kind of anomalies was discovered by Witten~\cite{WittenSU(2)} 
and affect gauge theories with an odd number of left-handed~$\mbox{SU}(2)$ doublets. 

The basic mathematical condition for the existence of Witten's anomaly is
that the fourth homotopy group of the gauge group is non-trivial. 
Among semisimple groups, this is only the case 
for~$\mbox{Sp}(n)$, for which~$\pi_{4}[\mbox{Sp}(n)]=\mathbb{Z}_{2}$. The
most interesting case for phenomenology is indeed $\mbox{Sp}(1)=\mbox{SU}(2)$.
Let us consider four-dimensional Euclidean space and impose 
boundary conditions to effectively compactify it to~$S^{4}$. Gauge trasformations~$U(x)$ are 
then mappings~$U:S^{4}\rightarrow \mbox{SU}(2)$ classified 
by~$\pi_{4}[\mbox{SU}(2)]=\mathbb{Z}_{2}$, implying the existence of gauge
transformations that cannot be continuously
deformed to the identity. 

To fix ideas, let us consider the theory of 
a single $\mbox{SU}(2)$ left-handed doublet. Since for a
doublet of Dirac fermions we have
\begin{align}
\int {\cal D}\psi {\cal D}\overline{\psi} \exp\left(-\int d^{4}x\,\overline{\psi}\,i\Dsl 
\psi\right)=\det\,(i\Dsl),
\end{align}
the value of the relevant functional integral for Weyl fermions is given by the square root
of the determinant of the Dirac operator
\begin{align}
\int ({\cal D}\psi {\cal D}\overline{\psi})_{\rm Weyl}  e^{-\int
d^{4}x\,\overline{\psi}i{\not D}\psi}=[\det\,(i\Dsl)]^{\frac{1}{2}}.
\end{align}
The problem however is the sign to the square root.
Since we are only interested in relative signs we may, for
example, choose the positive branch of the square root. However, for
this choice to be consistent, the sign must be preserved by
gauge transformations. This is evidently the case for infinitesimal gauge
transformations, but not for transformations that do not lie in the connected
component of the identity
\begin{align}
[\det\,i\Dsl(A_{\mu})]^{\frac{1}{2}}=-
[\det\,i\Dsl(A_{\mu}^{U})]^{\frac{1}{2}}.
\label{Sign-change}
\end{align}
This means trouble, because $A_{\mu}$ and $A_{\mu}^{U}$
are not continuously connected and there is no consistent way of excluding
$A_{\mu}^{U}$ from the path integral. As a consequence,
the partition function vanishes because the contribution
from $A_{\mu}$ is cancelled by that from $A_{\mu}^{U}$.

Equation~(\ref{Sign-change}) can be understood in terms of spectral flow.
We have seen that the non-zero modes of the Dirac operator~$i\Dsl$ are paired. 
We may then define $[\det\,i\Dsl]^{\frac{1}{2}}$
as the regularized product of the positives eigenvalues of the Dirac
operator
\begin{align}
[\det i\Dsl]^{\frac{1}{2}}=\prod_{\lambda_{n}>0}\lambda_{n}.
\label{det}
\end{align}
Consider now a family of connections~$A_{\mu}^{t}=tA_{\mu}^{U}+(1-t)A_{\mu}$, with $t\in [0,1]$, 
continuously connecting $A_{\mu}$ with
$A_{\mu}^{U}$.
We can study how the eigenvalues of~$i\Dsl(A_{\mu}^{t})$ flow as a
function of~$t$. The condition that~$\det i\Dsl$ changes
sign means that as~$t$ runs from~0 to~1 there is an odd number of 
eigenvalues~$\lambda_{n}(t)$ crossing zero. This result
follows from a slightly exotic version
of the Atiyah-Singer index theorem, the mod two index theorem, applied to
the five-dimensional Dirac operator
\begin{align}
{\cal D}=i\overline{\Gamma}\frac{\partial}{\partial \tau}+\Dsl(A),
\end{align}
where the parameter~$\tau$
is the fifth coordinate of a space~$S^{4}\times\mathbb{R}$. The fifth component of
the gauge field vanishes, $A_{\tau}(x,\tau)=0$, while the other ones satisfy boundary conditions
$A_{\mu}(x,\tau=-\infty)=A_{\mu}(x)$ and
$A_{\mu}(x,\tau=+\infty)=A_{\mu}^{U}(x)$, adiabatically interpolating between the 
four-dimensional field configurations~$A_{\mu}$ and~$A_{\mu}^{U}$. It can be seen that 
the operator ${\cal D}$ is real
and antisymmetric, so its non-zero eigenvalues are purely imaginary
and come in complex conjugate pairs. In fact, the number of zero modes modulo two is
a topological invariant that, using the mod $2$
index theorem, Witten computed to be one. This means that as we interpolate between  $A_{\mu}$ 
and~$A_{\mu}^{U}$ there is an odd number of eigenvalues of the four-dimensional Dirac
operator crossing zero. It is therefore not possible to assign a
definite sign to the square root in eq.~\eqref{det}. 

The upshot is that~$\mbox{SU}(2)$ gauge theory with an odd number of fermion
doublets has a global gauge anomaly and its physical Hilbert space is empty. Luckily, the 
SM contains an even number of weak~$\mbox{SU}(2)$ doublets per family, so disaster is avoided.

\section{String theory I. Bosonic string}
\label{sec:bosonic_string}

A basic lesson to be learned from our journey in sec.~\ref{sec:QFTapproach} through the
various approaches to quantize gravity is that all of them are
afflicted with problems. The situation with classical GR is somewhat reminiscent of Fermi's
theory of the weak interactions. It was able to explain a variety of low
energy phenomena, but it failed to be renormalizable and 
violated unitarity at high energies. The cure
came with the SM, based on completely different dynamical principles. Similarly,
it does not make much sense to regard Einstein's gravity as a
fundamental theory valid all the way from cosmological scales down to the
Planck length, spanning more than sixty orders of magnitude.
In view of the difficulties encountered it seems natural
to think of GR as an effective theory, so carrying out the program of 
quantizing gravity should start with identifying the appropriate framework 
capable of producing a quantum theory of space-time which in the long distance 
limit agrees with Einstein's theory. As it should be clear by now, such a
theory would provide profound insights into many of the riddles of the
SM like the origin of mass, symmetry breaking, and the chiral nature of
quarks and leptons. In our brief study of anomalies we have seen how when the space-time
dimension goes beyond four there exist very restrictive constraints to be 
satisfied by a chiral theory. For instance, in $d=10$ we are essentially left
with the low-energy spectrum of the heterotic
string~\cite{Pricenton-quartet}.

So far, the only candidate unifying at high energies gravity with the other
interactions is string theory. To be sure, there are still many
puzzles and difficulties to be clarified in string theory, but it should be 
also mentioned that many of the problems encountered in other approaches to
quantum gravity based on local field theories disappear in this
context. There are strong arguments (as well as explicit computations) showing
the ultraviolet finiteness of superstring theories. There are quite
interesting results such as duality or its finite temperature properties 
(see sec.~\ref{sec:R-duality})
indicating that string theory defined on spaces of different sizes,
and even different topologies, are completely equivalent. For instance, 
strings do not distinguish a circle of radius~$R$ from one with 
radius~$\alpha'/R$, where~$\sqrt{\alpha'}$ defines certain string fundamental length scale.

Although string theory contains a fundamental length it still has
local interactions. This is something that cannot be achieved in the
framework of local field theory. The heterotic
string also shows that string theory can in principle accommodate the
quantum numbers and interactions of the SM. Unfortunately, our knowledge of string
theory is too rudimentary to have a glimpse on its physical basis. 
Nevertheless, the results achieved so far give reasons to be optimistic.
It is likely that many of the new ingredients brought about by string
theory will appear in the correct theory of quantum gravity.

In this part of the lectures we begin with an exploration  
of string theory focused on its general properties and
the formalism developed for the computation of string scattering
amplitudes to any order of perturbation theory. Along the way we are going
to find how the Einstein equations emerge in string theory, 
study the simplest version of duality, and describe the
geometrical interpretation of string infinities and physical states. 
For the time being, to make the discussion simpler, our attention will be 
centered on the bosonic string. We will deal with fermionic strings later on
in the lectures.

\subsection{Bosonic string theory}
\label{sec:bosonic_string_theory}

We review next the basic features of bosonic string 
theory~\cite{Green-Schwarz-Witten,Kaku,Lust-Theisen}. To get some inspiration on
how to describe a one-dimensional object, we begin with the well-known
case of a free relativistic particle of mass~$m$ propagating in Minkowski space-time. Its action is
built from the simplest Lorentz invariant we can
associate
with a particle's trajectory, the length of its world-line
\begin{align}
S[x^{\mu}(s)]=-m\int_{a}^{b}
ds\,\sqrt{\eta_{\mu\nu}\dot{x}^{\mu}(s)\dot{x}^{\nu}(s)}\;.
\label{particle}
\end{align}
where~$x^{\mu}(s)$ parametrizes the world-line. 

Strings, on the other hand, are one-dimensional
objects in a $d$-dimensional Minkowski space
(called its target space) whose points are parametrized by a coordinate~$\sigma$
running from~$0$ to~$\pi$. If a particle's time evolution is described by its one-dimensional
world-line, strings sweeps a two-dimensional surface in Minkowski space-time, known as 
the string's world-sheet, parametrized
by two coordinates~$(\sigma,\tau)$. The first one is the one already introduced
to label the points on the string, while $\tau$ plays a role analog to the world-line
parameter of the point particle. Thus, the string world-sheet is described by a 
set of~$d$ functions,~$X^{\mu}(\sigma,\tau)$, transforming as a vector under Lorentz
transformations in the target space.

To construct an action functional for the bosonic string, we do as with the 
relativistic particle and take the action to be proportional to the area
of the world-sheet swept out by the string. To write it we notice that the metric induced on the
world-sheet by the target space Minkowski metric (its pullback) is
\begin{align}
\Gamma_{ab}=\eta_{\mu\nu}\frac{\partial X^{\mu}}{\partial \sigma^{a}}
\frac{\partial X^{\nu}}{\partial \sigma^{b}},
\label{eq:pullback_WS}
\end{align}
where $a=0,1$ and $(\sigma^{0},\sigma^{1})=(\tau,\sigma)$. The world-sheet area is
then obtained by integrating
square root of minus the determinant of $\Gamma_{ab}$, so the string action is given by
\begin{align}
S[X^{\mu}]=T\int d^{2}\sigma
\sqrt{-\det\left[\eta_{\mu\nu}\frac{\partial
X^{\mu}}{\partial \sigma^{a}}\frac{\partial X^{\nu}}{\partial
\sigma^{b}}\right]},
\label{eq:nambu-goto}
\end{align}
where $T$ is
called the string tension and measures its mass/energy per unit length.
The action~\eqref{eq:nambu-goto} was first proposed by Y. Nambu and T. Goto
\cite{Nambu-Goto} in the early 1970s and it is known as the
Nambu-Goto action. By using the expression of the Minkowski metric, together with
the definitions
\begin{align}
\dot{X}^{\mu}&=\frac{\partial X^{\mu}}{\partial \tau}, \nonumber \\[0.2cm]
X'^{\mu}&=\frac{\partial X^{\mu}}{\partial \sigma},
\end{align}
we can rewrite the Nambu-Goto action as
\begin{align}
S[X^{\mu}(\tau,\sigma)]=\int d\tau d\sigma \sqrt{\dot{X}^{2} X'^{2}-
(\dot{X}\cdot X')^{2}}.
\label{Nambu-Goto}
\end{align}

Although eq.~(\ref{Nambu-Goto}) has a very simple
physical interpretation, its form is not very
convenient because of the presence of the square root. We would
like to find another action classically equivalent to
(\ref{Nambu-Goto}) but without its unpleasant features. The idea of
using auxiliary fields to simplify the Nambu-Goto action was proposed by Brink,
Di Vecchia, and Howe \cite{Brink} and by Deser and Zumino
\cite{Deser-Zumino}. The idea is is to introduce, besides the embeddings $X^{\mu}(\tau,\sigma)$,
a new independent non-dynamical field~$g_{ab}(\tau,\sigma)$ which is the intrinsic 
metric on the world-sheet and write
\begin{align}
S_{P}[X^{\mu},g_{ab}]=-\frac{T}{2}\int
d^{2}\sigma\sqrt{-g}\,g^{ab}\partial_{a}X^{\mu}\partial_{b}X^{\nu}
\eta_{\mu\nu}.
\label{Polyakov}
\end{align}
This form of the action was used by Polyakov~\cite{Polyakov} to
formulate string theory in terms of path integrals and for this reason is known as the Polyakov action.

It is worth-noticing the existence of a particle analog of the
action (\ref{Polyakov}). We introduce an
auxiliary field~$e(s)$, playing the role of the intrinsic one-dimensional metric on
the world-line, to write the action
\begin{align}
S=\frac{1}{2}\int ds\left[\frac{\dot{x}^{2}(s)}{e(s)}-m^{2}e(s)\right],
\end{align}
which is also classically equivalent to~(\ref{particle}). This action has the
additional advantage that it can be applied to the massless case, where~(\ref{particle}) is ill-defined.

The string tension~$T$ is usually rewritten in terms of a new constant~$\alpha'$
called the Regge slope, a relic from the times of the dual models of hadronic resonances~\cite{GVeneziano}
\begin{align}
T=\frac{1}{2\pi\alpha'}.
\end{align}
Neither~$T$ nor~$\alpha'$ have any effect on classical string
solutions, since they are overall factors in the action.
Not so in the quantum
theory. In the path integral approach, each string configuration is weighted by a phase
\begin{align}
\exp{\left(i\frac{T}{\hbar}S'\right)},
\end{align}
where $S'$ is the string action without the~$T$ prefactor.
From~$\alpha'$ and~$\hbar$ it is possible to construct a quantity with
dimensions of length (in units of $c=1$)
\begin{align}
\ell_{S}=\sqrt{\alpha'\hbar},
\label{eq:string_length}
\end{align}
setting a fundamental length in string theory. Together
with the vacuum expectation value of the dilaton (determining the
string coupling constant) it fixes all couplings in string
theory.

The action (\ref{Polyakov}) has a number of symmetries.
It is invariant under infinitesimal world-sheet
diffeomorphism $\sigma^{a}\rightarrow
\sigma^{a}+\xi^{a}$
\begin{align}
\delta X^{\mu}&=\xi^{a}\partial_{a}X^{\mu}, \nonumber \\[0.2cm]
\delta g_{ab}&=\nabla_{a}\xi_{b}+\nabla_{b}\xi_{a}
\end{align}
as well as under Weyl rescalings of the metric
\begin{align}
\delta g_{ab}&=2\Omega(\sigma)g_{ab}, \nonumber \\[0.2cm]
\delta X^{\mu}&=0.
\label{eq:Weyl_bosonic}
\end{align}
In addition to this, the action is also invariant under Poincar\'e transformations
in Minkowski target space, in infinitesimal form
\begin{align}
\delta X^{\mu}&=a^{\mu}_{\,\,\,\nu}X^{\nu}+b^{\mu}, \hspace{1cm}
a_{(\mu\nu)}=0, \nonumber \\[0.2cm]
\delta g_{ab}&=0.
\end{align}
Invariance under Weyl transformations
implies that the two-dimensional classical field theory described by
the action~(\ref{Polyakov}) is a conformal field theory. As we will see below, the
preservation of this invariance after quantization imposes strong
constraints on the theory.

The classical equations of motion from~(\ref{Polyakov}) give the two-dimensional
wave equation for $X^{\mu}$
\begin{align}
\nabla_{a}\nabla^{a} X^{\mu}=0,
\label{eq:bosonic_string_wave2d}
\end{align}
The world-sheet
energy-momentum tensor~$T_{ab}$ is given by
\begin{align}
T_{ab}=-\frac{2}{T}\frac{1}{\sqrt{g}}\frac{\delta S}{\delta g^{ab}},
\end{align}
so the equation of motion for~$g_{ab}$ yields
\begin{align}
T_{ab}\equiv\partial_{a}X\cdot\partial_{b}X-
\frac{1}{2}g_{ab}g^{cd}\partial_{c}X\cdot\partial_{d}X=0,
\label{constrainT}
\end{align}
where the dot indicates the contraction of the target space indices with the
Minkowski metric.
A look at eq.~\eqref{constrainT} shows that the trace of the
energy-momentum tensor is identically zero, even without applying the constraint~$T_{ab}=0$.
This is a consequence of the invariance of the
action under Weyl rescalings of the metric~\eqref{eq:Weyl_bosonic}.
Moreover, from the constraint~\eqref{constrainT} we obtain
\begin{align}
g_{ab}={2\Gamma_{ab}\over g^{cd}\Gamma_{cd}} \hspace*{1cm} \Longrightarrow \hspace*{1cm}
\sqrt{-g}={\sqrt{-\det\Gamma_{ab}}\over g^{cd}\Gamma_{cd}},
\end{align}
where~$\Gamma_{ab}$ is the pullback of the target space Minkowski metric on the world-sheet 
shown in eq.~\eqref{eq:pullback_WS}. Plugging this expression back into the 
action~\eqref{Polyakov} we recover the Nambu-Goto action, thus showing that both actions are 
classically equivalent. This, however, does not imply their equivalence at the quantum level.

The invariances of the theory allows a wide freedom in choosing a gauge for the
intrinsic metric $g_{ab}$. For example, using the fact that two-dimensional manifolds
are conformally flat, it is possible to find suitable coordinates on the word-sheet
such that the metric is brought locally to the form
\begin{align}
g_{ab}=e^{\phi(\sigma)}\eta_{ab},
\end{align}
with $\eta_{ab}$ the flat Minkowski metric on the world-sheet. This is known as
the conformal gauge.

Let us analyze now the propagation of a free closed string.
We begin by imposing the periodicity condition of the string
coordinates
\begin{align}
X^{\mu}(\tau,\sigma+\pi)=X^{\mu}(\tau,\sigma).
\end{align}
The world-sheet in this case is a cylinder~$S^{1}\times \mathbb{R}$ parametrized 
by~$0\leq\sigma< \pi$ and~$-\infty<\tau<\infty$.
Choosing the conformal gauge and taking light-cone coordinates in the
world-sheet, $\sigma^{\pm}=\tau\pm\sigma$, the action can be written as
\begin{align}
S=-\frac{T}{2}\int_{S^{1}\times \mathbb{R}} d^{2}\sigma\, \partial_{+}X^{\mu}\partial_{-}X^{\nu}
\eta_{\mu\nu}.
\end{align}
It is convenient now to make a Wick rotate the world-sheet time,~$\tau \rightarrow -i\tau$, and define
complex coordinates
\begin{align}
w&=2(\tau-i\sigma), \nonumber \\[0.2cm]
\overline{w}&=2(\tau+i\sigma).
\end{align}
The cylinder can be mapped now into the punctured complex
plane~$\mathbb{C}^{*}=\mathbb{C}-\{ 0\}$ by the conformal transformation
\begin{align}
z=e^{w}.
\end{align}
The field equations for $X^{\mu}$ take the simple form
\begin{align}
\partial_{z}\partial_{\overline{z}}X^{\mu}(z,\overline{z})=0,
\label{embedding}
\end{align}
while the constraint~\eqref{constrainT} reads
\begin{align}
\partial_{z}X\cdot \partial_{z}X&=0, \nonumber \\[0.2cm]
\partial_{\overline{z}}X\cdot \partial_{\overline{z}}X&=0,
\end{align}
It is easy now to write the general solution to eq.~(\ref{embedding}) as the
sum of a holomorphic and antiholomorphic function
\begin{align}
X^{\mu}(z,\overline{z})=X^{\mu}(z)+X^{\mu}(\overline{z}),
\end{align}
Writing the two functions as a Laurent series around the origin of~$\mathbb{C}^{*}$, we arrive
at the general solution describing the free propagation of a classical closed string
\begin{align}
X^{\mu}(z,\overline{z})=q^{\mu}-\frac{i}{2}\alpha'p^{\mu}\log{|z|^{2}}+i\sqrt{\alpha'\over 2}
\sum_{n\neq 0}
\frac{\alpha_{n}^{\mu}}{n}z^{-n}+i\sqrt{\alpha'\over 2}\sum_{n\neq
0}\frac{\overline{\alpha}_{n}^{\mu}}{n}\overline{z}^{-n}.
\label{expansion}
\end{align}
It is instructive to write this solution in the original~$(\tau,\sigma)$ coordinates, after
undoing the Wick rotation on~$\tau$
\begin{align}
X^{\mu}(\tau,\sigma)&=q^{\mu}+\alpha'p^{\mu}\tau
+i\sqrt{\alpha'\over 2}\sum_{n\neq 0}{\alpha_{n}\over n}e^{-in(\tau-\sigma)}
+i\sqrt{\alpha'\over 2}\sum_{n\neq 0}{\overline{\alpha}_{n}\over n}e^{-in(\tau+\sigma)}.
\label{eq:bosonic_expansion_sigma_tau}
\end{align}
It is not difficult to see that~$q^{\mu}$ is the position at~$\tau=0$ of the string center, while~$p^{\mu}$ 
is its momentum of the string. The two sums represent the vibrational modes travelling along the string
in opposite directions that we respectively label as right- and left-movers.

So far we have been concerned with classical strings. It is time now to turn to its quantization. 
These are three main strategies.

\paragraph{a) Light-cone gauge quantization.} We use light-cone coordinates in the target space
defined by~$(X^{\pm},X^{i})$, with $X^{\pm}=X^{0}\pm
X^{d-1}$ and $i=1,\ldots,d-2$. Then, using the residual freedom left after fixing the
conformal gauge, we set (see ref.~\cite{Green-Schwarz-Witten} for details)
\begin{align}
X^{+}=q^{+}+\sqrt{2\alpha'}p^{+}\tau.
\end{align}
In this gauge the constrains can be explicitly solved, so we can substitute them in the solution of
eq.~(\ref{embedding}) and eliminate all unphysical degrees of
freedom. The theory is then described only in terms of transverse modes.
It is necessary however to check that the quantum theory preserves
Lorentz invariance. To verify this we check
whether the Lorentz generators~$J^{\mu\nu}$ satisfy the right algebra. 
Classically we would have for example the Poisson bracket~$\{J^{i-},J^{j-}\}=0$, while
the corresponding commutator in the quantum theory gives~\cite{Green-Schwarz-Witten}
\begin{align}
[J^{i-},J^{j-}]=-\frac{1}{(p^{+})^{2}}\sum_{m=1}^{\infty}\Delta_{m}
\big(\alpha^{i}_{-m}\alpha^{j}_{m}-\alpha^{j}_{-m}\alpha^{i}_{m}\big),
\end{align}
where the coefficients $\Delta_{m}$ are
\begin{align}
\Delta_{m}=m\left(\frac{26-d}{12}\right)+\frac{1}{m}\left(\frac{26-d}{12}
+2(1-b)\right),
\end{align}
and~$b$ is a normal-ordering constant appearing because of the
ambiguity in the order of the operators (see also below). Hence, in order to recover
Lorentz invariance we are led to set~$d=26$ and~$b=1$.
This means that a consistent quantization of the bosonic string is only possible when 
the target space-time is $26$-dimensional.

\paragraph{b) Old covariant quantization.} An obvious drawback of light-cone quantization is
the lack of target space covariance. To avoid this problem, let us consider the components of the
world-sheet energy-momentun tensor in complex coordinate basis $(z,\overline{z})$.
Taking into account that~$T_{z\overline{z}}=T_{\overline{z}z}=0$ due to tracelesness, its conservation
equation can be written as
\begin{align}
\partial_{z}T_{\overline{z}\overline{z}}=\partial_{\overline{z}}T_{zz}=0.
\end{align}
This means that~$T_{zz}$ and~$T_{\overline{z}\overline{z}}$ are respectively holomorphic and
anti-holomorphic functions, so we can Laurent expand them around $z=\overline{z}=0$
\begin{align}
T_{zz}(z)=\sum_{n\in \mathbb{Z}}L_{n}z^{-n-2}, \nonumber \\
T_{\overline{z}\overline{z}}(\overline{z})=\sum_{n\in \mathbb{Z}}\overline{L}_{n}z^{-n-2},
\label{eq:TzzTbzbz_virasoro_gen}
\end{align}
where $L_{n},\overline{L}_{n}$ are called the Virasoro generators and
satisfy~$L_{n}^{*}=L_{-n}$ and~$\overline{L}_{n}^{*}=\overline{L}_{-n}$. Implementing the 
expansion~(\ref{expansion}), as well as the canonical
Poisson brackets of the classical theory,
we find the algebra of Virasoro generators
\begin{align}
\{L_{n},L_{m}\}&=-i(n-m)L_{n+m}, \nonumber \\[0.2cm]
\{\overline{L}_{n},\overline{L}_{m}\}&=-i(n-m)\overline{L}_{n+m}, 
\label{eq:classical_virasoro}\\[0.2cm]
\{L_{n},\overline{L}_{m}\}&=0.
\nonumber 
\end{align}
After quantization,~$L_{n}$ and~$\overline{L}_{n}$ become operators. However, the 
algebra~\eqref{eq:classical_virasoro} is not
exactly recovered. Instead, we get the central extension
\begin{align}
[L_{m},L_{n}]&=(m-n)L_{m+n}+\frac{d}{12}m(m^{2}-1),
\nonumber \\[0.2cm]
{[}\overline{L}_{m},\overline{L}_{n}]&=(m-n)\overline{L}_{m+n}+
\frac{d}{12}m(m^{2}-1),
\label{eq:virasoro_quantum} \\[0.2cm]
{[}L_{m},\overline{L}_{n}]&=0.
\nonumber
\end{align}
The next step is to study the spectrum of the quantum theory.

On general grounds, whenever we have a classical system with a set of first-class 
constrains~$\{\phi_{i}\}$, physical states are identified as those 
annihilated by the corresponding operators~$\{\widehat{\phi}_{i}\}$, i.e.~$\widehat{\phi}_{i}
|\mbox{phys}\rangle=0$ for all~$i$.
However, in the case of the bosonic
string this does not work because of the two-dimensional gravitational anomaly
discussed in sec.~\ref{eq:expect} makes it impossible to impose the condition
$\widehat{T}_{ab}|{\rm phys}\rangle=0$. The best we can do is
to use an analog of the Gupta-Bleuler method employed in the quantization of
the electromagnetic field~\cite{Itzykson-Zuber}. In fact, it can be shown 
that the presence of the central term in the algebra~\eqref{eq:virasoro_quantum} 
is a consequence of the two-dimensional gravitational anomaly.

We therefore define physical states as those annihilated only by the 
positive frequency part of the energy-momentum tensor, namely
\begin{align}
L_{n}|{\rm phys}\rangle&=\overline{L}_{n}|{\rm phys}\rangle=0, \hspace{2cm} n>0,
\nonumber \\[0.2cm]
(L_{0}-b)|{\rm phys}\rangle&=(\overline{L}_{0}-b)|{\rm phys}\rangle=0,
\label{eq:bosonic_phys_cond}
\end{align}
where $b$ is a normal ordering constant that has to be included because
of the ambiguity in the order of the operators $a_{n}$, $\overline{a}_{n}$ in
the definition of $L_{0}$ and $\overline{L}_{0}$. We have to make sure that the
resulting spectrum does not include ghosts, i.e. physical states of negative norm. Brower, Goddard,
and Thorn \cite{Brower-Goddard-Thorn} proved a no-ghost theorem 
showing that ghost pop up in the spectrum for~$d>26$. The theory is ghost-free, however, if
$d=26$ and $b=1$ or $d<26$ and $b<1$, thus setting an upper bound on the dimensionality of the
target space.

\paragraph{c) Modern covariant quantization.} Let us begin with
the path integral \cite{Polyakov,KPZ}
\begin{align}
Z=\int{\cal D}g{\cal D}X \exp{\left(\frac{1}{4\pi\alpha'}\int
d^{2}\sigma\sqrt{-g}\,g^{ab}\partial_{a}X^{\mu}\partial_{b}X^{\nu}
\eta_{\mu\nu}\right)},
\end{align}
where, in order to carry the path integration avoiding overcounting, we need
to fix the gauge. We may use the conformal gauge
\begin{align}
g_{ab}=e^{\phi(\sigma)}\widehat{g}_{ab},
\label{conformal-factor}
\end{align}
where $\widehat{g}_{ab}$ is some fiducial metric. Since the action is also
diffeomorphism-invariant, we have to integrate only over those metric~$\widehat{g}_{ab}$ 
not related by world-sheet diffeomorphisms. 
Luckily, it can be shown that the class of gauge-equivalent metrics is a
finite-dimensional space~\cite{D'Hoker-Phong}, the moduli space of
Riemann surfaces.

Gauge fixing with respect to diffeomorphisms is implemented using Faddeev-Popov ghosts.
After factoring out the divergent volume of the group of diffeomorphisms,~${\rm Vol\,}({\rm Diff})$ 
we find the expression
\begin{align}
Z=\int d\mu\,{\cal D}_{e^{\phi}\widehat{g}}X {\cal D}_{e^{\phi}\widehat{g}}\phi
{\cal D}_{e^{\phi}\widehat{g}} b {\cal D}_{e^{\phi}\widehat{g}}c\,
\exp{\Big(-S_{P}[X,\widehat{g}]-S_{\rm gh}[b,c,\widehat{g}]\Big)},
\end{align}
where ${\cal D}_{e^{\phi}\widehat{g}}$ indicates that the integration
measure is computed for the original metric $g_{ab}$, $d\mu$ is the
measure in the space of fiducial metrics (i.e., the moduli space of Riemann surfaces), 
and~$S_{P}$ and~$S_{\rm gh}$ are respectively the string action~\eqref{Polyakov} 
and the action for the ghost fields
\begin{align}
S_{\rm gh}[b,c,\widehat{g}]=\frac{i}{4\pi\alpha'}\int
d^{2}\sigma\sqrt{\widehat{g}} \,\widehat{g}^{ab} b_{bc}\nabla_{a}c^{c}.
\end{align}
In order to make explicit the dependence in the field
$\phi(\sigma)$, it is convenient to recast the functional integral in terms of 
integration measures referred to the fiducial metric~$\widehat{g}_{ab}$. 
For the $X^{\mu}$ fields this is done using~\cite{Alvarez-Gaume-91}
\begin{align}
{\cal D}_{e^{\phi} \widehat{g}}X=e^{(d/48\pi)S_{L}[\phi,\widehat{g}]}
{\cal D}_{\widehat{g}}X,
\end{align}
while for the ghost fields we have
\begin{align}
{\cal D}_{e^{\phi} \widehat{g}}b{\cal D}_{e^{\phi}
\widehat{g}}c=e^{(-26/48\pi)S_{\rm L}[\phi,\widehat{g}]}
{\cal D}_{\widehat{g}}b {\cal D}_{\widehat{g}}c.
\end{align}
In these expressions $S_{L}[\phi,\widehat{g}]$ is the Liouville action, given by
\begin{align}
S_{L}[\phi,\widehat{g}]=\int d^{2}\sigma
\sqrt{\widehat{g}}\left(\frac{1}{2}\widehat{g}^{ab}\partial_{a}\phi
\partial_{b}\phi+R\phi+\mu e^{\phi}\right).
\end{align}
Combining all these results, we finally arrive at the following form of the partition function
\begin{align}
Z &=
\int d\mu\,{\cal D}_{e^{\phi}\widehat{g}}\phi {\cal D}_{\widehat{g}}X  {\cal
D}_{\widehat{g}} b {\cal D}_{\widehat{g}}c  \exp{\left(-S_{\rm P}[X,\widehat{g}]-
S_{\rm gh}[b,c,\widehat{g}]-
\frac{26-d}{48\pi}S_{L}[\phi,\widehat{g}]\right)}.
\label{eq:liouville1}
\end{align}

This results shows that the realization of conformal
symmetry is very different for~$d=26$ and $d\neq 26$. In the latter case we are dealing
with non-critical strings and the Liouville field is necessary to preserve conformal 
invariance in the quantum theory. 
Under some plausible assumptions we can extract the dependence of the
measure~${\cal D}_{e^{\phi}\widehat{g}}$ on the conformal factor~$\phi$
\begin{align}
Z&=\int {\cal D}_{\widehat{g}}\phi {\cal D}_{\widehat{g}}X  {\cal
D}_{\widehat{g}} b {\cal D}_{\widehat{g}}c 
\exp{\left(-S_{\rm P}[X,\widehat{g}]-
S_{\rm gh}[b,c,\widehat{g}]-
\frac{25-d}{48\pi}S_{L}'[\phi,\widehat{g}]\right)},
\end{align}
where $S_{L}'[\phi,\widehat{g}]$ is given by
\begin{align}
S_{L}'[\phi,\widehat{g}]=\int d^{2}\sigma \sqrt{\widehat{g}}\left(
\frac{1}{2}\widehat{g}^{ab}\partial_{a}\phi\partial_{b}\phi+ R\phi+
\mu e^{\gamma \phi}\right),
\end{align}
with
\begin{align}
\gamma=\frac{25-d-\sqrt{(25-d)(1-d)}}{12}.
\label{eq:gamma_DKK}
\end{align}
This expression was first obtained by David and Distler and Kawai~\cite{DDK} 
imposing a general ansatz for the Jacobian and demanding
the resulting theory to be well-defined. The result~\eqref{eq:gamma_DKK} 
seems reasonable only for $d\leq 1$, while for $1<d<25$ the naive arguments
of ref.~\cite{DDK} break down. For $d\geq 25$ the theory might still make sense
if we Wick rotate the Liouville field,~$\phi\rightarrow -i\phi$, so its
kinetic term has the right sign. Interestingly, in the particular
case of
an Euclidean $25$-dimensional target space, the Liouville field can be interpreted as
the time coordinate,~$X^{0}=-i\phi$, rendering the target theory effectively both Minkowskian
and $26$-dimensional.

Going back to eq.~\eqref{eq:liouville1}, we see that 
for critical strings ($d=26$) 
the Liouville field decouples from the action
and the integration over $\phi$ is absorbed in the global normalization of
the path integral. The only remnant of the integration over metrics is then the residual
integration over the moduli~$\mu$
\begin{align}
Z=\int d\mu\,{\cal D}_{\widehat{g}}X  {\cal
D}_{\widehat{g}} b {\cal D}_{\widehat{g}}c
\,\exp{\Big(-S_{\rm P}[X,\widehat{g}]-S_{\rm gh}[b,c,\widehat{g}]\Big)}.
\end{align}
We are therefore left with a two-dimensional
quantum field theory of~$26$ scalar fields,~$X^{\mu}$, and the reparametrization
ghosts preserving conformal invariance at the quantum level. This can be seen by
constructing the Virasoro generators for the ghost fields,
satisfying the algebra
\begin{align}
[L_{m}^{\rm gh},L_{n}^{\rm gh}]=(m-n)L_{m+n}^{\rm gh}+\frac{1}{6}m(1-13m^{2})
\delta_{m+n,0},
\end{align}
and similarly for~$\overline{L}_{m}^{\rm gh}$. Introducing the total
(matter+ghosts) Virasoro generators
\begin{align}
L_{m}^{\rm tot}&=L_{m}+L_{m}^{\rm gh}-b\delta_{m,0}, \nonumber \\[0.2cm]
\overline{L}_{m}^{\rm tot}&=\overline{L}_{m}+\overline{L}_{m}^{\rm gh}-b\delta_{m,0},
\end{align}
we find that they satisfy the classical Virasoro algebra without central
term
\begin{align}
[L_{m}^{\rm tot},L_{n}^{\rm tot}]&=(m-n)L_{m+n}^{\rm tot}, \nonumber \\[0.2cm]
{[}\overline{L}_{m}^{\rm tot},\overline{L}_{n}^{\rm tot}]&=(m-n)\overline{L}_{m+n}^{\rm tot},
\end{align}
with~$[L_{m}^{\rm tot},\overline{L}_{n}^{\rm tot}]=0$,
and as a consequence the quantum theory is invariant under the two-dimensional conformal
group.

The presence of ghosts fields makes BRST formalism~\cite{Green-Schwarz-Witten}
specially well suited. The BRST operator is defined from the matter and
ghost fields satisfying $Q^{2}=0$. Physical states are defined as those in the cohomolgy of
the $Q$ operator with minimum ghost number.

We have seen how the two-dimensional world-sheet field theory defined by 
the embedding fields~$X^{\mu}$ is a conformal field theory also at the quantum level. 
So far we have considered strings propagating on a flat target
space, but we would like to consider more 
general metrics~$G_{\mu\nu}$. Since consistency of the theory 
requires that it defines a two-dimensional conformal field
theory, in this case we need to demand that the theory's beta function
vanishes identically,~$\beta_{\mu\nu}(G)=0$. As we will see in sec.~\ref{sec:stringy_einstein_eq}
in more detail, at 
leading order in~$\alpha'$ these conditions are precisely the Einstein equations.

\subsection{Conformal field theory}
\label{sec:CFT}

Two-dimensional conformal field theories (CFT) are the basic building blocks
to construct string models and for this reason we are now reviewing 
some basic properties of these theories. They were
studied in the seminal paper by Belavin, Polyakov, and 
Zamolodchikov~\cite{Belavin-Polyakov-Zamolodchikov}, one of the basic
references on the subject (for a general review see also~\cite{Ginsparg}).

We begin, however, with a more general discussion and
consider a field theory on $d$-dimensional Minkowski space-time with
conserved energy-momentum
tensor~$T_{\mu\nu}$. Given a vector field $\xi^{\mu}(x)$ we can
construct the current $j_{\nu}=\xi^{\mu}(x)T_{\mu\nu}$ which will also be automatically
conserved, $\partial_{\mu}j^{\mu}=0$, provided~$\xi^{\mu}(x)$ is a Killing vector 
field,~$\partial_{\mu}\xi_{\nu}+\partial_{\nu}\xi_{\mu}=0$. This is seen from the identity
\begin{align}
\partial_{\mu}j^{\mu}&=\partial_{\mu}\xi_{\nu}T^{\mu\nu} \nonumber \\[0.2cm]
&=\frac{1}{2}(\partial_{\mu}\xi_{\nu}+\partial_{\nu}\xi_{\mu})T^{\mu\nu}.
\label{current-conserv}
\end{align}
If we take, however, the vector field generating dilatations,
$\xi^{\mu}(x)=x^{\mu}$, the previous equation shows that the energy-momentum
tensor would still be conserved if it is traceless,~$T^{\mu}_{\mu}=0$.
In this case the theory is classically invariant under scale
transformations.

A conformal trasformation is a change of coordinates $x^{\mu}\rightarrow
x'^{\mu}$ characterized by the relation
\begin{align}
g_{\mu\nu}(x')dx'^{\mu}dx'^{\nu}=
\Omega(x)\eta_{\mu\nu}dx^{\mu}dx^{\nu}.
\label{ct}
\end{align}
By considering an infinitesimal transformation $\Omega(x)=1+\lambda(x)$
and $x'^{\mu}=x^{\mu}+\xi^{\mu}(x)$, we arrive at the
relation
\begin{align}
\partial_{\mu}\xi_{\nu}+\partial_{\nu}\xi_{\mu}-
\frac{2}{d}\eta_{\mu\nu}\partial_{\sigma}\xi^{\sigma}=0.
\label{killing}
\end{align}
Vector fields $\xi_{\mu}(x)$ generating conformal transformations are
called conformal Killing vectors. For $d>2$ the
general solution to eq.~\eqref{killing} is 
\begin{align}
\xi_{\mu}=b_{\mu}+\omega_{\mu\nu}x^{\nu}+(c_{\nu}x^{\nu})x_{\mu}-
\frac{1}{2}c_{\mu}x^{2},
\end{align}
with $\omega_{(\mu\nu)}=0$. This shows that the conformal group 
is finite-dimensional when~$d>2$, since conformal transformations are labeled by a
finite number of parameters, $(b_{\mu},\omega_{\mu\nu},c_{\mu})$. For a
space-time with signature $(t,s)$, the conformal group is~$\mbox{SO}(t+1,s+1)$.

The situation is quite different in two dimensions. 
Using light-cone coordinates~$x^{\pm}=x^{0}\pm x^{1}$,
the conformal Killing condition~(\ref{killing}) reads
\begin{align}
\partial_{+}\xi^{-}=\partial_{-}\xi^{+}=0,
\end{align}
whose general solution is given by $\xi^{+}=\xi^{+}(x^{+})$ and
$\xi^{-}=\xi^{-}(x^{-})$. Conformal transformations are then given by
\begin{align}
x^{+}\longrightarrow f(x^{+}), \hspace{1cm}
x^{-}\longrightarrow g(x^{-}),
\end{align}
with~$f$ and~$g$ arbitrary functions.
We can now Wick-rotate and use complex coordinates $z$, $\overline{z}$,
so conformal transformations are holomorphic and
antiholomorphic trasformations for $z$ and $\overline{z}$
\begin{align}
z'=f(z), \hspace{1cm} \overline{z}'=\overline{f}(\overline{z}).
\label{conformal-trans}
\end{align}
In complex coordinates the energy-momentum tensor $T_{\mu\nu}(\sigma)$
only has two non-vanishing components, $T_{zz}(z)\equiv T(z)$ and
$T_{\overline{z}\overline{z}}(\overline{z})\equiv \overline{T}(\overline{z})$ while
$T_{z\overline{z}}=T_{\overline{z}z}=0$ as a consequence of the tracelessness
of $T_{\mu\nu}$. 
From the previous discussion it follows that the conformal group in two
dimensions is infinite-dimensional and its generators are
\begin{align}
\ell_{n}&=-z^{n+1}\frac{d}{dz}, \nonumber \\[0.2cm]
\overline{\ell}_{n}&=-\overline{z}^{n+1}{d\over d\overline{z}},
\end{align}
satisfying the Virasoro algebra
\begin{align}
[\ell_{m},\ell_{n}]&=(m-n)\ell_{m+n}, \nonumber \\[0.2cm]
[\overline{\ell}_{m},\overline{\ell}_{n}]&=(m-n)\overline{\ell}_{m+n},
\end{align}
and~$[\ell_{m},\overline{\ell}_{n}]=0$.

Next, we have to classify fields according to their behavior under the two-dimensional conformal group.
We define primary fields $\Phi_{h\overline{h}}(z,\overline{z})$ with conformal
weights $(h,\overline{h})$ as those fields such that
$\Phi_{h\overline{h}}(z,\overline{z})(dz)^{h}(d\overline{z})^{\overline{h}}$ is invariant
under conformal transformations~\eqref{conformal-trans}
\begin{align}
\Phi_{h\overline{h}}(z,\overline{z})(dz)^{h}(d\overline{z})^{\overline{h}}=
\Phi_{h\overline{h}}\big(f(z),\overline{f}(\overline{z})\big)(dz')^{h}
(d\overline{z}')^{\overline{h}}.
\end{align}
This implies the finite transformation
\begin{align}
\Phi_{h,\overline{h}}(z,\overline{z})
=f'(z)^{h}\overline{f}'(\overline{z})^{\overline{h}}
\Phi_{h,\overline{h}}\big(f(z),\overline{f}(\overline{z})\big),
\end{align}
or, infinitesimally
$z'=z+\epsilon(z)$
\begin{align}
\delta_{\epsilon}\Phi_{h,\overline{h}}(z,\overline{z})=
\left(\epsilon\partial_{z}+h\partial_{z}
\epsilon\right)\Phi_{h\overline{h}}(z,\overline{z}),
\end{align}
and similarly for the antiholomorphic part. 

We now turn, to the quantum theory. The operator
\begin{align}
T(\epsilon)=\oint_{|z|=1}\frac{dz}{2\pi i}\epsilon(z)T(z),
\end{align}
is the generator of conformal transformations, so we can write
\begin{align}
\delta_{\epsilon}\Phi_{h\overline{h}}=[T(\epsilon),\Phi_{h\overline{h}}(z,\overline{z})].
\label{commutator}
\end{align}
For simplicity, from now on we consider the
holomorphic parts, the antiholomorphic expressions being analogous. Any holomorphic function can be
expressed as a power series in $z$, so the whole conformal group can
be constructed from the generators associated with the monomial~$\epsilon_{n}=z^{n+1}$
\begin{align}
L_{n}=\oint_{|z|=1}\frac{dz}{2\pi i} z^{n+1}T(z).
\end{align}
Inverting this relation, we rewrite~$T(z)$ as
\begin{align}
T(z)=\sum_{n\in\mathbb{Z}}L_{n}z^{-n-2},
\end{align}
where $L_{n}$ are the Virasoro generators introduced in eq.~\eqref{eq:TzzTbzbz_virasoro_gen}
in the context of string theory.
Quantum mechanically, they satisfy the central extension of the
Virasoro algebra
\begin{align}
[L_{m},L_{n}]=(m-n)L_{m+n}+\frac{c}{12}m(m^{2}-1)\delta_{m+n,0},
\end{align}
where $c$ is called the central charge. It is easy to see that~$L_{\pm 1}$ and~$L_{0}$ 
generate an $\mbox{SL}(2,\mathbb{C})$ subalgebra. From
eq.~(\ref{commutator}) we also determine the commutator of a Virasoro
generator with a primary field
\begin{align}
[L_{n},\Phi_{h}(z)]=h(n+1)z^{n}\Phi_{h}(z)+
z^{n+1}\partial_{z}\Phi_{h}(z).
\label{comm-L-Phi}
\end{align}

Associated with
every primary field~$\Phi_{h}(z)$ we have a state in the theory's Fock
space defined by the action of this field on the vacuum at~$z=0$
\begin{align}
|h\rangle =\Phi_{h}(0)|0\rangle.
\end{align}
To have a well-defined action of the energy-momentum tensor operator on the vacuum
it is necessary to impose
\begin{align}
L_{n}|0\rangle =0, \hspace{1cm} n\geq-1,
\end{align}
so, in particular, the vacuum state~$|0\rangle$ is~$\mbox{SL}(2,\mathbb{C})$-invariant.
Moreover, from eq.~(\ref{comm-L-Phi}) we see that the state~$|h\rangle$ associated 
with the primary field~$\Phi_{h}(z)$ satisfies
\begin{align}
L_{0}|h\rangle &=h|h\rangle, \\[0.2cm]
L_{n}|h\rangle &=0 \hspace*{0.5cm} \mbox{for} \hspace{0.5cm} n>0.
\end{align}
Due to the second property, states created by primary fields are
called highest-weight states. 
On the other hand, applying~$L_{-n}$ with $n>0$ to a highest-weight state gives the
so-called descendent states, labeled by the eigenvalue of
$L_{0}$. Taking into account that~$[L_{0},L_{-n}]=nL_{-n}$, we find
\begin{align}
|h\rangle, &\hspace{1cm} L_{0}=h, \nonumber \\[0.2cm]
L_{-1}|h\rangle, &\hspace{1cm} L_{0}=h+1, \nonumber \\[0.2cm]
L_{-1}^{2}|h\rangle , L_{-2}|h\rangle, &\hspace{1cm} L_{0}=h+2,
\\[0.2cm]
&\vdots \nonumber
\label{family}
\end{align}
The representation of the Virasoro algebra spanned by these states is
called a Verma module~$V(h,c)$. It is labelled by the conformal weight of the primary field
and the central charge of the Virasoro algebra.

In the same way as we associated a
primary field~$\Phi_{h}(z)$ with a highest weight state $|h\rangle$ using~$|h\rangle=\Phi_{h}(0)|0\rangle$,
we can define descendent fields
associated with the states~$L_{-n_{1}}\ldots
L_{-n_{k}}|0\rangle$, namely
\begin{align}
\Phi^{n_{1}+\ldots +n_{k}}(z)=L_{-n_{1}}(z)\ldots
L_{-n_{k}}(z)\Phi_{h}(z),
\end{align}
where the operators $L_{n}(z)$ are defined by
\begin{align}
L_{n}(z)=\oint_{C_{n}}\frac{dw}{2\pi i}(w-z)^{n+1}T(w),
\end{align}
and the contours are taken in such a way that $C_{n_{k}}\subset \ldots
\subset C_{n_{1}}$. This defines one-to-one correspondence
between fields and states.

\begin{figure}[t]
\centerline{\includegraphics[scale=0.4]{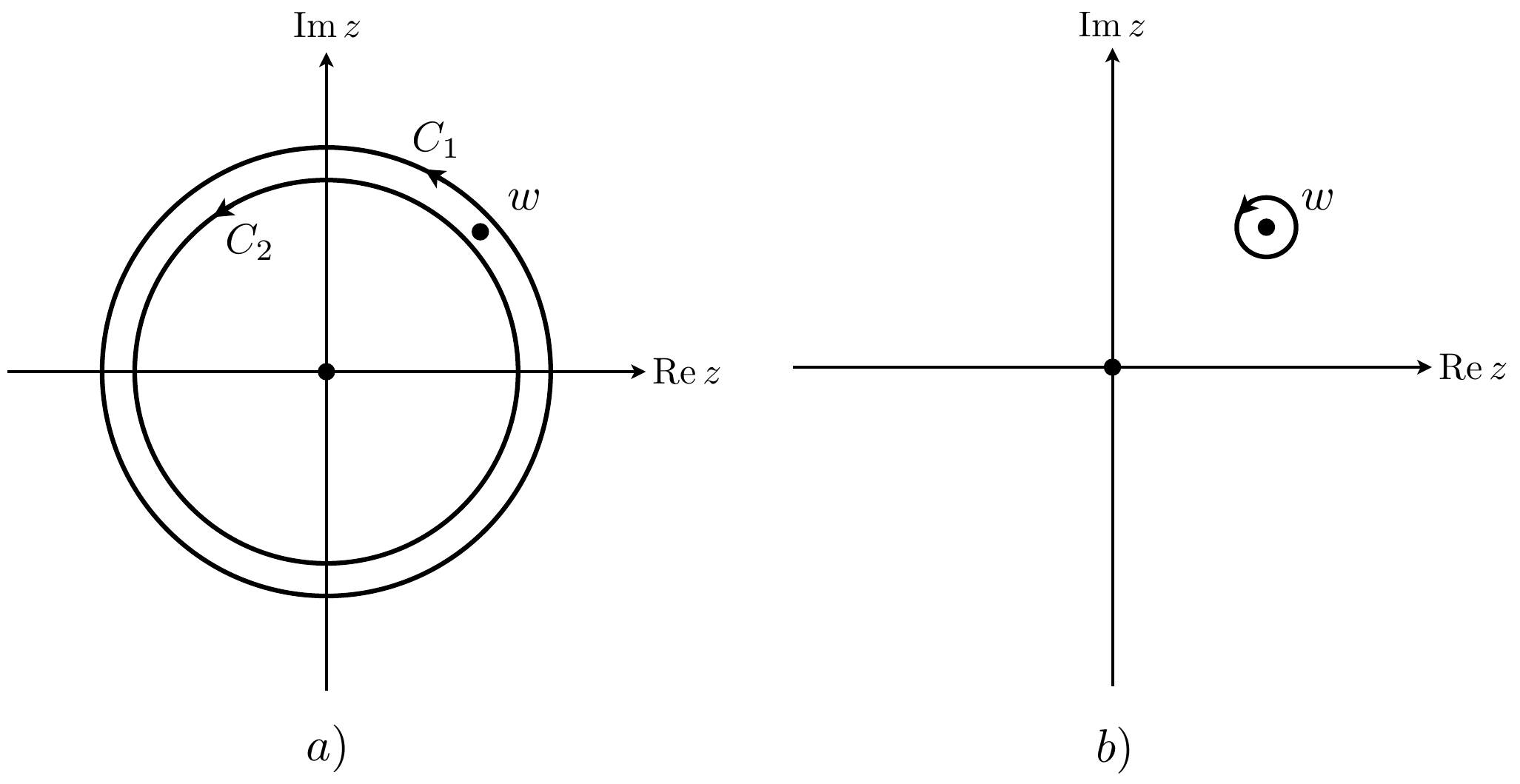}}
\caption{Integration contours to compute the equal-radius commutator}
\label{contours}
\end{figure}

Equal time (i.e., radius) commutators are related to the operator product
expansion (OPE) of the operators involved. Let us
consider a complete set of operators $\{\Phi_{k}(z)\}$. On general grounds, the
product of any two operators in this set, $\Phi_{i}(z)\Phi_{j}(w)$, diverges 
in the coincidence limit $z\rightarrow w$. However, since the
operators~$\Phi_{k}(z)$ are complete, we can write
\begin{align}
\Phi_{i}(z)\Phi_{j}(w)=\sum_{k}C_{ij}^{k}(z,w) \Phi_{k}(w),
\label{OPE}
\end{align}
which is called the operator product expansion of~$\Phi_{i}(z)$ and~$\Phi_{j}(w)$.
The divergence of the product in the coincidence limit 
is given by the behavior of the coefficients~$C_{ij}^{k}(z,w)$.
Let us take two analytic fields $A(z)$ and $B(z)$ and two
aarbitrary functions $f(z)$ and $g(z)$, and construct the operators
\begin{align}
A(f)&=\oint_{|z|=1}\frac{dz}{2\pi i}f(z)A(z), \nonumber \\[0.2cm]
B(g)&=\oint_{|w|=1}\frac{dw}{2\pi i}g(w)B(w).
\end{align}
We compute then the equal-time commutator as
\begin{align}
[A(f),B(g)]&=\oint_{C_{1}}\frac{dz}{2\pi i}f(z)A(z)
\oint_{C_{2}}\frac{dw}{2\pi i}g(w)B(w) \nonumber \\[0.2cm]
&-\oint_{C_{2}}\frac{dw}{2\pi
i}g(w)B(w)\oint_{C_{1}}\frac{dz}{2\pi i}f(z)A(z),
\label{comm1}
\end{align}
where the contour~$C_{2}$ lies inside~$C_{1}$, as schematically represented in fig. \ref{contours}a. By deforming
the contour as in fig. \ref{contours}b, the difference between the integrals in eq.~\eqref{comm1} is recast as
\begin{align}
[A(f),B(g)]=\oint_{|w|=1}\frac{dw}{2\pi i}g(w)\oint_{w}\frac{dz}{2\pi i}
A(z)B(w)f(z),
\label{OPE-commutator}
\end{align}
where the integral to the right is performed along a little circumference
around $w$ and is therefore
determined by the singularities in the OPE of~$A(z)B(w)$. 
Applying this result to the commutator 
of~$T(z)$ with a primary field~$\Phi_{h}(z)$, and using
eq.~(\ref{comm-L-Phi}), we arrive at the OPE
\begin{align}
T(z)\Phi_{h}(w)=\frac{h}{(z-w)^{2}}\Phi_{h}(w)+
\frac{1}{z-w}\partial_{w}\Phi_{h}(w)+ \mbox{regular terms},
\end{align}
where the regular terms are analytic in the limit~$z\rightarrow w$.
Using again eq.~(\ref{OPE-commutator}), the Virasoro algebra is represented 
in terms of the OPE
\begin{align}
T(z)T(w)&=\frac{c/2}{(z-w)^{4}}+\frac{2}{(z-w)^{2}}T(w) \nonumber \\[0.2cm]
&+\frac{1}{z-w}\partial_{w}T(w)+\mbox{regular terms},
\label{eq:TTOPE}
\end{align}
where $c$ is the central charge. The
vacuum expectation value of the product~$T(z)T(w)$ is computed from~\eqref{eq:TTOPE}
to be
\begin{align}
\langle T(z)T(w) \rangle = \frac{c/2}{(z-w)^{4}}.
\end{align}
from where the gravitational anomaly calculated in eq.~\eqref{eq:2D_grav_anomaly} is retrieved
\begin{align}
\int d^{2}x \langle T_{++}(x)T_{++}(0)\rangle e^{ipx} \sim
\frac{p_{+}^{3}}{p_{-}}.
\end{align}

Due to the presence of the central charge in~\eqref{eq:TTOPE}, 
the energy-momentum tensor is not a primary field. Indeed,
$T(z)$ can be written as
\begin{align}
T(z)=L_{-2}(z)\mathbb{1}(z)\;,
\end{align}
where $\mathbb{1}(z)$ is the identity operator, which is a trivial
primary
field of zero conformal weight. The energy momentum
tensor is then a descendant field of the identity operator. If instead of
infinitesimal we consider finite conformal transformations,~$z \rightarrow f(z)$, 
we find the following transformation for~$T(z)$
\begin{align}
T(z)\rightarrow [\partial_{z}f(z)]^{2}T(z)+\frac{c}{12}\{f,z\},
\label{T-transform}
\end{align}
where $\{f,z\}$ is the Schwarzian derivative
\begin{align}
\{f,z\}\equiv\frac{\partial_{z}f(z)\partial_{z}^{3}f(z)-
\frac{3}{2}[\partial_{z}^{2}f(z)]^{2}}{[\partial_{z}f(z)]^{2}}.
\end{align}

In sec.~\ref{sec:bosonic_string_theory} we
learned that the cylinder is mapped onto~$\mathbb{C}^{*}$ 
by means of the conformal transformation~$z=e^{w}\equiv e^{\tau+i\sigma}$.
Time ordering in the original description is converted into
radial ordering on~$\mathbb{C}^{*}$, with~$\tau=-\infty$ corresponding
to~$z=0$ and~$\tau=\infty$ to~$z=\infty$. 
Using the transformation
of the energy-momentum tensor given in eq.~(\ref{T-transform}), we get 
a relation between the energy-momentum tensor on the cylinder
and on~$\mathbb{C}^{*}$
\begin{align}
T(w)_{\rm cyl}=z^{2}T(z)-\frac{c}{24}.
\end{align}
This implies that all Virasoro generators remain the same apart from~$L_{0}$
which changes according to
\begin{align}
(L_{0})_{\rm cyl}=(L_{0})_{\rm plane}-\frac{c}{24},
\label{cyl-plane}
\end{align}
although in the following we omit the ``plane'' subscript. 
This expression gives the key to interpret physically the transformations induced by
$L_{0}$ and $\overline{L}_{0}$. From their definition in terms of the
energy-momentum tensor,
\begin{align}
L_{0}&=\oint_{|z|=1}\frac{dz}{2\pi i}zT(z), \nonumber \\[0.2cm]
\overline{L}_{0}&=\oint_{|\overline{z}|=1}\frac{d\overline{z}}{2\pi
i}\overline{z}\overline{T}(\overline{z}),
\end{align}
we see that $L_{0}+\overline{L}_{0}$ is the generator of dilatations
$z\rightarrow e^{\lambda} z$ in $\mathbb{C}^{*}$, with $\lambda$ a real number. 
In the cylinder coordinates~$(\tau,\sigma)$
these dilatations correspond to
time translations, $\tau \rightarrow \tau+\lambda$, so 
$L_{0}+\overline{L}_{0}$ plays the role of the Hamiltonian.
Incidentally, Fock space states generated on the vacuum by operators evaluated 
at the origin are in the cylinder picture created at~$\tau\rightarrow-\infty$.

The interpretation of $L_{0}-\overline{L}_{0}$ is also straighforward: 
it generates transformations $z\rightarrow e^{i\lambda} z$
for $\lambda\in\mathbb{R}$, which are rotations centered at the origin $z=0$. 
In the cylinder they correspond to shifts in the
coordinate $\sigma$. 
Thus,~$L_{0}-\overline{L}_{0}$ is interpreted as a momentum operator.

\begin{figure}[t]
\centerline{\includegraphics[scale=0.40]{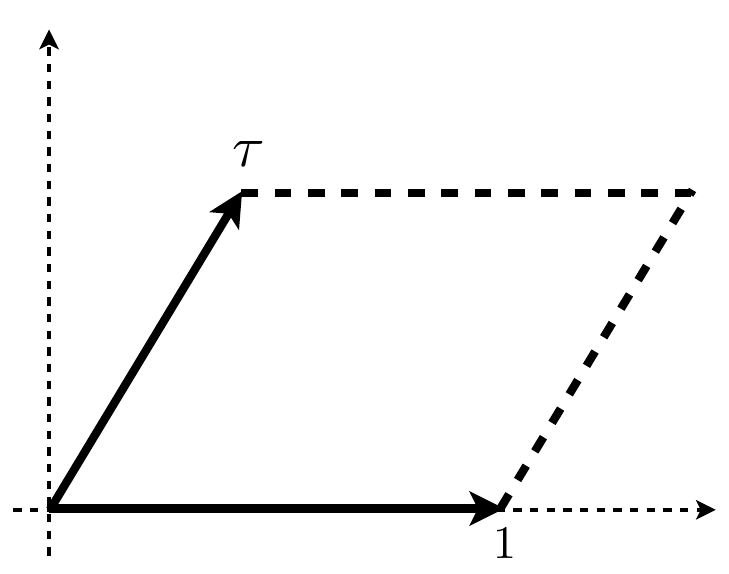}}
\caption{Torus with modular parameter $\tau$}
\label{torus}
\end{figure}

The partition function of a CFT counts the number of states per energy level. 
For later use, it is interesting to work not on the cylinder or the punctured complex plane, but
on a Riemman surface of arbitrary genus. A CFT on the torus can be constructed 
starting from the formulation on the cylinder just by identifying its two end (see \cite{Ginsparg} for
details)..
In this case we have two periods, one associated with time translations
and a second one with the shifts generated by~$(L_{0})_{\rm cyl}\pm(\overline{L}_{0})_{\rm cyl}$.
As a first step, we redefine the complex coordinate $w\rightarrow iw$, so we have the 
periodicity~$w \equiv w+2\pi$. The other period
is defined in terms of a complex number~$\tau=\tau_{1}+i\tau_{2}$, 
the modular parameter of the torus, as $w
\equiv w+2\pi \tau$ (see fig. \ref{torus}).
Implementing the relation~(\ref{cyl-plane}), we find the CFT partition function~\cite{Ginsparg}
\begin{align}
Z={\rm tr}\,\big(q^{L_{0}-c/24}\overline{q}^{\overline{L}_{0}-c/24}\big),
\label{eq:CFTpartition_function}
\end{align}
where $q\equiv e^{2\pi i\tau}$. The modular parameter~$\tau$ counts conformally inequivalent tori.

The partition function counts the number
of states associated with a given conformal family. Let $V(h,c)$ be the
Verma module associated with a primary field $\Phi_{h}(z)$
\begin{align}
V(h,c)=\big\{|h\rangle, L_{-1}|h\rangle, L_{-2}|h\rangle,
L_{-1}^{2}|h\rangle,\ldots \big\}
\end{align}
To count the number of states per level in~$V(h,c)$ we 
define the character
\begin{align}
\chi_{h}(q)={\rm tr}_{V(h,c)}q^{L_{0}-c/24},
\end{align}
which is the holomorphic part of the partition function~\eqref{eq:CFTpartition_function}, 
restricted to the states in $V(h,c)$. The trace can be written as
\begin{align}
\chi_{h}=\sum_{m} d_{m} q^{m},
\end{align}
where $d_{m}$ is the number of states in $V(h,c)$ with $L_{0}=m+c/24$.
For example, if $V(h,c)$ contains a single highest-weight vector the
character can be explicitly evaluated to give
\begin{align}
\chi_{h}=q^{h-c/24}\prod_{n=1}^{\infty}\frac{1}{1-q^{n}}.
\end{align}

To illustrate the previous discussion, we
analize the case of a massless free scalar field in two dimensions.
In complex coordinates $(z,\overline{z})$ we write the action
\begin{align}
S=\frac{1}{4\pi}\int d^{2}z\,
\partial\phi(z,\overline{z})\overline{\partial}\phi(z,\overline{z}),
\label{action-scalar}
\end{align}
where here and in the following we set~$\partial\equiv\partial_{z}$ and
$\overline{\partial}\equiv\partial_{\overline{z}}$.
The field equations read
\begin{align}
\partial\overline{\partial}\phi(z,\overline{z})=0,
\end{align}
and its general solution is given
by~$\phi(z,\overline{z})=\phi(z)+\overline{\phi}(\overline{z})$. 
Since~$\partial\phi$ and~$\overline{\partial}\overline{\phi}$ are respectively holomorphic 
and antiholomorphic fields, we write the Laurent expansions
\begin{align}
\partial\phi(z)&=\sum_{n\in \mathbb{Z}}\alpha_{n}z^{-n-1}, \label{partialphihol}\\[0.2cm]
\overline{\partial}\phi(\overline{z})&=\sum_{n\in
\mathbb{Z}}\overline{\alpha}_{n}\overline{z}^{-n-1}.
\label{partialphiantihol}
\end{align}
The action~(\ref{action-scalar}) gives the propagator
\begin{align}
\langle \phi(z,\overline{z})\phi(w,\overline{w})\rangle=-\log|z-w|^{2},
\end{align}
and the energy-momentum tensor
\begin{align}
T(z)=-\frac{1}{2}:\!\partial\phi(z)\partial\phi(z)\!:,
\end{align}
and similarly for the antiholomorphic component.
In order to calculate the central charge we compute the OPE for
two energy-momentum tensors. Applying Wick's theorem for radial
quantization, we obtain
\begin{align}
T(z)T(w)=\frac{1/2}{(z-w)^{4}}+ \mbox{other terms},
\end{align}
so the central charge is $c=1$. From the OPE
\begin{align}
T(z)\partial\phi(w) =
\frac{1}{(z-w)^{2}}\partial\phi(w)+\frac{1}{z-w}\partial^{2}\phi(w),
\end{align}
we see that $\partial\phi(z)$ is a primary field with conformal weight
$(h,\overline{h})=(1,0)$. A similar calculation shows that~$\overline{\phi}(\overline{z})$
is also a primary field with~$(h,\overline{h})=(0,1)$.
Other primary fields are the vertex operators
\begin{align}
V_{k}(z,\overline{z})= :\!e^{ik\phi(z,\overline{z})}\!:,
\end{align}
with conformal weights
\begin{align}
h(V_{k})=\overline{h}(V_{k})=\frac{k^{2}}{2},
\end{align}
as can be checked by computing the relevant OPE.

\subsection{Quantization of the bosonic string}
\label{sec:quantization_bosonic}

This example of the two-dimensional massless scalar field provides a
natural connection with the quantization of the closed bosonic string. 
We work in the critical dimensions~$d=26$, 
where the Liouville field decouples.
Using the conformal gauge and mapping the
cylindrical world-sheet onto $\mathbb{C}^{*}$, the action~\eqref{Polyakov}
takes the form
\begin{align}
S_{P}[X^{\mu}]=-\frac{1}{4\pi \alpha'}\int dzd\overline{z} \,\partial X^{\mu}
\overline{\partial} X^{\nu} \eta_{\mu\nu},
\label{eq:Polyakov_action_holantihol}
\end{align}
describing~$d$ massless scalar fields $X^{\mu}$. 
As with the scalar field analyzed at the end of the previous section, 
we can also construct vertex operators of definite conformal weight, for example
\begin{align}
V(k)=\,:\!e^{ip_{\mu}X^{\mu}(z,\overline{z})}\!:,
\label{eq:vertex_tachyon}
\end{align}
with~$h=\overline{h}={1\over 2}p_{\mu}p^{\mu}$. Geometrically, vertex operators can 
be interpreted as the Fourier transform of the operators~\cite{Polyakov92}
\begin{align}
V(X)\sim \int d^{2}\sigma \sqrt{g} \delta\big(X^{\mu}-X^{\mu}(\sigma)\big),
\end{align}
pinning a string at the location~$X$. Since only fields with conformal 
weight~$(h,\overline{h})=(1,1)$
can be integrated to give conformally invariant operators, 
physical states should satisfy
\begin{align}
L_{0}|{\rm phys}\rangle&=|{\rm phys}\rangle, \nonumber \\[0.2cm]
\overline{L}_{0}|{\rm phys}\rangle&=|{\rm phys}\rangle.
\end{align}
These conditions fix the value of the normal ordering constant
introduced in sec.~\ref{sec:bosonic_string_theory} to~$b=1$.

We now proceed to study the spectrum of the closed bosonic string. 
The solution to the equations
of motion~$\partial\overline{\partial}X^{\mu}(z,\overline{z})=0$ derived 
from the action~\eqref{eq:Polyakov_action_holantihol}  
admit the expansions written in eq.~\eqref{expansion}.
The canonical commutation relations between the field~$X^{\mu}(z,\overline{z})$
and its conjugate momentum~$\Pi^{\mu}$ give rise to the following commutation relations
between the operators appearing on the right-hand side of the expansion~\eqref{expansion}
\begin{align}
[q^{\mu},p_{\nu}]&=i\eta^{\mu\nu}, \nonumber \\[0.2cm]
{[}\alpha_{m}^{\mu},\alpha_{n}^{\nu}]&=m\eta^{\mu\nu}\delta_{m+n,0},
\label{commutation-rel} \\[0.2cm]
{[}\overline{\alpha}_{m}^{\mu},\overline{\alpha}_{n}^{\nu}]&=
m\eta^{\mu\nu}\delta_{m+n,0},
\nonumber
\end{align}
while all other commutators vanish. Equation~(\ref{expansion}) 
can be split into right- (holomorphic) and left-moving (antiholomorphic)
parts according to
\begin{align}
X_{\rm R}^{\mu}(z)&=
x_{\rm R}^{\mu}-\frac{i\alpha'}{2}p_{\rm R}^{\mu}\log{z}+ i\sqrt{\frac{\alpha'}{2}}\sum_{n\in\mathbb{Z}}
\frac{\alpha_{n}^{\mu}}{n}z^{-n}, \nonumber \\[0.2cm]
X_{\rm L}^{\mu}(\overline{z})&=
x_{\rm L}^{\mu}-\frac{i\alpha'}{2}p_{\rm L}^{\mu}\log{\overline{z}}+i\sqrt{\frac{\alpha'}{2}}
\sum_{n\in\mathbb{Z}} \frac{\overline{\alpha}_{n}^{\mu}}{n}\overline{z}^{-n}\;,
\label{eq:XLXRexpansions}
\end{align}
where $x_{\rm R}^{\mu}=x_{\rm L}^{\mu}=q^{\mu}/2$ and $p_{\rm R}^{\mu}=p_{\rm L}^{\mu}=
p^{\mu}$.
The operators~$L_{0}$ and~$\overline{L}_{0}$ are then written as
\begin{align}
L_{0}&=\frac{\alpha'}{2}p^{2}+\sum_{n\geq 1}\alpha_{-n}\cdot
\alpha_{n}, \nonumber \\[0.2cm]
\overline{L}_{0}&=\frac{\alpha'}{2}p^{2}+\sum_{n\geq 1}\overline{\alpha}_{-n}\cdot
\overline{\alpha}_{n},
\label{eq:L0_L0bar_modeexpansion}
\end{align}
where we used a dot to represent the contraction of target space indices. 
Setting $L_{0}=\overline{L}_{0}=1$ and taking
into account that the mass of a physical state is defined by
$m^{2}=-p_{\mu}p^{\mu}$ we arrive at the mass formula
\begin{align}
\frac{1}{2}\alpha'm^{2}=\sum_{n\geq 1} \alpha_{-n}\cdot\alpha_{n}+
\sum_{n\geq 1} \overline{\alpha}_{-n}\cdot\overline{\alpha}_{n}-2.
\label{eq:bosonic_mass_spectrum}
\end{align}
The condition $L_{0}-\overline{L}_{0}$, which now states that the theory is
invariant under redefinitions of the origin in the coordinate~$\sigma$,
gives the so-called level matching condition
\begin{align}
\sum_{n\geq 1}\alpha_{-n}\cdot\alpha_{n}=
\sum_{n\geq 1}\overline{\alpha}_{-n}\cdot\overline{\alpha}_{n}.
\label{eq:level_matching_bos}
\end{align}
Looking back to the commutation relations in eq.~(\ref{commutation-rel}), we find
that $N_{R}\equiv\sum_{n\geq 1}\alpha_{-n}\cdot\alpha_{n}$ can be interpreted as
the ocupation number
associated to the right-moving modes, and similarly for left-movers.
This allows us to rewrite the mass formula~\eqref{eq:bosonic_mass_spectrum} 
and the level matching
condition~\eqref{eq:level_matching_bos} in a more compact form as
\begin{align}
\frac{1}{2}\alpha'm^{2}&=N_{R}+N_{L}-2, \nonumber \\[0.2cm]
N_{R}&=N_{L}.
\label{eq:bos_mass_form_LMC}
\end{align}

We are ready now to construct the spectrum of the closed bosonic string.
We start with an~$\mbox{SL}(2,\mathbb{C})$-invariant vacuum state $|0,p\rangle$ defined by
\begin{align}
\alpha_{n}^{\mu}|0,p\rangle&=\overline{\alpha}_{n}^{\mu}|0,p\rangle=0,
\hspace{1cm} n\geq 1, \nonumber \\[0.2cm]
\widehat{p}^{\mu}|0,p\rangle&=p^{\mu}|0,p\rangle,
\end{align}
where, for the time being, we write~$\widehat{p}^{\mu}$ to distinguish the momentum operator from
its eigenvalue. Since~$N_{R}|0,p\rangle=N_{L}|0,p\rangle=0$, the vacuum state 
has~$m^{2}<0$ and is a tachyon. This state is created by the
action of the tachyon vertex operator~\eqref{eq:vertex_tachyon}.

In the first excited level there is only one possibility satisfying the
level matching condition, $N_{L}=N_{R}=1$. A general state will then be of
the form
$\zeta_{\mu\nu}(p)\alpha_{-1}^{\mu}\overline{\alpha}_{-1}^{\nu}|0,p\rangle$,
with $\zeta_{\mu\nu}(p)$ the polarization vector. Its associated vertex operator
is given by
\begin{align}
V(p)=\zeta_{\mu\nu}(p):\!\partial X^{\mu}
\overline{\partial}X^{\nu}e^{ip\cdot X}\!:.
\end{align}
The states in this first level are massless, as it can be seen from either
the mass formula or the condition that the previous
vertex operator has conformal weights $(1,1)$. This also results in the
transversality relation~$p^{\mu}\zeta_{\mu\nu}(p)=0$. 
Decomposing the polarization tensor into irreducible representations of~$\mbox{SO}(1,25)$,
we find
\begin{itemize}
\item[-]
$\zeta_{\mu\nu}=\zeta_{\nu\mu}$ with~$\eta^{\mu\nu}\zeta_{\mu\nu}=0$.

\item[-]
$\zeta_{(\mu\nu)}=0$.

\item[-]
$\zeta_{\mu\nu}=\eta_{\mu\nu}$.

\end{itemize}
corresponding respectively to the graviton $g_{\mu\nu}$, the antisymmetric tensor
$B_{\mu\nu}$, and the scalar dilaton $\Phi$.

Maybe the most remarkable feature of this analysis is that, 
without much effort, we have found that the spectrum of the bosonic string
contains a massless spin-2 particle. Its identification with the physical
graviton might seem a bit too hasty. However, once we introduce string interaction
using the prescription to be presented in the next section,  
it is possible to show that scattering amplitudes of these spin-2
states are described in the low energy limit by the linearized
Einstein-Hilbert action, so they can be consistently identified with the
graviton. This is one of the most relevant features of string theory: every
free closed string theory, and all theories of interacting strings, contain
a massless spin-2 state.

It should be stressed that the existence of a graviton state is  a
consequence of working in the critical dimension, as gravitons seem
to be excluded from the spectrum of non-critical strings. Thus, if we
want to use string theory to unify gravity with other interactions
we are forced to take~$d=26$. It is clear that the notion of dimension in 
string theory is encoded in the value of the central
extension of the Virasoro algebra. We need to have $c=26$,
although the contribution from ``geometrical'' dimensions may be smaller
than this number, since the critical value of~$c$ can be reached by adding 
up different CFTs without a geometrical interpretation.

\subsection{String interactions and the characterization
of the moduli space}
\label{sec:int+moduli}

\begin{figure}[t]
\centerline{\includegraphics[scale=0.50]{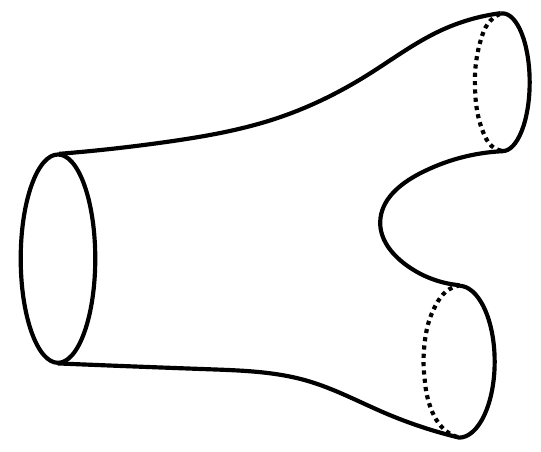}}
\caption{Fundamental vertex in string perturbation theory}
\label{vertex}
\end{figure}

\begin{figure}[t]
\centerline{\includegraphics[scale=0.45]{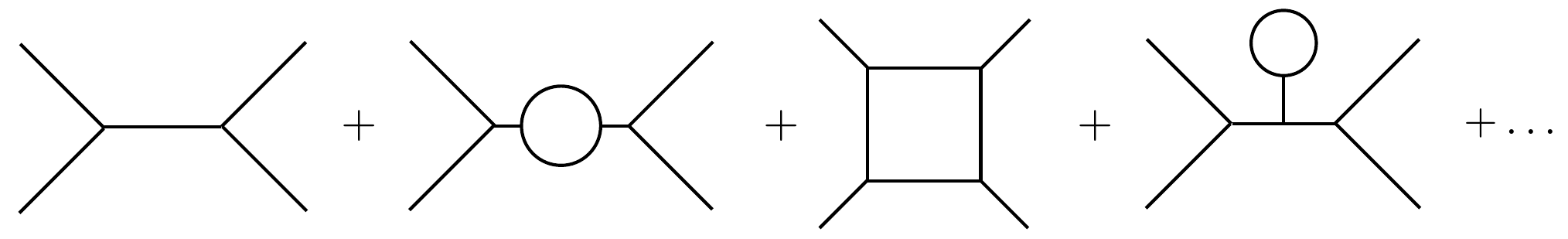}}
\caption{Feynman diagrams contributing to the four-point function in
$\phi^{3}$ scalar theory}
\label{phi3}
\end{figure}

\begin{figure}[t]
\centerline{\includegraphics[scale=0.45]{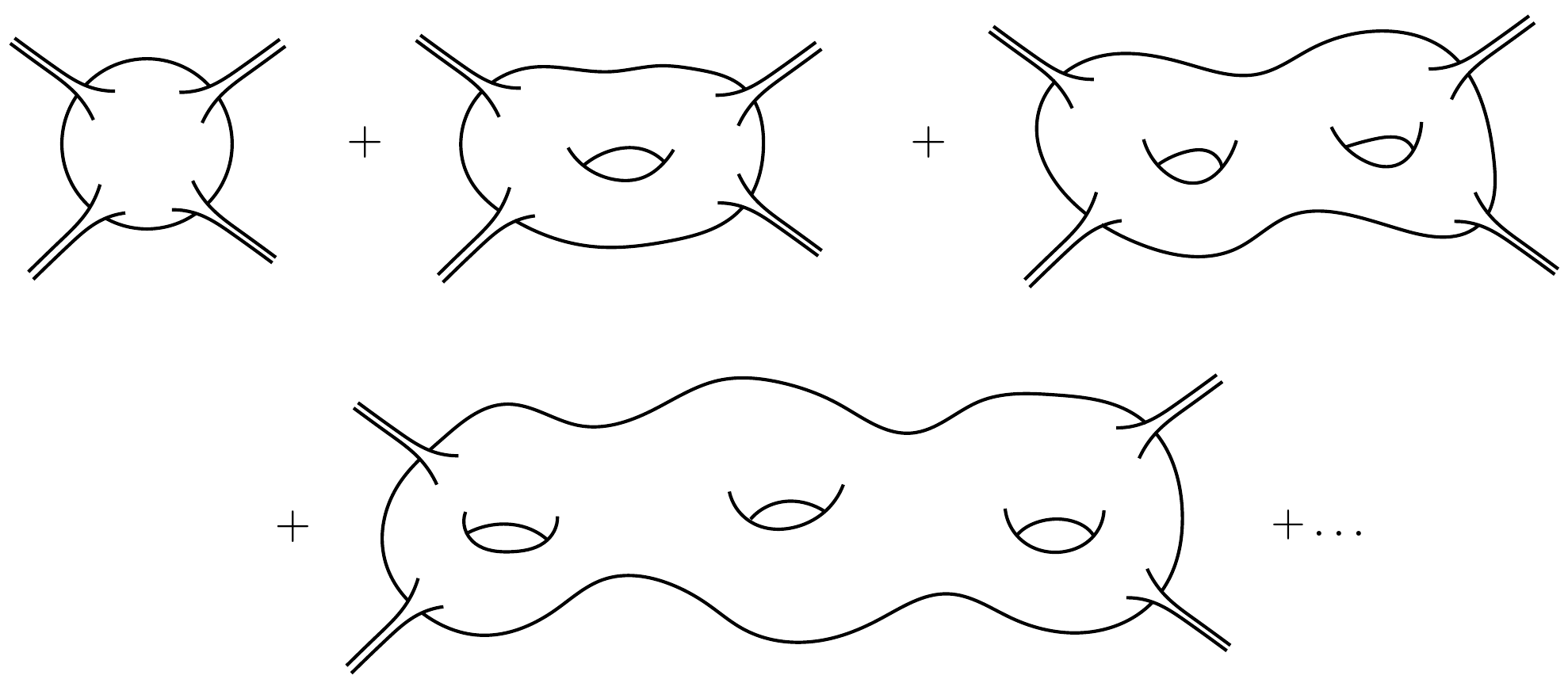}}
\caption{String graphs contributing to the scattering of four strings}
\label{strings-graphs}
\end{figure}

It is about time to
consider string interactions. Perturbatively, this is done
by introducing the basic vertex in fig.~\ref{vertex}, describing
the splitting of a string into two more.
Using a particle analog, this would
correspond to the vertex in a~$\phi^{3}$ scalar field theory.
The situation in string theory is however
quite different from that in QFT. In the scalar cubic theory, in
order to compute a given amplitude, we have to sum over all
Feynman graphs with a given number of
external states. For the four point amplitude, for example, we need to
add the contributions of all graphs with four external legs shown in 
fig.~\ref{phi3}.
By contrast, in string theory we should consider diagrams with a fixed number of
external string states, summing over all world-sheets
connecting ``in'' with ``out'' strings (see fig.
\ref{strings-graphs}),
all of them constructed from the fundamental vertex in fig. \ref{vertex}.
In fact, the number of vertices needed to construct a
given surface only depends on the genus~$g$ of the surface and the
number~$n$ of external strings and it is equal to~$2g+n-2$. Denoting by~$g_{\rm st}$
the string coupling constant associated with the fundamental vertex,
the $n$-string amplitude can be morally written as
\begin{align}
A(1,\ldots,n)=
\sum_{\rm topologies}g_{\rm st}^{2g+n-2}\sum_{\rm metrics}
\frac{1}{{\rm Vol}(\mbox{Diff})\times {\rm Vol}(\mbox{Weyl})}\sum_{\rm embeddings}
e^{-S_{\rm P}[X,g]},
\label{amplitude}
\end{align}
where the first sum indicates that we sum over all two-dimensional
surfaces of genus $g$ and with~$n$ external strings. We also divide out
by the volume of the group of diffeomorphisms and of the group of Weyl
transformations because these are, in the critical dimension, the gauge
invariances of the quantum theory. 

The Weyl invariance of
the theory can also be exploited to avoid considering two-dimensional surfaces 
with ``tubes'' extending all the way to
infinity. These cylinders are cut off the surface and, using a conformal transformation,
mapped into the punctured unit circle ${\cal C}^{*}
=\{z\in \mathbb{C}\,|\,0<|z|\leq 1\}$, before being attached back to 
each boundary. External string states are then inserted in the punctures by
means of vertex operators carrying the quantum
numbers of the corresponding external strings. When all this is properly done, 
the amplitude~(\ref{amplitude}) can be rewritten in mathematical precise terms as
\begin{align}
A(1,\ldots,n)=
\sum_{g=1}^{\infty} g_{\rm st}^{-\chi} \int_{\Sigma_{g,n}}
\frac{{\cal D}X{\cal D}g}{{\rm Vol}(\mbox{Diff})\times {\rm Vol}(\mbox{Weyl})}
\prod_{i=1}^{n} V_{\Lambda_{i}}(k_{i}) e^{-S_{\rm P}[X,g]},
\label{amplitude2}
\end{align}
where $V_{\Lambda_{i}}(k_{i})$ are the vertex operators representing
external strings with momenta $k_{i}$ and quantum numbers $\Lambda_{i}$,
while~$\chi$ is the Euler characteristic of a genus-$g$ manifold with~$n$
punctures,~$\chi=2-2g-n$.

What we have described is known as the Polyakov prescription for the computation of
string theory amplitudes: summing over all two-dimensional surfaces of genus $g$
and~$n$ punctures non-equivalent under the joined action of
diffeomorphisms and Weyl transformations. For every surface
$\Sigma_{g,n}$ the metric can be written, up to diffeomorphisms, in the form
\begin{align}
g_{ab}=e^{\phi(x)}\widehat{g}_{ab},
\end{align}
where~$\widehat{g}_{ab}$ is a fiducial metric. Since we are in critical
dimension, the dependence over the conformal factor $\phi(x)$ drops out
and the integration over $\phi(x)$ is cancelled
by the ${\rm Vol}(\mbox{Weyl})$ in the denominator. This means that we are left with the
problem of the classification of the fiducial metrics $\widehat{g}_{ab}$. As we will
see, the set $\{\widehat{g}_{ab}\}/\mbox{Diff}(\Sigma_{g,n})$ of conformally equivalent
metrics modulo diffeomorphism is finite-dimensional.

Taking complex coordinates, the problem of classifying  
compact orientable two-dimensional surfaces of
genus $g$ with punctures is equivalent to the classification  
of their complex structures. The space of
parameters labelling different classes is the moduli space of the
surface,~${\cal M}_{g,n}$. 
To characterize the moduli space of a genus~$g$ Riemann surface without
puntures, we apply the uniformization theorems
due to Klein, Poincar\'e, and Koebe: any Riemann surface~$\Sigma$
can be constructed from one of the following simply connected Riemann
surfaces, their universal cover~$\widehat{\Sigma}$:
\begin{itemize}
\item[-] 
The sphere $S^{2}$ with the round metric
\begin{align}
ds^{2}=\frac{dzd\overline{z}}{(1+z\overline{z})^{2}}.
\end{align}

\item[-] 
The complex plane $\mathbb{C}$ with the flat metric
\begin{align}
ds^{2}=dzd\overline{z}.
\end{align}

\item[-] 
The upper half plane $\mathbb{H}$ with the constant
negative curvature metric
\begin{align}
ds^{2}=\frac{2dzd\overline{z}}{(z-\overline{z})^{2}}.
\end{align}

\end{itemize}
The sphere is the covering space for genus zero Riemann surfaces, the
complex plane for $g=1$, and the upper half plane for any surface with
$g>1$. In any case, 
the corresponding Riemann surface is obtained from its universal cover~$\widehat{\Sigma}$ as the 
quotient~$\Sigma=\widehat{\Sigma}/\pi_{1}(\Sigma)$, where~$\pi_{1}(\Sigma)$
is the fundamental group of~$\Sigma$. Thus, to classify Riemann surfaces
into diffeomorphism classes, we need to find the action of
$\pi_{1}(\Sigma)$ on~$\widehat{\Sigma}$.

We also have the Gauss-Bonnet theorem
\begin{align}
\frac{1}{2\pi}\int_{\widehat{\Sigma}} d^{2}z R =2(1-g).
\end{align}

Let us begin with a genus zero Riemann surface. A theorem exits
stating that any Riemann surface homeomorphic to the sphere is also
isomorphic to it. This means that the moduli space for a genus zero
Riemann surface,~${\cal M}_{0}$, has a single point. For~$g=1$ 
the situation is not so trivial. Since the
only group acting on the complex plane without fixed points is the group of
translations, to get a torus we
have to divide~$\mathbb{C}$ by a subgroup of
discrete translations isomorphic to  its fundamental group~$\mathbb{Z}\oplus\mathbb{Z}$. 
We thus take two
complex numbers~$\omega_{1}$ and~$\omega_{2}$ and construct the
lattice generated by them. In fact, without loss of
generality, we can take one of the complex numbers equal to one
(e.g.,~$\omega_{1}=1$), so we are left only with a single complex modular
parameter,~$\omega_{2}=\tau$. Moreover, since the lattices
generated by $\tau$ and $-\tau$ are isomorphic, we can take~$\mbox{Im\,}\tau\geq 0$.

Tori parametrized by~$\tau$ represent equivalence classes under transformations
lying in the connected component of the identity.
We have however to restrict the values of $\tau$ in order to avoid
overcounting, because different $\tau$'s may correspond to
tori equivalent under diffeomorphisms not connected with the identity. 
This is in fact the case for two tori
characterized by modular parameters related by a M\"obius transformation
\begin{align}
\tau'=\frac{a\tau+b}{c\tau+d}, \hspace{1cm} ad-bc=1, \hspace{1cm}
a,b,c,d \in \mathbb{Z},
\end{align}
since the lattices defined by $\tau$ and $\tau'$ only differ in the choice of the fundamental cell. 
This means that we should consider only values of $\tau$ lying on the fundamental region of
the group~$\mbox{PSL}(2,\mathbb{Z})$, which we choose to be
\begin{align}
{\cal F}=\left\{\tau\in \mathbb{C} \,\left|\, \tau_{2}>0,\;
-\frac{1}{2}<\tau_{1}\leq\frac{1}{2},\;|\tau|>1\right\}\right.,
\label{eq:fundamental_region_F}
\end{align}
where~$\tau_{1}\equiv{\rm Re\,}\tau$ and~$\tau_{2}\equiv{\rm Im\,}\tau$.

\begin{figure}[t]
\centerline{\includegraphics[scale=0.45]{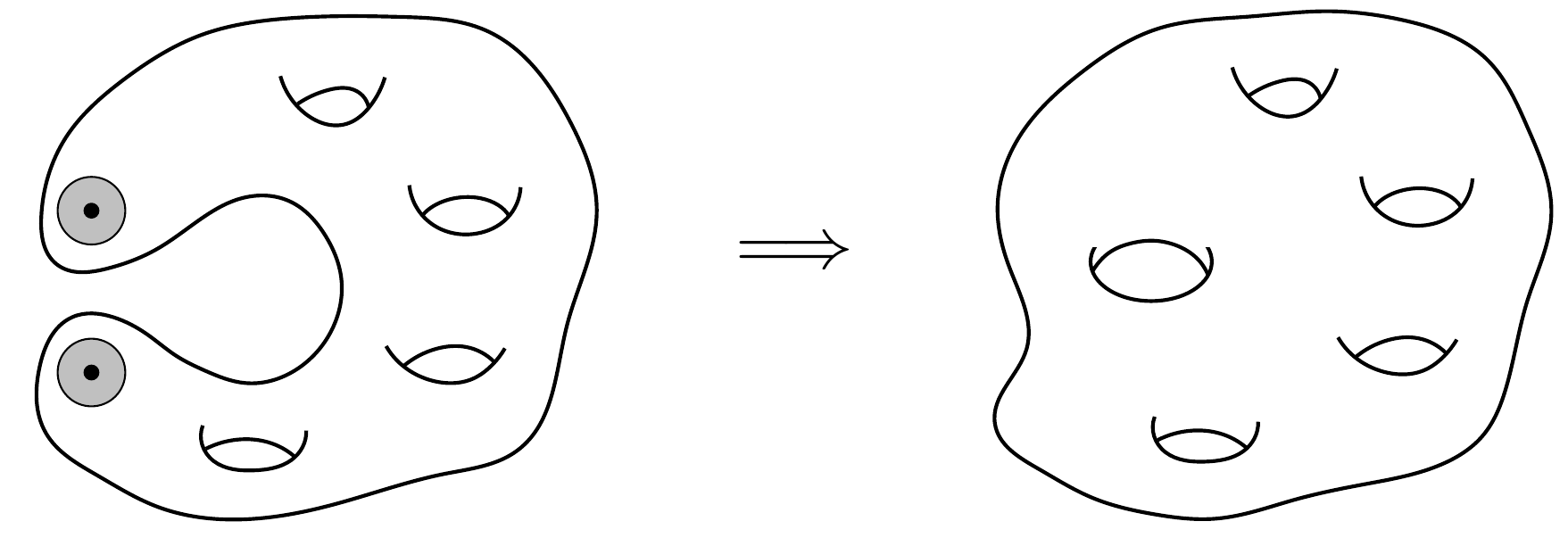}}
\caption{Sewing of two punctures in a genus $g$ Riemann surface to
obtain a genus $g+1$ Riemann surface}
\label{sewing}
\end{figure}

We have found that the modular space for a genus zero Riemann surface
${\cal M}_{0}$ consists of a single point while for genus
one ${\cal M}_{1}={\cal F}$. The case of higher genus requires more care. 
To construct the moduli space
${\cal M}_{g}$ with $g>1$ we are going to use the sewing technique.
Let be a genus-$g$ Riemann surface with two punctures
located at $P_{1}$ and $P_{2}$, together with
two annuli around these points with complex coordinates $z_{1}$ and $z_{2}$
such that $z_{1}(P_{1})=z_{2}(P_{2})=0$~(see fig.~\ref{sewing}). We can sew the Riemann surface
by identifying the two points on the annuli through the transformation
\begin{align}
z_{1}z_{2}=q,
\end{align}
with $q\in \mathbb{C}-\{0\}$ the sewing parameter. The result is
a Riemann surface with genus~$g+1$.

This construction allow us to establish a recursive relation between the
dimensions of the moduli spaces ${\cal M}_{g}$ and ${\cal M}_{g+1}$. 
The parameters characterizing the Riemann surface after
sewing two of its points are the moduli parameters of the
original Riemann surface of genus $g$, plus the three complex parameters
($z_{1},z_{2},q$) labelling the position of the points 
and the sewing parameter. This leads to the
relation
\begin{align}
{\rm dim\,}{\cal M}_{g+1}={\rm dim\,}{\cal M}_{g}+3.
\end{align}
We start with a genus-one surface. Due to the isometries of the
flat metric on the torus, one point can be located at~$z=0$, 
so the number of moduli parameters for a genus-two surface is three. 
Applying the sewing technique recursively,  
we find that dimension of the moduli space for a
genus-$g$ surface
\begin{align}
{\rm dim\,}{\cal M}_{0}&=0, \nonumber \\[0.2cm]
{\rm dim\,}{\cal M}_{1}&=1, \nonumber \\[0.2cm]
{\rm dim\,}{\cal M}_{g}&=3g-3, \hspace{1cm} g>1.
\end{align}

\begin{figure}[t]
\centerline{\includegraphics[scale=0.45]{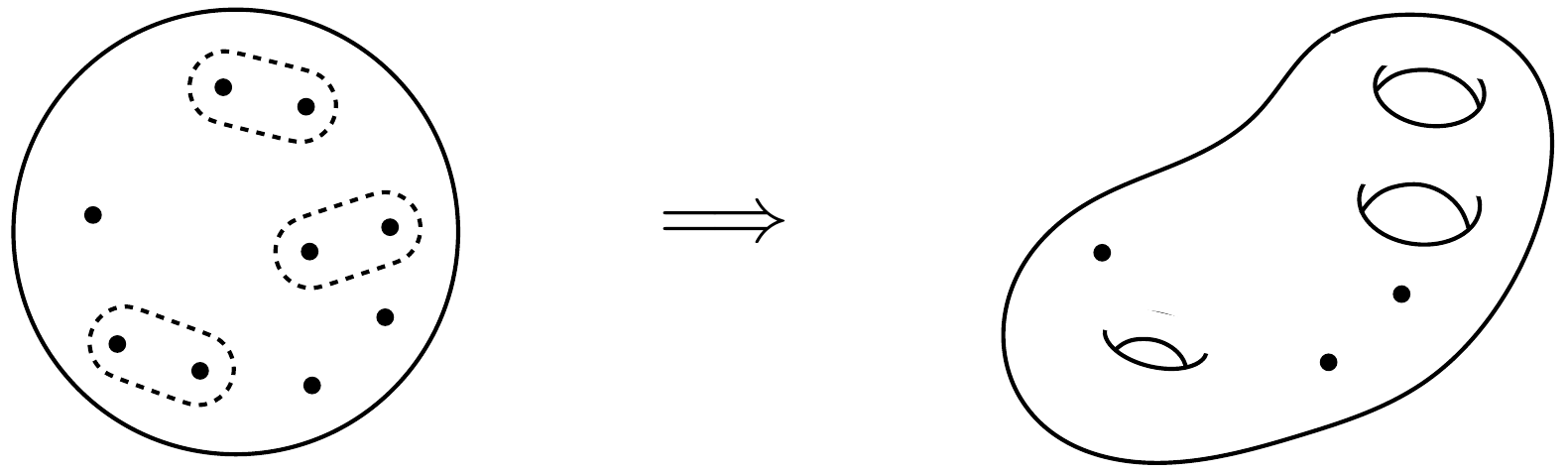}}
\caption{Construction of a genus $g$ Riemann surface with $n$ puctures
from a sphere with $2g+n$ punctures}
\label{construction}
\end{figure}

To compute string amplitudes we also need to find
the dimension of the moduli space ${\cal M}_{g,n}$ of genus-$g$ Riemann surfaces
with~$n$ punctures. Here we
use a slightly different method from the one employed to 
determine the dimension of~${\cal M}_{g}\equiv{\cal M}_{g,0}$.
We start with a sphere with $2g+n$ punctures and consider first the
case when~$2g+n>3$. By conformal invariance~\cite{Ginsparg}, 
three of the punctures can be located at~$z=0$,~$z=1$, and~$z=\infty$,
so we only need to specify~$2g+n-3$ complex parameters to
locate the remaining ones. Sewing~$2g$ punctures together
with~$g$ sewing parameters~$q_{1},\ldots,q_{g}$ increases this number
by~$g$, so we find that the
number of independent complex parameters in~$\mathcal{M}_{g,n}$ 
is~$3g+n-3$ for $g>1$ (see fig.
\ref{construction}). If~$2g+n\leq 2$, on the other hand,
all the punctures on the sphere can be brought to fixed positions.
The dimension of the moduli space equals then the number of
sewing parameters, which is equal to $g$.

We have seen that the space of two-dimensional metrics modulo
differmorphisms and Weyl rescalings is finite-dimensional. This means
that the path integration over metrics reduces to an
integration over the moduli space of the corresponding surface
\begin{align}
\int_{\Sigma_{g,n}}{{\cal D}g\over {\rm Vol}(\mbox{Diff})\times {\rm Vol}(\mbox{Weyl})}
\longrightarrow \int_{{\cal M}_{g,n}}
d\mu(g,n)
\end{align}
where $d\mu(g,n)$ is the integration measure over the moduli space. To
compute string amplitudes is necessary to characterize
this measure (see sec.~\ref{sec:operator_formalism}).

\subsection{Bosonic strings with background fields. ``Stringy''
corrections to the Einstein equations}
\label{sec:stringy_einstein_eq}

Up to now we have considered bosonic strings propagating in Minkowski
space. We are also interested in studying the propagation of
strings in general manifolds with metric $G_{\mu\nu}(X)$. This requires 
modifying the string action~\eqref{Polyakov} to include the
effect of the background metric
\begin{align}
S=-\frac{1}{4\pi \alpha'}\int d^{2}\sigma
\sqrt{g}g^{ab}\partial_{a}X^{\mu}\partial_{b}X^{\nu}G_{\mu\nu}(X).
\label{graviton-background}
\end{align}
In sec.~\ref{sec:quantization_bosonic} we also learned that, besides the graviton, 
the massless sector of the bosonic string contains also an antisymmetric tensor field 
and the scalar dilaton. It is natural to include also background values
for these two fields, that we denote by~$B_{\mu\nu}(X)$ and~$\Phi(X)$. The
general form of the string action is then given by
\begin{align}
S&=-\frac{1}{4\pi\alpha'}\int d^{2}\sigma
\sqrt{g}g^{ab}\partial_{a}X^{\mu}\partial_{b}X^{\nu}G_{\mu\nu}(X)
\label{action-background}
\\[0.2cm]
&-\frac{1}{4\pi\alpha'}\int d^{2}\sigma
\epsilon^{ab}\partial_{a}X^{\mu}\partial_{b}X^{\nu}B_{\mu\nu}(X)+
\frac{1}{4\pi}\int d^{2}\sigma \sqrt{g}R^{(2)}(g)\Phi(X), \nonumber
\end{align}
where~$\epsilon^{ab}$ is the
two-dimensional Levi-Civita tensor and~$R^{(2)}(g)$ is the world-sheet curvature.

For the string theory described by the action~(\ref{action-background}) to
be consistent, we have to impose that it defines a CFT at the quantum level.
This imposes constraints on the background fields. A very convenient way to
study the conformal invariance of (\ref{action-background}) is to write
the metric~$g_{\mu\nu}=e^{\phi}\widehat{g}^{\mu\nu}$ in $2+\epsilon$
dimensions, looking then for the conditions required for the conformal factor 
to drop out of the effective action in the limit~$\epsilon\rightarrow 0$. 
Doing this, we find the following equations for
the metric, antisymmetric tensor, and dilaton background fields, to lowest order in $\alpha'$
\cite{CFMP,FradTseyt}
\begin{align}
R_{\mu\nu}+\frac{1}{4}H_{\mu}^{\,\lambda\rho}H_{\nu\lambda\rho}-
2D_{\mu}D_{\nu}\Phi &= 0, \nonumber \\[0.2cm]
D_{\lambda}H^{\lambda}_{\,\mu\nu}-2D_{\lambda}\Phi\,H^{\lambda}_{\mu\nu}
&=0, \label{background-eqs}\\[0.2cm]
4(D_{\mu}\Phi)^{2}-4D_{\mu}D^{\mu}\Phi+R+\frac{1}{12}H_{\mu\nu\lambda}
H^{\mu\nu\lambda} &= 0,
\nonumber
\end{align}
where $H_{\mu\nu\lambda}$ is the field strength associated with the
antisymmetric tensor $B_{\mu\nu}$
\begin{align}
H_{\mu\nu\lambda}=\partial_{\mu}B_{\nu\lambda}+\partial_{\lambda}
B_{\mu\nu}+\partial_{\nu}B_{\lambda\nu}.
\end{align}
Remarkably, eq.~(\ref{background-eqs}) are
the field equations derived from the effective action functional
\begin{align}
\hspace{-.5cm} S_{d=26}=-\frac{1}{2\kappa^{2}}\int d^{26}x \sqrt{G}
e^{-2\Phi}\left( R-4D_{\mu}\Phi D^{\mu}\Phi+
\frac{1}{12}H_{\mu\nu\lambda}H^{\mu\nu\lambda}\right).
\end{align}

Alternatively, the equations to be satisfied by the background fields can be
obtained by imposing the vanishing of the beta-function of the two-dimensional
world-sheet theory. To simplify things, we forget about the antisymmetric tensor field and the
dilaton, considering the action~(\ref{graviton-background}). A computation of its
beta-function at one loop order gives (for the evaluation of
beta-functions in non-linear sigma-models, see ref.~\cite{Friedan})
\begin{align}
\beta_{\mu\nu}(X)=-\frac{1}{2\pi}R_{\mu\nu}(X).
\end{align}
Thus, conformal invariance requires~$\beta_{\mu\nu}(X)=0$, so the background metric
has to satisfy the vacuum Einstein equations. ``Stringy'' corrections to this result
are obtained by including higher order terms in~$\alpha'$. 
For example, 
computing the two-dimensional beta-function at two-loops we arrive at~\cite{Friedan}
\begin{align}
\beta_{\mu\nu}(X)=-\frac{1}{4\pi}\left[R_{\mu\nu}(X)+
\frac{\alpha'}{2}R_{\mu\lambda\sigma\tau}(X)
R_{\nu}^{\;\;\lambda\sigma\tau}(X) \right],
\end{align}
leading to the following order~$\alpha'$ corrections to the vacuum Einstein equations 
\begin{align}
R_{\mu\nu}(X)+\frac{\alpha'}{2}
R_{\mu\lambda\sigma\tau}(X)R_{\nu}^{\;\;\lambda\sigma\tau}(X)= 0.
\end{align}
The form of the beta-function to all orders in $\alpha'$ is not known.

\subsection{Toroidal compactifications. $R$-duality}
\label{sec:R-duality}

We discuss now duality, a distinct property of strings
with no analogue in QFT.
All known critical string theories are formulated in dimension higher
that four: $d=26$ for the bosonic string and, as we will see below, $d=10$ for 
supersymmetric strings. This means that, in order to connect with real world phenomenology, 
we have to go from~$26$ or~$10$ dimensiones to four. One way is implementing 
a Kaluza-Klein-style dimensional reduction.
Starting with, say, a $26$-dimensional manifold we compactify~$22$  
dimensions on a torus~$T^{22}$. Assuming that the typical radii of these 
compact dimensions is of the order
of the Planck length, the existence of extra dimensions would be
completely unobservable at the energies available in
current high-energy experiments. 

This
does not mean, however, that the compactification does not leave some imprint
in low-energy phenomenology, such as the isometries of the
internal manifold reflecting in the gauge group at low energies. 
In the last few years a great effort has been devoted
to the search of string models with acceptable four-dimensional
phenomenological properties, either using toroidal compactifications of the
type we discuss here or
more sophisticated internal manifolds such as orbifolds or Calabi-Yau manifolds.

To be more precise, we consider closed bosonic 
strings~\cite{Lust-Theisen} compactified on a $D$-dimensional torus $T^{D}=S^{1}\times
\stackrel{(D)}{\ldots}\times S^{1}$ and let us denote~$X^{i}$ the $26-D$ open 
dimensions and~$X^{I}$ the compact ones. The~$D$-dimensional torus is defined through
the identification
\begin{align}
X^{I}\sim X^{I}+\pi\sqrt{2}\sum_{i=1}^{D}n_{i}R_{i}e_{i}^{I}
\equiv X^{I}+2\pi L^{I}, \hspace{1cm} n_{i}\in \mathbb{Z},
\label{identification}
\end{align}
where~${\bf e}_{i}\equiv(e_{i}^{1},\ldots,e_{i}^{D})$ 
are~$D$ linearly independent vectors, normalized 
such that~${\bf e}_{i}^{2}=2$, and~$R_{i}$ are the radii of the compact dimensions.
The vectors $L^{I}$ define a $D$-dimensional lattice $\Lambda$ so the
torus is represented as the quotient
\begin{align}
T^{D}=\mathbb{R}^{D}/(2\pi \Lambda).
\end{align}
Translations of the string center~$x^{I}$ along the compact dimension 
are generated by the momenta~$p^{I}$. For the wave 
function~$\exp{(ix\cdot p)}$ to be single valued, the vector~$p^{I}$ 
must lie on~$(\Lambda^{D})^{*}$, the lattice dual to $\Lambda^{D}$,
generated by vectors~${\bf e}_{i}^{*}\equiv(e_{i}^{*1},\dots,e_{i}^{*D})$
satisfying
\begin{align}
{\bf e}_{i}\cdot {\bf e}_{j}^{*}=\delta_{ij} \hspace*{0.5cm} 
\mbox{with} \hspace*{0.5cm} {\bf e}_{i}^{* 2}=1/2. 
\end{align}
Compact momenta are then expanded as
\begin{align}
p^{I}=\sqrt{2}\sum_{i=1}^{D} \frac{m_{i}}{R_{i}}e^{*I}_{i}, 
\hspace*{1cm}m_{i}\in\mathbb{Z}.
\end{align}

Let us now study the propagation of a closed bosonic string on~$T^{D}$.
The basic difference with respect to the flat space case studied above is that
now the periodicity conditon~$X^{\mu}(\sigma+\pi,\tau)=
X^{\mu}(\sigma,\tau)$ can also be satisfied modulo the identification~(\ref{identification}),
namely
\begin{align}
X^{I}(\tau,\sigma+\pi)=X^{I}(\tau,\sigma)+2\pi L^{I}=
X^{I}(\tau,\sigma)+\pi\sqrt{2}\sum_{i=1}^{D}n_{i}R_{i}e_{i}^{I}.
\end{align}
Physically, this means that the string wraps~$n_{i}$ times around the $i$th compact
dimension and the integers $n_{i}$ is therefore known as winding numbers. To implement these
new boundary conditions, the expansion~\eqref{eq:bosonic_expansion_sigma_tau}
for the compact coordinates~$X^{I}(\tau,\sigma)=X_{L}^{I}(\tau+\sigma)+X_{R}^{I}(\tau-\sigma)$ 
has to be modified to
\begin{align}
X^{I}_{L}(\tau+\sigma)&=\frac{1}{2}x^{I}+
{1\over\sqrt{2\alpha'}}\big(\alpha'p^{I}+L^{I}\big)(\tau+\sigma)+ i\sqrt{\frac{\alpha'}{2}}\sum_{n\neq
0}\frac{\overline{\alpha}^{I}_{n}}{n} e^{-2in(\tau+\sigma)}, \nonumber \\[0.2cm]
X^{I}_{R}(\tau-\sigma)&=\frac{1}{2}x^{I}+
{1\over\sqrt{2\alpha'}}\left(\alpha'p^{I}-L^{I}\right)(\tau-\sigma)+ i\sqrt{\frac{\alpha'}{2}}\sum_{n\neq
0}\frac{\alpha^{I}_{n}}{n} e^{-2in(\tau-\sigma)}.
\end{align}
Thus, the mass formula and the level matching condition in eq.~\eqref{eq:bos_mass_form_LMC} 
now take the form
\begin{align}
\frac{1}{2}\alpha'm^{2}&={1\over 2\alpha'}\sum_{I=1}^{D}\Big(\alpha'^{2}
p^{I}p^{I}+
L^{I}L^{I}\Big)
+N_{R}+N_{L}-2, \nonumber \\[0.2cm]
N_{L}-N_{R}&=\sum_{I=1}^{D}p^{I}L^{I}.
\end{align}
It is important to keep in mind that the spectrum of the compactified bosonic
string still contains all the states of the
uncompactified theory in the so-called untwisted sector (with~$p^{I}=L^{I}=0$). 
This means that, in general, tachyons are not removed
by a naive toroidal compactification (they can still be removed by compactifying 
in the presence of Wilson lines, see ref.~\cite{Narain-Sarmadi}).
In addition, the theory contains two massless vector 
states~$\alpha^{\mu}\overline{\alpha}^{I}|0\rangle$ and
$\alpha^{I}\overline{\alpha}^{\mu}|0\rangle$, as well as $D^{2}$ massless
scalars~$\alpha^{I}\overline{\alpha}^{J}|0\rangle$. The vector
states are gauge bosons associated with the Abelian~$\mbox{U}(1)^{D}\times \mbox{U}(1)^{D}$ 
gauge invariance generated by the torus isometries.

Let us consider the simpler case of a single compactified
dimension with radius~$R$
\begin{align}
X^{25}\sim X^{25}+2\pi R\ell,
\end{align}
with $\ell\in\mathbb{Z}$ the winding number. The mass formula reads
\begin{align}
\frac{1}{2}\alpha'm^{2}=\frac{\alpha'M^{2}}{2R^{2}}+{\ell^{2}R^{2}\over 2\alpha'}+N_{L}+N_{R}-2,
\end{align}
with $N_{L}-N_{R}=M\ell$ and~$M\in\mathbb{Z}$. A very important
property of
the spectrum of this theory is the invariance of
the mass formula under the replacement~\cite{duality}
\begin{align}
M\longrightarrow \ell, \hspace{2cm} \ell\longrightarrow -M, \hspace{2cm}
R \longrightarrow \frac{\alpha'\hbar}{R},
\label{replace}
\end{align}
where we have restored the powers of~$\hbar$.
Interesting as it is, the symmetry of the mass formula does
not guarantee that the whole interacting theory remains invariant 
under~(\ref{replace}). It should be checked that this property is
also shared by string amplitudes.

Using the low energy field theory, it was shown in ref.~\cite{Ginsparg-Vafa} 
that for the theory to be invariant under~(\ref{replace}) the string
coupling constant~$g_{\rm st}$, determined by the vacuum expectation value of the dilaton
field~$\Phi$, should also transform according to
\begin{align}
g_{\rm st} \longrightarrow	g_{\rm st}\frac{\sqrt{\alpha'\hbar}}{R},
\end{align}
The induced transformation of the dilaton vacuum expectation value 
can also be obtained in a sigma-model
approach imposing that the transformed sigma-model
also preserves conformal invariance at zeroth order in~$\alpha'$~\cite{Buscher}. 
It was finally proved in ref.~\cite{Alvarez-Osorio-3} that
the duality transformation
\begin{align}
R\longrightarrow \frac{\alpha'\hbar}{R},\, \hspace{1cm}
g_{\rm st}\longrightarrow g_{\rm st}\frac{\sqrt{\alpha'\hbar}}{R},
\label{g}
\end{align}
is a symmetry of the whole string perturbation theory for the
bosonic and the heterotic string. This symmetry can also be shown to
hold in $c=1$~non-critical strings for the whole perturbative
expansion, once the matrix model is truncated to the sector without
vortices~\cite{Gross-Klebanov}.

The property that
the theory at radius~$R$ is equivalent to the one at radius~$\alpha'\hbar/R$, with an
appropriate transformation of the string coupling, is called~$R$-duality (for a
general review see \cite{J-Schwarz}).
This is a characteristic ``stringy'' property stemming for the extended nature of the
string and therefore has no counterpart in
field theory.

$R$-duality is not exclusive of the bosonic string, being present as well in 
fermionic string models like the heterotic and closed
superstrings. In recent years this property has been considered as
lending support to the idea that string theory has a fundamental minimal
length scale. In fact, the study of string scattering at high energies has led
to the formulation of a generalized uncertainity principle~\cite{Veneziano}
\begin{align}
\Delta x \sim \frac{\hbar}{\Delta E}+\alpha'\Delta E,
\label{Veneziano}
\end{align}
setting a lower bound for any measurable distance at~$\sqrt{\alpha'\hbar}$,
the self-dual radius under~\eqref{replace}. 
This makes sense, since~$R$-duality implies that the string does no
distinguish between dimensions of size~$R$ and~$(\alpha'\hbar)/R$. 
Moreover, the fact that this minimum distance also equals the string
length defined in eq.~\eqref{eq:string_length} indicates that
strings cannot probe distances shorter than
their own characteristic size.

\subsection{Operator formalism}
\label{sec:operator_formalism}

Our study of string theory has been carried out so far at a minimum of 
mathematical sophistication. We are changing this to discuss
the operator formalism for bosonic strings~\cite{Alvarez-Gaume-Gomez-Moore-Vafa},
a framework that allows addressing many
features of string theory at arbitrary orders of perturbation
theory, among them the origin of potential
infinities in string amplitudes and the geometrical interpretation
of the physical state conditions. Before presenting the operator formalism,
however, we review some basic features of Riemann surfaces 
that will turn out be useful later~\cite{Alvarez-Gaume-Moore-Vafa}. 

\begin{figure}[t]
\centerline{\includegraphics[scale=0.45]{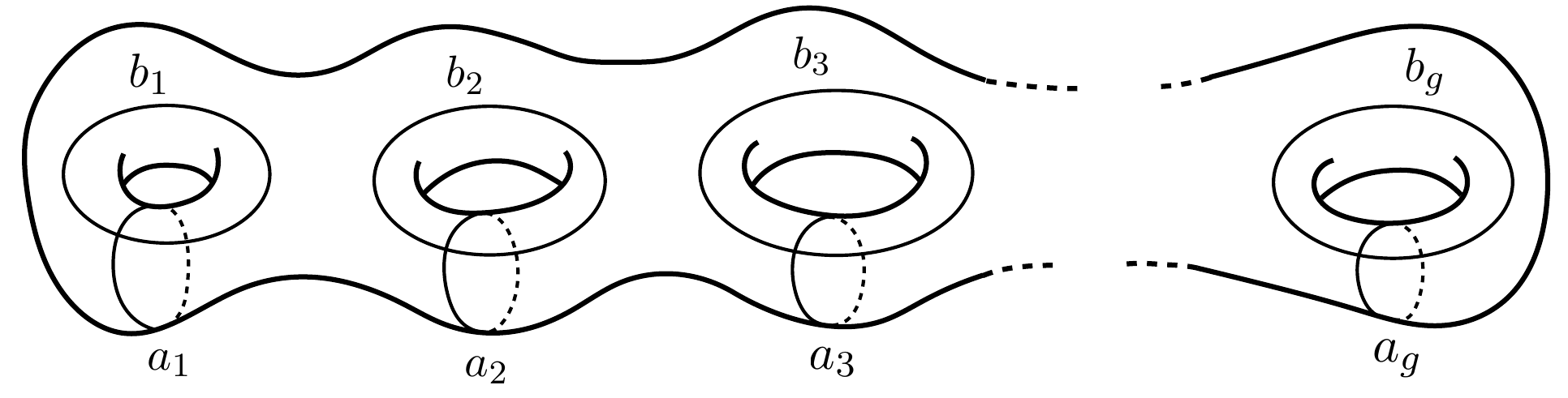}}
\caption{Basis of the first homology group $H_{1}(\Sigma_{g})$}
\label{cycles}
\end{figure}

\begin{figure}[t]
\centerline{\includegraphics[scale=0.45]{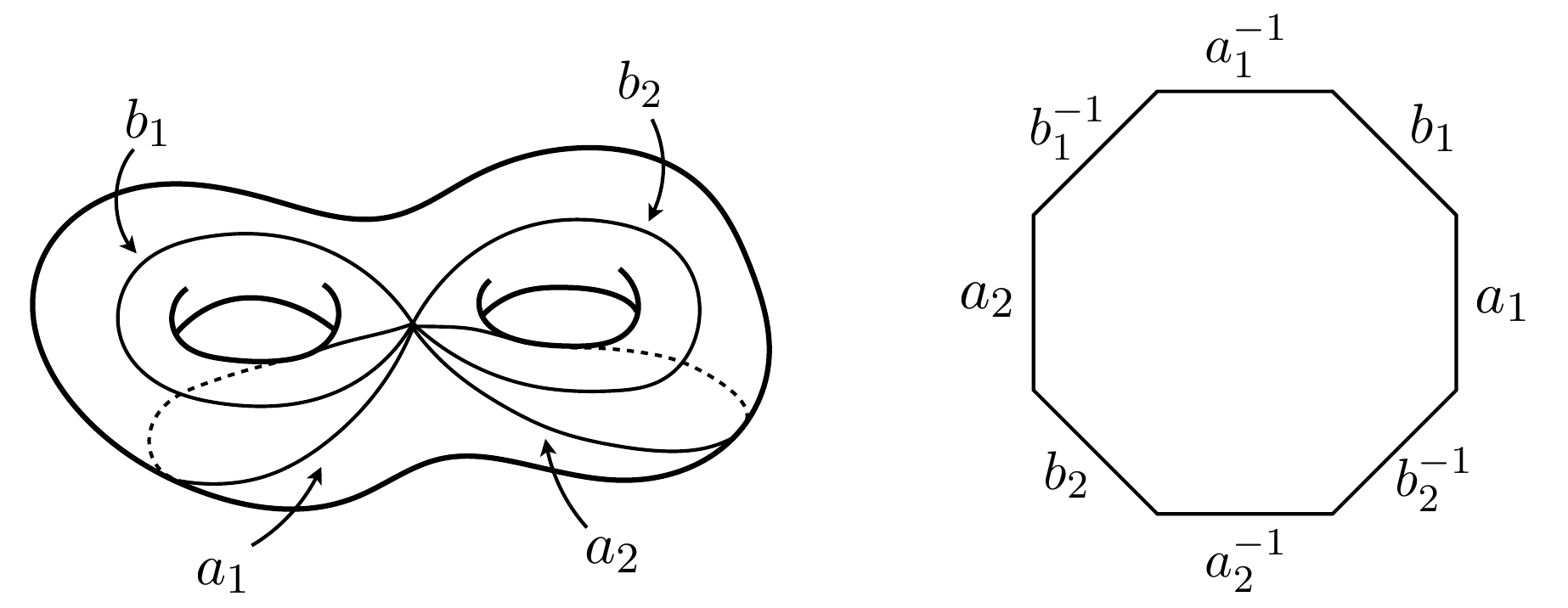}}
\caption{Octagon obtained by cutting a genus-$2$ Riemann surface along
the elements of the basis of $H_{1}(\Sigma_{2})$}
\label{g=2}
\end{figure}

Every closed, orientable two-dimensional surface~$\Sigma_{g}$ is topologically characterized by its
genus~$g$ or, equivalently, by its Euler characteristic $\chi=2-2g$.
Its homology groups have dimensions
\begin{align}
{\rm dim\,}H_{0}(\Sigma_{g})=1, \hspace{1cm} {\rm dim\,}H_{1}(\Sigma_{g})=2g,
\hspace{1cm} {\rm dim\,}H_{2}(\Sigma_{g})=1,
\end{align}
and in the case of the first homology group~$H_{1}(\Sigma_{g})$ a basis can be chosen
considering the~$2g$ cycles~$(a_{i},b_{i})$, with~$1\leq i \leq g$, shown
in fig. \ref{cycles},
so any closed curve on~$\Sigma_{g}$ can be uniquely decomposed
in terms of curves belonging to their homology classes.
Moreover, once a homology basis is chosen, any
genus-$g$ surface can be seen as a $4g$-sided polygon with a proper identifications
of its sides. For example, in fig.~\ref{g=2} we show how a
genus-$2$ surface is constructed from an octagon by gluing together the 
sides~$a_{1}a_{1}^{-1}$,~$a_{2}a_{2}^{-1}$,~$b_{1}b_{1}^{-1}$, and $b_{2}b_{2}^{-1}$.

Associated with the homology basis~$(a_{i},b_{i})$ there is a set of~$2g$ one-forms
$(\alpha_{i},\beta_{i})$, with~$1\leq i \leq g$, satisfying
\begin{align}
\int_{a_{i}}\alpha_{j}&=\delta_{ij}, \hspace{2cm}
\int_{a_{i}}\beta_{j}=0, \nonumber \\[0.2cm]
\int_{b_{i}}\alpha_{j}&=0, \hspace{2.3cm}
\int_{b_{i}}\beta_{j}=\delta_{ij}.
\end{align}
Introducing a complex structure on the two-dimensional surface
we have a Riemann surface whose cotangent space is decomposed
as the direct sum
\begin{align}
T^{*}\Sigma_{g}=T^{*(1,0)}\Sigma_{g}\oplus T^{*(0,1)}\Sigma_{g},
\end{align}
where the elements of~$T^{*(1,0)}\Sigma_{g}$ and~$T^{*(0,1)}\Sigma_{g}$ can be locally written
as~$f(z,\overline{z})dz$ and $f(z,\overline{z})d\overline{z}$, respectively. In particular there exists
a set of $g$ holomorphic $(1,0)$ differentials $\omega_{i}$,
locally of the form $\omega_{i}=f_{i}(z)dz$, 
normalized according to
\begin{align}
\int_{a_{i}}\omega_{j}=\delta_{ij}.
\end{align}
This condition completely determines the holomorphic differentials~$\omega_{i}$,
called Abelian differentials of the first kind. Integrating~$\omega_{i}$ 
along the cycle~$b_{j}$ gives the period matrix of the
Riemann surface
\begin{align}
\int_{b_{i}} \omega_{j}=\Omega_{ij}.
\end{align}
It can be shown that this matrix is symmetric,
$\Omega_{ij}=\Omega_{ji}$, and satisfies~${\rm Im\,}\Omega_{ij}>0$. The matrices
satisfying these properties define the Siegel upper half plane $\mathbb{H}_{g}$,
which contains the Teichm\"{u}ller space of a genus-$g$ Riemann surface
\begin{align}
{\cal
T}_{g}=\frac{\mbox{Metrics}}{\mbox{Weyl}\times
\mbox{Diff}_{0}^{+}(\Sigma_{g})},
\end{align}
where $\mbox{Diff}_{0}^{+}(\Sigma_{g})$ is the set of orientation-preserving
diffeomorphisms on $\Sigma_{g}$ in the connected component of the identity. The moduli 
space~$\mathcal{M}_{g}$ of inequivalent complex structures on~$\Sigma_{g}$
introduced in sec.~\ref{sec:int+moduli} is related to the Teichm\"{u}ller
space by 
\begin{align}
{\cal M}_{g}=\frac{{\cal T}_{g}}{\Omega(\Sigma_{g})}.
\end{align}
Here the mapping class group~$\Omega(\Sigma_{g})$ is defined as 
the set of orientation-preserving diffeomorphism
modulo those connected to the identity,~$\Omega(\Sigma_{g})
=\mbox{Diff}^{+}(\Sigma_{g})/\mbox{Diff}_{0}^{+}(\Sigma_{g})$.

Another important issue in the theory of Riemann surfaces is the
definition of spin structures. Given a genus-$g$ Riemann surface there are~$2^{2g}$
spin structures, roughly speaking corresponding to the choices of periodic or
antiperiodic boundary conditions for fermions along each generator
of the homology group~$H_{1}(\Sigma_{g})$. These spin structures are
classified into two groups depending on whether the number of
zero modes of the Weyl operator is even or odd.
In the case of the torus, for example, we have four spin structures
corresponding to $(P,P)$,~$(P,A)$,~$(A,P)$, and~$(A,A)$, with~$P$ and~$A$
respectively indicating periodic and antiperiodic boundary conditions along the
corresponding element of the homology basis. Choosing a flat metric on the torus
it can be seen that that~$(P,P)$ is the only odd spin structure, while the other three are even.

\begin{figure}[t]
\centerline{\includegraphics[scale=0.45]{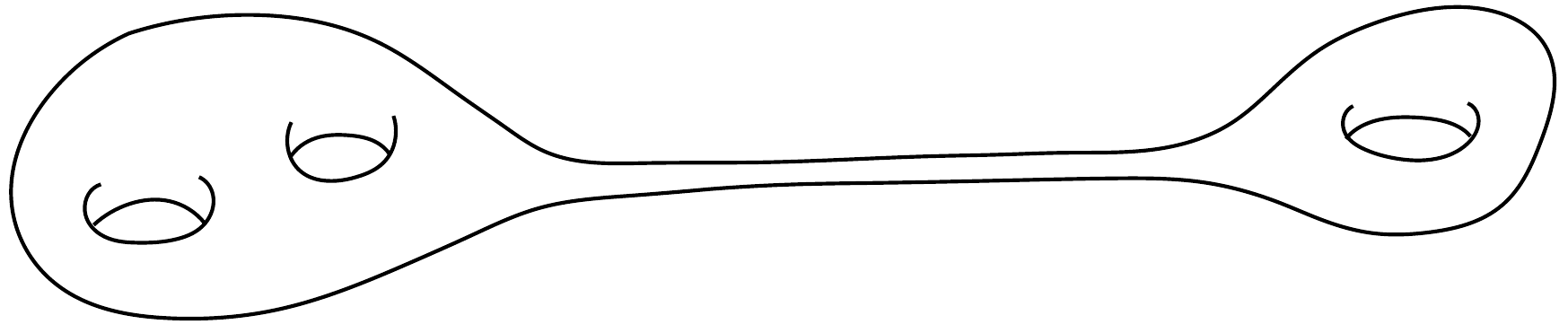}}
\caption{Riemann surface with a long narrow tube. Divergences in
string amplitudes appear when summing over the contributions from
this kind of surfaces, in the limit in which the length of the tube
tends to infinity}
\label{tubes}
\end{figure}

After these preliminaries, we focus on the analysis of
the origin of the divergences appearing in the partition function of the
bosonic string~\cite{Nelson} (the analysis of fermionic
strings proceeds along similar lines). The key observation is that divergences
in string perturbation theory come from the contribution of surfaces lying on the 
boundary of moduli space~${\cal M}_{g}$. In the case of the bosonic string, 
the presence of the tachyon renders the contribution coming from surfaces with long narrow
tubes, like the one showed in fig.~\ref{tubes}, divergent.
In the case of the bosonic string, the origin of this divergence can 
be traced back to the cylinder partition function
constructed in sec.~\ref{sec:CFT}. The contribution of the tachyon in 
eq.~\eqref{eq:CFTpartition_function} goes like~$q^{-1}$, diverging in
the limit of large cylinder lengths~$\tau\rightarrow i\infty$. 

\begin{figure}[t]
\centerline{\includegraphics[scale=0.5]{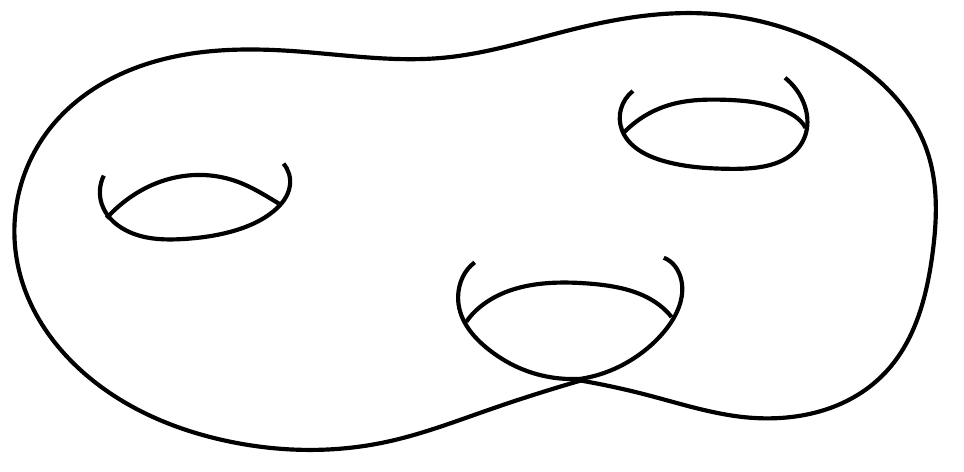}}
\caption{Riemann surface with a cycle pinched off}
\label{pinched}
\end{figure}

\begin{figure}[t]
\centerline{\includegraphics[scale=0.45]{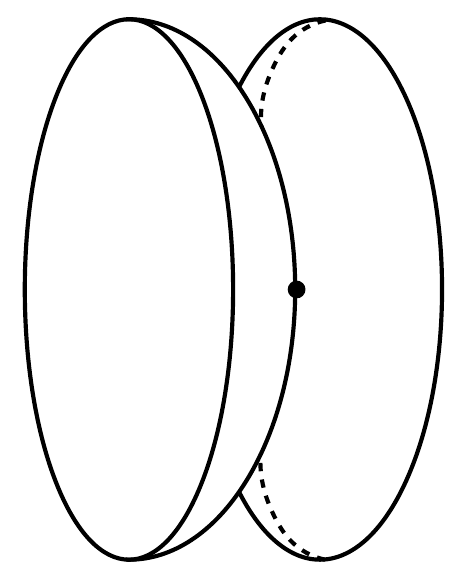}}
\caption{Local aspect of a node}
\label{node}
\end{figure}

For this reason, 
in order to study divergences of string diagrams, it is convenient to
consider a compactified moduli space~$\overline{\cal M}_{g}$ including 
all the points on the boundary of~${\cal M}_{g}$.
Points in~$\overline{\cal M}_{g}-{\cal M}_{g}$ correspond to Riemann
surfaces with infinitely long narrow tubes. These are conformally related 
to surfaces with nodes
with some cycle is pinched off (see fig.~\ref{pinched}).
The neighborhood of these nodes are topologically
equivalent to two disks joined by their centers, a situation depicted in fig.~\ref{node}.

\begin{figure}[t]
\centerline{\includegraphics[scale=0.45]{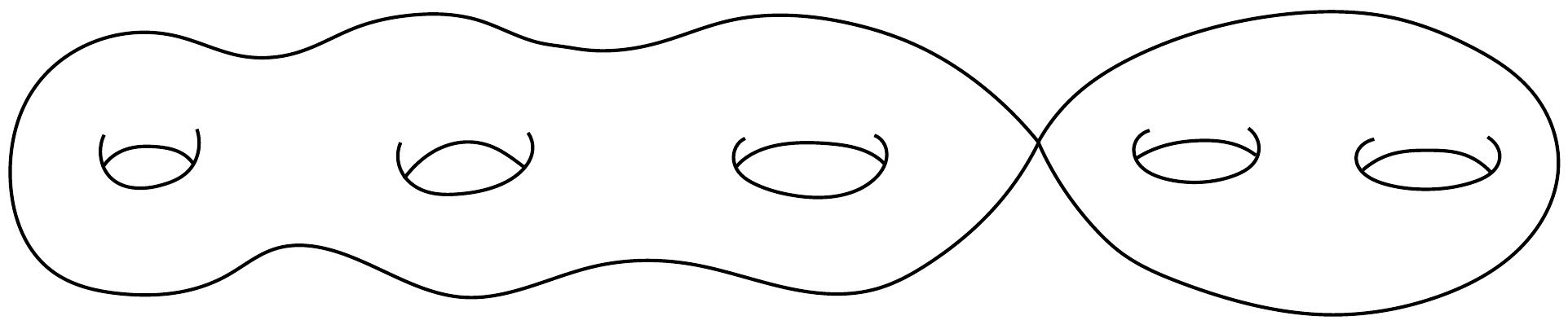}}
\caption{Splitting of a genus $g$ Riemann surface into two Riemann
surface by pinching off a homologically trivial cycle}
\label{Delta}
\end{figure}

Let us introduce some notation. 
We say that a surface in the boundary of the moduli space is of 
type~$\Delta_{i}$ if the degenerate surface splits a genus-$g$ surface into
two joined surfaces of genus~$i$ and~$g-i$~(see fig. \ref{Delta}). By
$\Delta_{0}$ we denote, on the other hand,
those surfaces where a homologically non-trivial cycle
is pinched
off, as in fig.~\ref{pinched}, and the result is a surface of genus~$g-1$ where two points
are identified\footnote{These two kinds of nodes are respectively known as 
splitting and non-splitting.}. 
This latter type of surfaces can be obtained by applying the sewing 
technique discussed in sec.~\ref{sec:int+moduli} to a Riemann surface
and taking the sewing parameter~$q=0$.

We now present the construction of the operator formalism on a
genus-$g$ Riemann surfaces. In sec.~\ref{sec:quantization_bosonic} 
we learned how scattering amplitudes on the sphere could be written
as correlation function on the punctured complex plane~$\mathbb{C}^{*}$
of the type
\begin{align}
\langle 0|\Phi_{1}(z_{1})\ldots \Phi_{N}(z_{N})|0\rangle.
\end{align}
For higher genus surfaces, we would like to construct a
state~$|W\rangle$ containing the details about the topology and
complex structure of the Riemann surface, and such that
\begin{align}
\frac{\langle 0|\Phi_{1}(z_{1})\ldots
\Phi_{N}(z_{N})|W\rangle}{\langle 0|W\rangle},
\end{align}
coincides with the result of the path integral computation of the
corresponding scattering amplitude. Let us define the surface's augmented
moduli space~${\cal P}(g,n)$ as
the moduli space of genus-$g$ surfaces with~$n$ parametrized boundaries,
i.e. $n$ distincted points~$P_{i}$ and~$n$ local complex coordinates~$z_{i}(P_{i})$
on their neighboorhood. Its fundamental group~$\pi_{1}\big({\cal P}(g,n)\big)$
equals the mapping class group of a genus-$g$ Riemann surface with~$n$ 
distinguished points,~$\Omega(\Sigma_{g,n})
=\mbox{Diff}(\Sigma_{g,n})/\mbox{Diff}_{0}(\Sigma_{g,n})$. One advantage of
using~${\cal P}(g,n)$ instead of~${\cal M}_{g,n}$ is that in order
to define oscillators and creation-annihilation operators we have to implement
Laurent expansions of the various fields in the local parameters~$z_{i}$ around
the points~$P_{i}$. In addition, ${\cal P}(g,n)$
resolves the orbifold singularities of ${\cal M}_{g,n}$.

\begin{figure}[t]
\centerline{\includegraphics[scale=0.45]{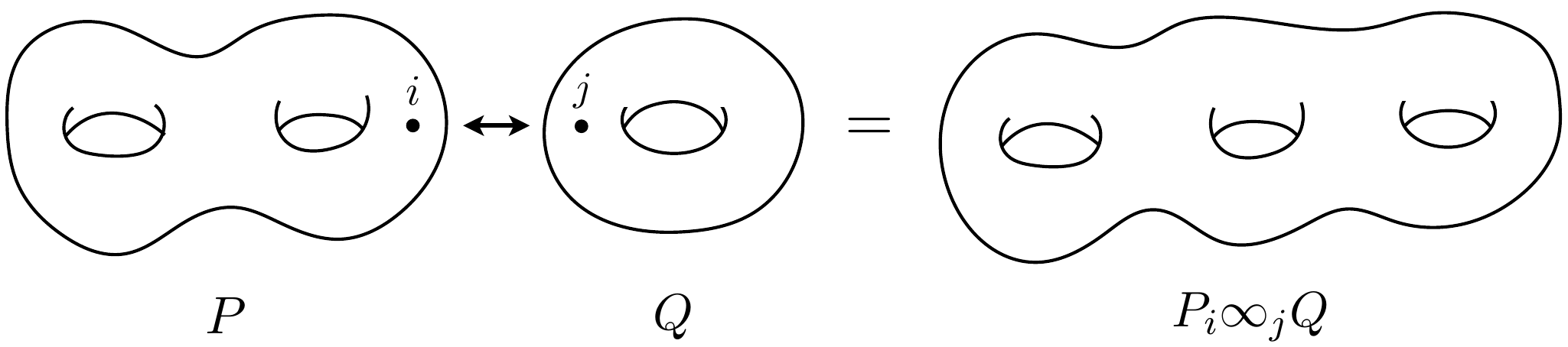}}
\caption{Construction of a genus-$(g_{1}+g_{2})$ Riemann surface by sewing
together two surfaces with genera $g_{1}$ and $g_{2}$}
\label{operations}
\end{figure}

\begin{figure}[t]
\centerline{\includegraphics[scale=0.45]{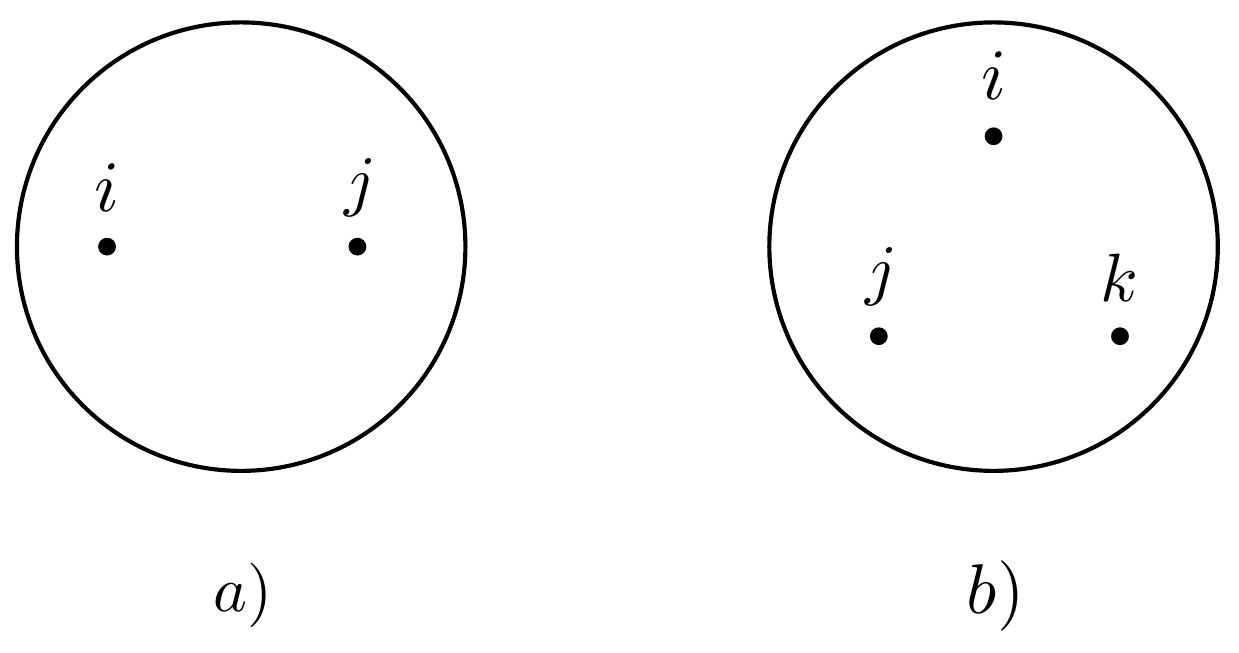}}
\caption{Buildings blocks for the construction of any element of ${\cal
P}(g,n)$}
\label{buildings}
\end{figure}

To construct the operator formalism for a general Riemann surface
we need a map between~${\cal P}(g,n)$
and~${\cal H}\otimes\ldots\otimes{\cal H}$, where ${\cal H}$ is the Hilbert
space of the theory. This means that to every CFT we associate
a ray in ${\cal H}^{\otimes n}$ for every point of~${\cal P}(g,n)$. 
Sewing points on Riemann surfaces defines a 
``composition''~${\cal P}(g_{1},n_{1})\times {\cal P}(g_{2},n_{2})
\longrightarrow {\cal P}(g_{1}+g_{2},n_{1}+n_{2}-2)$ such that
\begin{align}
P\in {\cal P}(g_{1},n_{1}),\,\, Q\in {\cal P}(g_{2},n_{2}) &\longrightarrow
P_{i}\infty_{j} Q \in {\cal P}(g_{1}+g_{2},n_{1}+n_{2}-2), 
\label{eq:first_operation_OF}
\end{align}
corresponding to sewing together the $i$th puncture of the
first surface with the $j$th puncture of the second 
(see fig.~\ref{operations}). A second 
operation~${\cal P}(g,n) \longrightarrow  {\cal P}(g+1,n-2)$ is defined by
\begin{align}
P \in {\cal P}(g,n) &\longrightarrow  P8^{i}_{j} \in {\cal P}(g+1,n-2),
\label{eq:second_operation_OF}
\end{align}
in which the $i$th and $j$th punctures are sewn together, as shown in fig.
\ref{sewing}.
Using these two operations, the moduli space~${\cal P}(g,n)$
can be iteratively constructed from~${\cal P}(0,3)$ and ${\cal P}(0,2)$
corresponding to the Riemann surfaces in fig.~\ref{buildings}.

The operations~\eqref{eq:first_operation_OF} 
and~\eqref{eq:second_operation_OF} admit representations in the space of states.
Taking an orthonormal basis~$\{|n\rangle\}$ of~${\cal H}$,
we associate with the Riemann surface in fig.~\ref{buildings}-{\it a}) the
state
\begin{align}
|S_{ij}\rangle=\sum_{n}|n\rangle_{i}|n\rangle_{j}.
\end{align}
Thus, given two Riemann surfaces~$P\in {\cal P}(g_{1},n_{1})$ and~$Q \in{\cal
P}(g_{2},n_{2})$ whose corresponding states are~$|P\rangle \in {\cal H}^{\otimes
n_{1}}$ and~$|Q\rangle \in {\cal H}^{\otimes n_{2}}$, we associate
with~$P_{i}\infty_{j}Q$ the state
\begin{align}
|R\rangle=\langle S_{ij}|P\rangle\otimes |Q\rangle.
\end{align}
This shows how starting with~${\cal P}(0,3)$ and using the state~$|S_{ij}\rangle$ 
we can in principle reach a Riemann surface with any genus and any number of punctures. 
In a sense, 
the procedure just described is quite similar to how Feynman diagram
are drawn by joining the legs on the theory 
vertices\footnote{This method has been explicitly employed in the
construction of a closed string field theory by B. Zwiebach (see his 
lecture~\cite{Zwiebach_lectures} at this school and ref.~\cite{SFT}).}.

Once we have a correspondence between elements in~${\cal P}(g,n)$ and
states in~${\cal H}^{\otimes n}$ we can reformulate Polyakov's
prescription to compute string scattering amplitudes
in the operator language. For example, 
the contribution of a Riemann surface $W\in {\cal P}_{g,n}$ to
the amplitude of $n$~external strings in the on-shell 
states~$|\chi_{i}\rangle \in {\cal H}$ is given by
\begin{align}
\langle\chi_{1}|\ldots\langle\chi_{n}|W\rangle.
\end{align}
Thus, the total amplitude $A(1,\ldots,n)$ can be formally written as
\begin{align}
A(1,\ldots,n)=\sum_{g=0}^{\infty}g_{\rm st}^{-\chi}\sum_{W}
\langle\chi_{1}|\ldots\langle\chi_{n}|W\rangle,
\label{amp.}
\end{align}
where the sum over Riemann surfaces is implemented by an integral over
moduli parameters.

It is worth-stressing that a ray is associated with
a point of~${\cal P}(g,n)$ only for CFTs. Furthermore, for $g>1$ the first homology
group~$H_{1}[{\cal P}(g,n)]$ vanishes, a result known as Harer theorem \cite{Harer}. 
This is very useful
to characterize the Polyakov measure, since it implies
that any flat line bundle is necessarily trivial. From eq.~\eqref{amp.} it
is clear that the aim of the string operator formalism is to obtain an
operator representation of ``scattering'' measures on~${\cal M}_{g,n}$.

A problem still pending is to determine the action of the Virasoro
generators on the space of states.
Let be~$R\in {\cal P}(g,n)$ and let $P_{i}$ be a point on $R$ with
local parameter $z_{i}$~(see fig.~\ref{disk}). We can
cut an annulus off the disk around this point and transform it by the action of a
meromorphic vector field~$v(z_{i})$
\begin{align}
z_{i}\rightarrow z_{i}+\epsilon v(z_{i}),
\label{vector}
\end{align}
taking into account that~$v(z_{i})$ may have poles at $P_{i}$.
Once this is done, we fill the inside of the annulus to get a disk and
glue it back on the surface. This is a deformation of the
original Riemann surface, so we may wonder about the relation of the
new surface with the original one. Here, there are three possibilities:  
\begin{figure}[t]
\centerline{\includegraphics[scale=0.45]{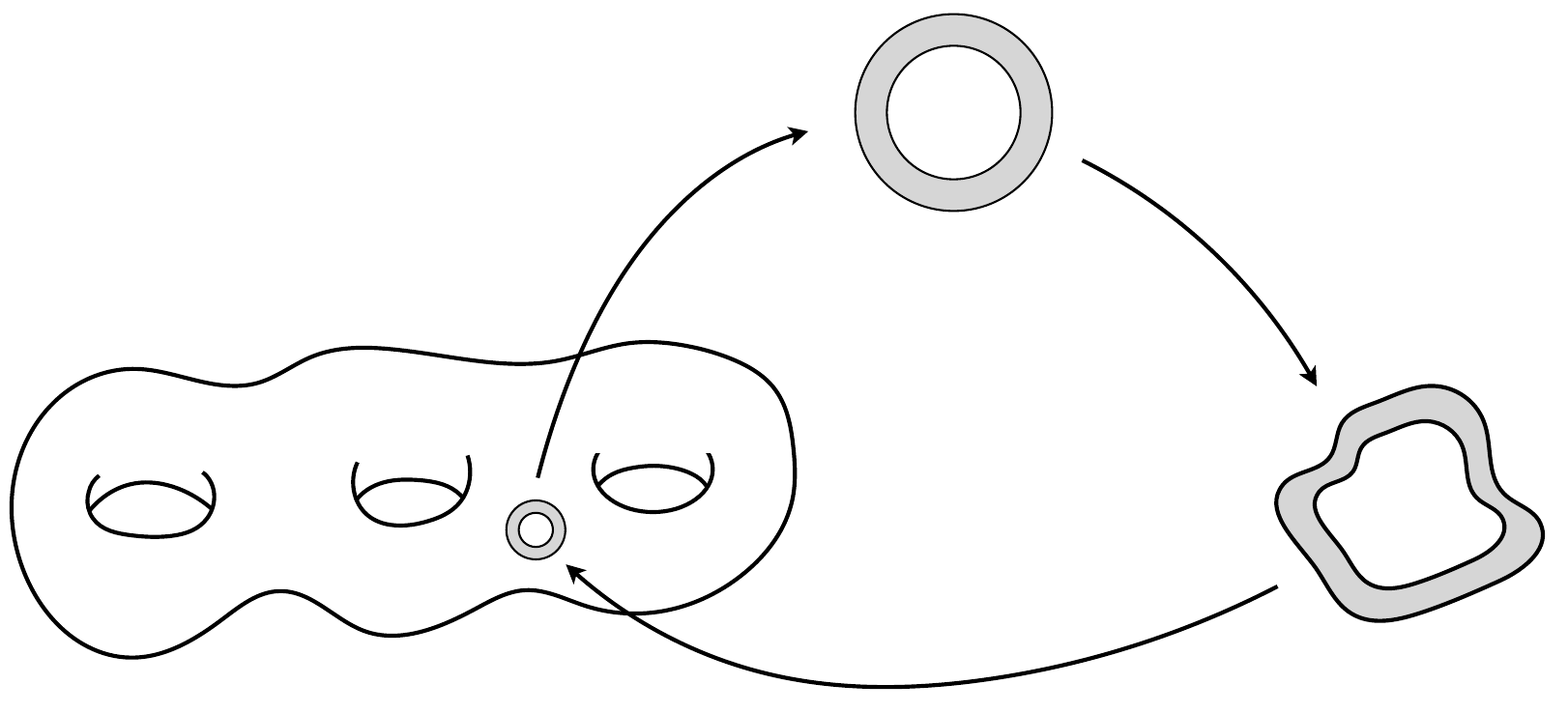}}
\caption{Deformation of the Riemann surface by cutting out a disk and
gluing it back after deforming it}
\label{disk}
\end{figure}

\begin{itemize}
\item[-] The vector field $v(z_{i})$ is holomorphic on the disk and
vanishes at $P_{i}$.
In this case the transformation given in eq.~(\ref{vector}) is equivalent to choosing
different coordinates at $P_{i}$. If the vector field
is holomorphic on the disk but does not vanish at $P_{i}$, what we have
is an infinitesimal translation of the point~$P_{i}$.

\item[-] The vector field can be extended to a holomorphic vector field on
the remaining of the surface~$R-P_{i}$. The
transformed surface would be identical to the original one, since any 
transformation~(\ref{vector})
can be undone by a transformation generated by~$v$ on~$R-P_{i}$.

\item[-] None of the above two possibilities are true. That is,~$v(z)$ is holomorphic
in the annulus but does not extend holomorphically either to the disk or
to~$R-P_{i}$. If this happens, the transformtion~(\ref{vector})
is an infinitesimal moduli deformation, i.e. an infinitesimal motion in
the space~${\cal M}_{g,n}$.

\end{itemize}

In view of the last possibility, we have to deal with the
problem of representing moduli deformations, or any
change induced by~$v(z)$ on~${\cal H}^{\otimes n}$. This is implemented by an 
operator~$O(v)$ acting on the states~$|W\rangle$. Because of the fundamental role played
by the Virasoro algebra in any CFT, we consider the Laurent expansion of 
the energy-momentum tensor in a neighborhood of~$P_{i}$
\begin{align}
T(z)=\sum_{n\in\mathbb{Z}} L_{n}z^{-n-2},
\end{align}
and construct the operator~$T(v)$ associated with the deformation
induced by~$v(z)$
\begin{align}
T(v)=\frac{1}{2\pi i}\oint_{P_{i}}dz\,T(z)v(z),
\label{eq:T(v)_definition}
\end{align}
so the change on the state~$|W\rangle$ is given by
\begin{align}
\delta_{v}|W\rangle=[T(v)+\overline{T}(\overline{v})]|W\rangle.
\label{delt-v}
\end{align}
This expression should be interpreted as defining the action of the
Virasoro algebra on the theory's Hilbert space. Since for matter fields
the Virasoro algebra has a central extension, eq.~(\ref{delt-v})
only makes sense in terms of rays.

A final consistency condition in the construction of $|W\rangle$ for
$P\in {\cal P}(g,n)$ requires that~$(L_{0}-\overline{L}_{0})|W\rangle=k|W\rangle$,
with~$k\in \mathbb{Z}$. This
condition has to be imposed since we always have the freedom to make a Dehn twist
around the point $P$, which in the local
holomorphic coordinate reads~$z \rightarrow \exp{(2\pi i\theta)}z$, with $0\leq \theta
\leq 1$. On the Hilbert space, this transformation is implemented by the
operator~$\exp{[2\pi i\theta(L_{0}-\overline{L}_{0})]}$ and for~$\theta=1$ 
invariance of the state requires~$|W\rangle$ 
to be an eigenstate of $L_{0}-\overline{L}_{0}$ with integer eigenvalue. 
This condition is necessary
to preserve modular invariance.

\paragraph{A pair of spin-$\mathbf{1\over 2}$ fermions.}
We now leave our general discussion in order to discuss a few useful examples
to clarify the general formalism presented so far.
We study in the first place the operator formalism for 
a pair of spin-$\frac{1}{2}$ fields. They admit the expansion
\begin{align}
b(z)dz^{\frac{1}{2}}&=\sum_{n\in\mathbb{Z}
+\frac{1}{2}}b_{n}z^{-n-\frac{1}{2}}dz^{\frac{1}{2}}, \nonumber \\[0.2cm]
c(z)dz^{\frac{1}{2}}&=\sum_{n\in\mathbb{Z}
+\frac{1}{2}}c_{n}z^{-n-\frac{1}{2}}dz^{\frac{1}{2}},
\end{align} 
whose coefficients satisfy the anticommutation relations
\begin{align}
\{b_{m},c_{n}\}&=\delta_{m+n,0}, \nonumber \\[0.2cm]
\{b_{m},b_{n}\}&=\{c_{m},c_{n}\}=0.
\label{eq:anticomm_rel_bc_fermions}
\end{align}
The field~$b(z)$ can be interpreted as the conjugated ``translation''
operator for~$c(z)$, namely
\begin{align}
b(z)\sim \frac{\delta}{\delta c(z)}.
\label{eq:bsimfdelta/deltac}
\end{align}
\begin{figure}[t]
\centerline{\includegraphics[scale=0.45]{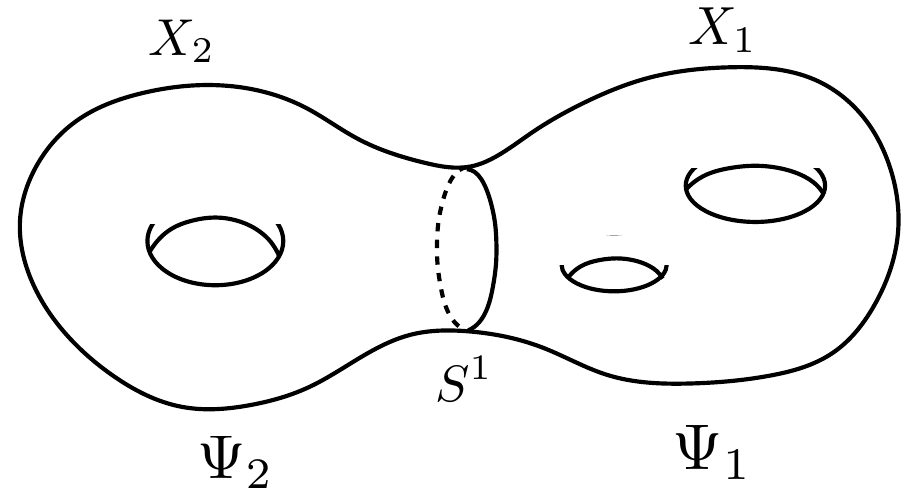}}
\caption{Riemann surface divided into two parts with a common boundary with topology~$S^{1}$}
\label{surface}
\end{figure}

The fields~$b(z)$,~$c(z)$ are defined on the Riemann surface~$X$ that,  
as shown in fig.~\ref{surface}, 
can be split into two 
pieces~$X_{1}$ and~$X_{2}$ joined along their common 
boundary~$\partial X_{1}=-\partial X_{2}=S^{1}$, the minus sign in the second term
reflecting the different orientation of the two boundaries. 
To each piece of the surface we associate states $\Psi_{1}$ and~$\Psi_{2}$,
whose inner product is given by the partition function on~$X$
\begin{align}
Z\equiv\int {\cal D}b{\cal D}c \exp{\left[-\int_{X}
\Big(b\overline{\partial}c+c\overline{\partial}b\Big)\right]}=\langle\Psi_{2}|\Psi_{1}\rangle.
\end{align}
The states $\Psi_{1},\Psi_{2}$ themselves admit the wave functional 
representation\footnote{Since the spinors $b(z)$ and $c(z)$ are canonically conjugate
fields, this expression is a Hamiltonian functional integral
representation of the wave functional~\cite{Faddeev_Slavnov}.}
\begin{align}
\Psi_{1}[f]=\int_{c\big|_{S^{1}}=f}{\cal D}b{\cal D}c\exp{\left[\,
-\int_{X_{1}}\Big(b\overline{\partial}c+c\overline{\partial}b\Big) + \oint_{S^{1}} cb
\right]},
\label{psi-1}
\end{align}
and similarly for~$\Psi_{2}[f]$, 
where no boundary conditions
are imposed on the spinor~$b(z)$,
while the boundary integral in the exponential accounts for the flux of the
fermionic current through $S^{1}$ (this term  should be added to the classical 
action in order to 
cancel boundary the contribution arising when computing the field equations~$\overline{\partial}b
=\overline{\partial}c=0$). 
Moreover, its presence also guarantees the invariance of the
wave functional~\eqref{psi-1} under shifts~$f(z)\rightarrow f(z)+w_{n}(z)$
\begin{align}
\Psi_{1}[f+w_{n}]=\Psi_{1}[f],
\label{Qinvariance}
\end{align}
where~$w_{n}(z)$ admits a meromorphic extension to~$X_{1}$ with poles at the punctures. Indeed, 
shifting~$c(z)\rightarrow c(z)-w_{n}(z)$ on~$X_{1}$ and using
the Stokes theorem, we find
\begin{align}
\int_{X_{1}}\overline{\partial}(w_{n}b)-\oint_{S^{1}}w_{n}b=0,
\end{align}
while no Jacobian is generated by the shift in the functional integration measure.
Since, as we have seen,~$b(z)$ plays the role of the translation
operator for~$c(z)$, defining the conserved charge
\begin{align}
Q_{n}\equiv Q(w_{n})=\frac{1}{2\pi i}\oint_{S^{1}}dz b(z)w_{n}(z).
\end{align}
the invariance~(\ref{Qinvariance}) is recast as
\begin{align}
Q_{n}\Psi_{1}=0.
\label{c-1}
\end{align}
Applying the Stokes theorem it is possible to show 
that~$Q(w_{n})$ only depends on the homology class of
the boundary.

The state~$\Psi_{1}$ associated with~$X_{1}$ also admits a representation in terms
of the ``Fourier transformed''
wave functional 
\begin{align}
\Psi_{1}[\widetilde{f}]=\int_{b\big|_{S^{1}}=\widetilde{f}}{\cal D}b{\cal D}c\exp{\left[\,
-\int_{X_{1}}\Big(b\overline{\partial}c+c\overline{\partial}b\Big) + \oint_{S^{1}} cb
\right]},
\label{psi-1_FT}
\end{align}
where the boundary condition on~$S^{1}$ is now imposed on
the conjugate field~$b(z)$. Repeating the previous analysis,
we find that this functional is 
invariant under the translation
of boundary value~$\widetilde{f}(z)$ by a holomorphic section~$\widetilde{w}_{n}(z)$.
Again, this is implemented by the operator
\begin{align}
\widetilde{Q}_{n}\equiv\widetilde{Q}(\widetilde{w}_{n})=\frac{1}{2\pi i}\oint_{S^{1}}dz
\,c(z)\widetilde{w}_{n}(z),
\end{align}
so the state~$\Psi_{1}$ satisfies
\begin{align}
\widetilde{Q}_{n}\Psi_{1}=0.
\label{c-2}
\end{align}
In fact, the two charges $Q_{n}$ and $\widetilde{Q}_{m}$
verify the anticommutation relation
\begin{align}
\{Q_{n},\widetilde{Q}_{m}\}=\frac{1}{2\pi i}\oint_{S^{1}} dz\,w_{n}(z)
\widetilde{w}_{m}(z).
\label{eq:anticommQ_nQ_m_full}
\end{align}
Since~$w_{n}$ and~$\widetilde{w}_{n}$ are holomorphic on~$S^{1}$ and 
admit a holomorphic extension to~$X_{1}$,
by deforming the integration contour across~$X_{1}$ the integral on the right-hand side
of eq.~\eqref{eq:anticommQ_nQ_m_full} can be seen to vanish. We thus find
\begin{align}
\{Q_{n},\widetilde{Q}_{m}\}=0.
\end{align}
Equations~\eqref{c-1} and~\eqref{c-2}
provide a maximal set of conditions for the state~$\Psi_{1}$. 
This means that to associate a quantum state to a Riemann surface~$W\in{\cal P}(g,n)$ 
we have to search for all the meromorphic sections~$w_{n}$,~$\widetilde{w}_{m}$ 
of the spinor bundle with poles only at the punctures. Using them, we construct the corresponding conserved charges~$Q_{n}$ 
and~$\widetilde{Q}_{n}$, the state being defined by
eqs.~(\ref{c-1}) and~(\ref{c-2}). 

As a first example to illustrate this, let us look at 
the sphere with a single puncture at~$z=0$. The meromorphic
sections with poles at $z=0$ are
\begin{align}
w_{n}(z)=z^{-n}
\hspace*{1cm} \mbox{with} \hspace{1cm} n>0,
\end{align}
so using the mode expansion for~$b(z)$ and~$c(z)$ we
arrive at the operators
\begin{align}
Q_{n}&=\int_{P}\frac{dz}{2\pi i}b(z)z^{-n}=b_{-n-\frac{1}{2}}, \nonumber
\\[0.2cm] 
\widetilde{Q}_{m}&=\int_{P}\frac{dz}{2\pi
i}c(z)z^{-m}=c_{{-m-\frac{1}{2}}},
\end{align}
with $n,m>0$. The state associated to the sphere with
a single puncture is then defined by
\begin{align}
b_{-n-\frac{1}{2}}|W\rangle=c_{{-n-\frac{1}{2}}}|W\rangle=0,
\hspace{1cm} n>0,
\label{0-rangle}
\end{align}
that is the~$\mbox{SL}(2,\mathbb{C})$-invariant vacuum state.

\begin{figure}[t]
\centerline{\includegraphics[scale=0.5]{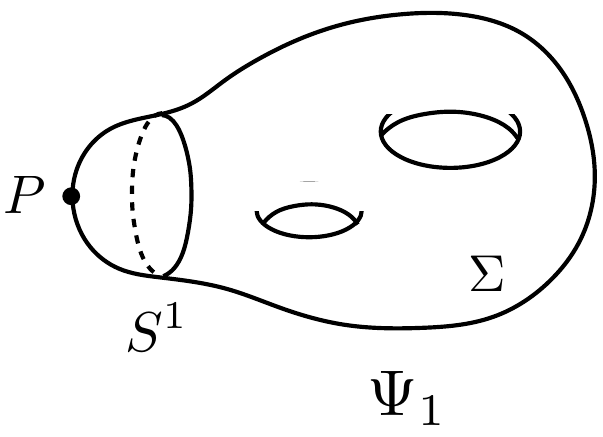}}
\caption{Riemann surface~$\Sigma$ with a single puncture at $P$}
\label{P}
\end{figure}

We now proceed to construct the state associated with a genus-$g$
Riemann surface with a single puncture at $P$ (see fig.
\ref{P}), searching again for meromorphic sections with poles only at~$P$. The Riemann-Roch
theorem ensures that these sections exist with poles of arbitrary order. 
In order to find them, we consider the Szeg\"{o} kernel for 
spin~$\frac{1}{2}$, which is the fermion two-point function~$S(z,w)=\langle c(z)b(w)\rangle$ 
on~$\Sigma$. Labelling the spin structures 
by~$(\alpha,\beta)$, where
\begin{align}
({\scriptstyle {1\over 2}},{\scriptstyle {1\over 2}})\rightarrow (P,P), \hspace*{0.5cm}
({\scriptstyle {1\over 2}},{\scriptstyle 0})\rightarrow (P,A), \hspace*{0.5cm}
({\scriptstyle 0},{\scriptstyle {1\over 2}})\rightarrow (A,P), \hspace*{0.5cm}
({\scriptstyle 0},{\scriptstyle 0})\rightarrow (A,A),
\end{align}
the Szeg\"o kernel takes the form (for a constructive derivation, 
see ref.~\cite{Alvarez-Gaume-Gomez-Reina})
\begin{align}
S(z,w)=\frac{\vartheta\left[
\begin{array}{c}
\alpha \\
\beta
\end{array}
\right](z-w|\Omega)}{\vartheta\left[
\begin{array}{c}
\alpha \\
\beta
\end{array}
\right](0|\Omega)E(z,w)},
\label{eq:Szego}
\end{align}
where~$\Omega$ is the period
matrix of~$\Sigma$ and
we have introduced the Jacobi theta function~\cite{Mumford,Alvarez-Gaume-Moore-Vafa}
\begin{align}
\vartheta\left[
\begin{array}{c}
\alpha \\
\beta
\end{array}
\right](z|\tau)=\sum_{n\in\mathbb{Z}}e^{i\pi\tau(n+\alpha)^{2}+
2\pi i(n+\alpha)(z+\beta)}.
\label{vartheta-function}
\end{align} 
The prime form~$E(z,w)$ in the denominator of  
eq.~\eqref{eq:Szego} is defined as the 
unique~$(-\frac{1}{2},-\frac{1}{2})$ differential 
satisfying~$E(z,w)=-E(w,z)$ and such that it only vanishes at~$z=w$. 
To find its explicit expression,
we take an odd non-singular spin
structure~$(\alpha',\beta')$, i.e. which has just one zero mode, and define the function
\begin{align}
f(z,w)=\vartheta\left[
\begin{array}{c}
\alpha' \\
\beta'
\end{array}
\right]\left.\left(\int_{w}^{z}\omega\right|\Omega\right).
\label{eq:pre_prime_form}
\end{align}
As a consequence of the Riemann vanishing theorem \cite{Alvarez-Gaume-Moore-Vafa},
this function vanishes at $g-1$ points~$P_{1},\ldots,P_{g-1}$ as a function of~$z$ 
with~$w$ fixed. Since it also vanish when $z\rightarrow w$,  
~$f(z,w)=0$ at the same points when
regarded as a function of $w$ with $z$ fixed. Moreover, if~$z\sim w\sim P_{i}$ 
we have~$f(z,w)\sim
(z-w)(z-P_{i})(w-P_{i})$, so differentiating with respect to~$z$ we
can define the holomorphic form
\begin{align}
h(z)^{2}=\sum_{j=1}^{g}\omega_{i}(z)\frac{\partial}{\partial u_{i}}
\left.\vartheta \left[
\begin{array}{c}
\alpha' \\
\beta'
\end{array}
\right]\left.\left(\int_{z}^{w}\omega\right|\Omega\right)\right|_{z=w},
\end{align}
where~$u_{i}$ stands for the first argument of the theta function.
The function~$h(z)^{2}$
has second order zeros at~$P_{i}$, so we can take
the square root without producing branch cuts. The prime form is then given by
\begin{align}
E(z,w)=\frac{\vartheta\left[
\begin{array}{c}
\alpha' \\
\beta'
\end{array}
\right](z-w|\Omega)}{h(z)h(w)}.
\label{eq:prime_form}
\end{align}

We not return to the problem of finding the holomorphic sections~$w_{n}(z)$. 
They are obtained by differentiating the
Szeg\"{o} kernel with respect to $w$ and then setting~$w=0$. If~$z$ and~$w$
are in a neighborhood of $P\in \Sigma$, we have
\begin{align}
w_{n}(z)&=\left.\frac{1}{(n-1)!}\frac{\partial^{n-1}}{\partial w^{n-1}}
S(z,w)\right|_{w=0} \nonumber \\[0.2cm]
&=
\frac{1}{z^{n}}+\sum_{m=1}^{\infty}B_{nm}z^{m-1}\;,
\end{align}
where the coefficients~$B_{nm}$ are given by
\begin{align}
B_{nm}=\frac{1}{(n-1)!(m-1)!}\frac{\partial^{m-1}}{\partial z^{m-1}}
\frac{\partial^{n-1}}{\partial
w^{n-1}}\left.\left[S(z,w)-\frac{1}{z-w}\right]\right|_{z=w=0}.
\end{align}
The anticommutation 
relations~\eqref{eq:anticomm_rel_bc_fermions} admit a representation 
in terms of differential operators in which $b_{-n}$
and $c_{-n}$, for $n>0$, act by multiplication while
\begin{align}
b_{n}=\frac{\partial}{\partial c_{-n}}, \hspace{1cm}
c_{n}=\frac{\partial}{\partial b_{-n}}, \hspace{0.5cm} \mbox{when $n>0$}.
\label{eq:rep_cb}
\end{align}
Given the expression of the charges $Q_{n}$, $\widetilde{Q}_{m}$
\begin{align}
Q_{n}&=b_{n-\frac{1}{2}}+\sum_{m=1}^{\infty}B_{nm}b_{-m+\frac{1}{2}},
\nonumber \\[0.2cm]
\widetilde{Q}_{n}&=
c_{n-\frac{1}{2}}+\sum_{m=1}^{\infty}B_{nm}c_{-m+\frac{1}{2}},
\label{Bogoliubov}
\end{align}
and using the representation~\eqref{eq:rep_cb}, 
we arrive at a system of differential equations to be satisfied 
by the state~$\Psi(c_{-n},b_{-n})$. Its solution
\begin{align}
|\Psi\rangle=C\,\exp{\left(-\sum_{n,m=1}^{\infty}B_{nm}c_{-n+\frac{1}{2}}
b_{-m+\frac{1}{2}}\right)}|0\rangle,
\end{align}
is a Bogoliubov transformation on
the standard $\mbox{SL}(2,\mathbb{C})$~vacuum state~$|0\rangle$ annihilated by~$b_{n-{1\over 2}}$
and~$c_{n-{1\over 2}}$, with~$n>0$ [see eq.~(\ref{0-rangle})]. The
charge operators $Q_{n}$, $\widetilde{Q}_{m}$, which are linear combinations
of creation and annihilation operators with respect to
$|0\rangle$, are themselves annihilation operators with respect to the new
vacuum state~$|\Psi\rangle$. All the geometric properties of the state are neatly
summarized by this Bogoliubov transformation.

For a genus-$g$ Riemann surface with~$n$ punctures, $\Sigma \in {\cal P}(g,n)$, we need to find
meromorphic sections of the spinor bundle with poles at
$P_{1},\ldots,P_{n}$. The simplest example is the sphere with two
punctures that, without loss of generality, can be located at $z=0$ and $z=\infty$. This
is the surface we associated to the ``sewing'' state~$|S_{12}\rangle$.
Now, the meromorphic sections on the sphere with poles at~$z=0$ and~$z=\infty$ are generated by
\begin{align}
w_{n}=z^{n}dz^{\frac{1}{2}}\;, \hspace{1cm} n \in \mathbb{Z},
\end{align}
and the conserved charges~$Q_{n}$ and~$\widetilde{Q}_{n}$ are easily obtained by integrating on 
patches located around
each puncture
\begin{align}
Q_{n}&=\oint_{P_{1}}\frac{dz_{1}}{2\pi i}b^{(1)}(z_{1})w_{n}(z_{1})+
\oint_{P_{2}}\frac{dz_{2}}{2\pi i}b^{(2)}(z_{2})w_{n}(z_{2})
\equiv Q_{n}^{(1)}+Q_{n}^{(2)},
\nonumber \\[0.2cm]
\widetilde{Q}_{n}&=\oint_{P_{1}}\frac{dz_{1}}{2\pi
i}c^{(1)}(z_{1})w_{n}(z_{1})+
\oint_{P_{2}}\frac{dz_{2}}{2\pi i}c^{(2)}(z_{2})w_{n}(z_{2})
=\widetilde{Q}_{n}^{(1)}+\widetilde{Q}_{n}^{(2)},
\end{align}
where the superscripts on the last expressions indicate the puncture.
Solving the corresponding set of differential equations, we find the sewing state
\begin{align}
|S_{12}\rangle=\prod_{m=1}^{\infty}
\exp{\left(c^{(1)}_{-m+\frac{1}{2}}b^{(2)}_{-m+\frac{1}{2}}+
c^{(2)}_{-m+\frac{1}{2}}b^{(1)}_{-m+\frac{1}{2}}\right)}
|0\rangle_{1}\otimes|0\rangle_{2},
\end{align}
satisfying
\begin{align}
(Q_{n}^{(1)}+Q_{n}^{(2)})|S_{12}\rangle=(\widetilde{Q}_{n}^{(1)}+\widetilde{Q}_{n}^{(2)})|S_{12}\rangle
=0,
\end{align}
with~$n>0$.

\begin{figure}[t]
\centerline{\includegraphics[scale=0.45]{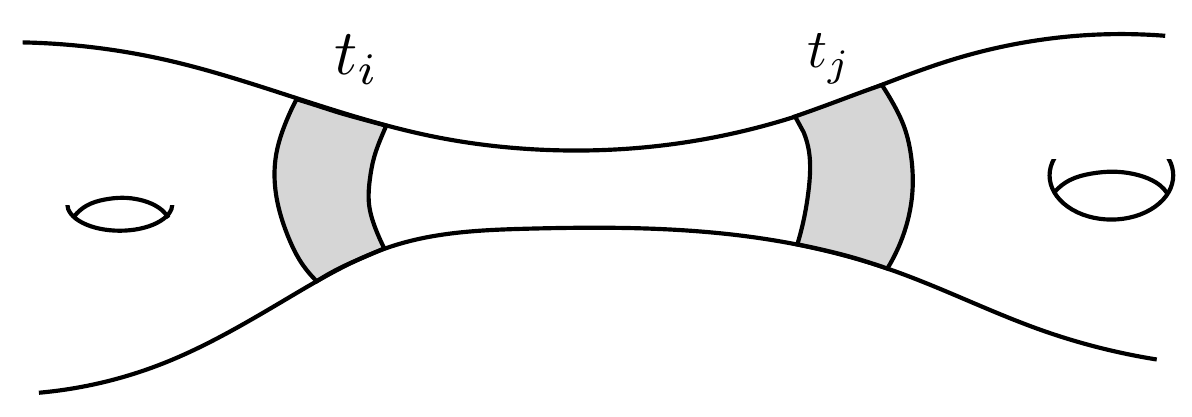}}
\caption{Remnants of the two disks located around punctures~$P_{i}$ and~$P_{j}$
with local parameters~$t_{i}$,~$t_{j}$ after sewing them together}
\label{remnants}
\end{figure}

We know that sewing together two Riemann surfaces~$P\in {\cal P}(g_{1},n_{1})$, $Q\in {\cal P}(g_{2},n_{2})$ 
with sewing parameter $q=1$ a new Riemann surface~$R=P_{i}\infty_{j}Q
\in{\cal P}(g_{1}+g_{2},n_{1}+n_{2}-2)$ is obtained. As an example of
how the operator formalism works, we verify that the state associated with the surface~$R$
\begin{align}
|R\rangle=\langle S_{ij}|P\rangle\otimes |Q\rangle,
\end{align}
satisfies the condition 
\begin{align}
\big(Q_{n}^{(1)}+\ldots+Q_{n}^{(i-1)}+Q_{n}^{(j+1)}+\ldots+Q_{n}^{(r)}\big)|R\rangle=0,
\nonumber \\[0.2cm]
\big(\widetilde{Q}_{n}^{(1)}+\ldots+\widetilde{Q}_{n}^{(i-1)}
+\widetilde{Q}_{n}^{(j+1)}+\ldots+\widetilde{Q}_{n}^{(r)}\big)|R\rangle=0,
\end{align}
where~$1,\ldots,i-1$ label punctures on~$P$ and~$j+1,\ldots,r$ on~$Q$.
Sewing the $i$th and $j$th punctures together, a remnants of the two
discs is left in the form of two annuli with local coordinates~$t_{i}$ and~$t_{j}$ 
respectively (see fig.~\ref{remnants}). Using them to expand the holomorphic spinors, 
we construct the operators~$Q^{(i)}_{n}$ and~$Q^{(j)}_{n}$ ($n>0$) satisfying
\begin{align}
\langle S_{ij}|(Q_{n}^{(i)}+Q_{n}^{(j)})=0.
\end{align}
With this, we write
\begin{align}
\big(Q_{n}^{(1)}+&\ldots+Q_{n}^{(i-1)}+Q_{n}^{(j+1)}+\ldots+Q_{n}^{(r)}\big)\langle
S_{ij}|P\rangle\otimes|Q\rangle \\[0.2cm]
&=\langle
S_{ij}|\big[\big(Q_{n}^{(1)}+\ldots+Q_{n}^{(i-1)}
+Q_{n}^{(i)}\big)+\big(Q_{n}^{j}+Q_{n}^{j+1}+\ldots+Q_{n}^{(r)}\big)\big]
|P\rangle\otimes |Q\rangle=0,   
\nonumber
\end{align}
as we wanted to prove.

\begin{figure}[t]
\centerline{\includegraphics[scale=0.45]{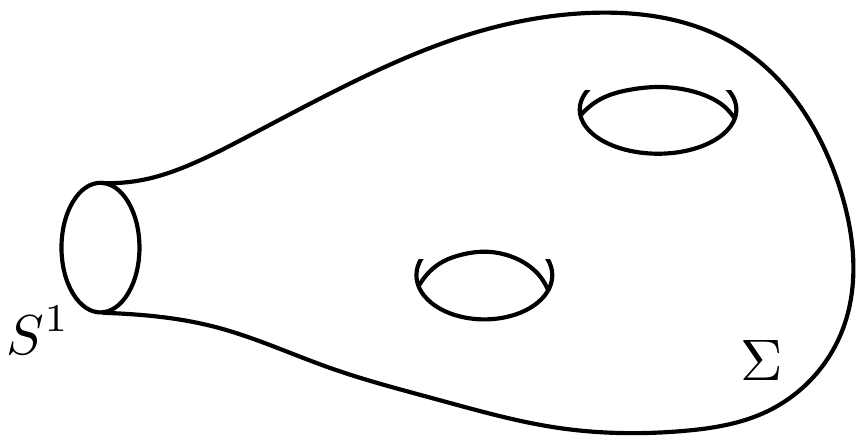}}
\caption{Riemann surface~$\Sigma$ with boundary $S^{1}$}
\label{scalar-field}
\end{figure}

\paragraph{Real scalar field.}
The second example to be discussed is a single-valued real scalar field~$\phi$. 
Proceeding along the lines of the spin-$\frac{1}{2}$ case, 
we construct the wave functional associated with a Riemann surface~$\Sigma$ 
with boundary~$S^{1}$ (see fig.~\ref{scalar-field})
\begin{align}
\Psi[f]=\int_{\phi|_{S^{1}}=f}{\cal D}\phi
\exp{\left(-\frac{1}{2}\int_{\Sigma_{1}}d\phi\wedge \star d\phi
\right)},
\label{eq:Psi[f]_scalar}
\end{align}
where~$f$ is a real function on $S^{1}$ which
can be uniquely extended to a harmonic
function,~$d\star df=0$, on~$\Sigma$. Introducing the new field~$\phi'$
\begin{align}
\phi=\phi'+f,
\label{change-phi}
\end{align}
satisfying the boundary condition~$\phi'|_{S^{1}}=0$, we change variables in the
functional integral on the rigt-hand side of eq.~\eqref{eq:Psi[f]_scalar}
\begin{align}
\Psi[f]=
\int_{\phi'|_{S^{1}}=0}{\cal D}\phi'\,e^{-S[\phi'+f]}.
\end{align}
Evaluating now $S[\phi'+f]$, we find
\begin{align}
S[\phi'+f]&=\frac{1}{2}\int_{\Sigma_{1}} d(\phi'+f)\wedge
\star d(\phi'+f)
\nonumber \\[0.2cm]
&= S[\phi']+\frac{1}{2}\int_{\Sigma_{1}} df\wedge \star df+\int_{\Sigma_{1}} d\phi'\wedge \star df
\\[0.2cm]
&=S[\phi']+\frac{1}{2}\oint_{S^{1}} f\star df,
\nonumber
\end{align}
where we have used the Stokes theorem, the boundary condition 
of~$\phi'$, and the fact that~$f$ is a harmonic function. Finally, carrying out the functional integration
over~$\phi'$, the state functional is expressed as
\begin{align}
\Psi[f]=(\det{}\!'\,\Delta)_{\phi'|_{S^{1}}=0}^{-\frac{1}{2}}
\exp{\left(-\frac{1}{2}\oint_{S^{1}} f\star df\right)},
\end{align}
with~$\Delta$ the two-dimensional Laplacian.
The conserved charges in this case can be written using Green's
identity for harmonic functions
\begin{align}
Q(h_{n})=\oint_{S^{1}}(h_{n}\star d\phi-\phi\star dh_{n}),
\label{charge-scalar}
\end{align}
where~$h_{n}$ are harmonic functions analytic on~$\Sigma$.
It is easy to see that since the field equations for~$\phi$ 
read~$d\star d\phi=0$, the operator $Q(h_{n})$ does not change
under deformations of $S^{1}$ and only depends on the homology class of
the boundary. As in the fermionic case, the state $\Psi$ is constructed
by requiring it to be annihilated by the conserved charges.
Splitting~$h_{n}$
into its holomorphic and antiholomorphic pieces, $h_{n}^{H}+h_{n}^{A}$, 
the charge~(\ref{charge-scalar}) can be derived from the
conserved current
\begin{align}
j(h_{n})=\partial\phi h_{n}^{H}-\overline{\partial}\phi h_{n}^{A}.
\end{align}

We still need to find an explicit expression for the functions~$h_{n}$. 
Considering, for example, a genus-$g$ Riemann surface with a single
puncture at~$P$, the simplest solutions to our problem would be the real 
and imaginary parts of a holomorphic function on~$\Sigma-P$ with 
poles at~$P$. Unfortunately,
the Weierstrass gap theorem states that in this case the order of the pole
at~$P$ must be larger than~$g$, so we would need~$g$ extra conditions in order to determine completely
the state~$\Psi$. As a consequence, it is not possible to restrict to holomorphic
objects and we have to consider functions with holomorphic and
antiholomorphic parts. 

We start by considering a holomorphic
function with poles at~$P$ of order less than~$g$ and in order to
bypass Weierstrass' theorem, is allowed to be multivalued
around the cycles. Then we add some antiholomorphic function
to get a single-valued harmonic function.
Let us consider the multivalued meromorphic differentials
\begin{align}
\eta_{n}(t)=\frac{1}{(n-1)!}\left.\frac{\partial}{\partial t}
\frac{\partial^{n}}{\partial y^{n}} \log{E(t,y)}\right|_{y=0}.
\end{align}
where $E(t,y)$ is the prime form defined in eq.~\eqref{eq:prime_form}
and~$t$ is the local coordinate at~$P$. We
define then the functions
\begin{align}
h_{n}(t)=
\int^{t}\Big[\eta_{n}(t)
-\sum_{i}A_{ni}({\rm Im\,}\Omega)^{-1}(\omega_{i}-\overline{\omega}_{i})\Big],
\label{hn-func}
\end{align}
where~$\omega_{i}$ are the Abelian differentials and the coefficients~$A_{ni}$ 
are defined in terms of the local coordinate~$t$ through the
expansion
\begin{align}
\omega_{i}(t)=\sum_{n=1}^{\infty}A_{ni}t^{n-1}dt,
\end{align}
and the differentials~$\eta_{n}(t)$ satisfy
\begin{align}
\oint_{b_{i}}\eta_{n}(t)=A_{ni}.
\end{align}
Using the properties of the prime form~$E(t,y)$~\cite{Mumford}, it 
can be shown that the functions~$h_{n}(t)$ defined in~(\ref{hn-func}) are
single-valued.

Next we use the familiar expansion of the field $\phi(t)$
\begin{align}
\phi(t)=q+ip\log{t}+ip\log{\overline{t}}+\sum_{n\neq
0}\left(\frac{a_{n}}{n}t^{n}+\mbox{c.c.}\right),
\end{align}
where the coefficients satisfy the non-vanishing commutation relations
\begin{align}
[q,p]&=i, \nonumber \\[0.2cm] [a_{n},a_{m}]&=n\delta_{n+m,0},
\label{eq:creation_annihilation_scalarOF} \\[0.2cm]
[\overline{a}_{n},\overline{a}_{m}]&=n\delta_{n+m,0}.
\nonumber
\end{align}
Using these expressions, together with Green's theorem for harmonic
functions, it can be shown that the charges $Q(h_{n})$ defined in eq.~\eqref{charge-scalar}
commute
\begin{align}
[Q(h_{n}),Q(h_{m})]=0.
\end{align}
Moreover, we can construct an operator representation of the creation-annihilation
algebra in eq.~\eqref{eq:creation_annihilation_scalarOF} by
introducing variables~$x_{n},\overline{x}_{n}$ and defining
\begin{align}
a_{n}=\frac{\partial}{\partial x_{n}}, \hspace{1cm}
a_{-n}=nx_{n}, \hspace{1cm} \mbox{for $n\geq0$},
\label{eq:a_rep_x_OF}
\end{align}
with the corresponding expressions for barred quantities. We then impose
the conditions on~$|\Psi\rangle$
\begin{align}
Q(h_{n})|\Psi\rangle=\overline{Q}(\overline{h}_{n})|\Psi\rangle=0,
\end{align}
and using the representation~\eqref{eq:a_rep_x_OF} arrive at a set of differential
equations on $\langle x,\overline{x}|\Psi\rangle$ that upon integration
yields
\begin{align}
\langle x,\overline{x}|\Psi\rangle=\exp{\left[(x,\overline{x})M\left(
\begin{array}{c}
x \\
\overline{x}
\end{array}
\right)\right]}.
\end{align}
The matrix $M$ is given by
\begin{align}
M=\left(
\begin{array}{cc}
{\displaystyle Q_{nm}+\frac{\pi}{2}A_{n}({\rm Im\,}\Omega)^{-1}A_{m}} &
{\displaystyle -\frac{\pi}{2}A_{n}({\rm Im\,}\Omega)^{-1}\overline{A}_{m}} \\[0.4cm]
{\displaystyle -\frac{\pi}{2}\overline{A}_{n}({\rm Im\,}\Omega)^{-1}A_{m}}&
{\displaystyle \overline{Q}_{nm}+\frac{\pi}{2}\overline{A}_{n}({\rm Im\,}\Omega)^{-1}\overline{A}_{m}}
\end{array}
\right),
\end{align}
with
\begin{align}
Q_{nm}=\frac{1}{2(n-1)!(m-1)!}\frac{\partial^{n}}{\partial t^{n}}
\frac{\partial^{m}}{\partial y^{m}}
\log{\left[\frac{E(t,y)}{t-y}\right]}.
\end{align}

The sewing state $|S_{12}\rangle$, associated to the sphere with 
two punctures, can be computed now with the result
\begin{align}
|S_{12}\rangle=\int dp \prod_{n=1}^{\infty}
\exp{\big(a_{n}^{(1)+}a_{n}^{(2)+}+\overline{a}_{n}^{(1)+}\overline{a}_{n}^{(2)+}
\big)}|p\rangle_{1}\otimes|p\rangle_{2},
\end{align}
where the state $|p\rangle$ carries momentum $p$. In the case of a genus-$g$
Riemann surface with~$n$ punctures,~$\Sigma_{g,n}$, we introduce the
scalar Green function
\begin{align}
G(z,w)=-2\pi\left({\rm Im\,}\int_{z}^{w}\omega\right)({\rm Im\,}\Omega)^{-1}
\left({\rm Im\,}\int_{z}^{w}\omega\right)-\log{|E(z,w)|^{2}},
\end{align}
and define
\begin{align}
g(z_{i},w_{j})=\left\{
\begin{array}{ll}
G(z_{i},w_{j})+\log{|z_{i}-w_{j}|^{2}} & \hspace*{0.5cm}\mbox{if
$z_{i},w_{j}$ lie in the same patch} \\[0.3cm]
G(z_{i},w_{j}) &\hspace*{0.5cm} \mbox{otherwise}
\end{array}
\right..
\end{align}
Splitting~$g(z_{i},w_{j})$ into four parts corresponding to the
holomorphic or antiholomorphic contribution in each of the two variable,
and constructing with these four pieces the matrix $M(z_{i},w_{j})$, we
obtain the unique ray associated with~$\Sigma_{g,n}$
\begin{align}
|\Psi_{g}^{n}\rangle&=\int dp_{1}\ldots dp_{n} \delta\left(\sum_{i}p_{i}\right)
 \\[0.2cm]
&\times \exp{
\left[-\frac{1}{(2\pi i)^{2}}\sum_{i,j}\oint_{P_{i}}\oint_{P_{j}}
\big(\partial\phi(z_{i}),\overline{\partial}\phi(z_{i})\big)M(z_{i},z_{j})\left(
\begin{array}{c}
\partial\phi(z_{i}) \\[0.2cm]
\overline{\partial}\phi(z_{j})
\end{array}
\right)\right]}
|p_{1}\rangle\otimes\ldots\otimes |p_{n}\rangle.
\nonumber 
\end{align}

\paragraph{Ghost system.}
We consider now a ghost system consisting of a pair of
anticommutating fields:~$b(z)$, with spin~$2$, and
$c(z)$, with spin~$-1$. We begin with the sphere with a single puncture at~$P$ and local
coordinates such that~$z(P)=0$. Expanding the fields as
\begin{align}
b(z)&=\sum_{n\in \mathbb{Z}}b_{n}z^{n-2}, \nonumber \\[0.2cm] 
c(z)&=\sum_{n\in
\mathbb{Z}}c_{n}z^{n+1},
\end{align}
the coefficients satisfy the anticommutation relations
\begin{align}
\{b_{n},c_{m}\}=\delta_{m+n,0}.
\end{align}
We have to look for vector fields holomorphic off $P$ and quadratic
differentials with the same property.
Using the Riemann-Roch theorem \cite{Griffiths-Harris} on the sphere, it
is
possible to show that the only vector fields with the desired properties
are
\begin{align}
z^{2-n}\partial, \hspace*{1cm} n>0,
\end{align}
while for the quadratic differentials we have
\begin{align}
z^{-4-n}dz^{2}, \hspace*{1cm} n>0.
\end{align}
In the case of
$b(z)$, the associated conserved charges are
\begin{align}
\oint_{P}\frac{dz}{2\pi i}b(z)z^{k}=b_{-k+1}, \hspace{1cm}
k=2,1,0,-1,\ldots,
\end{align}
while for $c(z)$ we have
\begin{align}
\oint_{P}\frac{dz}{2\pi i}c(z)z^{-k}=c_{k-2}\;, \hspace{1cm}
k=4,5,\ldots
\end{align}
Thus, the state $|\phi_{0}\rangle$ associate with the sphere with one
puncture and the ghost system is defined by
\begin{align}
b_{n}|\phi_{0}\rangle&=0, \hspace{1cm} n>-2, \nonumber \\[0.2cm]
c_{n}|\phi_{0}\rangle&=0, \hspace{1cm} n>1,
\end{align}
which is a $\mbox{SL}(2,\mathbb{C})$-invariant state,
because it is annihilated by $L_{0},L_{\pm 1}$, where $L_{n}$ are the
Virasoro generators associated with the ghost system. We define the
ghost current
\begin{align}
j(z)_{\rm gh}=:\!c(z)b(z)\!:,
\end{align}
and the associated charge
\begin{align}
Q_{\rm gh}=\oint_{P}\frac{1}{2\pi i}j(z).
\end{align}
Our first task will be to determine the ghost charge of the vacuum
$|\phi_{0}\rangle$. Evaluating the commutator of $L_{0}$ with the
ghost fields we get
\begin{align}
[L_{0},b_{n}]&=-nb_{n}, \nonumber \\[0.2cm]
{[}L_{0},c_{n}]&=-nc_{n},
\end{align}
showing that the state $|\phi_{0}\rangle$ is not a highest-weight
state of the oscillators algebra, since it is not annihilated by
negative energy modes.
Moreover, the zero modes $b_{0}$, $c_{0}$ form a Clifford algebra since
$\{b_{0},c_{0}\}=1$. Then we can define the highest-weight states
$\{|+\rangle, |-\rangle\}$ building a representation of the
Clifford algebra
\begin{align}
c_{n}|+\rangle&=b_{n}|-\rangle=0, \hspace{1cm} n\geq 0, \nonumber \\[0.2cm]
c_{0}|-\rangle&=|+\rangle, \\[0.2cm]
b_{0}|+\rangle&=|-\rangle.
\nonumber
\end{align}
It can be argued \cite{Green-Schwarz-Witten} that the natural assignment
of ghost number to these states is
$Q|\pm\rangle=\pm\frac{1}{2}|\pm\rangle$. These states are constructed
from the~$\mbox{SL}(2,\mathbb{C})$-invariant vacuum as
\begin{align}
|-\rangle=c_{1}|\phi_{0}\rangle, \hspace{1cm}
|+\rangle=c_{0}c_{1}|\phi_{0}\rangle,
\end{align}
where~$|\phi_{0}\rangle$ has charge~$-{3\over 2}$.
Now it is straighforward to verify that
$L_{0}|\pm\rangle=-|\pm\rangle$. This is in fact the origin of the bosonic string tachyon.

Next we construct the sewing state $|S_{12}\rangle$. For
a sphere with two punctures (at~$0$ and at~$\infty$) the quadratic differentials and vector fields
holomorphic off those points are respectively
\begin{align}
\mu=z^{n}dz^{2}, \hspace{1cm} v=z^{n+1}\frac{d}{dz}, \hspace{1cm}
n\in\mathbb{Z}.
\end{align}
We proceed as in previous cases. Imposing
the conditions
\begin{align}
(b^{(1)}_{n}-b^{(2)}_{-n})|S_{12}\rangle&=0, \nonumber \\[0.2cm]
(c^{(1)}_{n}-c^{(2)}_{-n})|S_{12}\rangle&=0,
\end{align}
we obtain the solution
\begin{align}
|S_{12}\rangle=\prod_{m=1}^{\infty}\exp{\left(-c_{-m}^{(1)}b_{-m}^{(2)}-
c_{-m}^{(2)}b_{-m}^{(1)}\right)}(b_{0}^{(1)}-b_{0}^{(2)})|+\rangle_{1}
\otimes |-\rangle_{2}.
\end{align}
Evaluating now the total ghost charge associated with the sewing state,
we obtain
\begin{align}
Q_{\rm gh}|S_{12}\rangle=(Q^{(1)}_{\rm gh}+Q^{(2)}_{\rm gh})|S_{12}\rangle=0.
\end{align}

To generalize the construction to a genus-$g$ Riemann surfaces with $n$
punctures we have to keep in mind that for~$g>1$ the Riemann-Roch
theorem states that there are $3g-3$ holomorphic quadratic
differentials. In the neighborhood of a point $P$ they can be
written using the local coordinate~$z$ as
\begin{align}
\psi_{n+1}=z^{n}+\sum_{m\geq q}C_{nm}^{(2)}z^{m}, \hspace{1cm}
q=3g-3,
\end{align}
with $n=0,\ldots,3g-4$. In a similar fashion, we write quadratic
differentials with poles at~$P$ as
\begin{align}
s_{n}=z^{-n}+\sum_{m\geq q}B_{nm}^{(2)}z^{m},
\end{align}
and the corresponding expressions for the vectors. The coefficients
$B^{(2)}_{nm}$ could in principle be written in terms of prime forms
and $\vartheta$-functions. It is worth noticing that since
\begin{align}
\oint_{P_{i}}\frac{dz}{2\pi i}c(z)\psi_{n+1}(z)=c_{-n-2}+\sum_{m\geq q}
C^{(2)}_{nm}c_{-m-2},
\end{align}
and
\begin{align}
\oint_{P_{i}}\frac{dz}{2\pi i}c(z)s_{1}(z)=c_{-1}+\sum_{m\geq q}
B^{(2)}_{1m}c_{-m-2},
\end{align}
only contain creation operators, the only way the associated
state~$|\phi^{n}_{g}\rangle$ is annihilated by these charges is
if they appear
explicitly in the state. For quadratic differentials with
poles of higher order we also get annihilation operators. As
above, we use a representation for the operators~$c_{n}$,~$b_{n}$~(with $n>0$)
\begin{align}
c_{n}=\frac{\partial}{\partial b_{-n}}, \hspace{1cm}
b_{n}=\frac{\partial}{\partial c_{-n}}, \hspace{1cm} n>0,
\end{align}
while~$c_{-n}$, $b_{-n}$ act as multiplication operators.
The conserved charges conditions lead to a differential equation for the state whose solution is
\begin{align}
|\phi_{g}^{n}\rangle&=C_{1}\ldots C_{3g-3}A^{(1)}_{1}\ldots A^{(n)}_{1}
 \exp{\left(-\sum_{m\geq q, n\geq 2} B^{(2)ij}_{mn}
c^{(i)}_{-m-2}b^{(j)}_{-n+2}\right)}
|+\rangle_{1}\otimes\ldots\otimes|+\rangle_{n},
\end{align}
where
\begin{align}
A_{1}^{(i)}=\sum_{j}\oint_{P_{j}}s_{-1}^{(i)}(z_{j})c^{(j)}(z_{j}),
\end{align}
with~$s_{-1}^{(i)}(z)$ the quadratic differential with a single
pole at $P_{i}$, and
\begin{align}
C_{i+1}=\sum_{j}\oint_{P_{j}}\psi_{i}(z_{j})c^{(j)}(z_{j}).
\end{align}
The ghost charge associated with the state~$|\phi_{g}^{n}\rangle$ can be
evaluated taking into account that~$Q_{\rm gh}(b)=-Q_{\rm gh}(c)=1$.
We thus obtain
\begin{align}
Q_{\rm gh}|\phi_{g}^{n}\rangle=\left(3g-3+\frac{3n}{2}\right)|\phi_{g}^{n}\rangle.
\end{align}
As a final check, we make sure that the ghost charge associated with a genus $g$
Riemann surface and $n$ punctures is consistent with the sewing rules.
In fact, taking two states $P\in{\cal P}(g_{1},n_{1})$ and
$Q\in{\cal P}(g_{2},n_{2})$ and keeping in mind that the sewing state
carry no ghost charge at all, we find the ghost charge associated with
$P_{i}\infty_{j}Q$
\begin{align}
\left(3g_{1}-3+\frac{3n_{1}}{2}\right)+\left(3g_{2}-3+\frac{3n_{2}}{2}\right)=
3(g_{1}+g_{2})-3+\frac{3(n_{1}+n_{2}-2)}{2},
\end{align}
which is consistent with the sewing prescription. Incidentally, the 
previous formalism can be also applied to the commuting ghosts emerging in the
quantization of superstrings (see ref.~\cite{LAG-Gomez-Nelson-Sierra-Vafa}
and references therein).

\paragraph{The bosonic string.}
As a final application of the operator formalism, we look into
the bosonic string. In this case we have a system of~$26$
bosons and a~$b$-$c$ ghost system, such that the total
central charge (matter+ghost) is equal to zero. This condition
guarantees that the energy-momentum tensor acts without a central extension,
so given two vectors~$v_{1}\equiv v_{1}(z)\partial$ and~$v_{2}\equiv v_{2}(z)\partial$ 
in the neighborhood of a point~$P$ we have
\begin{align}
[T(v_{1}),T(v_{2})]=T([v_{1},v_{2}]),
\label{eq:[Tv1,Tv2]=T[v1,v2]}
\end{align}
where $[v_{1},v_{2}]$ is the Lie bracket of the two vector fields,~$T(v)$ is 
defined in eq.~\eqref{eq:T(v)_definition}, and the 
identity follows from the OPE in eq.~\eqref{eq:TTOPE}
with~$c=0$.
Thus, the Hilbert space of the free bosonic string is constructed by 
taking the tensor product of 26 copies of the Hilbert space of the free boson
described above together with the one of the ghost system, so 
any point in~${\cal P}(g,n)$
is represented by a ray in the combined matter+ghosts Hilbert space. We need to
worry, however, about the relative normalization of these rays. 
Since infinitesimal motions in~${\cal P}(g,n)$ are generated by~$T(v)$, 
and for the total system~$c=0$, no normalization ambiguity 
arises at different points as long as we can compare rays by translating them 
along contractible paths. Problems may only arise along non-trivial elements 
of~$H_{1}[{\cal P}(g,n)]$. Luckily, Harer's theorem \cite{Harer}
guarantees that this group vanishes for~$g>2$ and state~$|\phi\rangle$
is globally well defined and free of
normalization ambiguities. For $g\leq 2$, on the other hand, one can reach the same
conclusion using sewing.

Since the state $|\phi\rangle$ is globally defined, we would like to obtain
the string measure in moduli space~${\cal M}_{g}$ using the operator formalism, a 
quantity needed in computing the bosonic string partition function.
Starting for simplicity with ${\cal P}(g,1)$, we have a projection
map
\begin{align}
\begin{array}{cc}
{\cal P}(g,1) & \\[0.2cm]
\Bigg\downarrow & \hspace*{-1.3cm}\pi \\[0.5cm]
{\cal M}_{g} &
\end{array}
\nonumber
\end{align}
by simply forgetting the puncture and the local parameter\footnote{This means that the
projection is not one-to-one, since elements of~${\cal P}(g,1)$ with the puncture located
at different positions get mapped to the same point in~$\mathcal{M}_{g}$.}.
To construct a measure in ${\cal M}_{g}$ we associate with a
basis of holomorphic tangent vectors
$\{V_{1},\ldots,V_{3g-3}\}$ at a point $P\in{\cal M}_{g}$ 
the volume spanned by them, that we denote by
\begin{align}
\mu(P)(V_{1},\ldots,V_{3g-3},\overline{V}_{1},\ldots,\overline{V}_{3g-3}),
\end{align}
where, for simplicity, we analyze the case~$g>1$.
Thus, for every point $P\in{\cal M}_{g}$ and a basis of tangent 
vectors~$\{V_{i}\}$, we can find at some point in~$\pi^{-1}(P)\subset {\cal
P}(g,1)$ a set of vector fields $\{v_{i}\}$ in the neighborhood of the
puncture getting mapped into~$\{V_{i}\}$ under the projection 
map~$\pi$. For any such set~$\{v_{i}\}$ we define the measure
\begin{align}
\mu(P)(V_{1},\ldots,V_{3g-3},\overline{V}_{1},\ldots,\overline{V}_{3g-3})&=
\langle 0|\prod_{i=1}^{3g-3}b(v_{i})
\prod_{i=1}^{3g-3}\overline{b}(\overline{v}_{i})|\phi\rangle_{P},
\label{measure}
\end{align}
where the operator $b(v)$ is given by
\begin{align}
b(v)=\oint_{P}\frac{dz}{2\pi i}b(z)v(z),
\end{align}
and~$|0\rangle$ is the $\mbox{SL}(2,\mathbb{C})$-invariant vacuum, while
$|\phi\rangle_{P}\in {\cal F}_{\rm matter}\otimes{\cal F}_{bc}
\otimes{\cal F}_{\overline{b}\overline{c}}$ is the state associated with the point in~${\cal
P}(g,1)$. A basic check is
that the ghost numbers match on the right-hand side of eq.~(\ref{measure}), so the 
correlator is non-zero.

It can be shown that eq.~(\ref{measure}) is a proper measure on
moduli space. First, let us change the representative~$\{v_{i}\}$ 
in~${\cal P}(g,1)$ of the basis~$\{V_{i}\}$ by $v_{i}\rightarrow v_{i}+\epsilon_{i}$.
This does not induce any change in the measure provided~$\epsilon_{i}$ 
extends homomorphically away from~$P$, since then~$b(\epsilon_{i})|\phi\rangle=0$.
The same happens changing the local parameter or moving~$P$
infinitesimally, since $L_{n}|0\rangle=0$ for $n\geq -1$ and the total
central charge vanishes.
A last condition that~$\mu(P)$ should verify in order to be
identified with the bosonic string measure is the
Belavin-Knizhnik theorem~\cite{Belavin-Knizhnik}
\begin{align}
\partial\overline{\partial}\log{\mu}=-13\,\partial\overline{\partial}
\log{({\rm Im\,}\Omega)},
\label{B-K}
\end{align}
from where it follows that
\begin{align}
\mu=\frac{\rho\wedge\overline{\rho}}{(\det\,{\rm Im\,}\Omega)^{13}},
\label{eq:belavin_knizhnik}
\end{align}
where $\rho$ is a holomorphic $3g-3$ form in the moduli space. To prove
eq.~(\ref{B-K}) we make use of the relation~(\ref{delt-v}) to write
\begin{align}
\partial\overline{\partial}\log{\mu (\ldots, V,\ldots)}&=
\frac{\langle0|\ldots T(\,\,\,)\overline{T}(\,\,\,)
\ldots|\phi\rangle}{\langle0|\phi\rangle} 
-\frac{\langle0|\ldots T(\,\,\,)|\phi\rangle\langle0|\ldots
\overline{T}(\,\,\,)|\phi\rangle}{\langle0|\phi\rangle^{2}},
\end{align}
where the ellipsis stands for~$b(v_{i})$ insertions.
Since the only obstruction to holomorphic factorization comes from the
matter
sector, we can restrict the calculation to the matter states. Using
Rohrlich's formula, with~$\nu$ the Beltrami differential,~\cite{Fay}
\begin{align}
\delta\Omega_{ij}=-\oint_{P}\omega_{i}\omega_{j}\nu,
\end{align}
we obtain
\begin{align}
\partial\overline{\partial}\log{\mu(\ldots,
V,\ldots)}=-13\,\partial\overline{\partial} \log{\det\,{\rm Im\,}\Omega}.
\end{align}
This shows that~$\mu(P)$ is a well-defined measure over the moduli space~$\mathcal{M}_{g}$.

As an application of our analysis we show that, for splitting nodes, the measure has second-order
poles at the boundary of the moduli space\footnote{For non-splitting
nodes, on the other hand, the measure has first-order poles~\cite{Gava_Iengo_poles}.}. Let us consider a surface
pinched off along a non-contractible loop (see sec.~\ref{sec:int+moduli}), 
obtained by sewing two surfaces together with sewing parameter~$q\rightarrow 0$. 
In this region the measure looks like
\begin{align}
\mu\sim \left(\frac{dq}{q}\frac{d\overline{q}}{\overline{q}}\right)
\langle
S_{ij}|b_{0}\overline{b}_{0}q^{L_{0}}\overline{q}^{\overline{L}_{0}}|\phi\rangle.
\end{align}
Thus, since the lowest eigenvalue of $L_{0}$ and $\overline{L}_{0}$ is $-1$ 
(corresponding to the tachyon), in the~$q\rightarrow 0$ limit we find
\begin{align}
\mu\sim\left|\frac{dqd\overline{q}}{q^{2}\overline{q}^{2}}\right|.
\end{align}
This shows that~$\rho$ in eq.~\eqref{eq:belavin_knizhnik} 
has a second order pole at the boundary of
moduli space corresponding to the equivalence class of Riemann surfaces
with a node. This completes the identification of
(\ref{measure}) with the Polyakov measure. It also exhibits in a
rigorous way the intuitive expectation that string infinities originate
in the exchange of tachyons along thin and long tubes connecting different
elements of the surface.

As a final application, we study string scattering amplitudes in order to obtain
a geometrical interpretation of the physical state conditions. In 
order to compute a scattering amplitude with $n$~external states at genus~$g$, we need 
to construct a measure on ${\cal M}_{g,n}$. Using the operator formalism we find a
well-defined state $|\phi\rangle_{P}$ for any $P\in{\cal P}(g,n)$ and once
again we have a projection map
\begin{align}
\begin{array}{cc}
{\cal P}(g,n) & \\[0.2cm]
\Bigg\downarrow & \hspace*{-1.3cm} \pi \\[0.5cm]
{\cal M}_{g,n} &
\end{array}
\nonumber
\end{align}
by simply erasing the information about the punctures and the local parameters. We
can construct then a measure similar to the one in eq.~(\ref{measure})
\begin{align}
\mu\sim\langle\chi_{1}|\ldots\langle\chi_{n}|b(\,\,\,)\ldots\overline{b}(\,\,\,)
|\phi\rangle_{P}.
\label{measure-n}
\end{align}
For this, however, to be a well-defined
we need a~$(3g-3+n,3g-3+n)$-form, which requires the insertion of~$3g-3+n$ $b$ operators. 
Since~$|\phi\rangle_{P}$ has ghost number~$3g-3+3n/2$, 
the external states must have ghost number~$-\frac{n}{2}$. Distributing this ghost
number democratically among the external states, we conclude that~$\langle\chi_{i}|$ 
should have ghost number~$-\frac{1}{2}$.
Imposing the conditions to have a well-defined measure
over~${\cal M}_{g,n}$, we arrive at
\begin{align}
L_{n}|\chi_{i}\rangle&=\overline{L}_{n}|\chi_{i}\rangle=0\;, \nonumber \\[0.2cm]
b_{n}|\chi_{i}\rangle&=\overline{b}_{n}|\chi_{i}\rangle=0\;, \hspace{1cm}
n\geq 0\;,
\end{align}
with no constraint coming from~$L_{-1}$, since the measure
depends on the puncture position. These conditions, together
with
those over the ghost number of $|\chi_{i}\rangle$, are solved by the state
\begin{align}
|\chi\rangle=\sum |\psi\rangle_{\rm matter}\otimes|-\rangle,
\end{align}
where the state~$|\psi\rangle_{\rm matter}$ satisfies
\begin{align}
L_{n}|\psi\rangle_{\rm matter}&=\overline{L}_{n}|\psi\rangle_{\rm matter}=0, \hspace{1cm} n>0\;,
\nonumber \\[0.2cm]
L_{0}|\psi\rangle_{\rm matter}&=\overline{L}_{0}|\psi\rangle_{\rm matter}=|\psi\rangle_{\rm matter}.
\end{align}
The standard physical state conditions are thus retrieved.

We have seen how the operator formalism provides a natural framework to
analyze many qualitative and quantitative questions in string
theory. From a geometrical point of view, the computation of string
amplitudes is related to the construction of measure on~${\cal M}_{g,n}$
satisfying the Belavin-Knizhnik condition~\eqref{B-K}. 
This measure can be computed in the
operator formalism using the same techniques as in a CFT on the
plane. All the geometrical information is encoded into the Bogoliubov
transformation relating the standard vacuum~$|0\rangle$ to the 
state~$|\phi\rangle_{P}$ associated with~$P\in{\cal P}(g,n)$. 
Although the operator formalism can also be extended to the supersymmetric
case, we will not discuss the details here. The interested reader can
find then in ref.~\cite{LAG-Gomez-Nelson-Sierra-Vafa}.

\section{String theory II. Fermionic strings}
\label{sec:fermionic_strings}

We now begin our brief study of fermionic strings with the aim of showing how
many of the undesirable features of the
bosonic string are eliminated in a supersymmetric theory. We will also emphasize how some simple
consistency conditions, such as absence of global anomalies on the world-sheet,
impose strong constraints on the spectrum of the theory. It is quite
remarkable that once the modular anomalies (i.e. global diffeomorphism
anomalies) on the world-sheet are cancelled, and after we make the
theory invariant under the mapping class group (see sec.~\ref{sec:operator_formalism}),
all space-time anomalies also cancel~\cite{Schellekens-Warner}. At the
end of the present section we also discuss some special features of
superstring theory, such as their finite temperature behavior, and offer a brief status
report on finiteness of string perturbation theory.
A brief introduction to string black holes (covered in more detail in
A.~Polyakov's lectures~\cite{Polyakov92}) is deferred to the next section.

\subsection{Fermionic String}
\label{sec:fermionic_string}

Despite all the beautiful features of the bosonic
string, the model is far from satisfactory. Its first and more basic problem 
is the presence of a tachyon in the spectrum, itself the source
of many difficulties. Furthermore, the theory only contains 
bosonic degrees of freedom propagating in target
space, so there is no hope of getting interesting phenomenology from the
model. As we will see, one can get rid of these two problems at once by
adding fermions to the two-dimensional world-sheet field theory.
This lead to fermionic string models or, more properly,
supersymmetric strings. Now, in addition to the bosonic 
coordinates~$X^{\mu}(\tau,\sigma)$ ($\mu=1,\ldots,d$), we also have 
$d$~Majorana-Weyl
world-sheet spinors~$\Psi^{\mu}(\tau,\sigma)$ transforming as
a $d$-dimensional vector under Lorentz transformations in target
space. 

In the bosonic string the negative-norm states in the spectrum were
disposed of by using reparametrization invariance,
giving rise to the Virasoro constraints. The fermionic string also
contains negative-norm states coming from the action of the time component of the spinor field
$\Psi^{0}(\tau,\sigma)$ on the vacuum, so we need to find a new local invariance
to get rid of them: two-dimensional supergravity. Let us
introduce two new fields, the {\it zweibein}~$e_{\alpha}^{\;\;a}(\tau,\sigma)$ defined by
\begin{align}
h_{\alpha\beta}=e_{\alpha}^{\;\;a}e_{\beta}^{\;\;b}\eta_{ab},
\label{eq:vierbein_2d}
\end{align}
and a Majorana gravitino~$\chi_{\alpha}$ (in both cases $a,b=0,1$
and~$\alpha,\beta=0,1$ are respectively two-dimensional Lorentz and Einstein indices;~$\mu,\nu
=0,\ldots,d-1$ are reserved for target space indices). 
The complete world-sheet action we consider is~\cite{Brink,Deser-Zumino}
\begin{align}
S&=-\frac{1}{4\pi\alpha'} \int d^{2}\sigma
(\det\;e)\eta_{\mu\nu}\Bigg( h^{\alpha\beta} \partial_{\alpha}
X^{\mu}\partial_{\beta}X^{\nu}-ie^{\alpha}_{\;\;a}\overline{\Psi}^{\mu}
\rho^{a}\partial_{\alpha}\Psi^{\nu}  \nonumber \\[0.2cm]
&+ 2e^{\alpha}_{\;\;a}e^{\beta}_{\;\;b}\overline{\chi}_{\alpha}
\rho^{b}\rho^{a}
\Psi^{\mu}\partial_{\beta}X^{\nu}-\frac{1}{2}\overline{\Psi}^{\mu}\Psi^{\nu}
e^{\alpha}_{\;\;a}e^{\beta}_{\;\;b}
\overline{\chi}_{\alpha}\rho^{b}\rho^{a}\chi_{\beta}\Bigg),
\label{eq:WSferm_action}
\end{align}
where~$\rho^{a}$ are the two-dimensional Dirac matrices, satisfying the
Clifford algebra
\begin{align}
\{\rho^{a},\rho^{b}\}=-2\eta^{ab}.
\end{align}
The action~\eqref{eq:WSferm_action} is invariant under Weyl 
\begin{align}
\delta X^{\mu}&=0, \nonumber \\[0.2cm]
\delta \Psi^{\mu}&=-{1\over 2}\Omega(\sigma)\Psi^{\mu}, \nonumber\\[0.2cm]
\delta e_{\alpha}^{\,\,\,a}&= \Omega(\sigma) e_{\alpha}^{\,\,\,a}, \\[0.2cm]
\delta\chi_{\alpha} &={1\over 2}\Omega(\sigma)\chi_{\alpha},\nonumber 
\end{align}
as well as super-Weyl rescalings
\begin{align}
\delta\chi_{\alpha}= e_{\alpha a}\rho^{a}\lambda(\sigma),
\end{align}
where~$\lambda$ is a Majorana spinor and all remaining fields remain 
invariant. 
These symmetries can be fixed using the
superconformal gauge:
\begin{align}
e_{\alpha}^{\;\;a}&=e^{\phi}\delta_{\alpha}^{\;\;a}, \nonumber \\[0.2cm]
\hspace{1cm}
\chi_{\alpha}&=\rho_{a}\lambda.
\end{align}
so~$\phi$ and~$\lambda$
can be gauged away in the classical theory\footnote{As we will see below, there is a critical value of
the target space dimension for which these classical Weyl and super-Weyl
transformations are preserved as symmetries of the quantum theory.}. 
We thus arrive at the gauge-fixed classical action
\begin{align}
S=-\frac{1}{4\pi\alpha'}\int d^{2}\sigma \Big(
\partial_{\alpha}X^{\mu}\partial^{\alpha}X^{\nu}-i\overline{\Psi}^{\mu}
\rho^{\alpha}\partial_{\alpha}\Psi^{\nu}\Big)\eta_{\mu\nu},
\label{SST-conf-gauge}
\end{align}
which is still invariant under the supersymmetry transformations
\begin{align}
\delta_{\epsilon} X^{\mu}&=\overline{\epsilon}\Psi^{\mu}, \nonumber \\[0.2cm]
\delta_{\epsilon}\Psi^{\mu}&=-i\rho^{\alpha} \partial_{\alpha}
X^{\mu}\epsilon,
\end{align}
with $\epsilon$ a constant Majorana spinor. Associated with these,
there is a supercurrent
\begin{align}
J_{\alpha}=\frac{1}{2}\rho^{\beta}\rho_{\alpha}\Psi^{\mu}
\partial_{\beta} X_{\mu},
\end{align}
satisfying~$\partial_{\alpha}J^{\alpha}=0$.
Similarly, the energy momentum tensor
\begin{align}
T_{\alpha\beta}&=\partial_{\alpha}X^{\mu}\partial_{\beta}X_{\mu}+
\frac{i}{2}\overline{\Psi}^{\mu}\rho_{(\alpha}\partial_{\beta)}\Psi_{\mu}
\nonumber \\[0.2cm]
&-{1\over 2}\left(\eta^{\alpha\beta}\partial_{\alpha}X^{\mu}\partial_{\beta}X_{\mu}+
\frac{i}{2}\eta^{\alpha\beta}\overline{\Psi}^{\mu}
\rho_{\alpha}\partial_{\beta}\Psi_{\mu}\right)\times\eta_{\mu\nu},
\end{align}
is also conserved, $\partial_{\alpha}T^{\alpha\beta}=0$. With this, we arrive at the classical
equation of motion
\begin{align}
\partial_{\alpha}\partial^{\alpha}X^{\mu}&=0, \nonumber \\[0.2cm]
\rho^{\alpha}\partial_{\alpha}\Psi^{\mu}&=0,
\end{align}
supplemented by the constraints
\begin{align}
T_{\alpha\beta}&=0, \nonumber \\[0.2cm]
J_{\alpha}&=0,
\end{align}
following from the equations of motion of the non-dynamical zweibein~$e_{\alpha}^{\;\;a}$ and 
gravitino~$\chi_{\alpha}$ fields respectively.
Moreover, invariance under Weyl and super-Weyl transformations imply that
\begin{align}
T^{\alpha}_{\;\alpha}&=0, \nonumber \\[0.2cm]
\rho^{\alpha}J_{\alpha}&=0.
\end{align}

Let us study the free closed superstring. As in the
bosonic case, the world-sheet is a cylinder parametrized 
by~$-\infty<\tau<\infty$ and $0\leq\sigma< \pi$. After Wick
rotating
these coordinates and conformally mapping the
cylinder onto the punctured complex plane $\mathbb{C}^{*}$, the action
becomes
\begin{align}
S=-\frac{1}{4\pi\alpha'}\int d^{2}z \Big(\overline{\partial}X^{\mu}
\partial X_{\mu}-\psi^{\mu}\overline{\partial}\psi_{\mu}-
\overline{\psi}^{\mu}\partial \overline{\psi}_{\mu}\Big),
\end{align}
where the two-dimensional spinor $\Psi^{\mu}$ is decomposed in components according to
\begin{align}
\Psi^{\mu}\equiv \left(
\begin{array}{c}
\psi^{\mu} \\[0.1cm]
\overline{\psi}^{\mu}
\end{array}
\right),
\end{align}
and used the following representation for the two-dimensional
Dirac algebra
\begin{align}
\rho^{0}=\left(
\begin{array}{cc}
0  &  -i \\
i  &   0
\end{array}
\right), \hspace{2cm} \rho^{1}= \left(
\begin{array}{cc}
0  &   i \\
i  &   0
\end{array}
\right).
\end{align}
The equations of motion are
\begin{align}
\partial\overline{\partial}X^{\mu}(z,\overline{z})&=0, \nonumber \\[0.2cm]
\partial\overline{\psi}(z,\overline{z})&=\overline{\partial}\psi(z,\overline{z})=0,
\end{align}
so $\psi(z)$, $\overline{\psi}(\overline{z})$ are respectively holomorphic and
antiholomorphic fields. We can also compute the energy-momentum tensor and supercurrent as
\begin{align}
T(z)&\equiv T_{zz}(z)=-\frac{1}{2}\partial X^{\mu}\partial X_{\mu}-
\frac{1}{2}\partial\psi^{\mu}\psi_{\mu}, \nonumber \\[0.2cm]
T_{F}(z)&\equiv J_{z}(z)=-\frac{1}{2}\psi^{\mu}\partial X_{\mu},
\end{align}
and the corresponding expressions for barred quantities,
$\overline{T}(\overline{z})=T_{\overline{z}\overline{z}}(\overline{z})$,
$\overline{T}_{F}(\overline{z})=J_{\overline{z}}(\overline{z})$.

Superfield formalism is particularly well-suited to study the
supersymmetric string~\cite{Green-Schwarz-Witten,Kaku}.
Together with the commuting coordinates $z,
\overline{z}\in\mathbb{C}^{*}$, we introduce a
pair of anticommuting variables $\theta,\overline{\theta}$ to write the
supercoordinates
\begin{align}
\boldsymbol{z}=(z,\theta), \hspace{1cm} \overline{\boldsymbol{z}}=(\overline{z},\overline{\theta}).
\end{align}
Taking into account the anticommuting character of $\theta$ and
$\overline{\theta}$, any function $f(z,\overline{z},\theta,\overline{\theta})$ can be
expanded as
\begin{align}
f(z,\overline{z},\theta,\overline{\theta})=f_{0}(z,\overline{z})+
f_{1}(z,\overline{z})\theta+f_{2}(z,\overline{z})\overline{\theta}+
f_{3}(z,\overline{z})\theta\overline{\theta}.
\end{align}
As for calculus in superspace, the supersymmetric derivative is defined by
\begin{align}
D=\frac{\partial}{\partial\theta}+\theta\frac{\partial}{\partial z},
\end{align}
while integration over~$\theta$,~$\overline{\theta}$ is carried out using the Berezin rules~\cite{Berezin}
\begin{align}
\int d\theta=0, \hspace{1cm} \int d\theta\, \theta=1.
\end{align}

To reformulate the superstring action in the new
language, we introduce the superfield
\begin{align}
Y^{\mu}(z,\overline{z},\theta,\overline{\theta})=X^{\mu}(z,\overline{z})+
\theta\psi^{\mu}(z,\overline{z})+\overline{\theta}\,\overline{\psi}^{\mu}(z,\overline{z}),
\end{align}
so eq.~(\ref{SST-conf-gauge}) is recast as
\begin{align}
S=-\frac{1}{4\pi\alpha'}\int d^{2}z\,d\theta\,d\overline{\theta}\,
\overline{D}Y^{\mu}DY_{\mu},
\end{align}
invariant under superconformal transformations\footnote{Here we are not
dwelling on the broad subject of superconformal field theories. For 
details and results, as well as the formulation
of superstrings in this language, see ref.~\cite{FMS}.}.
We also define the components of the super-energy-momentum tensor components
${\cal T}(z,\theta)$, ${\overline{\cal T}}(\overline{z},\overline{\theta})$
\begin{align}
{\cal T}(z,\theta)&=T_{F}(z)+\theta T(z), \nonumber \\[0.2cm]
\overline{\cal T}(\overline{z},\overline{\theta})&=\overline{T}_{F}(\overline{z})+
\overline{\theta}\,\overline{T}(\overline{z}).
\end{align}
After quantization, we are led to the following OPE's for~$T(z)$ 
and~$T_{F}(z)$
\begin{align}
T(z)T(w)&=\frac{c/2}{(z-w)^{4}}+\frac{2T(w)}{(z-w)^{2}}+
\frac{1}{z-w}\partial T(w)+\mbox{regular terms}, \nonumber \\[0.2cm]
T(z)T_{F}(w)&=\frac{3/2}{(z-w)^{2}}T_{F}(w)+\frac{1}{z-w}\partial
T_{F}(w)+\ldots, \label{OPE-superconformal} \\[0.2cm]
T_{F}(z)T_{F}(w)&=\frac{c/6}{(z-w)^{3}}+\frac{1}{2(z-w)}T(w)
+\ldots
\nonumber
\end{align}
The second identity in particular implies that~$T_{F}(z)$ is a primary field 
with conformal weight~$\frac{3}{2}$.

To quantize the theory using the functional integral formalism,  
all local symmetries need to be gauge-fixed. Using the
Faddeev-Popov procedure, we add to the anticommuting
reparametrization
ghosts~$b_{\alpha\beta}$ and~$c^{\alpha}$ a pair of commuting
superconformal ghosts~$\beta_{\alpha}$
and~$\gamma$, with spins~$\frac{3}{2}$ and~$-\frac{1}{2}$
respectively. Their action reads
\begin{align}
S_{\rm gh}=\frac{1}{2\pi\alpha'}\int d^{2}z\Big(b_{zz}\overline{\partial}
c^{z}+\beta_{z}\overline{\partial}\gamma+\mbox{c.c.}\Big).
\end{align}
In the superspace formalism, the ghost fields are also combined into a pair of ghosts
superfields
\begin{align}
B(z,\theta)&=\beta(z)+\theta b(z), \nonumber \\[0.2cm]
C(z,\theta)&=c(z)+\theta\gamma(z),
\end{align}
and the ghost action takes the more compact form
\begin{align}
S_{\rm gh}=\frac{1}{2\pi\alpha'}\int d^{2}z d\theta d\overline{\theta}
\big(B\overline{D}C+\mbox{c.c.}\big).
\end{align}

To summarize, in the free quantum superstring we have to deal with the following fields:
\begin{align}
X^{\mu}(z), \hspace{.5cm} \psi^{\mu}(z), \hspace{.5cm} b(z),
\hspace{.5cm} c(z), \hspace{.5cm} \beta(z), \hspace{.5cm}
\gamma(z),
\end{align}
and their corresponding antiholomorphic components.
Their contributions to the central charge~$c$ are given by
\begin{align}
\begin{array}{ccc}
X^{\mu}    & \longrightarrow  & d, \\[0.2cm]
\psi^{\mu} & \longrightarrow  & \frac{d}{2}, \\[0.2cm]
b,\;c	 & \longrightarrow  & -26, \\[0.2cm]
\beta,\;\gamma& \longrightarrow  & 11,
\end{array}
\end{align}
so the total central charge is equal to 
\begin{align}
c_{\rm tot}=\frac{3d}{2}-15.
\end{align} 
This
means that conformal and superconformal anomalies cancel only if the
target space dimension is $d=10$. As with the
bosonic string, what distinguishes critical and non-critical superstrings is
the decoupling of the conformal modes. Since a graviton only appears
when~$d=10$ and we are interested in the fermionic string
as a theory of quantum gravity unifying of all fundamental
interactions, from now on we focus our attention on critical superstrings.

A crucial issue in dealing with fermionic string is the question of
fermion
boundary conditions. As in the closed bosonic string, we
impose the periodicity condition of the embedding bosonic
field,~$X^{\mu}(\tau,\sigma+\pi)=X^{\mu}(\tau,\sigma)$. Fermion fields, 
on the other hand, can be either periodic or antiperiodic.
These are respectively known as Ramond (R)
and Neveu-Schwarz
(NS) boundary conditions
\begin{align}
\Psi^{\mu}(\tau,\sigma+\pi)&=\Psi^{\mu}(\tau,\sigma), \hspace{2.4cm}
\mbox{(R)} \nonumber \\[0.2cm]
\Psi^{\mu}(\tau,\sigma+\pi)&=-\Psi^{\mu}(\tau,\sigma). \hspace{2cm}
\mbox{(NS)}
\end{align}
When mapping the cylinder onto the punctured complex plane
$\mathbb{C}^{*}$, we need to translate these boundary conditions. 
Since~$\psi^{\mu}(z)$ is a~$(\frac{1}{2},0)$ field and~$z\equiv e^{w}=\exp{[2(\tau-i\sigma)]}$,
we have
\begin{align}
\psi^{\mu}(z)dz^{\frac{1}{2}}=\psi^{\mu}(w)dw^{\frac{1}{2}}.
\end{align}
This means the periodicity of $\psi^{\mu}(z)$ under~$z\rightarrow e^{2\pi i}z$
is opposite to that of~$\psi^{\mu}(w)$ under~$w\rightarrow w+2\pi i$. Thus, on the plane
R and NS boundary conditions read
\begin{align}
\psi^{\mu}(e^{2\pi i}z)&=-\psi^{\mu}(z) \hspace{0.9cm} \mbox{(R)},
\nonumber \\[0.2cm]
\psi^{\mu}(e^{2\pi i}z)&=\psi^{\mu}(z) \hspace{1cm} \mbox{(NS)}.
\end{align}
For closed superstrings there are four possible sectors, corresponding to
the boundary conditions for left- and right-moving modes: (R,R), (R,NS),
(NS,R) and (NS,NS). Expanding the fermion fields~$\psi(z)$ in Fourier modes, we have
\begin{align}
\psi(z)&=\sum_{n\in\mathbb{Z}} d^{\mu}_{n}
z^{-n-\frac{1}{2}} \hspace{1cm} \mbox{(R)}, \nonumber \\[0.2cm]
\psi(z)&=\sum_{r\in\mathbb{Z}+\frac{1}{2}} b^{\mu}_{r}
z^{-r-\frac{1}{2}}  \hspace{0.5cm} \mbox{(NS)}.
\end{align}
The field~$X^{\mu}(z,\overline{z})$, on the other hand, still admits the expansion~\eqref{expansion}
as in the bosonic string.

The energy-momentum tensor, being a bosonic field, can be
expanded in the usual way
\begin{align}
T(z)=\sum_{n\in\mathbb{Z}}L_{n}z^{-n-2},
\end{align}
where~$L_{n}$ are again the Virasoro generators. For the supercurrent~$T_{F}(z)$,
however, the boundary conditions have to be consistent with those
of the fermion fields~$\psi^{\mu}(z)$, namely
\begin{align}
T_{F}(z)&=\frac{1}{2}\sum_{r\in\mathbb{Z}}G_{r}z^{-r-\frac{3}{2}}
\hspace{1.5cm} \mbox{(R)}, \nonumber \\[0.2cm]
T_{F}(z)&=\frac{1}{2}\sum_{r\in\mathbb{Z}+\frac{1}{2}}G_{r}
z^{-r-\frac{3}{2}} \hspace{1cm} \mbox{(NS)},
\end{align}
where $G_{r}$ are the generators of superconformal transformations.
The (anti)commutation relations of the generators follow from the OPE's
in eq.~(\ref{OPE-superconformal})
\begin{align}
[L_{m},L_{n}]&=(m-n)L_{m+n}+\frac{\widehat{c}}{8}m(m^{2}-1)\delta_{m+n,0}, \nonumber \\[0.2cm]
{[}L_{m},G_{r}]&=\left(\frac{m}{2}-r\right)G_{m+r}, \\[0.2cm]
\{G_{r},G_{s}\}&=2L_{r+s}+\frac{\widehat{c}}{2}
\left(r^{2}-\frac{1}{4}\right)\delta_{r+s,0},
\nonumber 
\end{align}
where~$\widehat{c}=\frac{2}{3}c$ and~$r\in\mathbb{Z}$ 
or~$r\in\mathbb{Z}+\frac{1}{2}$ depending on
whether we are in the (R) or (NS) sectors.

Let us study now the spectrum of the closed
supersymmetric string. Upon quantization of the two-dimensional supersymmetric 
field theory, we find the following non-zero
(anti)commutation relations among oscillators
\begin{align}
[\alpha_{m},\alpha_{n}]&=m\eta^{\mu\nu}\delta_{m+n,0}, \nonumber \\[0.2cm]
\{d^{\mu}_{m},d^{\nu}_{n}\}&=\eta^{\mu\nu}\delta_{m+n,0},  \\[0.2cm]
\{b^{\mu}_{r},b^{\nu}_{r}\}&=\eta^{\mu\nu}\delta_{r+s,0}.
\nonumber
\end{align}
As in the bosonic string, we have negative norm states that are eliminated
following the same steps explained in
sec.~\ref{sec:bosonic_string_theory}. Again, the constraints
cannot be ``strongly'' imposed because of the anomaly. Rather, we require physical
states to be annihilated by half of the Virasoro and super-Virasoro generators
\begin{align}
L_{n}|{\rm phys}\rangle&=0, \hspace{1cm} n>0, \nonumber \\[0.2cm]
G_{r}|{\rm phys}\rangle&=0, \hspace{1cm} r>0\;\mbox{(NS)},\hspace{0.5cm}
r\geq 0 \;\mbox{(R)},
\end{align}
together with the conditions
\begin{align}
(L_{0}-a)|{\rm phys}\rangle&=0, \nonumber \\[0.2cm]
(L_{0}-\overline{L}_{0})|{\rm phys}\rangle&=0,
\end{align} where the normal ordering
constant~$a$ takes the values~$a=\frac{1}{2},0$ in the NS and R sectors respectively
[cf. eq.~\eqref{eq:bosonic_phys_cond}].
To find the spectrum, we work in the light-cone gauge where
the constraints can be explicitly solved. Making use of the residual invariance
after implementing the superconformal gauge, we impose the light-cone
gauge conditions
\begin{align}
X^{+}&=q^{+}+\alpha'p^{+}\tau, \nonumber \\[0.2cm]
\psi^{+}&=0,
\end{align}
so the theory is completely described in terms of transverse
modes.

Let us begin with the NS sector, focusing on the left-movers. Their 
contribution to the mass formula in this sector is computed 
following the same steps as in sec.~\ref{sec:quantization_bosonic},
with the result 
\begin{align}
\frac{1}{2}\alpha'm_{L}^{2}=\sum_{n>0}\alpha_{-n}^{i}\alpha_{n}^{i}+
\sum_{r>0}rb_{-r}^{i}b_{r}^{i}-\frac{1}{2},
\label{m-super}
\end{align}
and a corresponding expression~$m_{R}^{2}$ for right-movers.
The vacuum state $|0\rangle_{L}$ is defined by the conditions
\begin{align}
\alpha_{n}^{i}|0\rangle_{L}=b_{r}^{i}|0\rangle_{L}=0, \hspace{1cm} n,r>0.
\label{vac-super}
\end{align}
and has~$m_{L}^{2}<0$ and it is a tachyon. This looks like the kind
of trouble we encountered in the bosonic string and were trying to escape from. Fortunately, as
it will be shown shortly, these tachyon states can be purged from the theory.
In the first excited level, on the other hand, we get the state
\begin{align}
b_{-\frac{1}{2}}^{i}|0\rangle_{L},
\end{align}
which is massless and transforms as a vector of~$\mbox{SO}(8)$, the little group for massless particles 
in ten dimensions. In addition, for the first massive level we have two possible states 
\begin{align}
\alpha_{-1}|0\rangle_{L} \hspace*{0.5cm} \mbox{and} \hspace*{0.5cm}
b_{-\frac{1}{2}}^{i}b_{-\frac{1}{2}}^{j}|0\rangle_{L},
\end{align}
both with~$\alpha'm_{L}^{2}=1$.

The contribution to the of the left-movers to the mass formula in the R sector 
is derived from~$L_{0}=0$ condition
\begin{align}
\frac{1}{2}\alpha'm_{L}^{2}=\sum_{n>0}\alpha^{i}_{-n}\alpha^{i}_{n}+
\sum_{n>0}nd^{i}_{-n}d^{i}_{n}.
\label{eq:massR}
\end{align}
Here, however, we meet with an additional complication. The
zero modes $d_{0}^{\mu}$ form a closed subalgebra
\begin{align}
\{d_{0}^{\mu},d_{0}^{\nu}\}=2\eta^{\mu\nu}.
\label{subalgebra}
\end{align}
This implies that given the massless vacuum state~$|{\rm vac}\rangle_{L}$
defined by
\begin{align}
\alpha_{n}|{\rm vac}\rangle_{L}=d_{n}|{\rm vac}\rangle_{L}=0, \hspace{1cm} n>0,
\end{align}
all the states $d_{0}^{\mu}|{\rm vac}\rangle_{L}$ are massless as well, since
$\{d_{0}^{\mu},d_{n}^{\nu}\}=0$ for~$n\neq 0$. These states furnish
a representation of the Clifford algebra~(\ref{subalgebra})
and we conclude the ground state of the R sector is a ten-dimensional
spinor~$|A\rangle_{L}$, with~$A=1,\ldots,8$ a space-time spinor index. In sec.~\ref{sec:spinors} we learned
that in ten
dimensions both the Weyl and
Majorana conditions can be simultaneously imposed 
or, in other words, our ground states can be chosen to have
definite chiralities that we denote 
by~$|A^{+}\rangle_{L}$, $|A^{-}\rangle_{L}$, with~$A^{+},A^{-}=1,\ldots,8$.
Thus, the spectrum of the R~sector of the fermionic string is constructed
from this massless spinor state. Modulo a choice of vacuum chirality, 
the first excited level contains two states
\begin{align}
d^{i}_{-1}|A^{+}\rangle_{L}, \hspace*{0.5cm} \mbox{and} \hspace*{0.5cm} 
\alpha^{i}_{-1}|A^{+}\rangle_{L},
\end{align} 
both with masses
$\alpha'm_{L}^{2}=4$. Higher excited levels are obtained in a similar fashion.

To obtain the spectrum of the closed fermionic string we
tensor-product right- and left-movers in the four sectors~(R,R),~(R,NS),~(NS,R),
and~(NS,NS). The associated mass formulae are obtained by adding the corresponding
expressions~\eqref{m-super} and~\eqref{eq:massR} for the left- ($m_{L}^{2})$
and right-moving ($m_{R}^{2})$ sectors, taking also into
account the level matching condition~$L_{0}=\overline{L}_{0}$. 
Notice, however, that we still need to deal with
the presence of the tachyon state in the sector with NS~boundary conditions. 
Historically, a projection was done onto a sector
of definite $G$-parity states~\cite{GSO}, thus eliminating the tachyon and also
rendering the
spectrum space-time supersymmetric. 

\paragraph{The GSO projection and modular invariance.}
We begin by presenting first how this projection is done, and we will see later how this 
projection is 
naturally understood in terms of modular invariance, the absence of
global diffeomorphism anomalies on the world-sheet. The $G$-parity
operator in the NS sector is defined by
\begin{align}
G=(-1)^{F+1}=(-1)^{\sum_{r\in\mathbb{Z}+\frac{1}{2}}
b^{i}_{-n}b^{i}_{n}+1},
\end{align}
so the dangerous NS vacuum state has $G|0\rangle=-|0\rangle$. In
the R sector, on the other hand, the $G$-parity operator is constructed from a string
generalization of the chirality matrix
\begin{align}
G&=(-1)^{\sum_{n>0}
d^{i}_{-n}d^{i}_{n}} \Gamma^{0}\ldots\Gamma^{8}.
\end{align}
By using these operators we can at the same time project out the unwanted tachyon 
and make the string spectrum supersymmetric. This can be accomplished by using the~GSO
projection~\cite{GSO} introduced by Gliozzi, Scherk, and Olive.
We project the string spectrum onto the states with positive
$G$-parity for each handedness separately. Since the fundamental state in
the NS sector has $G=-1$, this projection eliminates the tachyon from the
theory. With respect to the~$\Gamma$ operator in the R~sector, we have two 
different,~$\Gamma=\pm 1$. Since the action of~$\Gamma$ on the R~ground state~$|a\rangle$ 
gives us the chirality of this state,
different choices in the left- and right-moving sector lead to chiral and non-chiral spectra. 
For example, taking~$\Gamma=\overline{\Gamma}=1$ the left- and right-R ground 
states are spinors with the same chirality,~$|A^{+}\rangle$ and~$|B^{+}\rangle$. 
This type of string is called type-IIB
superstring~\cite{type-IIB} and its massless sector contains the
states\footnote{Apart from the GSO projection, the
level matching condition has also been implemented. The same applies to the derivation of
eq.~\eqref{IIA} below.}
\begin{align}
|A^{+}\rangle_{L}&\otimes |B^{+}\rangle_{R}, \nonumber \\[0.2cm]
\overline{b}_{-\frac{1}{2}}^{i}|0\rangle_{L}&\otimes 
b_{-\frac{1}{2}}^{j}|0\rangle_{R}, \nonumber \\[0.2cm]
|A^{+}\rangle_{L} & \otimes  b_{-\frac{1}{2}}^{i}|0\rangle_{R}, \label{IIB} \\[0.2cm]
\overline{b}_{-\frac{1}{2}}^{i}|0\rangle_{L} & \otimes  |B^{+}\rangle_{R}.
\nonumber
\end{align}
Decomposing these states into irreducible representations of the little
group $\mbox{SO}(8)$, we find the graviton,
two real scalars, two antisymmetric rank-two tensors, and a rank-four
antisymmetric tensor. In addition, we have two spin-$\frac{3}{2}$
and two spin-$\frac{1}{2}$ states, all with the same chirality. These states in the massless 
sector are those of chiral~$N=2$ SUGRA in~$d=10$
\cite{Nahm,SUGRA-d=10}.

By taking the choice $\Gamma=-\overline{\Gamma}=1$, on the other hand, we obtain the type-IIA
superstring~\cite{Green-Schwarz-Witten}. The states in the massless
sector are
\begin{align}
|A^{+}\rangle_{L} &\otimes |B^{-}\rangle_{R}, \nonumber \\[0.2cm]
\overline{b}_{-\frac{1}{2}}^{i}|0\rangle_{L} & \otimes 
b_{-\frac{1}{2}}^{j}|0\rangle_{R}, \nonumber \\[0.2cm]
|A^{+}\rangle_{L} &\otimes  b_{-\frac{1}{2}}^{i}|0\rangle_{R},
\label{IIA} \\[0.2cm]
\overline{b}_{-\frac{1}{2}}^{i}|0\rangle_{L} &\otimes  |B^{-}\rangle_{R}.
\nonumber
\end{align}
The theory has the same states as type-IIB superstring, but with the important
difference that fermionic states come in both chiralities.
Then, the particle content of type-IIA superstring is that of
non-chiral~$N=2$ SUGRA in~$d=10$~\cite{Nahm}, which is
also obtained by dimensional reduction from~$N=1$ SUGRA in~$d=11$.

It is rather surprising that the 
GSO projection renders the spectrum of the fermionic string  space-time supersymmetric. 
After all, supersymmetry was only implemented on the world-sheet and there
is no a priori reason to expect that it is also 
realized in target space. As we will see below, there exits a second formalism developed by
Green and Schwarz~\cite{Green-Schwarz-2,Schwarz} where space-time supersymmetry is
implemented on the string action from the start.

To compute amplitudes we need to consider
world-sheets with arbitrary genus~$g$ and we have to be specially
careful in properly handling the $2g$~spin structures of world-sheet fermion fields. 
As discussed in sec.~\ref{sec:operator_formalism}, spin structures are divided 
into even and odd depending on the number or zero modes of the Weyl operator or, equivalently,
on whether the number of holomorphic sections of the corresponding line
bundle is even or odd. There are $2^{g-1}(2^{g}+1)$ even spin structures
and $2^{g-1}(2^{g}-1)$ odd ones (for a detailed discussion of spin
structures on Riemann surfaces see ref.~\cite{Fay}). 

Summing 
over spin structures implements the GSO projection and is also equivalent to 
the cancellation of modular anomalies. This is well illustrated in 
the~$g=1$ case, the torus, whose spin line bundles are
flat line bundles that can be represented in terms of the boundary 
conditions of world-sheet fermions along the two homology cycles. The four possibilities
are schematically depicted as:
\begin{align*}
\centerline{\includegraphics[scale=0.40]{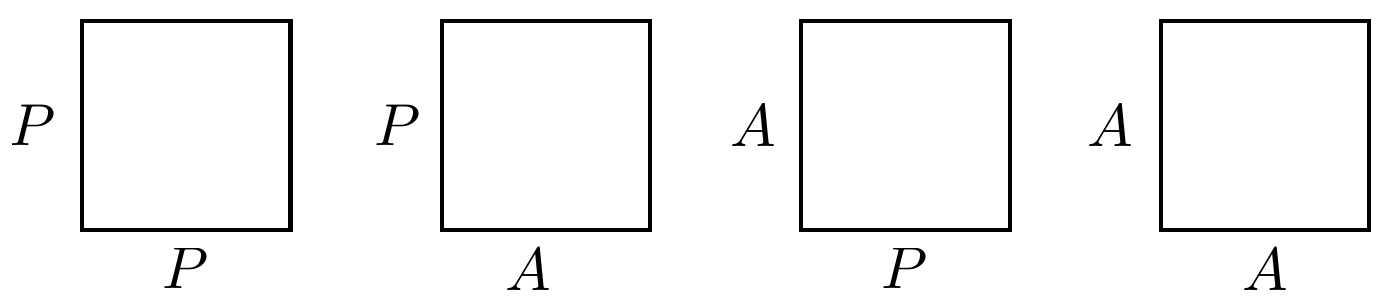}}
\end{align*}
where the horizontal line represents the world-sheet~$\sigma$-coordinate
with the vertical line showing the~$\tau$ direction.

The mapping class group, or modular group, on the torus 
is~$\mbox{PSL}(2,\mathbb{Z})$, acting on the complex modular parameter~$\tau\equiv\tau_{1}+i\tau_{2}$ as
\begin{align}
\tau\longrightarrow \frac{a\tau+b}{c\tau+d}, \hspace{1cm}
a,b,c,d \in \mathbb{Z}, \hspace{1cm} ab-cd=1.
\end{align}
It is generated by the two transformations
\begin{align}
T:\tau&\longrightarrow \tau+1 , \nonumber \\[0.2cm]
S:\tau&\longrightarrow -\frac{1}{\tau},
\end{align}
satisfying the relations
\begin{align}
S^{2}=(ST)^{3}=1.
\end{align}
The action of the modular group results in a change of the fermion boundary conditions. For
instance, acting with~$T$ on~$(A,A)$ the fundamental cell of the torus
\begin{align*}
\centerline{\includegraphics[scale=0.40]{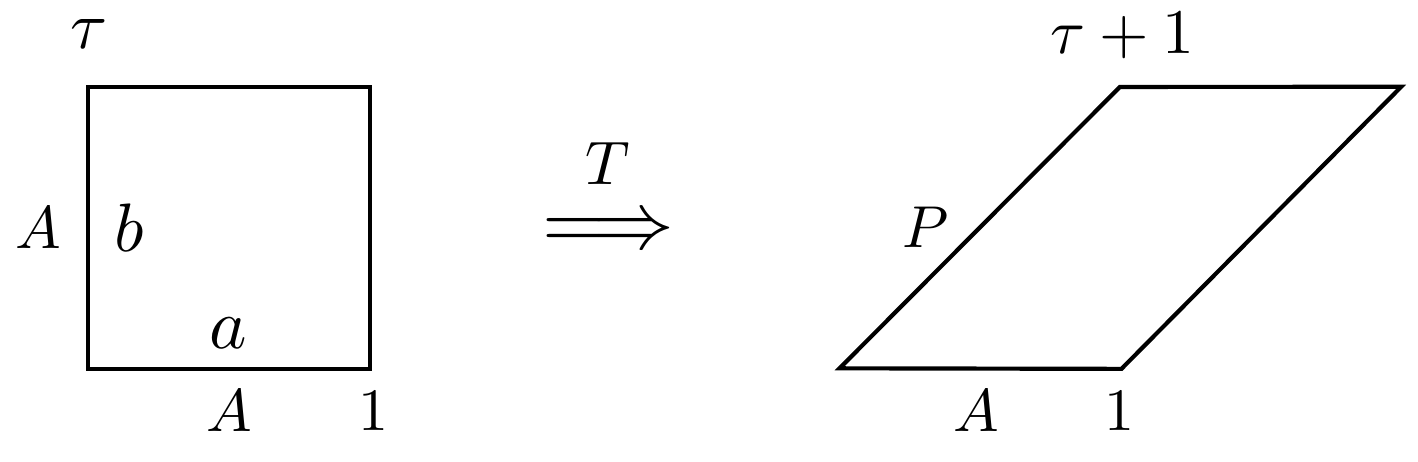}}
\end{align*}
maps the~$(A,A)$ spin structure into~$(P,A)$. 
Acting further with~$S$, which effectively exchanges the two cycles,
we obtain the~$(A,P)$ spin structure. On the other hand, it is not difficult to
see that the~$(P,P)$ spin structure is left invariant by both~$T$ and~$S$.
This shows that the
action of the group~$\mbox{PSL}(2,\mathbb{Z})$ on the spin structures
consists of two orbits: one comprising the even spin 
structures~$(A,A)$,~$(A,P)$, and~$(P,A)$ and the a second 
including the odd one~$(P,P)$. The reader
can easily check that in the former case the operator $\overline{\partial}$, the Dirac operator
in this case, has no zero modes, whereas it has one with~$(P,P)$ 
boundary conditions. Moreover, denoting  
by~$\Gamma_{2}$ the subgroup of $\mbox{SL}(2,\mathbb{Z})$ leaving the spin
structure invariant, it can be shown that the
quotient~$\mbox{SL}(2,\mathbb{Z})/\Gamma_{2}$ is a finite group of order 
six\footnote{The elements of the subgroup~$\Gamma_{2}\subset \mbox{PSL}(2,\mathbb{Z})$ are
\begin{align}
\left(
\begin{array}{cc}
a & b \\
c & d
\end{array}
\right) \equiv \left(
\begin{array}{cc}
1 & 0 \\
0 & 1
\end{array}
\right)
\;\;\mbox{mod}\;2.
\nonumber
\end{align}
It is generated by~$T^{2}$ and~$ST^{-2}S$.}. 

After this
brief mathematical interlude,
we address the problem of global anomalies on the world-sheet,
first was analyzed
by Witten (see first reference in \cite{Witten-anom}). When 
defining scattering measures in sec.~\ref{sec:operator_formalism}, we
needed to integrate over metrics modulo diffeomorphisms
and singled out the normal subgroup of orientation-preserving 
diffeomorphisms in the connected component of the 
identity,~$\mbox{Diff}_{0}^{+}(\Sigma)$, so the mapping class group was defined as
$\Omega(\Sigma)=\mbox{Diff}^{+}(\Sigma)/\mbox{Diff}_{0}^{+}(\Sigma)$.
Since all quantities involved in the analysis of the bosonic string were from the
beginning generally covariant, we were rather cavalier in the treatment
of~$\Omega(\Sigma)$. Now, however, we need to be more careful. The fields
in the fermionic string section are acted on non-trivially by
$\Omega(\Sigma)$, so we need to ensure that the quantum theory is invariant
under the mapping class group. If we fix the spin structures of the fermions,
the absence of global anomalies is equivalent to the condition that the theory does
not change under the action of~$\Gamma_{2}$, the
subgroup of $\Omega(\Sigma)$ preserving the spin structures introduced above. One can then 
classify all possible ways of summing over spin
structures which are invariant under $\Omega(\Sigma)/\Gamma_{2}$. This
should produce all modular invariant string theories with the same
world-sheet field content. 

Let us go back now to the case of the fermionic string, where the GSO projection
is implemented independently on left- and right-movers, and compute the contribution
of the four spin structures to the torus
partition functions. Writing again~$q\equiv \exp{(2\pi i\tau)}$, we
have
\begin{align}
Z_{(A,A)}&={\rm tr\,}q^{H_{\rm NS}}, \nonumber \\[0.2cm]
Z_{(A,P)}&={\rm tr\,}(-1)^{F}q^{H_{\rm NS}}, \nonumber \\[0.2cm]
Z_{(P,A)}&={\rm tr\,}q^{H_{\rm R}}, \label{Z-traces} \\[0.2cm]
Z_{(P,P)}&={\rm tr\,}(-1)^{F}q^{H_{\rm R}},
\nonumber
\end{align}
with~$H_{\rm NS}$ and~$H_{\rm R}$ the Hamiltonians
in the~NS and~R sectors
\begin{align}
H_{\rm NS}&=(L_{0})_{\rm plane}-\frac{1}{2}=
\sum_{n>0}\alpha_{-n}^{i}\alpha_{n}^{i}+\sum_{r>0}rb_{-r}^{i}b_{r}^{i}-
\frac{1}{2}, \nonumber \\[0.2cm]
H_{\rm R}&=(L_{0})_{\rm plane}-\frac{1}{2}=
\sum_{n>0}\alpha_{-n}^{i}\alpha_{n}^{i}+\sum_{n>0}nd_{-n}^{i}d_{n}^{i},
\end{align}
where we used that in the light-cone gauge~$c=8\times 1+8\times \frac{1}{2}=12$.
To implement periodic boundary conditions along 
the~$\tau$-direction we need to insert $(-1)^{F}$ in the traces in
eq.~(\ref{Z-traces}), what provides a first hint on the geometrical
interpretation of the GSO projection. Notice as well that in writing the corresponding traces, we
also dropped the contribution of the zero modes~$\alpha_{0}^{i}\equiv{1\over 2}p^{i}$ 
[cf.~\eqref{eq:L0_L0bar_modeexpansion}, keeping in mind that now we are working in the light-cone gauge]. 
After considering left- and right-movers, this results in a factor
proportional to~$\tau_{2}^{-4}$ 
\begin{align}
\int d^{8}p \,q^{{1\over 2}p^{i}p^{i}}\overline{q}^{{1\over 2}p^{i}p^{i}}
&=\int d^{8}p \,e^{-2\pi \tau_{2}p^{i}p^{i}} \nonumber \\[0.2cm]
&=(2\tau_{2})^{-4}.
\end{align}
In addition, an explicit evaluation of 
the trace of the bosonic non-zero modes $\alpha^{i}_{n}$, 
a common factor in all the contributions in~\eqref{Z-traces}, gives the result
\begin{align}
{\rm tr\,} q^{\sum_{n>0}\alpha_{-n}^{i}\alpha_{n}^{i}}=
q^{\frac{1}{3}}\eta^{-8}(\tau),
\label{eq:bosonic_zero_modes_ST}
\end{align}
where $\eta(\tau)$ is the Dedekind eta function
\begin{align}
\eta(\tau)=q^{\frac{1}{24}}\prod_{n=1}^{\infty}(1-q^{n}).
\end{align}
The remaining contributions to the partition function in the four different sectors can be expressed in
terms of Jacobi theta functions introduced in eq.~\eqref{vartheta-function}.
The corresponding partition functions are computed to be:
\begin{align}
Z_{(A,A)}&=\eta^{-8}(\tau)\left(\frac{\vartheta\left[
\begin{array}{c}
0 \\
0
\end{array}
\right](0|\tau)}{\eta(\tau)}\right)^{4}, \nonumber \\
Z_{(A,P)}&=\eta^{-8}(\tau)\left(\frac{\vartheta\left[
\begin{array}{c}
0 \\
\frac{1}{2}
\end{array}
\right](0|\tau)}{\eta(\tau)}\right)^{4}, \nonumber \\
Z_{(P,A)}&=\eta^{-8}(\tau)\left(\frac{\vartheta\left[
\begin{array}{c}
\frac{1}{2} \\
0
\end{array}
\right](0|\tau)}{\eta(\tau)}\right)^{4}, \label{theta-contrib} \\
Z_{(P,P)}&=\eta^{-8}(\tau)\left(\frac{\vartheta\left[
\begin{array}{c}
\frac{1}{2} \\
\frac{1}{2}
\end{array}
\right](0|\tau)}{\eta(\tau)}\right)^{4}=0,
\nonumber
\end{align}
where the factor $q^{\frac{1}{3}}$ in eq.~\eqref{eq:bosonic_zero_modes_ST}
cancels a similar factor coming from 
traces over the fermionic operators. 
We see that the odd spin structure does not contribute, as 
it is expected due to the presence of a fermionic zero mode. It 
should be mentioned that the fourth power of the $\vartheta$-functions
appearing in eq.~(\ref{theta-contrib}) is the minimal power guaranteeing
invariance of the partition function under~$\Gamma_{2}$, the subgroup of modular
transformations preserving the spin structures. This follows from the fact 
that the $\vartheta$-functions
just pick up a fourth root of unity under {\em any} transformation in this
subgroup. To understand the origin of these fourth powers in mathematical terms, 
let us recall that in the
light-cone gauge there are eight Weyl-Majorana world-sheet fermions within each
spin structure, which combine into {\em four} Weyl fermions. The
determinant of the Weyl operator within each spin structure is then proportional to 
the corresponding~$\vartheta$-function~\cite{Alvarez-Gaume-Moore-Vafa}. 

Once
global anomalies are cancelled, we can write a modular
invariant combination of the four elements in eq.~\eqref{theta-contrib} by
fixing their relative phases in the total partition function
\begin{align}
Z(\tau)&=Z_{(A,A)}(\tau)+w(A,P)Z_{(A,P)}(\tau) \nonumber \\[0.2cm]
&+w(P,A)Z_{(P,A)}(\tau)+w(P,P)Z_{(P,P)}(\tau).
\label{tor-partition}
\end{align}
Since the~$(A,A)$, $(P,A)$, $(A,P)$ spin structures
are in the same orbit under the action of~$T$ and~$S$, the 
phases~$w(A,P)$,~$w(P,A)$ are determined from the requirement~$w(A,A)=1$ to be
\begin{align}
w(A,P)=w(P,A)=-1.
\end{align}
With this, the one-loop partition function takes the form
\begin{align}
Z(\tau)={\rm tr\,}\big[1+(-1)^{F+1}\big]q^{H_{\rm NS}}
-{\rm tr\,}\big[1+w(P,P)(-1)^{F+1}\big]q^{H_{\rm R}},
\label{eq:Z(tau)GSO}
\end{align}
showing that, to preserve modular invariance, we need to
project onto the sector with~$(-1)^{F+1}=1$. This implements the~GSO projection. 

Introducing a more compact notation
\begin{align}
\vartheta\left[
\begin{array}{c}
0 \\
0
\end{array}
\right]\equiv\theta_{3}, \hspace{1cm}
\vartheta\left[
\begin{array}{c}
\frac{1}{2} \\
0
\end{array}
\right]\equiv\theta_{2}, \hspace{1cm}
\vartheta\left[
\begin{array}{c}
0 \\
\frac{1}{2}
\end{array}
\right]\equiv\theta_{4}, \hspace{1cm}
\vartheta\left[
\begin{array}{c}
\frac{1}{2} \\
\frac{1}{2}
\end{array}
\right]\equiv\theta_{1},
\end{align}
the partition function~\eqref{eq:Z(tau)GSO} is written as
\begin{align}
Z(\tau)={\theta_{3}(0|\tau)^{4}-
\theta_{4}(0|\tau)^{4}-\theta_{2}(0|\tau)^{4}\over \eta(\tau)^{12}}.
\label{aequatio}
\end{align}
The combination appearing in the numerator of the right-hand side vanishes
identically due to Jacobi's celebrated {\it aequatio identica  satis
abstrusa}\footnote{In English, ``a rather obscure identity''.}. Since the
full one-loop partition function of the type-II superstring is proportional 
to~$\tau_{2}^{-4}Z(\tau)Z(\overline{\tau})$,
we find that it 
is equal to zero. This is a consequence of supersymmetry, since the contribution
to of target space bosons is precisely
cancelled by the fermions.

As a final remark we notice that,
since~$Z_{(P,P)}(\tau)=0$, the phase~$w(P,P)$ remains
undetermined and so does the $G$-parity projection in the~R sector [see eq.~\eqref{eq:Z(tau)GSO}]. 
This is not surprising. From a target space point
of view, the states in the NS~sector transform under single-valued
representations of the little group~$\mbox{SO}(8)$ and they should be quantized
as bosons. In the R~sector, on the other hand, states are constructed 
by applying creation operators to the spinor ground state and
should be treated as space-time fermions. Since there is an
ambiguity in how to define space-time chirality, and $G$-parity is defined in terms of
the chirality matrix, it is not unreasonable
that~$w(P,P)$ is not determined. In fact, to fix the relative
phases between the NS and~R sectors one has to use the
factorization properties of two-loop amplitudes. 

The lesson to be
learned from this lengthy discussion is that the spectrum of fermionic
string theories is strongly constrained by a simple consistency
requirement: absence of global gravitational anomalies or, equivalently,
modular invariance. Playing with the
modular properties of both left- and right-movers independently opens up more possibilities, with
each modular invariant combination corresponding to theories with
different spectra. This way of constructing string theories has been widely
explored in the literature, specially in the context of the heterotic
string~\cite{string-construction}.

\paragraph{The Green-Schwarz spinor formalism.}
We already pointed out that in the analysis presented so far (also called the RNS~string formalism)
target space supersymmetry emerges as a surprise after implementing the GSO projection. 
An alternative analysis is provided by the Green-Schwarz formalism~\cite{Green-Schwarz-2}, where
one starts with a world-sheet action where space-time supersymmetry is explicit. A
difficulty of this formalism, however, is that it cannot be easily
quantized in a covariant way (it requires the implementation of Batalin-Vilkoviski
techniques). Here we use light-cone gauge quantization. In this approach we have, 
besides the~$X^{i}(\tau,\sigma)$ bosonic
coordinates, a field $S^{Aa}(\tau,\sigma)$ where $a=1,2$ 
and $A=1,\ldots,32$ are respectively world-sheet and space-time spinor 
indices\footnote{The reader is advised not to confuse the world-sheet spinor indices~$a,b=1,2$
introduced here with the world-sheet Lorentz indices~$a,b,=1,2$ used earlier in this same section
[see eq.~\eqref{eq:vierbein_2d}].}. 
Imposing the ten-dimensional Weyl and Majorana conditions, 
the $64$~complex components of~$S^{Aa}$ are reduced to $32$ real
components. Imposing the light-cone gauge condition on $S^{Aa}$
\begin{align}
(\Gamma^{+})^{AB}S^{Ba}=0,
\end{align}
with $\Gamma^{\mu}$ the ten-dimensional Dirac matrices,
we are left with $16$~real components. Since the Dirac equation further
reduces this number to $8$~real components, we have that, on-shell,
the number of fermionic space-time degrees of freedom equals the
number of transverse bosonic coordinates, and we can have in principle space-time supersymmetry. 

The light-cone gauge action is given
by
\begin{align}
S=-\frac{1}{4\pi\alpha'}\int d\tau d\sigma \Big(\partial_{\alpha}X^{i}
\partial^{\alpha}X^{i}-i\overline{S}^{A}\rho^{\alpha}
\partial_{\alpha}S^{A}\Big).
\end{align}
The equations of motion of the spinors $S^{Aa}$ are simply
\begin{align}
\partial_{+}S^{A1}=\partial_{-}S^{A2}=0.
\end{align}
This, together with the periodic boundary conditions
$S^{A}(\tau,\sigma+\pi)=S^{A}(\tau,\sigma)$ for the closed
superstring, lead to the mode expansion
\begin{align}
S^{A1}&=(2\alpha')^{1\over 4}\sum_{n\in\mathbb{Z}}S^{A}_{n}e^{2in(\tau-\sigma)}, \nonumber \\[0.2cm]
S^{A2}&=(2\alpha')^{1\over 4}\sum_{n\in\mathbb{Z}}\widetilde{S}^{A}_{n}e^{2in(\tau+\sigma)},
\end{align}
where the fermionic modes satisfy the anticommutation relations
\begin{align}
\{S_{m}^{A},\widetilde{S}_{n}^{B}\}=\delta^{AB}\delta_{m+n,0}.
\end{align}
To avoid confusions with the components of the conjugate spinor,
we use a tilde to denote left-moving modes.
In terms of these operators, the mass formula and level matching
condition for the closed superstring respectively read
\begin{align}
\frac{1}{2}\alpha'm^{2}&=\sum_{n>0}\Big(\alpha_{-n}^{i}\alpha_{n}^{i}+
\widetilde{\alpha}_{-n}^{i}\widetilde{\alpha}_{n}^{i}+nS_{-n}^{A}S_{n}^{A}+
n\widetilde{S}_{-n}^{A}\widetilde{S}_{n}^{A}\Big), \nonumber \\[0.2cm]
\sum_{n>0}\Big( \alpha_{-n}^{i} \alpha_{n}^{i}+nS_{-n}^{A}S_{n}^{A}\Big)&=
\sum_{n>0}\Big(\widetilde{\alpha}_{-n}^{i}\widetilde{\alpha}_{n}^{i}+
n\widetilde{S}_{-n}^{A}\widetilde{S}_{n}^{A}\Big).
\end{align}
Denoting by~$|i\rangle$ the eight physical degrees of freedom
of the massless vector state, we define their supersymmetric partners~$|A\rangle$
\begin{align}
|A\rangle=\frac{i}{8}(\Gamma_{j}S_{0})^{A}|j\rangle,
\label{partner}
\end{align}
where the states are normalized according to
\begin{align}
\langle i|j\rangle=\delta_{ij}, \hspace{1cm} \langle A|B\rangle=
\frac{1}{2}(h\Gamma^{+})^{AB},
\end{align}
and~$h$ is the ten-dimensional helicity operator. 

The massless
spectrum of the closed superstring contain $128$ bosonic states
\begin{align}
|i\rangle_{L}\otimes|j\rangle_{R},  \hspace{1cm}
|A\rangle_{L}\otimes|B\rangle_{R},
\end{align}
and the same number of fermionic states
\begin{align}
|i\rangle_{L}\otimes|B\rangle_{R}, \hspace{1cm}
|A\rangle_{L}\otimes|i\rangle_{R}.
\end{align}
As indicated above, if~$|A\rangle_{L}$ and~$|B\rangle_{R}$ have the
same helicity the result is type-IIB (chiral) superstring theory. 
Taking the states to have opposite helicities, on the other hand,
we obtain type-IIA (non-chiral) superstring theory.

As a closing remark let us indicate the spectrum of type-I (open)
superstring. Although now the fields do not satisfy any periodicity boundary condition
on the~$\sigma$ string coordinate, it is possible to show that the massless spectrum is
the same as in one of the sectors of type II superstring: we have 
eight transverse bosonic states~$|i\rangle$ and their
fermionic partners~$|A\rangle$ defined by eq.~\eqref{partner}. Thus,
the states in the massless sector of the open superstring are those of ten-dimensional
chiral $N=1$ SUGRA. In this model 
gauge invariance can be implemented by the Chan-Paton procedure~\cite{Chan-Paton}:
charges are attached to the string endpoints. 
However, using this method the only possible gauge groups are~$\mbox{USp}(N)$,~$\mbox{SO}(N)$, 
and~$\mbox{U}(N)$.
The resulting low energy field theory is ten-dimensional~$N=1$ SUGRA coupled
to~$N=1$ super-Yang-Mills. As explained in sec.~\ref{sec:GSandWSU2}, Green and 
Schwarz~\cite{Green-Schwarz} proved that this low energy field theory is
free from gauge and gravitational anomalies provided
the gauge groups are $\mbox{E}_{8}\times \mbox{E}_{8}$,
$\mbox{SO}(32)$ or~$\mbox{E}_{8}\times \mbox{U}(1)^{248}$. Combining this result with the restriction 
on the gauge groups imposed by the Chan-Paton procedure, we conclude
that the only consistent open string theory has $\mbox{SO}(32)$ as gauge
group~\cite{Marcus-Sagnotti}.

\subsection{The heterotic string}

Up to now we have studied two kinds of closed string models: the closed bosonic
string, that can be consistently quantized only when the target space dimension
is~$26$, and the type II superstring with
critical dimension~10. In both cases we have to deal with 
left- and the right-moving modes which do not mix among themselves.
In addition, strings are ``neutral''
objects, in the sense that charges cannot be added using the Chan-Paton procedure.

This independence between the left- and right-moving modes characteristic of closed string models 
provides the key to construct a hybrid theory, the heterotic
string \cite{Pricenton-quartet} (for a general review see also
\cite{Green-Schwarz-Witten,Lust-Theisen,Gross}). We combine the~$10$
right-moving modes of a type-II superstring with the $26$~left-movers
of a bosonic string, after compactifying $16$ of them into an internal
manifold. Working in the light-cone gauge, the physical
degrees of freedom in the right-moving sector are the~$8$ transverse
bosons~$X^{i}(\tau-\sigma)$ and one Majorana-Weyl ten-dimensional
fermion~$S^{A}(\tau-\sigma)$ (with $A=1,\ldots,8$). In the
left-moving sector, on the other hand, we
have~$24$ transverse bosonic coordinates $X^{i}(\tau+\sigma)$ and
$X^{I}(\tau+\sigma)$ (with~$i=1,\ldots,8$ and $I=1,\ldots,16$), where the latter~$16$ bosonic
fields are interpreted as parametrizing some internal manifold. Consistency of
the resulting theory imposes that this internal manifold should be a
$16$-dimensional torus ${\bf T}$
\begin{align}
{\bf T}=\mathbb{R}^{16}/\Lambda_{16},
\end{align}
with~$\Lambda_{16}$ a $16$-dimensional lattice. From the previous discussion, we
find that the action of the
free heterotic string is written in term of transverse world-sheet fields as
\begin{align}
S=-\frac{1}{4\pi\alpha'}\int d\tau d\sigma
\Big[\partial_{\alpha}X^{i}\partial^{\alpha}X^{i}+
\partial_{\alpha}X^{I}\partial^{\alpha}X^{I}+
i\overline{S}\Gamma^{-}(\partial_{\tau}+\partial_{\sigma})S\Big],
\label{het-action}
\end{align}
together with the constraints implementing that the~$X^{I}$ are left-movers
\begin{align}
\Phi\equiv (\partial_{\tau}-\partial_{\sigma})X^{I}=0, \hspace{1cm}
I=1,\ldots,16.
\label{Phi-constr}
\end{align}
The action (\ref{het-action}) is invariant under the supersymmetry
transformations
\begin{align}
\delta X^{i}&=\frac{1}{\sqrt{p^{+}}}\overline{\epsilon}\Gamma^{i}S, \nonumber
\\[0.2cm] 
\delta S &= \frac{i}{\sqrt{p^{+}}}\Gamma_{-}\Gamma_{\mu}
(\partial_{\tau}-\partial_{\sigma})X^{\mu}\epsilon,
\end{align}
where $\epsilon$ is a right-moving Majorana-Weyl light-cone spinor.

The spectrum for the heterotic string can be easily obtained, since
all we have to do is put together the results for the bosonic
and supersymmetric strings studied above. Expanding the different fields into 
their corresponding modes,
we get the set of operators
\begin{align}
\big\{\alpha_{n}^{i},S^{A}_{n},\widetilde{\alpha}_{n}^{i},
\widetilde{\alpha}^{I}_{n}\big\},
\end{align}
with $n\in\mathbb{Z}$ and $A,i=1,\ldots,8$. Besides these
operators, we also have the center of mass positions, $x^{i}$ and
$x^{I}$, as well as the momenta $p^{i}$ and $p^{I}$.
The commutation relations for these modes are
\begin{align}
[\alpha_{n}^{i},\alpha_{m}^{j}]&=
[\widetilde{\alpha}_{n}^{i},\widetilde{\alpha}_{m}^{j}]=
n\delta^{ij}\delta_{m+n,0}, \nonumber \\[0.2cm]
\{S_{n}^{a},S_{m}^{b}\}&=(\Gamma^{+}h)^{ab}\delta_{n+m,0}, \nonumber \\[0.2cm]
{[}\widetilde{\alpha}_{n}^{I},\widetilde{\alpha}_{m}^{J}]&=
n\delta^{IJ}\delta_{m+n,0}, \nonumber \\[0.2cm]
{[}q^{i},p^{j}]&=i\delta^{ij}, \\[0.2cm]
[q^{I},p^{J}]&=\frac{i}{2}\delta^{IJ},
\nonumber
\end{align}
where the commutation relations of the modes associated with the fields $X^{I}$ satisfying
the constraints (\ref{Phi-constr}) are obtained using Dirac brackets
(see ref.~\cite{Pricenton-quartet}). 
In terms of the normal-ordered number operators
\begin{align}
N&=\sum_{n>0}\left(\alpha_{-n}^{i}\alpha_{n}^{i}+\frac{n}{2}\overline{S}_{-n}
\Gamma^{-}S_{n}\right), \nonumber \\[0.2cm]
\widetilde{N}&=\sum_{n>0}\Big(\widetilde{\alpha}_{-n}^{i}\widetilde{\alpha}_{n}^{i}+
\widetilde{\alpha}_{-n}^{I}\widetilde{\alpha}_{n}^{I}\Big),
\end{align}
the mass formula is expressed as
\begin{align}
\frac{1}{2}\alpha'm^{2}=N+\widetilde{N}+
\alpha'\sum_{I=1}^{16}(p^{I})^{2}.
\end{align}
As in the case of bosonic and supersymmetric strings, we also have to
implement the level matching condition, that reads
\begin{align}
N=\widetilde{N}+\alpha'\sum_{I=1}^{16}(p^{I})^{2}-1.
\end{align}

We still need to find a suitable internal manifold to
compatify the~$16$ extra right-moving bosonic coordinates. As can be expected, 
modular invariance severely constraint the choice of~$\Lambda_{16}$ 
determining the spectrum. In particular, as it will be shown below,~$\Lambda_{16}$ 
must be an even integral Euclidean self-dual
lattice. In fact, there are only two such lattices: the weight lattice
of $\mbox{Spin}(32)/\mathbb{Z}_{2}$ [the modding by $\mathbb{Z}_{2}$ eliminates one of
the spin representations leaving the root lattice of~$\mbox{SO}(32)$] 
and the root lattice of $\mbox{E}_{8}\times \mbox{E}_{8}$.

Let us proceed to determine the spectrum of the heterotic string. The
ground state~$|0\rangle_{L}$ for left-moving modes is defined by
\begin{align}
\widetilde{\alpha}^{i}_{n}|0\rangle_{L}=\widetilde{\alpha}^{I}_{n}|0\rangle_{L}=0\;,
\hspace{1cm} n>0.
\end{align}
In the right-moving sector, on the other hand, we have the eight bosonic states $|i\rangle$
and the their fermionic partners~$|A\rangle$.
The mass formula, the level matching condition and the structure of~$\Lambda_{16}$ 
imply that there are no tachyons in the spectrum. In the massless sector we find the 
states
\begin{align}
\widetilde{\alpha}_{-1}^{i}|0\rangle_{L}&\otimes |i\rangle_{R},
\label{grav-multip-1} \\[0.2cm]
\widetilde{\alpha}_{-1}^{i}|0\rangle_{L}&\otimes |a\rangle_{R},
\label{grav-multip-2} \\[0.2cm]
\widetilde{\alpha}_{-1}^{I}|0\rangle_{L}&\otimes |i\rangle_{R},
\label{super-YM-1}\\[0.2cm]
\widetilde{\alpha}_{-1}^{I}|0\rangle_{L}&\otimes |a\rangle_{R},
\label{super-YM-2} \\[0.2cm]
|p^{I}\rangle_{L}&\otimes |i\rangle_{R}, \label{super-YM-3}\\[0.2cm]
|p^{I}\rangle_{L}&\otimes |a\rangle_{R}, \label{super-YM-4}
\end{align}
where~$\alpha'(p^{I})^{2}=1$. The states (\ref{grav-multip-1}) and
(\ref{grav-multip-2}) build a~$N=1$,~$d=10$ SUGRA
multiplet containing a graviton, a gravitino, 
a rank-two antisymmetric tensor, a dilaton, and their supersymmetric partners.
States (\ref{super-YM-1})-(\ref{super-YM-4}) form a $N=1$, $d=10$
super-Yang-Mills multiplet with gauge group $\mbox{E}_{8}\times \mbox{E}_{8}$ or
$\mbox{Spin}(32)/\mathbb{Z}_{2}$. The low energy field theory
of the heterotic string is therefore $N=1$ SUGRA
coupled to $N=1$ super-Yang-Mills in $d=10$ with gauge 
groups~$\mbox{E}_{8}\times \mbox{E}_{8}$ or $\mbox{SO}(32)$.
The heterotic string thus provides the first
example of a string theory with gauge group $\mbox{E}_{8}\times \mbox{E}_{8}$ which, as we saw above,
could not be implemented in the open string case.

Applying the Green-Schwarz cancellation mechanism, we see that the low
energy field theory for the heterotic string is free from both
gravitational and gauge anomalies. It can be shown moreover that the
theory is anomaly-free also at the string level if the gauge groups are
$\mbox{E}_{8}\times \mbox{E}_{8}$ or $\mbox{Spin}(32)/\mathbb{Z}_{2}$~\cite{Gross-Mende}, so
the anomaly cancellation with these gauge groups not only works
at the level of the low energy effective field theory but it can be
verified for the full string theory.

We conclude our discussion of the heterotic string with the computation of its partition function.
Had we formulated the heterotic string using the so-called~RNS
formalism (which is explicitly world-sheet supersymmetric) we would
have had to include the GSO-projected~NS and~R sectors for the right
movers. In addition to this we would also have to include the~$8$
transverse bosonic coordinates of the left-moving sector and the~$16$
right-moving internal degrees of freedom. Taking into account all these contributions,
the one-loop partition function for
the heterotic string is, up to a global numerical factor, given by
\begin{align}
Z(\tau,\overline{\tau})_{\rm het}= \tau_{2}^{-4}
\frac{\overline{\Theta}(\overline{\tau})}{\overline{\eta}(\overline{\tau})^{24}}
\eta(\tau)^{-12}\Big[\theta_{3}(0|\tau)^{4}-
\theta_{3}(0|\tau)^{4}-\theta_{2}(0|\tau)^{4}\Big],
\label{het-aequatio}
\end{align}
where $\overline{\Theta}(\overline{\tau})$ is the theta function associated with
the internal lattice $\Lambda_{16}$
\begin{align}
\overline{\Theta}(\overline{\tau})\equiv \sum_{{\bf r}\in \Lambda_{16}}
\overline{q}^{{1\over 2}{\bf r}^{2}}.
\end{align}
In order for~$Z(\tau,\overline{\tau})_{\rm het}$ to be
modular invariant,~$\overline{\Theta}(\overline{\tau})$ should have the following transformation
under the $T$ and $S$ generators
\begin{align}
T:\overline{\Theta}(\overline{\tau}) &\longrightarrow \overline{\Theta}(\overline{\tau}),
\nonumber \\[0.2cm]
S:\overline{\Theta}(\overline{\tau}) &\longrightarrow
\overline{\tau}^{8}\overline{\Theta}(\overline{\tau}).
\label{eq:Theta_int_het_ST}
\end{align}
It is easy to check that the first condition implies that
${\bf p}^{2}$ is an even integer, i.e.~$\Lambda_{16}$
is an even lattice. To implement the second condition,
we use the Poisson summation formula to write
\begin{align}
\overline{\Theta}\left(-\frac{1}{\overline{\tau}}\right)=
\frac{\overline{\tau}^{8}}{\sqrt{{\det\,g}}}
\sum_{{\bf p}\in \Lambda_{16}^{*}}\overline{q}^{{\bf p}^{2}},
\end{align}
where~$\Lambda_{16}^{*}$ indicates the lattice dual to $\Lambda_{16}$ and
$g_{ij}$ is the lattice metric. The conditions~\eqref{eq:Theta_int_het_ST} thus
imply that~$\Lambda_{16}$ is an even self-dual lattice. Let us also point out that, as
in the closed superstring case, the
partition function~\eqref{het-aequatio} vanishes due to Jacobi's
{\em aequatio} identity, again a consequence of space-time supersymmetry.

\subsection{Strings at finite temperature}

A basic prediction of the cosmological standard model (see, for example, ref.~\cite{Weinberg}) is that
at very early times the universe was
in a very hot state. Close to the Planck time its temperature should be
of the order of the Planck mass so,
if string theory is a correct theory of quantum gravity, it should have a lot to
say about the very early universe~\cite{EAlvarez-cosm,Tye,Brandenberger-Vafa,Tseytlin-Vafa}.
Thus, a first step towards string cosmology is
to study the thermal properties of strings, the aim of the present
section. 

When studying (critical) strings at finite temperature we are faced with two ``stringy''. 
The first one is 
that for every string theory there is a temperature where the
canonical partition function diverges,
the Hagedorn temperature \cite{Hagedorn}. This is a consequence of the fact
that in any critical string theory  the
number of states for a given mass grows
exponentially with the mass. In the bosonic string, for example, the asymptotic
density of states for large~$m$ is given by
\begin{align}
\rho(m)\sim m^{-\frac{25}{2}}e^{4\pi\sqrt{\alpha'}m}.
\label{asymptotic-density}
\end{align}
The canonical partition function~$Z(\beta)$ is defined as the
Laplace transform of the density of states~($\beta=1/T$)
\begin{align}
Z(\beta)=\int_{0}^{\infty}dE \rho(E)e^{-\beta E},
\end{align}
so in our case we find~\cite{Alvarez}
\begin{align}
Z(\beta)\sim \int_{\mu}^{\infty} dE
E^{-\frac{25}{2}}e^{(4\pi\sqrt{\alpha'}-\beta)E},
\end{align}
where $\mu$ is a cutoff in order to make (\ref{asymptotic-density})
applicable. It is now easy to see that~$Z(\beta)$
is well-defined provided
\begin{align}
\beta>4\pi\sqrt{\alpha'}.
\end{align}
The right-hand side of this inequality defines the Hagedorn (inverse) temperature
for the bosonic string
\begin{align}
\beta_{H}=4\pi\sqrt{\alpha'}.
\end{align}
In the case of the bosonic string, the presence of the tachion makes the
Helmholtz free energy ill defined (infinity) for all values of~$\beta$. 
The Hagedorn
temperature is nevertheless present in every critical string theory, including those
with a well-defined low temperature phase as the
superstring and the heterotic string (cf. for example~\cite{Kani-Vafa}). 
As already pointed out, this is a consequence of the exponential 
degeneracy of the string spectrum at large masses.

The second ``stringy'' feature is a duality relation, sometimes called~$\beta$-duality,
connecting the values of the partition function at low and high temperatures.
In the case of the heterotic string, it reads
\cite{O'Brien-Tan}
\begin{align}
Z(\beta)=Z\left(\frac{2\pi^{2}\alpha'}{\beta}\right).
\end{align}
Although formally this duality is reminiscent of the $R$-duality property discussed 
in sec.~\ref{sec:R-duality}, they are in
fact somewhat different, as they are the kind of compactifications giving rise to both kinds of duality.
In fact, although $\beta$-duality
is formally present in the bosonic (as a symmetry of the
integrand of the Helmholtz free energy)
and the heterotic string (see below), it is absent for type-II superstrings
In this latter case it is however possible to construct a kind of
``generalized'' duality relation for the one-loop free energy~\cite{Osorio-1}.

\begin{figure}[t]
\centerline{\includegraphics[scale=0.47]{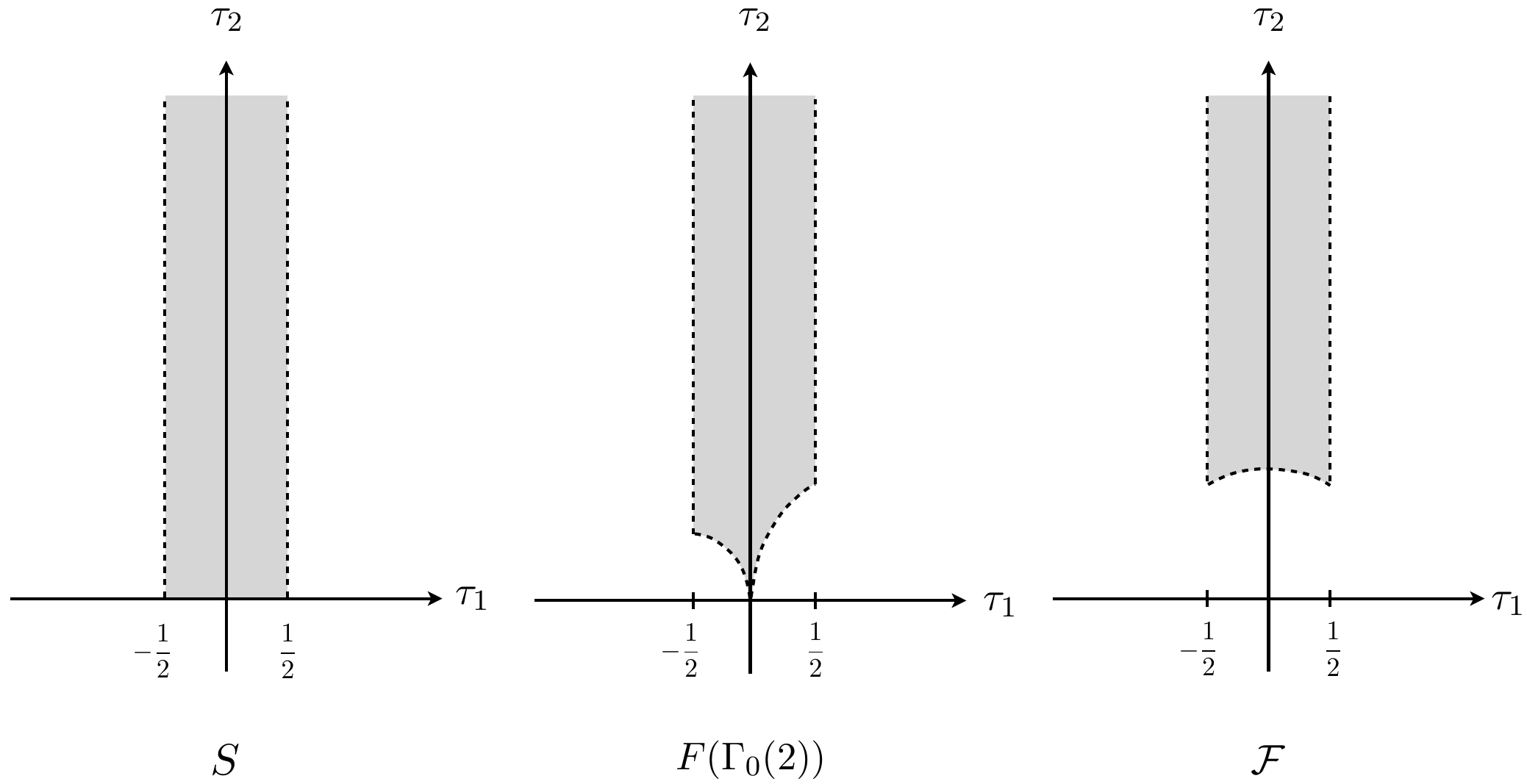}}
\caption{Fundamental regions of the translation subgroup generated by~$T$ (left), 
the congruence subgroup~$\Gamma_{0}(2)$ (center), and the full modular group~$\mbox{PSL}(2,\mathbb{Z})$
(right). For a detailed discussion see, e.g., the appendix of ref.~\cite{Alvarez-Osorio-2}.}
\label{fig:regions}
\end{figure}

As in ordinary statistical mechanics, it is convenient to work with the
Helmholtz free energy~$F(\beta)$
\begin{align}
F(\beta)=-\frac{1}{\beta}\log{Z(\beta)}.
\end{align}
In the imaginary time formalism, the canonical partition function of a thermal field
theory on~$\mathbb{R}^{1,d-1}$ is
given by the vacuum energy
of the same theory on $\mathbb{R}^{d-1}\times S^{1}$, with the
length of the compactified circle equal to the inverse temperature~$\beta$. Using
this method,
Polchinski computed the Helmholtz free-energy per unit volume for the
bosonic string at one-loop~\cite{Polchinski}
\begin{align}
F(\beta)=-{1\over 2^{27}\pi^{26}\alpha'^{13}}\int_{S}\frac{d^{2}\tau}{\tau^{2}_{2}}
\tau_{2}^{-12}|\eta(\tau)|^{-48}
\left[\theta_{3}\left(0\left|\frac{i\beta^{2}}{4\pi^{2}\alpha'\tau_{2}}
\right.\right) -1\right],
\label{bosonic-S}
\end{align}
where $\tau\equiv\tau_{1}+i\tau_{2}$ is again the torus modular parameter,
and~$\eta(\tau)$ and~$\theta_{3}(0|\tau)$ are the Dedekind eta
and Jacobi theta functions introduced in sec.~\ref{sec:fermionic_string}. The integral is evaluated
over the strip~
\begin{align}
S=\left\{\tau\in\mathbb{C}\,\left|\,-{1\over 2}<\tau_{1}<{1\over 2};\,\tau_{2}>0\right\}\right.,
\end{align} 
shown on the left panel 
of fig.~\ref{fig:regions}, which is the fundamental region of the subgroup of real translations 
generated by the transformation~$T$.

As already pointed out, eq.~\eqref{bosonic-S}
does not converge because of the infrared divergence caused by
the tachyon as $\tau_{2}\rightarrow \infty$. Besides this basic flaw, it should be said that
this result does not really correspond to the vacuum energy of bosonic
strings on $\mathbb{R}^{d-1}\times S^{1}$. The reason is that, unlike the Kaluza-Klein momenta
along the compactified Euclidean time, the expression does not
include the contribution of strings winding along this dimension. This shows
in the fact that eq.~\eqref{bosonic-S} is not manifestly invariant under the full 
modular group~$\mbox{PSL}(2,\mathbb{Z})$, but only under the subgroup of translation 
generated by~$T$. This is indeed surprising, since in a
a path integral evaluation of the vacuum energy the integral should render a modular invariant
result, where the integration over the modular
parameter coming from summing the contribution of inequivalent tori~\cite{McClain-Roth,O'Brien-Tan}.

It was observed by Polchinski~\cite{Polchinski} that the expression~\eqref{bosonic-S} 
coincides with the result of adding the contributions to the free energy of the 
different states in the full string spectrum. 
This method of computing the free energy, that
amounts to considering the string as a collection of quantum fields, is called
the analog model~\cite{EAlvarez-cosm,Sundborg,Bowick-W,Tye,Alvarez,Alvarez-Osorio-1}.

To be more precise, let us begin~\cite{Alvarez-Osorio-1} with the expression of the free
energy per physical degree
of freedom and per unit volume of a bosonic (fermionic)
quantum field in $d$ dimensions
\begin{align}
F(\beta)_{B,F}=\pm\frac{1}{\beta}\int \frac{d^{d-1}k}{(2\pi)^{d-1}}
\log(1\mp e^{-\beta \omega_{k}}),
\label{int-log}
\end{align}
where $\omega_{k}\equiv\sqrt{{\bf k}^{2}+m^{2}}$, with~$m$ the field mass. After some elementary
manipulations including all degrees of freedom, eq.~\eqref{int-log} can be recast as
\begin{align}
F(\beta)_{B}&=-\pi^{d/2}2^{d/2-1}
\int_{0}^{\infty}ds\,s^{-1-d/2}e^{-m^{2}s/2}
\left[\theta_{3}\left(0\left|\frac{i\beta^{2}}{2\pi
s}\right.\right)-1\right], \nonumber\\[0.2cm]
F(\beta)_{F}&=\pi^{d/2}2^{d/2-1}
\int_{0}^{\infty}ds\,s^{-1-d/2}e^{-m^{2}s/2}
\left[\theta_{4}\left(0\left|\frac{i\beta^{2}}{2\pi
s}\right.\right)-1\right],
\label{boson-fermion} 
\end{align}
for bosonic and fermionic fields respectively.
The string free energy is computed from these expressions adding the 
contribution of all states in the string spectrum, taking into account their statistics.
In addition, we have to impose the level
matching condition implemented by writing the~$\delta$-function
\begin{align}
\delta_{nm}=\int_{-\frac{1}{2}}^{\frac{1}{2}} d\tau_{1}
e^{2\pi i(n-m)\tau_{1}}.
\end{align}
It can be seen that the free
energy obtained in this way gives the free energy
of an ensemble of strings, since we are adding the contributions of second
quantized fields. Thus, using the analog model it is possible to evaluate
the one-loop free energy of any string theory. The extension of the analog model to 
$\ell$-loops can be done by plugging in eqs.~\eqref{boson-fermion} 
the renormalized
mass evaluated to~$(\ell-1)$-loops~\cite{Osorio} (see
also ref.~\cite{Moore}).

We look now at the thermal partition function of the heterotic since, since
this is the most interesting string model from many points of view. 
Using the analog model (or, equivalently, evaluating the path integral in the
light-cone gauge for a single string and exponentiating the result) we
obtain that the one loop free energy is given by~\cite{Alvarez-Osorio-1}
\begin{align}
F(\beta)=-\frac{1}{2^{11}\pi^{10}\alpha'^{5}}\int_{S}\frac{d^{2}\tau}{\tau_{2}^{2}}
\tau_{2}^{-4}\frac{\theta_{2}^{4}}{\eta^{12}}
\frac{\overline{\Theta}}{\overline{\eta}^{24}}
\theta_{2}\left(0\left|\frac{i\beta^{2}}{\pi^{2}\alpha'\tau_{2}}\right.\right).
\label{heterotic-T}
\end{align}
As in eq.~\eqref{bosonic-S}, the integral is extended over the
strip~$S$ instead of the fundamental region of the modular group~${\cal
F}$, defined in eq.~\eqref{eq:fundamental_region_F}
and depicted on the right of fig.~\ref{fig:regions}. 

Interestingly, it is possible to obtain an explicitly
modular invariant form for the
one-loop free energy given in eq.~\eqref{heterotic-T}~\cite{O'Brien-Tan,Alvarez-Osorio-2}. 
It can be checked that the integrand is invariant under the subgroup~$U\subset\mbox{PSL}(2,\mathbb{Z})$
generated by~$T$ leaving invariant the
spin structure~$(P,A)$. The fundamental
region of this subgroup, the strip~$S$, is obtained by acting with
transformations belonging to~$\Gamma\equiv\mbox{PSL}(2,\mathbb{Z})$ on ${\cal F}$.
Thus, the translation subgroup~$U$
can be written as the union of cosets of the modular group
\begin{align}
U=\bigcup_{(c,d)=1}\Gamma\gamma_{cd},
\label{coset-exp}
\end{align}
where $c$, $d$ are two co-prime integers and $\gamma_{cd}\in
\mbox{PSL}(2,\mathbb{Z})$ is given by
\begin{align}
\gamma_{cd}=\left(
\begin{array}{cc}
* & * \\
c & d
\end{array}
\right),
\label{eq:gammacd}
\end{align}
with the two upper entries of the matrix any integers satisfying the
condition on the determinant. Then, an integral over the
strip~$S$ can be rewritten as
\begin{align}
\int_{S}\frac{d^{2}\tau}{\tau_{2}^{2}}F(\tau)=
\int_{\bigcup\gamma_{cd}{\cal F}}\frac{d^{2}\tau}{\tau_{2}^{2}}F(\tau)=
\sum_{(c,d)=1}\int_{\gamma_{ab}{\cal F}}
\frac{d^{2}\tau}{\tau_{2}^{2}}F(\tau).
\end{align}
Carrying out a change of variables for each integral in the sum and commuting
the sum with the integral, we obtain
\begin{align}
\int_{S}\frac{d^{2}\tau}{\tau_{2}^{2}}F(\tau)=
\int_{\cal F} \frac{d^{2}\tau}{\tau_{2}^{2}}\sum_{(c,d)=1}
F(\gamma_{ab}\tau),
\label{f-m-i}
\end{align}
where we have used the modular invariance of the integration measure
$d^{2}\tau/\tau_{2}^{2}$. It can be seen that the
integrand on the right-hand side of~\eqref{f-m-i} is modular invariant, since any modular
transformation acting on the sum amounts to
changing representatives in the coset expansion~\eqref{coset-exp}.
The interest of this calculation is that the method can be applied to any integral
whose integrand is invariant under any subgroup of $\mbox{PSL}(2,\mathbb{Z})$
and is extended to the corresponding fundamental region.
Notice that the equivalence between the original integral and its modular
invariant extension depends on the legitimacy of performing the change of
variables and commuting the integral and the sum.

Using this method, we can write from~\eqref{bosonic-S} an explicit (although
divergent) modular
invariant expression of the bosonic string free energy. In the case of the heterotic string,
on the other hand,
to obtain the modular invariant
extension of eq.~\eqref{heterotic-T} we have to work a bit harder.
Instead of going from $U$ directly to $\mbox{PSL}(2,\mathbb{Z})$,
we first extend from~$U$ to the congruence subgroup
$\Gamma_{0}(2)\subset \mbox{PSL}(2,\mathbb{Z})$~\cite{Koblitz} and from 
there to the full modular group. In the first step we restrict~$\gamma_{cd}$
in~\eqref{eq:gammacd}
to~$\Gamma_{0}(2)$, with the result
\begin{align}
F(\beta)=-\frac{1}{2^{11}\pi^{10}\alpha'^{5}}\int_{F(\Gamma_{0}(2))}
\frac{d^{2}\tau}{\tau_{2}^{2}}\tau_{2}^{-4}
\frac{\overline{\Theta}}{\overline{\eta}^{24}}
\frac{\theta_{2}^{4}}{\eta^{12}} \theta \left[
\begin{array}{cc}
0  &  \frac{1}{2} \\
0  &  0
\end{array}
\right](0|4\Omega),
\end{align}
where $F\big(\Gamma_{0}(2)\big)$ is the fundamental region of the intermediary
subgroup (see the central panel of fig.~\ref{fig:regions}) 
and we introduced Riemann theta functions~\cite{Mumford}
with period matrix
\begin{align}
\Omega=\frac{i\beta^{2}}{4\pi^{2}\alpha'\tau_{2}}\left(
\begin{array}{cc}
|\tau|^{2} & -\tau_{1} \\
-\tau_{1}  &  1
\end{array}
\right).
\end{align}
Performing now the extension from $\Gamma_{0}(2)$ to~$\mbox{PSL}(2,\mathbb{Z})$, 
we arrive at the modular invariant expression for the
Helmholtz free energy for the heterotic string
\begin{align}
F(\beta)&=-\frac{1}{2^{11}\pi^{10}\alpha'^{5}}
\int_{\cal F}\frac{d^{2}\tau}{\tau_{2}^{2}}\tau_{2}^{-4}
\frac{\overline{\Theta}}{\overline{\eta}^{24}}\left\{
\frac{\theta_{2}^{4}}{\eta^{12}} \theta \left[
\begin{array}{cc}
0  &  \frac{1}{2} \\
0  &  0
\end{array}
\right](0|4\Omega)\right. \nonumber \\[0.2cm]
&+\left.\frac{\theta_{4}^{4}}{\eta^{12}} \theta \left[
\begin{array}{cc}
\frac{1}{2}  & 0  \\
0  &  0
\end{array}
\right](0|4\Omega)
-\frac{\theta_{3}^{4}}{\eta^{12}} \theta \left[
\begin{array}{cc}
\frac{1}{2}  & \frac{1}{2}  \\
0  &  0
\end{array}
\right](0|4\Omega)\right\}.
\label{eq:het_F(beta)_modinv}
\end{align}
This second step is quite straightforward because~$\Gamma_{0}(2)$ has
index~$3$ with respect to~$\mbox{PSL}(2,\mathbb{Z})$. Incidentally, 
a short calculation shows that taking the limit
$\beta\rightarrow\infty$ we find~$F(\beta)\rightarrow 0$. This is expected, since we are dealing with
a supersymmetric string theory whose vacuum energy vanishes [see eq.~\eqref{het-aequatio}].

From eq.~\eqref{eq:het_F(beta)_modinv} it can be seen
that there is a critical value of~$\beta$ below which 
the free energy diverges. This Hagedorn temperature of the heterotic string is 
given by~\cite{Pricenton-quartet}
\begin{align}
\beta_{H}\equiv 1/T_{H}=(2+\sqrt{2})\pi\sqrt{\alpha'}.
\end{align} 
Applying the Poisson summation
formula to the Riemann theta functions in~\eqref{eq:het_F(beta)_modinv} it can be
shown the $\beta$-duality relation~\cite{O'Brien-Tan}
\begin{align}
F(\beta)=\frac{2\pi^{2}\alpha'}{\beta^{2}}F\left(\frac{2\pi^{2}\alpha'}{\beta}\right).
\end{align}
Remarkably, this relation valid at one loop is in fact satisfied to all orders in string perturbation
theory~\cite{Alvarez-Osorio-3}.
Let us write the perturbation expansion of the Helmholtz free energy
\begin{align}
F(g_{st},\beta)=\sum_{g=1}^{\infty}g_{\rm st}^{2(g-1)}F_{g}(\beta),
\label{eq:pert_expan_F_het}
\end{align}
with~$g_{\rm st}$ the string coupling constant.
Using the modular properties of the thermal Riemann theta functions
appearing at genus-$g$, the following identity can be proved
\begin{align}
F_{g}(\beta)=\left(\frac{2\pi^{2}\alpha'}{\beta^{2}}\right)^{g}
F_{g}\left(\frac{2\pi^{2}\alpha'}{\beta}\right),
\end{align}
so the full perturbative series~\eqref{eq:pert_expan_F_het} satisfies the $\beta$-duality relation
\begin{align}
F(g_{\rm st},\beta)=\frac{2\pi^{2}\alpha'}{\beta^{2}}
F\left(\frac{2\pi^{2}\alpha'}{\beta}g_{\rm st},\frac{2\pi^{2}\alpha'}{\beta}\right).
\end{align}
Combining the existence of the Hagedorn temperature and the $\beta$-duality relation 
an interesting three phase structure for
the free energy emerges~\cite{O'Brien-Tan}, where~$F(\beta)$ is finite 
for~$\beta\leq 2\pi^{2}\alpha'/\beta_{H}$ and~$\beta\geq\beta_{H}$ and diverges
in the range~$2\pi^{2}\alpha'/\beta_{H}<\beta<\beta_{H}$.

The high-temperature phase,~$\beta<2\pi^{2}\alpha'/\beta_{H}$,
has singular
properties, the canonical entropy being negative.
Another most intriguing
feature is that evaluating the free energy in the limit~$T\rightarrow\infty$ we find
\begin{align}
\lim_{\beta\rightarrow 0}F(\beta)\sim \frac{2\pi^{2}\alpha'}{\beta^{2}}\Lambda,
\label{F-L}
\end{align}
with~$\Lambda$ the cosmological constant. Since in 
our case~$\Lambda=0$, this seems to imply that in the high temperature limit there are
no degrees of freedom at all.

Some authors have interpreted these facts as indication that the high
temperature phase is unphysical and the Hagedorn temperature must
be the maximum temperature of the universe~\cite{Brandenberger-Vafa}.
Others, however, have seen in the loss of gauge-invariant
degrees of freedom at high temperature evidence
that in this regime the theory could be described by a topological theory~\cite{Atick-Witten}. 
In this were the case, the Hagedorn
temperature would indicate a transition between a topological and a non-topological
phase. In~\cite{Kogan-Sathiapalan}
a close resemblance between the possible phase
transition at the Hagedorn temperature and a Kosterlitz-Thouless phase
transition~\cite{Kosterlitz-Thouless} taking place at the same temperature has been claimed. 
It should be pointed out, however, that these two phenomena seem to be disconnected
in the case of the~$c=1$
non-critical string, where there is a Kosterlitz-Thouless phase transition but no Hagedorn
temperature~\cite{Gross-Klebanov}. At any rate, the problem of what
happens at the Hagedorn temperature to the string gas remains open.

Using the canonical ensemble it can be seen that the string gas undergoes 
violent fluctuations when
approaching the Hagedorn temperature~\cite{Alvarez}. This may signal
the breakdown of the canonical description, so a microcanonical analysis should 
be appropriate to study the system beyond the Hagedorn 
temperature~\cite{Sundborg,Bowick-W,Alvarez,Bowick-Giddings,Deo-Jain-Tan}.
The microcanonical density of states~$\Omega(E)$ is defined by
\begin{align}
\Omega(E)={\rm tr\,}\delta(E-H),
\end{align}
and can be written as the inverse Laplace transform of the partition
function
\begin{align}
\Omega(E,V)=\int_{c-i\infty}^{c+i\infty}\frac{d\beta}{2\pi i}
Z(\beta,V)e^{\beta E},
\end{align}
where~$c\geq \beta_{H}$. By analytically continuing
the partition function~$Z(\beta)$ to the whole complex~$\beta$-plane
and deforming the contour of integration, a density of
states valid in regimes different from the Hagedorn one could then be obtained (cf., for example, 
refs.~\cite{Deo-Jain-Tan,Deo-Jain-Tan-2}). Looking at how
the energy distributes among the strings in the ensemble as
the Hagedorn temperature is approached from the low temperature phase,
the phase transition would correspond to the fact that, when the number
of noncompact dimensions is larger than two, a single string
is able to absorb all the available energy due to the exponential growth of
its number of state per energy level~\cite{Deo-Jain-Tan-2}. 
In the microcanonical description this
implies that the specific heat is negative, showing an instability in
the system that might be associated with a phase transition. Many important issues about
string theory thermodynamics are still to be clarified.

\subsection{Is string theory finite?}

One of the most appealing features of string theory is its promise of 
an ultraviolet finite theory of quantum gravity. We have seen
in sec.~\ref{sec:operator_formalism} that in the bosonic strings divergences arise
from the boundary of moduli space,~${\cal M}_{g}$
due to propagation of the ground state tachyon,
while the integration measure is finite and well behaved except near
the boundary. This might lead us to think that, since in the superstring and the
heterotic string the tachyon is projected out by the GSO projection,
no further divergences would arise and these theories would be automatically finite. 

This is nevertheless not necessarily correct because massless states could also give rise to
subleading divergences. Explicit calculations however have shown that one-loop amplitudes
are free of divergences~\cite{Green-Schwarz-Witten,Kaku}. Although 
we still lack a complete proof of finiteness at higher loops, a number of computations 
have been carried out at two-loop level, both for the bosonic~\cite{Several-bosonic,Moore} as well 
as for the supersymmetric string~\cite{Verlinde-Verlinde-SST}. In the case of the heterotic string
there exist various computations of the two-loop cosmological constant~\cite{Several-heterotic},
while in~\cite{Ortin} the existence of a modular invariant expression for the two-loop
cosmological constant was proved, vanishing after summing over all spin
structures. General arguments based on supersymmetry~\cite{Martinec} 
seem to point to the finiteness of superstring
perturbation theory. A detailed analysis of the many technical issues
involved can be found in
refs.~\cite{Atick-Moore-Sen,D'Hoker-Phong} and references therein.
Another approach to the proof of finiteness has been pursued by
Mandelstam \cite{Mandelstam}. 

There are several thorny obstacles
encountered in working out a rigorous proof of finiteness. First,
the moduli space of Riemann surfaces encountered in sec.~\ref{sec:operator_formalism}
is now replaced by supermoduli space of super-Riemann
surfaces which, unfortunately, is poorly understood. The moduli space of genus~$g$
supersurfaces is a superalgebraic variety of graded 
dimension~$(3g-3|2g-2)$. Denoting a coordinate system in a given patch 
by~$(m_{\alpha},\tau_{a})$, with~$\alpha=1,\ldots,3g-3$ and~$a=1,\ldots,2g-2$,
the transition function between two overlapping patches
can be expanded in powers of the odd moduli
\begin{align}
m'_{\alpha}&=f_{\alpha}(m_{\beta},\tau_{b})=
f^{(0)}_{\alpha}(m_{\beta})+f^{ab}_{\alpha}(m_{\beta})\tau_{a}\tau_{b}
\ldots
\label{1a} \\[0.2cm]
\tau'_{a}&=g_{a}(m_{\beta},\tau_{b})=g^{(0)b}_{a}(m_{\beta})
\tau_{b}+g_{a}^{\;\;bcd}(m_{\beta})\tau_{b}\tau_{c}\tau_{d}+\ldots
\label{1b}
\end{align}
A supervariety is said to be projected if a coordinate
cover exists such that in eq.~\eqref{1a} the only nonzero coefficient 
is~$f^{(0)}_{\alpha}(m_{\beta})$, whereas it is 
split if~$g_{a}^{\;\;bcd}$ and all other higher order terms in~\eqref{1b}
vanish. A split supermanifold would be equivalent to
a vector bundle. While the genus-one case is essentially trivial and at genus two it was
shown by Deligne \cite{Deligne} that the supermoduli space is split, not
much is known for~$g>2$. The integration measure for superstring
scattering amplitudes is a supermeasure, whose properties can be
studied using an extension of 
the operator formalism presented in sec.~\ref{sec:operator_formalism} 
to the fermionic string~\cite{LAG-Gomez-Nelson-Sierra-Vafa}. 
The supermoduli space also has a boundary built up by
surfaces with nodes, that now come classified according to
whether the punctures are identified as~NS or~R.

One would like to integrate over the odd moduli~$\tau_{a}$ to be left with a measure 
on the spin-moduli space
of Riemann surfaces with a fixed spin structure. Since the modular
group acts transitively on the even and odd spin structures, this space
has two components. Remember that the tachyon is eliminated after
adding all spin structures, so the integration over spin moduli space
should implement this sum. However, if the supermoduli space is not split 
there are ambiguities represented by total
derivatives that may give non-trivial contributions on the
moduli space boundary. Related to these integration ambiguities is the
existence of certain spurious poles found in the superstring measure~\cite{Verlinde-Verlinde-SST}. 
These brief and general remarks are just intended to give the reader
a flavor of the technical obstacles found in trying to proof
that superstring theory is finite. Many more details can be found in the reference quoted.

But even if superstring
perturbation theory is finite, the issue remains as to its summability. 
Non-summability would be an indication of the
existence of large non-perturbative effects which could play the role in addressing
many dynamical questions. For example, for critical strings compactification of
extra dimensions and breaking of various symmetries are expected
to result from non-perturbative effects. For the bosonic string 
it has been shown that the perturbative expansion is not Borel
summable~\cite{Gross-Periwal} (see also~\cite{Kaku2}): the genus~$g$
contribution grows as $(2g)!$ for large~$g$, a rate of
divergence even larger than in QFT. Since we are dealing with the bosonic string,
one might argue that this behavior is associated with the tachyon. This interpretation, however, 
is probably not
correct. In fact, some recent studies of strings in dimensions~$d<1$, which are tachyon-free, 
show the same growth as $(2g)!$~\cite{Shenker}. Most likely, this behavior
evidences the presence of important non-perturbative effects. 
Whether the same occurs for supersymmetric strings remains an open question.

\section{Other developments and conclusions}
\label{sec:developments}

\subsection{String phenomenology}

There are many topics in string theory not included in these lectures,
one of the most conspicuously absent being string phenomenology~\cite{phenomenology}. 
Many of the attempts to make contact with low-energy experiments are based on 
the~$\mbox{E}_{8}\times\mbox{E}_{8}$
heterotic string model. The reason is double. First, 
the gauge group~$\mbox{E}_{8}\times \mbox{E}_{8}$ is large enough to
contain the SM~group. But equally important is that, after compactification of the
extra dimensions, the model can naturally account for a chiral spectrum of low-energy fermions. 
Considering a classical
background of the form $M_{4}\times K$, with $M_{4}$ the four-dimensional
Minkowski space and~$K$ is a six-dimensional manifold or an appropriate CFT, 
the dynamics at low energies should look like a supersymmetric
extension of the SM. Moreover, its coupling
would be in principle computable from the knowledge of the internal manifold~$K$.

As we learned in sec. 3.5, the consistent propagation of the string in target space
requires that the world-sheet theory is a CFT. In the case of the superstring theory,
this two-dimensional field theory should be superconformal invariant. In addition,
a general argument~\cite{Banks} indicates that 
the requirement of space-time supersymmetry demands $N=2$~world-sheet superconformal invariance, 
with an integral lattice for the~$\mbox{U}(1)$ 
generator of the $N=2$ algebra. For backgrounds of the form~$M_{4}\times
K$, the dynamics of light
modes is described by a four-dimensional $N=1$ SUGRA Lagrangian~\cite{Cremmer-et-al}. 
This is characterized by
three functions: the K\"{a}hler potential~${\cal K}$, that determines the
kinetic term of the scalars in the theory, the superpotential~$W$,
responsible among other things for the Yukawa couplings, and the gauge
kinetic function~$f$, determining de gauge coupling constants.
To a large extend, the subject of string phenomenology is about the computation of the
three functions~${\cal K}$,~$W$, and~$f$. In ordinary model building the
form of these functions depends very much on the way the model is constructed.
In string theory, however, these functions are in principle
computable in terms from the world-sheet CFT on the corresponding string background.  

One of the massless scalars ubiquitous in string theory
is the dilaton. Its vacuum expectation value is related to the
tree-level gauge coupling constants by~$\langle\Phi\rangle=1/g_{\rm st}^{2}$. 
An important feature of four-dimensional string models is that the coupling
constants at low energies are determined by the expectation value of
the dilaton field, as well as of other scalars associated with the
compactification and collectively known as
moduli fields. An important problem faced in fixing
the expectation value for the dilaton is that it is a flat direction
in pertubation theory, so a non-trivial dilaton potential has to be
generated by non-perturbative effects. As we will see, there are some
interesting recent results in this direction.

Another important issue are the constraints imposed on the scale of space-time supersymmetry breaking. 
Although the basic scale in the problem is the Planck energy scale,~$M_{P}\sim 10^{19}$~GeV, 
to find satisfactory low energy models requires supersymmetry to be broken at
scales of the order of $10^{10}$-$10^{11}$~GeV. This energy hierarchy 
is difficult to implement naturally with the current model building technology. 

Some peculiar features of string theory have
been applied with encouraging success to address some the issues confronted by string phenomenology. 
One is $R$-duality, discussed in sec.~\ref{sec:R-duality}, a property without analogue in~QFT.
Although the structure of the duality group is only known in a few cases, this symmetry
has proved to be very useful in constraining the couplings of the low-energy
theory (see~\cite{dual-group} for reviews and references). Another
hallmark of superstring models is mirror
symmetry~\cite{mirror}, based on the general properties
of~$N=2$ superconformal field theories. Similarly to $R$-duality, it implies that a 
string theory defined on apparently different
geometries result in the same physics. This symmetry allows an explicit
computation of some Yukawa couplings in the
superpotential~$W$. The full physical implications of both $R$-duality and mirror symmetry
are now being vigorously explored.

To determine the gauge coupling constants we need to find the non-perturbative form of the dilaton
effective potential. In QFT non-perturbative effects display a
dependence on the coupling constant of the form $\exp{(-1/g^{2})}$, which
can be generated by the mechanism of gaugino condesation 
(for a review see, e.g.,~\cite{Nilles}). The resulting dilaton effective potential
has the form
\begin{align}
V_{\rm eff}\sim \sum_{a} c_{a}\exp{\left(-\frac{24\pi^{2}}{b_{a}g_{a}^{2}}
\right)},
\label{gaugino-cond}
\end{align}
where the sum runs over the different factors on the gauge groups
responsible for gaugino condensation, $b_{a}$ are the one-loop
coefficient of the $\beta$ functions, and the dilaton vacuum expectation value enters
in~\eqref{gaugino-cond} through the relation between~$g_{a}$ and the~$\langle\Phi\rangle$. 
This potential has a minimum at~$\langle\Phi\rangle=\infty$ corresponding to 
a vanishing coupling constant~$g_{a}=0$. This, however, changes
once loop threshold effects~$\Delta_{a}(T)$, resulting from integrating
out massive modes, are included~\cite{correct}
\begin{align}
\frac{1}{g_{a}^{2}(p^{2})}=\langle\Phi\rangle+b_{a}\log{\left(
\frac{M_{P}^{2}}{p^{2}}\right)}+\Delta_{a}(T),
\label{loop-corr}
\end{align}
where~$T$
collectively denotes the light scalar moduli fields. These
threshold corrections are crucial to reconcile the string model predictions
with the current value of the weak mixing angle. A combination of eqs.~\eqref{loop-corr} 
and \eqref{gaugino-cond} fixes the dilaton expectation value, 
opening up the possibility of finding vacua where supersymmetry is broken 
(for more details and references see the second entry in~\cite{phenomenology}).
So far this has been explored in a few models, but 
more systematic studies are expected to be carried out in the near future. 

\subsection{Black holes}

We finally turn to some of the exciting recent work on string black
holes. It is well known that in GR collapsing matter under certain
conditions generates black holes. Under physical conditions, singularities in space-time
are supposed to have always a horizon, a statement known as the cosmic censorship
hypothesis. An interesting property of classical black holes is that they come with no
hair~\cite{Hawking-Ellis}: the only parameters characterizing a
static black hole are
its mass, angular momentum, and charge, while any other attribute of the collapsing matter
is radiated away. This provides a rather
efficient mechanism to violate global symmetries like baryon number. 

The information problem in black holes presented in sec.~\ref{sec:QFT_curved_spaces}, 
has generated a big controversy in the
last fifteen years. The computations carried out by Hawking and others
involve quantizing fields in the presence of a classical gravitational background
while ignoring back reaction,
among other reasons because to properly understand it we would
need a well-defined theory of quantum gravity.
This last sentence should of course be qualified by string theory,
which is alleged to provide a consistent quantum theory of gravity.
Two approaches has been followed in analyzing this problem. On the one
hand, one can study some simple toy model based on two-dimensional dilaton gravity, a theory 
suggested by the equation of motion of string theory
(for details and references see \cite{Strominger}). These simple systems
provide a precise formulation of Hawking's paradoxes in a situation
where we have some control on the approximation used. 

The second
approach began after Witten \cite{Witten-BH} found an exact string
theory based on a~$\mbox{SL}(2,\mathbb{R})/\mbox{U}(1)$ coset model
describing a genuine two-dimensional string black hole. We know from
QFT that when translating a problem from four to two
dimensions frequently it loses most of its flavor. This does not seem to be the
case here. Many of the conceptual issues involved in the
four-dimensional problem continue to be non-trivial in two dimensions. 
Since Witten's solution is a perfectly sensible 
CFT and (more importantly) string theory, it has triggered an important
research program trying to understand what
answers it might provide to black hole paradoxes. Furthermore, recent revisions
of the no-hair theorem (see~\cite{Coleman-Preskill-Wilczek} for details
and references) also find a formulation within this two-dimensional scenario, in
a way possibly connected with the loss of quantum coherence. The authors
of~\cite{Ellis-Mavromatos-Nanopoulos}, for example,
have proposed on the basis of the properties of
Witten's model solutions to many of the riddles of
black holes and string physics.

The study of dilaton black hole solutions and of Witten's black
hole string theory is currently going at a fast pace. Details can be
found in the literature, as well as in Polyakov's lectures~\cite{Polyakov92}. 
The very fact that some of the deepest questions in quantum gravity are beginning
to be addressed by string theory is both exciting and encouraging. It is
however hard to guess at this moment what will be the outcome of these
investigations.

\vspace*{0.5cm}

\paragraph{Parting comments.}
With these general remarks, we have finally reached the end of these lectures, in the hope of having
conveyed the impression that string theory is an important and active
area of research, which may ultimately lead to a quantum theory of 
gravity and its unification with the other fundamental
interactions. We may only hope that some of the ideas and proposals put
forward in the last few years contain the seeds of this
grand synthesis.

\end{document}